
\font\tenmib=cmmib10
\font\sevenmib=cmmib10 scaled 800
\font\titolo=cmbx12
\font\titolone=cmbx10 scaled\magstep 2

\font\cs=cmcsc10
\font\sc=cmcsc10
\font\css=cmcsc8
\font\ss=cmss10

\font\ninerm=cmr9
\font\ottorm=cmr8
\textfont5=\tenmib
\scriptfont5=\sevenmib
\scriptscriptfont5=\fivei

\font\euftw=eufm9 scaled\magstep1

\font\msytw=msbm9 scaled\magstep1

\font\msytwww=msbm5 scaled\magstep1

\font\indbf=cmbx10 scaled\magstep2

\font\ottorm=cmr8\font\ottoi=cmmi8\font\ottosy=cmsy8%
\font\ottobf=cmbx8\font\ottott=cmtt8%
\font\ottocss=cmcsc8%
\font\ottosl=cmsl8\font\ottoit=cmti8%
\font\sixrm=cmr6\font\sixbf=cmbx6\font\sixi=cmmi6\font\sixsy=cmsy6%
\font\fiverm=cmr5\font\fivesy=cmsy5
\font\fivei=cmmi5
\font\fivebf=cmbx5%

\def\ottopunti{\def\rm{\fam0\ottorm}%
\textfont0=\ottorm\scriptfont0=\sixrm\scriptscriptfont0=\fiverm%
\textfont1=\ottoi\scriptfont1=\sixi\scriptscriptfont1=\fivei%
\textfont2=\ottosy\scriptfont2=\sixsy\scriptscriptfont2=\fivesy%
\textfont3=\tenex\scriptfont3=\tenex\scriptscriptfont3=\tenex%
\textfont4=\ottocss\scriptfont4=\sc\scriptscriptfont4=\sc%
\textfont5=\tenmib\scriptfont5=\sevenmib\scriptscriptfont5=\fivei
\textfont\itfam=\ottoit\def\it{\fam\itfam\ottoit}%
\textfont\slfam=\ottosl\def\sl{\fam\slfam\ottosl}%
\textfont\ttfam=\ottott\def\tt{\fam\ttfam\ottott}%
\textfont\bffam=\ottobf\scriptfont\bffam=\sixbf%
\scriptscriptfont\bffam=\fivebf\def\bf{\fam\bffam\ottobf}%
\setbox\strutbox=\hbox{\vrule height7pt depth2pt width0pt}%
\normalbaselineskip=9pt\let\sc=\sixrm\normalbaselines\rm}
\let\nota=\ottopunti%

\font\csss=cmcsc14



\newcount\mgnf
\ifnum\mgnf=0
   \magnification=\magstep0
   \hsize=16.5truecm\vsize=22.truecm
   \parindent=0.3cm\baselineskip=0.45cm\fi
\ifnum\mgnf=1
   \magnification=\magstep1\hoffset=0.truecm
   \hsize=16.5truecm\vsize=22.truecm
   \baselineskip=18truept plus0.1pt minus0.1pt \parindent=0.9truecm
   \lineskip=0.5truecm\lineskiplimit=0.1pt      \parskip=0.1pt plus1pt\fi

%
%
%
%
%
%
%

\global\newcount\numsec\global\newcount\numapp
\global\newcount\numfor\global\newcount\numfig\global\newcount\numsub
\global\newcount\numlemma\global\newcount\numtheorem\global\newcount\numdef
\global\newcount\appflag \numsec=0\numapp=0\numfig=1
\def\veroparagrafo{\number\numsec}\def\veraformula{\number\numfor}
\def\veraappendice{\number\numapp}\def\verasub{\number\numsub}
\def\verafigura{\number\numfig}
\def\verolemma{\number\numlemma}
\def\verotheorem{\number\numtheorem}
\def\veradef{\number\numdef}

\def\section(#1,#2){\advance\numsec by 1\numfor=1\numsub=1%
\SIA p,#1,{\veroparagrafo} %
\write15{\string\Fp (#1){\secc(#1)}}%
\write16{ sec. #1 ==> \secc(#1)  }%
\hbox to \hsize{\titolo\hfill \number\numsec. #2\hfill%
\expandafter{\alato(sec. #1)}}\*}

\def\appendix(#1,#2){\advance\numapp by 1\numfor=1\numsub=1%
\SIA p,#1,{A\veraappendice} %
\write15{\string\Fp (#1){\secc(#1)}}%
\write16{ app. #1 ==> \secc(#1)  }%
\hbox to \hsize{\titolo\hfill Appendix A\number\numapp. #2\hfill%
\expandafter{\alato(app. #1)}}\*}

\def\senondefinito#1{\expandafter\ifx\csname#1\endcsname\relax}

\def\SIA #1,#2,#3 {\senondefinito{#1#2}%
\expandafter\xdef\csname #1#2\endcsname{#3}\else
\write16{???? ma #1#2 e' gia' stato definito !!!!} \fi}

\def \Fe(#1)#2{\SIA fe,#1,#2 }
\def \Fp(#1)#2{\SIA fp,#1,#2 }
\def \Fg(#1)#2{\SIA fg,#1,#2 }
\def \Fl(#1)#2{\SIA fl,#1,#2 }
\def \Ft(#1)#2{\SIA ft,#1,#2 }
\def \Fd(#1)#2{\SIA fd,#1,#2 }

\def\etichetta(#1){(\veroparagrafo.\veraformula)%
\SIA e,#1,(\veroparagrafo.\veraformula) %
\global\advance\numfor by 1%
\write15{\string\Fe (#1){\equ(#1)}}%
\write16{ EQ #1 ==> \equ(#1)  }}

\def\etichettaa(#1){(A\veraappendice.\veraformula)%
\SIA e,#1,(A\veraappendice.\veraformula) %
\global\advance\numfor by 1%
\write15{\string\Fe (#1){\equ(#1)}}%
\write16{ EQ #1 ==> \equ(#1) }}

\def\getichetta(#1){Fig. \verafigura%
\SIA g,#1,{\verafigura} %
\global\advance\numfig by 1%
\write15{\string\Fg (#1){\graf(#1)}}%
\write16{ Fig. #1 ==> \graf(#1) }}

\def\etichettap(#1){\veroparagrafo.\verasub%
\SIA p,#1,{\veroparagrafo.\verasub} %
\global\advance\numsub by 1%
\write15{\string\Fp (#1){\secc(#1)}}%
\write16{ par #1 ==> \secc(#1)  }}

\def\etichettapa(#1){A\veraappendice.\verasub%
\SIA p,#1,{A\veraappendice.\verasub} %
\global\advance\numsub by 1%
\write15{\string\Fp (#1){\secc(#1)}}%
\write16{ par #1 ==> \secc(#1)  }}

\def\etichettal(#1){%
\ifnum\appflag=0{\veroparagrafo.\verolemma}%
\SIA l,#1,{\veroparagrafo.\verolemma} \fi%
\ifnum\appflag=1{A\veraappendice.\verolemma}%
\SIA l,#1,{A\veraappendice.\verolemma} \fi%
\global\advance\numlemma by 1%
\write15{\string\Fl (#1){\lm(#1)}}%
\write16{ lemma #1 ==> \lm(#1)  }}

\def\etichettat(#1){%
\ifnum\appflag=0{\veroparagrafo.\verotheorem}%
\SIA t,#1,{\veroparagrafo.\verotheorem} \fi%
\ifnum\appflag=1{A\veraappendice.\verotheorem}%
\SIA t,#1,{A\veraappendice.\verotheorem} \fi%
\global\advance\numtheorem by 1%
\write15{\string\Ft (#1){\thm(#1)}}%
\write16{ th. #1 ==> \thm(#1)  }}

\def\etichettad(#1){%
\inum\appflag=0{\veroparagrafo.\veradef}%
\SIA d,#1,{\veroparagrafo.\veradef} \fi%
\inum\appflag=1{A\veraappendice.\veradef}%
\SIA d,#1,{A\veraappendice.\veradef} \fi%
\global\advance\numdef by 1%
\write15{\string\Fd (#1){\defz(#1)}}%
\write16{ def. #1 ==> \defz(#1)  }}

\def\Eq(#1){\eqno{\etichetta(#1)\alato(#1)}}
\def\eq(#1){\etichetta(#1)\alato(#1)}
\def\Eqa(#1){\eqno{\etichettaa(#1)\alato(#1)}}
\def\eqa(#1){\etichettaa(#1)\alato(#1)}
\def\eqg(#1){\getichetta(#1)\alato(fig. #1)}
\def\sub(#1){\0\palato(p. #1){\bf \etichettap(#1).}}
\def\asub(#1){\0\palato(p. #1){\bf \etichettapa(#1).}}
\def\lemma(#1){\0\palato(lm #1){\cs Lemma \etichettal(#1)\hskip.3truecm}}
\def\theorem(#1){\0\palato(th #1){\cs Theorem \etichettat(#1)%
\hskip.3truecm}}
\def\definition(#1){\0\palato(df #1){\cs Definition \etichettad(#1)%
\hskip.3truecm}}

\def\equv(#1){\senondefinito{fe#1}$\clubsuit$#1%
\write16{eq. #1 non e' (ancora) definita}%
\else\csname fe#1\endcsname\fi}
\def\grafv(#1){\senondefinito{fg#1}$\clubsuit$#1%
\write16{fig. #1 non e' (ancora) definito}%
\else\csname fg#1\endcsname\fi}
\def\secv(#1){\senondefinito{fp#1}$\clubsuit$#1%
\write16{par. #1 non e' (ancora) definito}%
\else\csname fp#1\endcsname\fi}
\def\lmv(#1){\senondefinito{fl#1}$\clubsuit$#1%
\write16{lemma #1 non e' (ancora) definito}%
\else\csname fl#1\endcsname\fi}

\def\thmv(#1){\senondefinito{ft#1}$\clubsuit$#1%
\write16{th. #1 non e' (ancora) definito}%
\else\csname ft#1\endcsname\fi}

\def\defzv(#1){\senondefinito{fd#1}$\clubsuit$#1%
\write16{def. #1 non e' (ancora) definito}%
\else\csname fd#1\endcsname\fi}

\def\equ(#1){\senondefinito{e#1}\equv(#1)\else\csname e#1\endcsname\fi}
\def\graf(#1){\senondefinito{g#1}\grafv(#1)\else\csname g#1\endcsname\fi}
\def\secc(#1){\senondefinito{p#1}\secv(#1)\else\csname p#1\endcsname\fi}
\def\sec(#1){{\S\secc(#1)}}
\def\lm(#1){\senondefinito{l#1}\lmv(#1)\else\csname l#1\endcsname\fi}
\def\thm(#1){\senondefinito{t#1}\thmv(#1)\else\csname t#1\endcsname\fi}
\def\defz(#1){\senondefinito{d#1}\defzv(#1)\else\csname d#1\endcsname\fi}

\def\BOZZA{
\def\alato(##1){\rlap{\kern-\hsize\kern-1.2truecm{$\scriptstyle##1$}}}
\def\palato(##1){\rlap{\kern-1.2truecm{$\scriptstyle##1$}}}
}

\def\alato(#1){}
\def\galato(#1){}
\def\palato(#1){}


{\count255=\time\divide\count255 by 60 \xdef\hourmin{\number\count255}
        \multiply\count255 by-60\advance\count255 by\time
   \xdef\hourmin{\hourmin:\ifnum\count255<10 0\fi\the\count255}}

\def\oramin{\hourmin }

\def\data{\number\day/\ifcase\month\or gennaio \or febbraio \or marzo \or
aprile \or maggio \or giugno \or luglio \or agosto \or settembre
\or ottobre \or novembre \or dicembre \fi/\number\year;\ \oramin}
\setbox200\hbox{$\scriptscriptstyle \data $}
\footline={\rlap{\hbox{\copy200}}\tenrm\hss \number\pageno\hss}

%
%
\newcount\driver 
\newdimen\xshift \newdimen\xwidth
\def\ins#1#2#3{\vbox to0pt{\kern-#2 \hbox{\kern#1 #3}\vss}\nointerlineskip}

\newdimen\xshift \newdimen\xwidth \newdimen\yshift

\def\insertplotbm#1#2#3#4#5{\par%
\xwidth=#1 \xshift=\hsize \advance\xshift by-\xwidth \divide\xshift by 2%
\yshift=#2 \divide\yshift by 2%
\line{\hskip\xshift \vbox to #2{\vfil%
#3 \includegraphics{#4.pst}}\hfill \raise\yshift\hbox{#5} }}

\def\insertplot#1#2#3#4#5{\par%
\xwidth=#1 \xshift=\hsize \advance\xshift by-\xwidth \divide\xshift by 2%
\yshift=#2 \divide\yshift by 2%
\line{\hskip\xshift \vbox to #2{\vfil%
\ifnum\driver=0 #3
\special{ps: plotfile #4.ps} 
\fi
\ifnum\driver=1 #3 \includegraphics{#4.ps}\fi
\ifnum\driver=2 #3
\ifnum\mgnf=0\special{#4.ps 1. 1. scale} \fi
\ifnum\mgnf=1\special{#4.ps 1.2 1.2 scale}\fi
\fi }\hfill \raise\yshift\hbox{#5}}}

\def\insertplotttt#1#2#3#4{\par%
\xwidth=#1 \xshift=\hsize \advance\xshift by-\xwidth \divide\xshift by 2%
\yshift=#2 \divide\yshift by 2%
\line{\hskip\xshift \vbox to #2{\vfil%
\ifnum\driver=0 #3
\special{ps: plotfile #4.ps} 
\fi
\ifnum\driver=1 #3 \includegraphics{#4.ps}\fi
\ifnum\driver=2 #3
\ifnum\mgnf=0\special{#4.ps 1. 1. scale} \fi
\ifnum\mgnf=1\special{#4.ps 1.2 1.2 scale}\fi
\fi }\hfill}}

\newdimen\xshift \newdimen\xwidth \newdimen\yshift
\def\eqfig#1#2#3#4#5{
\par\xwidth=#1 \xshift=\hsize \advance\xshift
by-\xwidth \divide\xshift by 2
\yshift=#2 \divide\yshift by 2
\line{\hglue\xshift \vbox to #2{\vfil
\ifnum\driver=0 #3
\special{ps: plotfile #4.ps} 
\fi
\ifnum\driver=1 #3 \includegraphics{#4.ps}\fi
\ifnum\driver=2 #3 \special{
\ifnum\mgnf=0 #4.ps 1. 1. scale \fi
\ifnum\mgnf=1 #4.ps 1.2 1.2 scale\fi}
\fi}\hfill\raise\yshift\hbox{#5}}}

\let\a=\alpha \let\b=\beta  \let\g=\gamma     \let\d=\delta  \let\e=\varepsilon
\let\z=\zeta  \let\h=\eta   \let\th=\vartheta \let\k=\kappa   \let\l=\lambda
\let\m=\mu    \let\n=\nu    \let\x=\xi        \let\p=\pi      \let\r=\rho
\let\s=\sigma \let\t=\tau        \let\c=\chi
   \let\o=\omega 
\let\G=\Gamma \let\D=\Delta     \let\L=\Lambda  
           
\let\O=\Omega  

\def\\{\hfill\break} \let\==\equiv

\let\io=\infty 

\let\0=\noindent \def\pagina{{\vfill\eject}}

\def\ie{\hbox{\it i.e.\ }}\def\eg{\hbox{\it e.g.\ }}
\let\dpr=\partial 
\let\bs=\backslash
\def\defin{{\buildrel def\over=}}
\def\tende#1{\,\vtop{\ialign{##\crcr\rightarrowfill\crcr
 \noalign{\kern-1pt\nointerlineskip}
 \hskip3.pt${\scriptstyle #1}$\hskip3.pt\crcr}}\,}
\def\otto{\,{\kern-1.truept\leftarrow\kern-5.truept\to\kern-1.truept}\,}
\def\fra#1#2{{#1\over#2}}
\def\Pf{{\rm Pf}\,}

\def\PP{{\cal P}}\def\EE{{\cal E}}\def\VV{{\cal V}}
\def\HH{{\cal H}}\def\WW{{\cal W}}
\def\TT{{\cal T}}\def\NN{{\cal N}}\def\BB{{\cal B}}
\def\RR{{\cal R}}\def\LL{{\cal L}}\def\JJ{{\cal J}}\def\II{{\cal I}}
\def\DD{{\cal D}}\def\SS{{\cal S}}

\def\der{{\rm d}}
\def\T#1{{#1_{\kern-3pt\lower7pt\hbox{$\widetilde{}$}}\kern3pt}}
\def\VVV#1{{\underline #1}_{\kern-3pt
\lower7pt\hbox{$\widetilde{}$}}\kern3pt\,}
\def\W#1{#1_{\kern-3pt\lower7.5pt\hbox{$\widetilde{}$}}\kern2pt\,}

\def\lis{\overline}
 \def\acapo{\hfill\break}
  \def\sign{{\rm sign}\,}
\def\indica{\leaders \hbox to 0.5cm{\hss.\hss}\hfill}
\def\guida{\leaders\hbox to 1em{\hss.\hss}\hfill}
\mathchardef\oo= "0521

\def\bS{{\bf S}}
\def\V#1{{\bf #1}}
\def\pp{{\bf p}}\def\qq{{\bf q}}\def\ii{{\bf i}}\def\xx{{\bf x}}
\def\yy{{\bf y}}\def\kk{{\bf k}}\def\nn{{\bf n}}
\def\dd{{\bf d}}\def\zz{{\bf z}}\def\uu{{\bf u}}\def\vv{{\bf v}}
 \def\bP{{\bf P}}
\def\tt{{\bf t}}\def\jj{{\bf j}}\def\ff{{\bf f}}\def\ggg{{\bf g}}
\def\ss{{\underline \sigma}}\def\oo{{\underline \omega}}
\def\ee{{\underline \varepsilon}}\def\aa{{\underline \alpha}}
\def\un{{\underline \nu}}\def\ul{{\underline \lambda}}
\def\ux{{\underline x}}

\def\qed{\raise1pt\hbox{\vrule height5pt width5pt depth0pt}}

\def\indic{\hbox{\raise-2pt \hbox{\indbf 1}}}
\def\bk#1#2{\bar\kk_{#1#2}}

\def\virg{\quad,\quad}
\def\bT{{\bf T}}
\def\proof{\0{\cs Proof - } }
\def\Halmos{\hfill\vrule height6pt width4pt depth2pt \par\hbox to \hsize{}}

\def\MMM{\hbox{\euftw M}}\def\BBB{\hbox{\euftw B}}
\def\RRR{\hbox{\msytw R}} 
 \def\CCC{\hbox{\msytw C}}

 \def\ZZZ{\hbox{\msytw Z}}
 \def\zzz{\hbox{\msytwww Z}}


\mgnf=0   
\driver=1 
\openin14=\jobname.aux \ifeof14 \relax \else
\input \jobname.aux \closein14 \fi
\openout15=\jobname.aux

\def\openone{\leavevmode\hbox{\ninerm 1\kern-3.3pt\tenrm1}}%



\advance\voffset by 1.5truecm

\font\sc=cmcsc10
\newcount\firstpage

\def\setcap#1{\null\def\titlecap{#1}\global\firstpage=\pageno}
\def\titletesi{Universality and non--universality 
in the Ashkin--Teller model}

\footline={\hfil}

\def\ppagina{\ifodd\pageno\pagina\null\pagina\else\pagina\fi}

\def\nopagenumbers{\headline={\hfil}}

\def\pagenumbers{\headline={%
\ifodd\pageno\hfill{\sc\titlecap}~~{\bf\folio}%
\else{\bf\folio}~~{\sc\titletesi}\hfill\fi}}


\newwrite\indiceout

\def\inizioindicecorrente#1{
   \def\nomeindice{#1.ind}
   \immediate\openout \indiceout = \nomeindice
   }
\def\fineindicecorrente{ 
   \immediate\write\indiceout{ end }
   \immediate\closeout\indiceout
   \write16{ indice scritto nel file \nomeindice }
   }

\def\flushindex#1#2#3{
   \write\indiceout{ #1 }
   \write\indiceout{ #2 }
   \write\indiceout{ #3 } 
   \write\indiceout{ \the\pageno }
   }

\def\prtindex#1{\flushindex{prt}{0}{#1} }
\def\capindex#1#2{\flushindex{cap}{#1}{#2} }

\def\appindex#1#2{\flushindex{app}{#1}{#2} }

\def\bibindex#1{\flushindex{bib}{0}{#1} }

\def\leaderfill{\leaders\hbox to 1em{\hss . \hss} \hfill }

\inizioindicecorrente{t}
\null
\vskip4.truecm
\nopagenumbers
\insertplotttt{100pt}{100pt}{}{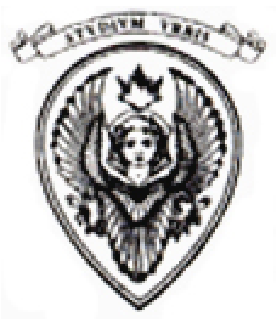}
\centerline{\csss UNIVERSIT\`A DEGLI STUDI DI ROMA}
\vskip.05truecm
\centerline{\csss LA SAPIENZA}
\vskip1.truecm
\centerline{\cs FACOLT\`A DI SCIENZE MATEMATICHE, FISICHE E NATURALI}
\vskip1.5truecm
\centerline{\cs PhD thesis in Physics}
\centerline{ A.Y. 2003/2004}
\vskip1.5truecm
\centerline{\titolone Universality and non--universality}
\vskip.05truecm
\centerline{\titolone in the Ashkin--Teller model}
\prtindex{Universality and non--universality
in the Ashkin--Teller model}
\vskip2.truecm

\leftline{\css Advisors:}
\leftline{{\css Prof.}{\it Giovanni Gallavotti}}
\leftline{{\css Prof.}{\it Vieri Mastropietro}}
\vskip1.truecm
\leftline{\css Author:}
\leftline{\it Alessandro Giuliani}
\ppagina
\null
\vskip1.truecm
\centerline{\titolo Index:}
\acapo\acapo
\hbox to 6.3 in{\bf 1. Introduction.\dotfill 7}\\
\hbox to 6.3 in{\qquad 1.1. The Ashkin--Teller model.\dotfill 9}\\
\hbox to 6.3 in{\qquad 1.2. Results.\dotfill 11}\\
\hbox to 6.3 in{\qquad 1.3. Outline of the proof.\dotfill 14}\\
\hbox to 6.3 in{\qquad 1.4. Summary.\dotfill 16}\\
\hbox to 6.3 in{\bf 2. The Ising model exact solution.\dotfill 18}\\
\hbox to 6.3 in{\qquad 2.1. The multipolygon representation.\dotfill 18}\\
\hbox to 6.3 in{\qquad 2.2. The Grassmann integration rules.\dotfill 18}\\
\hbox to 6.3 in{\qquad 2.3. The Grassmann representation of the 2D Ising model 
with open boundary conditions.\dotfill 19}\\
\hbox to 6.3 in{\qquad 2.4. The Grassmann representation of the 2D Ising model 
with periodic boundary conditions.\dotfill 27}\\
\hbox to 6.3 in{\qquad 2.5. The Ising model's free energy.\dotfill 28}\\
\hbox to 6.3 in{\bf 3. The Grassmann formulation of Ashkin--Teller.\dotfill 30}\\
\hbox to 6.3 in{\qquad 3.1. The Grassmann representation for a pair of 
Ising models with multi spin interactions.\dotfill 30}\\
\hbox to 6.3 in{\qquad 3.2. The Grassmann representation for the 
Ashkin--Teller model.\dotfill 33}\\
\hbox to 6.3 in{\bf 4. The ultraviolet integration.\dotfill 37}\\
\hbox to 6.3 in{\qquad 4.1. The effective interaction on scale 1.\dotfill 37}\\
\hbox to 6.3 in{\qquad 4.2. The integration of the $\c$ fields.\dotfill 38}\\
\hbox to 6.3 in{\qquad 4.3. Symmetry properties.\dotfill 43}\\
\hbox to 6.3 in{\bf 5. Renormalization Group for light fermions. 
The anomalous regime.\dotfill 48}\\
\hbox to 6.3 in{\qquad 5.1. Multiscale analysis.\dotfill 48}\\
\hbox to 6.3 in{\qquad 5.2. The localization operator.\dotfill 49}\\
\hbox to 6.3 in{\qquad 5.3. Renormalization.\dotfill 51}\\
\hbox to 6.3 in{\qquad 5.4. Analyticity of the effective potential.\dotfill 53}\\
\hbox to 6.3 in{\qquad 5.4. Proof of Theorem 5.1.\dotfill 54}\\
\hbox to 6.3 in{\bf 6. The flow of the running coupling constants.\dotfill 61}\\
\hbox to 6.3 in{\qquad 6.1. The flow equations.\dotfill 61}\\
\hbox to 6.3 in{\qquad 6.2. Proof of Theorem 6.1.\dotfill 61}\\
\hbox to 6.3 in{\qquad 6.2. The flow of the renormalization constants.\dotfill 64}\\
\hbox to 6.3 in{\qquad 6.4. The scale $h^*_1$.\dotfill 65}\\
\hbox to 6.3 in{\bf 7. Renormalization Group for light fermions. 
The non anomalous regime.\dotfill 67}\\
\hbox to 6.3 in{\qquad 7.1. Integration of the $\psi^{(1)}$ field.\dotfill 67}\\
\hbox to 6.3 in{\qquad 7.2. The localization operator.\dotfill 68}\\
\hbox to 6.3 in{\qquad 7.3. Renormalization for $h\le h^*_1$.\dotfill 69}\\
\hbox to 6.3 in{\qquad 7.4. The integration of the scales $\le h^*_2$.\dotfill 70}\\
\hbox to 6.3 in{\qquad 7.5. Keeping $h^*_2$ finite.\dotfill 71}\\
\hbox to 6.3 in{\qquad 7.6. The critical points.\dotfill 71}\\
\hbox to 6.3 in{\qquad 7.7. Computation of $h^*_2$.\dotfill 72}\\
\hbox to 6.3 in{\bf 8. The specific heat.\dotfill 74}\\
\hbox to 6.3 in{\bf 9. Conclusions and open problems.\dotfill 78}\\
\hbox to 6.3 in{\bf Appendix A1. Grassmann integration. Truncated expectations.
\dotfill 79}\\
\hbox to 6.3 in{\qquad A1.1 Truncated expectations and some more rules.\dotfill 79}\\
\hbox to 6.3 in{\qquad A1.2 Graphical representation for truncated expectations.\dotfill
81}\\ 
\hbox to 6.3 in{\bf Appendix A2. The Pfaffian expansion.\dotfill 83}\\
\hbox to 6.3 in{\bf Appendix A3. Gram--Hadamard inequality.\dotfill 90}\\
\hbox to 6.3 in{\bf Appendix A4. Proof of Lemma 5.3.\dotfill 94}\\
\hbox to 6.3 in{\bf Appendix A5. Proof of {(5.42)}.\dotfill 95}\\
\hbox to 6.3 in{\bf Appendix A6. Vanishing of the Beta function.\dotfill 96}\\
\hbox to 6.3 in{\qquad A6.1. The reference model.\dotfill 96}\\
\hbox to 6.3 in{\qquad A6.2. The Dyson equation.\dotfill 97}\\
\hbox to 6.3 in{\qquad A6.3. Ward identities and the first addend of 
\equ(A6.2.11a).\dotfill 99}\\
\hbox to 6.3 in{\qquad A6.4. The first correction identity.\dotfill 102}\\
\hbox to 6.3 in{\qquad A6.5. Ward identities and the second addend in 
\equ(A6.2.11a).\dotfill 108}\\
\hbox to 6.3 in{\qquad A6.6. Proof of \equ(A6.2.26).\dotfill 112}\\
\hbox to 6.3 in{\bf Appendix A7. The properties of 
{$D_\o ({\fam \bffam \tenbf p})^{-1} C_\o ({\fam \bffam \tenbf k},
{\fam \bffam \tenbf k}-{\fam \bffam \tenbf p}) $}.\dotfill 127}\\
\hbox to 6.3 in{\bf Appendix A8. Proof of Lemma 7.3.\dotfill 129}\\
\hbox to 6.3 in{\bf Appendix A9. Independence from boundary conditions.\dotfill
131}\\
\hbox to 6.3 in{\bf Acknowledgments.\dotfill 132}\\
\hbox to 6.3 in{\bf References.\dotfill 133}\\ 

\pagina\null
\ppagina\null

\pagenumbers


\setcap{1. Introduction.}
\capindex{1}{Introduction.}
\vskip1.truecm
\section(1,Introduction.)
\capindex{1}{Introduction.}

Two dimensional classical spin systems play a very special role 
in statistical mechanics, in providing the simplest non trivial examples
of systems undergoing a phase transition.

The first of these model to be extensively studied was the Ising model, 
[Pe][O][Ka49][KO][Ya]
whose importance relies in the fact that it first gave firm and quantitative
indications that a microscopic short range interaction
can produce phase transitions which deeply differ from that
described by mean field approximation. 

In the Ising model many detailed informations about the 
microscopic structure of the phases in the low or high 
temperature regime can be obtained by perturbative techniques 
(cluster expansion [Ru63][GM68][D] [Ru69]), by correlation inequalities
[Gr][FKG][Le74] and
by probabilistic methods (\eg the ``infinite
cluster'' method [Ru79][Ai80][Hi])
and some of the critical properties can be deduced 
by combination of the previous techniques together with the use of ``infrared
bounds'' [Fr][Ai82]. However, 
most of the results about the 
behaviour of thermodynamic functions near the critical temperature 
rely on the exact solution, 
first obtained by Onsager and after him reproduced in  
many different independent ways [KO][KWa][SML][H][S]. 

The Ising model in zero magnetic field is solvable in a very strong sense:
it can be exactly mapped into a system of free fermions [SML][H][S]
and, as a consequence,
not only one can calculate the free energy and the magnetization, 
but exact formulae for many important spin correlation functions
can be derived, and the asymptotic behaviour for 
large distances of some of them can be exactly computed
\footnote{${}^1$}{\nota
It must be stressed that 
these informations cannot be trivially derived from the 
exact expression of the free energy, and hard work together with
amazing algebraic cancellations are needed for the computation of the 
asymptotics of correlation functions, even for the ``simple''
spin--spin correlation function along the same horizontal line, see [MW].}.
For istance the energy--energy correlation functions can be computed,
as well as the asymptotic behaviour of the spin--spin correlation function
[MPW][BMW][TM][WMTB][MW]
and of some multi--spin correlation functions, when we let the relative
distances of the positions of the spins diverge in some special way 
and directions (\eg along the same horizontal line [Ka69]). 

These results allow to calculate the critical exponents, as defined 
in the usual scaling theory of critical phenomena, and to verify that, even if
the scaling laws 
are all satisfied, as expected, the 2D Ising model
exponents are {\it different} from those expected from Curie--Weiss theory:
one says that the Ising model belongs to a different {\it universality class}.
\\

The development of Renormalization Group [Ka66][DJ][C][Sy][W1][W2]
[WF], starting from the end of the $60$'s,
clarified the concept of universality class, and gave a fundamental 
explanation to the fenomenological expectation that different models,
even describing completely different physical situations,
could show the same critical behaviour, in the sense that
their critical exponents are the same (if one suitably identifies the 
corresponding thermodynamic functions in the two systems).
In the context of statistical mechanics, 
it became clear that two systems, with the same symmetries and with
interactions differing only by {\it irrelevant} terms 
have correlation functions
that, at the critical point, show the same asymptotic behaviour
in the limit of large distances; that is 
the two systems have the same critical exponents.

Independently from Renormalization Group, and approximatively at the same time,
a new important branch of statistical mechanics arose, 
that of exactly solvable models, for a review see [Ba82]. In this context,
and more specifically in that of 2D spin
systems, many explicit examples were constructed
of new and unexpected universality classes, different from Ising's.
We refer in particular to two dimensional 6 vertex (6V) and 8 vertex (8V)
models
\footnote{${}^2$}{\nota
The vertex models are defined by associating a direction to each of the bonds 
linking the sites of a 2D lattice; and by allowig only a few configurations
of the arrows entering or exiting a lattice site. In the 6V (8V) models only 
6 (8) different configuration are allowed at each site, 
and different 
energies are assigned to each allowed configuration. 
The 8V model
can be easily mapped into 2D spin models, described by two 
Ising layers, coupled by a 4 spin interaction. The 6V models can 
be obtained from the spin description of 8V by letting the 
coupling constants tending to infinity in some specific way (they can be
considered as Ising models ``with constraints'').}.
The class of 6V models includes the ice model, first solved by
Lieb [L1], the F--model and the KDP--model, solved in rapid succession
after the ice--model exact solution [L2][L3][Su]; see [LW] for a review
on the 6V models. The Lieb's solution was a breakthrough in statistical
mechanics, both because first showed the existence of 
exactly solvable models other than Ising itself, and because 
concretely showed the existence of many new universality classes 
different from Ising's
in the context of 2D spin systems. The latter point was of great 
importance for the development of the theory of critical phenomena:
in fact at the time of the solution of the ice model the universality theory
of critical point singularity was not yet developed in its final form.
So, when Renormalization Group approach arose around 1969, the 6V models 
appeared as a counterexample to the universality that Renormalization Group
was supposed to predict: depending on the 
specific choices of the energies assigned to the different vertex 
configurations one could find different values for the critical exponents.

This fact, not well understood at the beginning by
a fundamental point of view, was dismissed by the 
theoretical physics community on the grounds that the 6V models
are spin model ``with constraints'' (see footnote 2), 
that is too pathological to be well described by the universality theory 
of critical phenomena. 

However the deep meaning of Lieb's counterexamples was 
made clear by Baxter's exact solution of the 8V model, contained in 
a series of papers from 1971 to 1977 [Ba]\footnote{${}^3$}{\nota
Baxter's solution represents one of the major 
achievements of mathematical physics in the 1970's: it
introduced for the first time in theoretical physics the use
of triangle--star equations and of corner transfer matrix, 
which are nowadays fundamental tools for the study of quantum groups and 
integrable systems.}: it
made clear to everybody that
the 6V models could not be considered as 
pathological counterexamples. As remarked in footnote 2, 8V models
are genuine
short range Ising models with finite interaction and one can 
for instance consider a path in the parameters space 
continuously linking two 6V models defined by different choices 
of the energies associated
with the vertex configurations. The remarkable result
following by the 8V solution is that along this path the 8V critical 
exponents change {\it continuously}, and continuously connect those
of the two different 6V models. 

This observation was crucial and led to a much better understanding of 
the theories that were put forward to explain critical phenomena,
first among all Renormalization Group itself. In modern
language the solution of the above ``paradoxes'' relies
on the fact that the 6V and 8V models with different choices of parameters
differ by {\it marginal} terms: however this fact is not appearent
in the original spin variable, and in order to realize this one has
to reformulate all this models as suitable field theory models (that is not
an easy task).\\

Even if many important informations about the thermodynamics
of vertex models can be found from their exact solution, 
these models are exactly solvable in a sense much
weaker than that of Ising.

The 6V models are solvable by Bethe ansatz, that is by assuming that the
eigenvector of the transfer matrix with largest eigenvalue is a 
linear combination of plane
waves; and calculating the coefficients of the linear combination by solving
a (complicated) integral equation. This allows to find an exact
expression for the free energy $f(\b,E)$, 
as a function of the temperature $\b^{-1}$ and of
an external electric field $E$ (so that by computing the derivatives of 
$f$ w.r.t. $E$ one can study the critical behaviour
of the electric response function); but nothing can be said about
more complicated correlation functions, it is not even possible to write 
formal expression for them. 

The solution of the 8V model is even more involved and sophisticated
and is based on a reformulation of the problem of calculating the free energy
into the problem of solving a set of coupled elliptic integral 
equations (the so called Yang--Baxter triangle--star equations). Also in this
case it is not possible to find (even formal) expressions for generic
correlation functions, but only the free energy as a function of some
thermodynamic parameters can be calculated, so that only informations
about special low order correlation functions can be obtained. 
 
Relying the solutions of 6V and 8V on the explicit analytic solution
of special integral equations, it is not 
surprising that even small and apparently harmless 
modifications (from the Renormalization Group point of view)
of these models completely destroy their integrability. Also, the exact
solutions do not give any information about the thermodynamic behaviour
of systems obtained as small perturbations of 6V and 8V. 

On the other hand one can hope that many relevant properties of the integrable
models are quite robust under perturbations. Indeed, on the basis of operator
algebra and scaling theory, it was conjectured since a long time
that a universality property holds for Ising, in the sense that by adding 
to it, for instance, a next to nearest neighbor interaction, the critical 
indexes remain unchanged.
A similar universality property was conjectured for the 8V model.
By scaling theory arguments,
Kadanoff [Ka77] found evidence that 8V is in the same class of universality of
the Ashkin--Teller model
\footnote{${}^4$}{\nota
Ashkin--Teller (AT) is defined as a pair of Ising layers coupled via a 
four spin plaquette interaction, different from that of 8V; 
AT is not integrable and, in correspondence of some special 
choices of its parameters, it reduces to Ising and to the 4--states Potts 
model.}, 
in the sense that the critical exponents are the 
same, if one suitably identifies the coupling constants. 
Further evidence for this conclusion was given in [PB], by 
second order Renormalization Group, and in [LP][N],
by a heuristic mapping of both the 8V and the Ashkin--Teller 
models into the massive Luttinger model, a not integrable model describing 
massive interacting fermions on the continuum in 1+1 dimensions.\\

As suggested by the previous discussion, the natural method to relate
non--integrable models to integrable ones is given by Renormalization Group 
(RG). This was realized long ago, but the main open problem in 
this context was to implement RG in a rigorous way; and, even at a 
heuristic level, to understand in a detailed and quantitative way 
from the RG point of view how the crossovers 
between the different universality classes are realized, when one let 
continuously vary the strength of the coupling constants defining
the interaction among spins. 

In this dissertation we want to describe a constructive method
for studying thermodynamic and correlation functions at the critical
point for a wide class of two dimensional classical spin systems, 
obtained as perturbations of the Ising model, including
the next to nearest neighbor Ising, the 8V model and Ashkin--Teller.
The method was first introduced in [PS] and [M] and  
is based on an exact mapping of the spin model into a 
model of interacting spinless fermions in 1+1 dimensions and
on the implementation of constructive fermionic Renormalization Group methods
for the construction of the effective potential and of the correlation 
functions. The constructive fermionic Renormalization Group methods
we apply were developed by the Roma's school in the last decade
[BG1][BGPS][BoM][GS][BM] and are technically based on the so--called 
functional renormalization group, developed in the 1980's starting from [Po]
[GN], see [G1][BG] for reviews.

We will apply the method to the analysis of the critical behaviour
of the specific heat $C_v$ in the {\it Ashkin--Teller} model
and we will rigorously prove an old conjecture by Baxter and Kadanoff
about the critical behaviour of
Ashkin--Teller (AT), in correspondence of different
choices of the parameters defining the model (the inter--layer interaction 
$\l$ and the anisotropy $J^{(1)}-J^{(2)}$, see \equ(1.1)).
We shall study in detail how the crossover between the different universality
classes is realized when we let $J^{(1)}-J^{(2)}\to 0$
and how the location of the critical points
is renormalized by the interaction $\l$, in the region of small $\l$. 



\\
\sub(1.1){\bf The Ashkin--Teller model.}\\
The Ashkin--Teller model [AT] was introduced
as a generalization
of the Ising model to a four 
component system; in each site of a bidimensional
lattice there is a spin which can take four values,
and only nearest neighbor spins interact.
The model can be also considered a generalization
of the four state Potts model to which
it reduces for a suitable choice of the parameters.

A very convenient representation
of the Ashkin--Teller model is 
in terms of Ising spins [F]: given a
square sublattice 
$\L_M\subset\ZZZ^2$ of side $M$, one associates 
at each site $\xx\in\L_M$ two kinds of Ising spins,
$\s^{(1)}_\xx$, $\s^{(2)}_\xx$, assuming two possible values $\pm 1$. 
The AT Hamiltonian is assumed to be:
$$H^{AT}_{\L_M}=-\sum_{<\xx,\yy>\in\L_M}\Big[ J^{(1)}
\s^{(1)}_\xx\s^{(1)}_{\yy}+
J^{(2)}
\s^{(2)}_\xx\s^{(2)}_{\yy}
+\l\s^{(1)}_\xx\s^{(1)}_{\yy}
\s^{(2)}_\xx\s^{(2)}_{\yy}\Big]\=\sum_{\xx\in\L_M}H^{AT}_\xx\;,\Eq(1.1)$$
where $\xx,\yy$ are nearest neighbor sites and the last
identity is a definition for $H^{AT}_\xx$. Periodic boundary
conditions will be assumed throughout the work. 
\\
The case in which the two Ising subsystems are 
identical $J^{(1)}=J^{(2)}$ is called
{\it isotropic}, the opposite case {\it anisotropic}. 
\\
When the coupling $\l$ is $=0$, 
Ashkin--Teller (AT) reduces to two independent Ising models
and it has of course {\it two} critical temperatures if $J^{(1)}\not= J^{(2)}$.
\\
When $J^{(1)}=J^{(2)}=\l$, AT reduces to the four states Potts model.
\\
We shall study the case $J^{(i)}>0$, $i=1,2$, that is the case in which
the two Ising subsystems are {\it ferromagnetic}.\\

AT is a model for a number of 2d magnetic compounds: for instance
layers of atoms and molecules adsorbed on clean surfaces,
like selenium on nichel, molecular 
oxygen on graphite, atomic oxygen on tungsten; 
and layers of oxygen atoms in the basal Cu--O plane of 
some cuprates, like YBa$_2$Cu$_3$O$_z$, are believed
to constitute physical realizations
of the AT model [DR][Bak][Bar]. 
Theoretical results on AT can give detailed informations on the critical 
behaviour and the phase diagrams of such systems, which can 
be experimentally measured 
by means of electron diffraction techniques. 

Also, as explained in previous section, 
the importance of AT is in providing
a conceptual laboratory in which the higly non trivial
phenomenon of phase transitions
can be understood quantitatively in a relatively manegeable
model; in particular it has attracted great theoretical interest 
because is a simple and non trivial generalization of the Ising
and four-state Potts models, showing a rich variety of 
critical behaviours, depending on the choices of 
the parameters $J^{(i)}$ and $\l$ in \equ(1.1). 
AT is not exactly solvable, except in the trivial 
$\l=0$ case, and it has great theoretical
interest to develop techniques that, {\it without any use of exact
solutions}, could allow to understand the AT critical
behaviour. In fact exact solutions are
quite rare and generally peculiar of low dimensions, while RG methods
are expected to work in much more general situations: then it is  
important to refine RG tecniques in a simple but non trivial
playground, as that offered by AT.\\

The thermodynamic behaviour of the anisotropic AT model is 
not well understood even at a heuristic level. What is ``known''
is mainly based on conjectures, suggested by scaling theory,
and on numerics. 

A first conjecture, proposed by Wu and Lin [WL], concerns the 
critical points: from the symmetries of the model,
it is expected that AT, even in the interacting case 
(\ie $\l\not=0$),
has {\it two} critical temperatures for $J^{(1)}\not= J^{(2)}$ which coincide
at the isotropic point $J^{(1)}= J^{(2)}$. However nothing
has been proposed about the location of the critical points, even 
at a conjectural level.

Kadanoff [Ka77] and Baxter [Ba82] conjectured that 
the critical properties in the anisotropic 
and in the isotropic case are completely different; in
the first case the critical behaviour should be described
in terms of {\it universal} critical indices (identical
to those of the 2D Ising model)
while in the isotropic case the critical behaviour should be 
{\it nonuniversal} and described in terms
of indexes which are non trivial functions of $\l$.
In other words, the AT model should exhibit a 
{\it universal--nonuniversal} crossover when
the isotropic point is reached.

The general anisotropic case was studied numerically by 
Migdal--Kadanoff Renormalization Group [DR],
Mean Field Approximation and Monte Carlo [Be], 
real--space Renormalization Group [Bez]
Transfer Matrix Finite--Size--Scaling [Bad]; such results give evidence of
the fact that, far away from the isotropic point, 
AT has two critical points and belongs to the same universality class
of the Ising model but give essentially no informations
on the critical behaviour when the anisotropy
is small. The problem of how 
the crossover from universal to nonuniversal
behaviour is realized in the isotropic limit
remained for years completely unsolved, even at a heuristic level.\\
\\
\sub(1.5) {\bf Results.}
\\
Our main results concern the analytical properties of the free energy
in an interval of temperatures around the critical temperatures; 
and the critical behaviour of the specific heat. These thermodynamic
quantities are defined in the usual way: if 
$\b$ is the inverse temperature, the partition function
at finite volume is:
$$\Xi_{\L_M}\defin \sum_{\s^{(1)}_{\L_M},\ \s^{(2)}_{\L_M}}
e^{-\b H^{AT}_{\L_M}}
\;,\Eq(Z)$$
where $\s^{(i)}_{\L_M}\=\{\s^{(i)}_\xx\, |\, \xx\in \L_M\}$;
correspondingly, 
the free energy and the specific heat are defined as:
$$f=-{1\over \b}\lim_{M\to\io}{1\over M^2}\log\Xi_{\L_M}\virg 
C_{v}=\lim_{M\to\io}{\b^2\over M^2}\sum_{\xx,\yy\in\L_M}<H_\xx^{AT}
H_\yy^{AT}>_{\L_M,T}\;,\Eq(cv)$$
where $<\cdot>_{\L_M,T}$ denotes
the truncated expectation w.r.t. the Gibbs distribution
with Hamiltonian \equ(1.1). 

We find convenient to introduce the variables 
$$t={t^{(1)}+t^{(2)}\over 2}
\;,\quad u={t^{(1)}-t^{(2)}\over 2}\Eq(1.2)$$
with $t^{(j)}=\tanh\b J^{(j)}$, $j=1,2$. 
The parameter $t$ has the role of a {\it reduced temperature}
and $u$ measures the {\it anisotropy}
of the system. We shall consider the free
energy or the specific heat as functions of $t,u,\l$.
When $\l=0$ the model \equ(1.1) reduces to a pair
of decoupled Ising models and the specific heat $C_v$
can be immediately computed from the Ising model
exact solution; the system admits two critical points, defined by
$$\tanh\b J^{(i)}=\sqrt2 -1,\qquad i=1,2\;,\Eq(1.2az)$$
or, in terms of the parameters $t,u$ defined in \equ(1.2):
$$t_c^\pm=\sqrt2-1\pm|u|\;.\Eq(1.2za)$$
As it is well--known from Ising's exact solution, 
near the two critical temperatures the specific heat
shows a logarithmic divergence:
$C_v\simeq -C\log|t-t_c^\pm|$, where $C>0$.

Consider now the $\l\not=0$ case. If the anisotropy is strong 
the two Ising subsystems 
have very different critical temperatures:
so, if the temperature of the coupled system
is near to the critical temperature of one of the 
Ising subsystems, one can expect that AT is essentially equivalent to a 
single critical Ising model, perturbed by a small 
``random noise'', produced by the non--critical fluctuations 
of the second Ising subsystem; in such a case one expects
that the effect of the coupling
is at most that of changing the value
of the critical temperatures [PS]\footnote{$^5$}{\nota Note that,
because of the structure of the Hamiltonian \equ(1.1) 
(in which the interaction has 
the form of a product of bond interactions), this heuristic picture 
applies both to the case the non critical Ising model is well inside
the paramagnetic phase and to the case it is well inside the
magnetized phase: in both cases, if the system 2 is the system far
from criticality, we can rewrite $\s^{(2)}_\xx$ as
$\s^{(2)}_\xx=m^*_2+\d\s^{(2)}_\xx$, where $m^*_2$ is the (unperturbed)
magnetization of system 2, and $\d\s^{(2)}_\xx$ is the field associated with
the non critical fluctuations of $\s^{(2)}_\xx$ around
its average value; one can then expect that
the effect of the interaction of system 1 with system 2
is just that of changing the coupling $J^{(1)}$ into an effective coupling 
$J^{(1)}+\l(m^*_2)^2+\d J$, where $\d J$ is a small random noise, generated
by the non-critical fluctuations of $\s^{(2)}$ around
its average value. Since we shall assume
$J^{(1)}$ to be $O(1)$, it makes no qualitative difference whether $m^*_2$ 
is vanishing or not.}.
On the other hand if the anisotropy is small
the two system will become critical almost at the same 
temperature and the properties of the system could change drastically.

With the notations introduced above
and calling $D$ a sufficiently small $O(1)$ interval 
(\ie with amplitude independent of $\l$)
centered around $\sqrt2-1$, 
we can express our main result as follows [GM1][GM2].
\vskip.5cm
\0{\cs Theorem.}\ {\it There exists $\e>0$ such that, if $|\l|\le \e$ 
and $t\pm u\in D$, 
the AT model admits 
two critical points of the form:
$$t_c^\pm(\l,u)=\sqrt2-1+\n(\l)\pm |u|^{1+\h}(1+\d(\l,u))\;.\Eq(1.3)$$
Here $\n$ and $\d$ are $O(\l)$ corrections and $\h=\h(\l)=-b\l+O(\l^2)$,
$b>0$, is an analytic function of $\l$.
If $|\l|\le \e$, $t\pm u\in D$ and 
$t\not=t_c^\pm$,
the free energy and the specific heat of the model are 
analytic in $\l,t,u$; in the same region of parameters,
the specific heat $C_v$ can be written as:
$$C_v=F_1\D^{2\h_c}\log{|t-t_c^-|\cdot|t-t_c^+|\over \D^2}+F_2{1-\D^{2\h_c}
\over \h_c}+F_3\;,\Eq(1.4)$$
where: $2\D^2=(t-t_c^-)^2+(t-t_c^+)^2$; $\h_c=a\l+O(\l^2)$, $a\not=0$;
and $F_1$, $F_2$, $F_3$ are functions of $t,u,\l$, bounded above and below
by $O(1)$ constants.}
\\

A first interesting result that can be read from the Theorem 
is that the location of the critical points is dramatically changed 
by the interaction, see \equ(1.3). The difference of
the interacting critical temperatures normalized with the free one
$G(\l,u)\=(t_c^+(\l,u)-t_c^-(\l,u))/(t_c^+(0,u)-t_c^-(0,u))$
rescales with the anisotropy parameter as
a power law $\sim|u|^{\h}$, and in the limit $u\to 0$
it vanishes or diverges, depending on the sign of $\l$ 
(this is because $\h=-b\l+O(\l^2)$, 
with $b>0$). In Fig. 1 we plot the qualitative behaviour
of $G(\l,u)$ as a function of $u$, for two different values of $\l$
(\ie we plot the function $u^\h$, with $\h=0.3,-0.3$ respectively).


\midinsert
\*
\insertplotttt{330pt}{210pt}{}{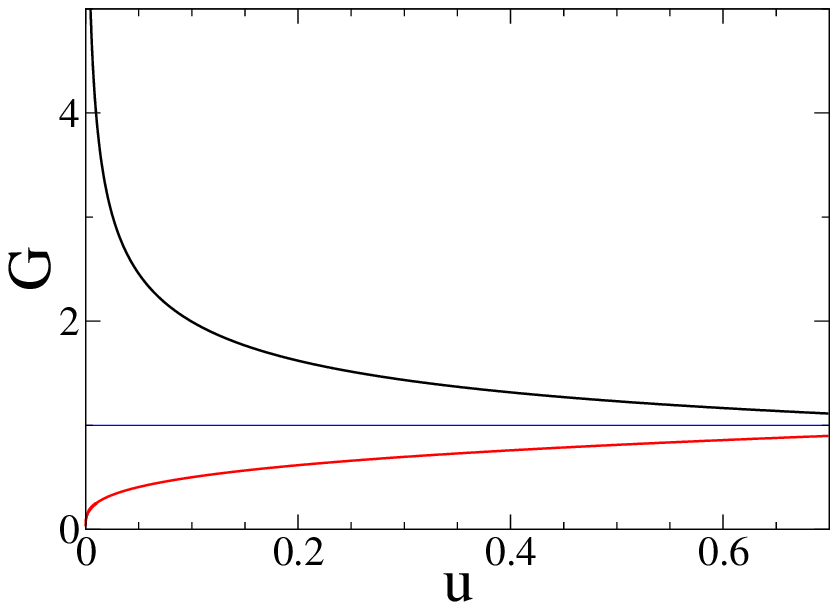}
\vskip.4truecm
\line{\vtop{\line{\hskip1.3truecm\vbox{\advance\hsize by -2.0 truecm
\0{\css Fig. 2.}
{\ottorm  The behaviour of the difference $G$
between the interacting critical temperatures normalized
to the free one, for two different values of $\l$;
depending on the sign of the interaction, it diverges
or vanishes in the isotropic limit. 
} \hfill} }}}
\*
\endinsert

As far as we know, the existence of
the critical index $\h(\l)$ was not known in the literature,
even at a heuristic level.\\

From \equ(1.4) it follows that there is universality for the 
specific heat, in the sense that it diverges logarithmically
at the critical points, as in the Ising model.
However the coefficient of the log
is {\it anomalous}: in fact if $t$ is near to one of the 
critical temperatures 
$\D\simeq\sqrt{2}|u|^{1+\h}$ so that the coefficient in front of the logarithm
behaves like $\sim |u|^{2(1+\h)\h_c}$, with 
$\h_c$ a new anomalous exponent $O(\l)$; 
in particular it is vanishing or diverging as 
$u\to 0$ depending on the sign of $\l$.
We can say that the system shows an {\it anomalous universality}
which is a sort of new paradigmatic behaviour: 
the singularity at the critical 
points is described in terms of universal critical indexes and
nevertheless, in the isotropic limit $u\to 0$, 
some quantities, like the difference
of the critical temperatures and the constant in front of the logarithm 
in the specific heat, scale with anomalous critical indexes, and they
vanish or diverge, depending on the sign of $\l$.

Eq\equ(1.4) clarifies how the universality--nonuniversality
crossover is realized as $u\to 0$.
When $u\not=0$ only the first term in eq\equ(1.4) can be log--singular
in correspondence of the two critical points; 
however
the logarithmic term dominates on the second one only if $t$ varies 
inside an extremely small region $O(|u|^{1+\h}e^{-c/|\l|})$
around the 
critical points (here $c$ is a positive $O(1)$ constant). 
Outside such region the power law behaviour 
corresponding to the second addend dominates. When $u\to 0$ one
recovers the power law decay first found by Mastropietro [M]
in the isotropic case:
$$C_v\simeq F_2{1-|t-t_c|^{2\h_c}\over\h_c}\Eq(1.5)$$
In Fig. 2
we plot the qualitative behaviour of $C_v$ 
as a function of $t$. 
The three graphs are 
plots of eq\equ(1.4), with $F_1=F_2=1$, $F_3=0$, 
$u=0.01$, $\h=\h_c=0.1,0,-0.1$ respectively; the central curve 
corresponds to the case $\h=0$, the upper one to $\h<0$ and 
the lower to $\h>0$.

\midinsert
\*
\insertplotttt{330pt}{180pt}{}{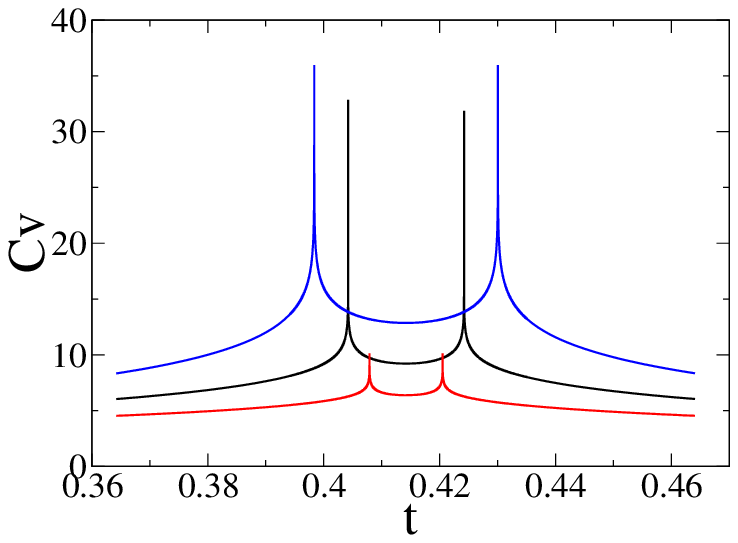}
\vskip.3truecm
\line{\vtop{\line{\hskip1.3truecm\vbox{\advance\hsize by -2.0 truecm
\0{\css Fig. 2.}
{\ottorm The behaviour
of the specific heat $C_v$ 
for three different values of $\l$, showing the log--singularities
at the critical points;
in the isotropic limit the two critical points tend to coincide,
the lower curve becomes continuous while the upper 
develops a power law divergence.
} \hfill} }}}
\*
\endinsert
\\

It now worths to make some technical remarks
about the Theorem above. 

The first is about the range of parameters
where the Theorem holds. The key hypothesis 
for the validity of the Theorem is the 
smallness of $\l$. When $\l=0$ the critical points
correspond to $t\pm u=\sqrt2-1$: hence for simplicity
we restrict $t\pm u$ in a sufficiently small $O(1)$
interval around $\sqrt2-1$. A possible explicit choice for $D$, 
convenient for our proof, 
could be $D=[{3(\sqrt2-1)\over 4},{5(\sqrt2-1)\over 4}]$.  
We expect that our technique would allow us to prove the above theorem, 
at the cost of a lengthier discussion, 
for any $t^{(1)},t^{(2)}>0$: of course in that case we should 
distinguish different regions of parameters and treat in a different
way the cases of low or high temperature or the case of big anisotropy
(\ie the cases $t<<\sqrt2-1$ or $t>>\sqrt2-1$ or $|u|>>1$).

The second remark is about the analyticity of the specific heat.
It is claimed that $C_v$ is analytic in
$\l,t,u$ outside the critical line. However, this is not appearent 
from \equ(1.4), because $\D$ is non analytic in $u$ at $u=0$ (of course the
bounded functions $F_j$ are non analytic in $u$ also, in a suitable
way compensating the non analyticity of $\D$). We get to \equ(1.4)
by interpolating two different asymptotic behaviours of $C_v$ in the regions
$|t-\lis t_c|<2|u|^{1+\h}$ and $|t-\lis t_c|\ge 2|u|^{1+\h}$,
where $\lis t_c$ is the average point between $t_c^+$ and $t_c^-$; then, the
non analyticity of $\D$ is introduced ``by hands'' by our estimates
and it is not intrinsic for $C_v$. \equ(1.4) is simply a convenient 
way to describe the crossover between different critical behaviours of $C_v$.

Finally, it must be stressed that we do not study the free energy 
directly at $t=t_c^\pm(\l,u)$,
therefore in order to show that
$t=t_c^\pm(\l,u)$ is a critical point we must study 
some thermodynamic property like 
the specific heat by evaluating
it at $t\ne t_c^\pm(\l,u)$ and $M=\io$ and then verify that it has a
singular behavior as $t\to t_c^\pm$. The case $t$ precisely equal to
$t_c^\pm$ cannot be discussed at the moment 
with our techniques, in spite of the uniformity
of our bounds as $t\to t_c^\pm$. The reason is that 
we write the AT partition function as a sum
of 16 different partition functions, differing
for boundary terms. Our estimates on each single term 
are uniform up to the critical point;
however, in order to show that the free energy computed with one of the 
16 terms is the same as the complete free energy,
we need to stay at $t\ne t_c^\pm$: in this case boundary terms are suppressed
as $\sim e^{-\k M|t-t_c^\pm|}$, $\k>0$, as $M\to\io$. If we stay exactly
at the critical point cancellations between the 16 terms can be present
(as it is well known already from the Ising model exact solution [MW])
and we do not have control on the behaviour of the free energy, as 
the infinite volume limit is approached. We believe that
this is a purely technical difficulty and that it could
be solved by a more detailed analysis of the cancellations among
the different terms appearing in the Ising's partition function. Another
possibility to study AT directly at the critical point would be 
to adapt our method to the case of open boundary conditions
(where even in the fermionic
representation the free energy can be written as the logarithm of a single
partition function). The interest 
of studying the model directly at criticality is linked to the possibility
of explicitly studying the finite size corrections to the correlation 
functions and the approach to their conformal limit.
\\
\\
\sub(1.6) {\bf Outline of the proof.}
\\
The proof of the Theorem above is based on a multiscale analysis
of the free energy and of the generating function of the energy--energy 
correlation functions. 

The first step to set up the Renormalization Group machinery is finding
a convenient field theory which gives an equivalent description of our 
spin system. We give a fermionic representation of the theory, following
the same strategy of [PS][M].
We start from the well known 
representation of the Ising model 
free energy in terms of a sum of {\it Pfaffians} [MW]
which can be equivalently written (see Ref.
[ID][S]) 
as {\it Grassmann functional integrals}, formally describing
massive non interacting Majorana fermions 
$\psi,\lis\psi$ on a lattice with action
$$\eqalign{&\sum_\xx{t\over 4}\Big[
\psi_\xx(\partial_1-i\partial_0)\psi_\xx+
\lis\psi_\xx(\partial_1+i\partial_0)\lis\psi_\xx-2
i\lis\psi_\xx(\partial_1+\dpr_0)\psi_\xx\Big]
+i(\sqrt{2}-1-t)\lis\psi_\xx\psi_\xx\;,\cr}\Eq(1.6)$$
where $\dpr_j$ are discrete derivatives; criticality
corresponds to the massless case. If
$\l=0$ the free energy and specific
heat of the AT model
can be written as sum of Grassmann integrals describing {\it two}
kinds of Majorana fields, with masses $m^{(1)}=t^{(1)}-\sqrt2+1$
and $m^{(2)}=t^{(2)}-\sqrt2+1$.

If $\l\not=0$ again the free energy and the specific heat
can be written as Grassmann integrals, but the Majorana
fields are {\it interacting} with a short range potential.
By performing a suitable change
of variables [ID][PS][M] and integrating out the ultraviolet
degrees of freedom, the effective action 
can be written as
$$\eqalign{&
Z_1\sum_{\xx,\o,\a}\Big[
\psi^+_{\o,\xx}(\dpr_1-i\o\dpr_0)\psi^-_{\o,\xx}
-i\o\s_1\psi^+_{\o,\xx}\psi^-_{-\o,\xx}
+i\o\m_1\psi^\a_{\o,\xx}
\psi^\a_{-\o,-\xx}+
\l_1\psi^+_{1,\xx}\psi^-_{1,\xx}\psi^+_{-1,\xx}\psi^-_{-1,\xx}\Big] +
{\cal W}_1\cr}\Eq(1.7)$$
where $\a=\pm$ is a {\it creation--annihilation} index
and $\o=\pm 1$ is a {\it quasi--particle} index.
$\s_1$ and $\m_1$ have the role of two {\it masses}
and it holds
$\s_1=O(t-\sqrt2+1)+O(\l)$, $\m_1=O(u)$. ${\cal W}_1$
is a sum of monomials of $\psi$ of arbitrary
order, with kernels which
are {\it analytic functions} 
of $\l_1$; analyticity is a very nontrivial property
obtained exploiting anticommutativity
properties of Grassman variables via {\it Gram inequality}
for determinants [Le][BGPS]. The 
$\psi^\pm$ are {\it Dirac} fields,
which are
combinations of the Majorana variables $\psi^{(j)},\lis\psi^{(j)}$, $j=1,2$,
associated with the two Ising subsystems.

One can compute the partition function by expanding 
the exponential of the action 
in Taylor series in $\l$ and naively
integrating term by term the Grassmann monomials, using the Wick rule;
however such a procedure gives poor bounds for the coefficients
of this series that, in the thermodynamic
limit, can converge 
only far from the critical points.

In order to study the 
critical behaviour of the system
we perform a multiscale analysis 
involving non trivial resummations of the perturbative series.
The first step is to decompose the 
propagator $\hat g(\kk)$ as a sum of propagators more and more singular 
in the infrared 
region, labeled by an integer $h\le 1$, so that $\hat g(\kk)=
\sum_{h=-\io}^1\hat g^{(h)}(\kk)$,
$\hat g^{(h)}(\kk)\sim\g^{-h}$. 
We compute the Grassmann integrals defining the partition function
by iteratively integrating
the propagators
$\hat g^{(1)},\hat g^{(0)},\ldots$
After each integration step we rewrite the partition function in a way 
similar to the last equation,
with $Z_h,\s_h,\m_h,\l_h,{\cal W}_h$
replacing $Z_1,\s_1,\m_1,\l_1,{\cal W}_1$, in particular 
the masses $\s_h\pm\m_h$ 
and the wave function renormalization $Z_h$ are modified through 
the iterative scheme;
the structure of 
the action is preserved because of
symmetry properties; moreover
${\cal W}_h$ is shown to be a sum of monomials of $\psi$ of arbitrary
order, with kernels decaying in real space on scale $\g^{-h}$, which
are {\it analytic functions} 
of $\{\l_h,\ldots,\l_1\}$,
if $\l_k$ are small enough, $k\ge h$, and $|\s_k|\g^{-k},|\m_k|\g^{-k}\le 1$;
again analyticity follows from Gram--Hadamard type of bounds.

All the above construction is based 
on the crucial property that the effective interaction
at each scale does not increase:
$|\l_h|\le 2|\l|$. This property is highly non trivial 
and at a first naive analysis it even seems false. In fact 
the effective coupling constants $\l_h$ obey a complicated set of
recursive equations, whose right hand side is called, as usual,
the {\it Beta function}. The Beta function can be written as sum
of two terms; the first term is common to a wide class of models, including 
the {\it Luttinger model}, the {\it Thirring model}, 
the {\it Holstein--Hubbard model}
for spinless fermions, the {\it Heisenberg XYZ spin chain}, the {\it 
8 vertex model}; the other term is model dependent. 
The first term is dimensionally {\it marginal}, that is it tends
to let the effective coupling constants grow logarithmically. 
But, if one could show that it is {\it exactly vanishing}, 
than the flow of the running coupling constants in all
the above models could be controlled just by dimensional bounds, and the
expansion would be convergent; the observables would then be expressed by
explicit convergent series from which all the physical information
can be extracted.

In the years two different strategies have been followed to prove the vanishing
of the Beta function in the above sense. 
The first one, proposed by Benfatto and Gallavotti [BG1] and proved in 
[BoM][BM2], consists 
of an indirect argument, based on the fact that the first term of the Beta 
function (the one that is common to the class of models listed above)
is the same as that one obtaines from a multiscale analysis of the 
Luttinger model, that is an exactly solvable model [ML]; by contradiction,
one shows that the Beta function must be vanishing, otherwise the 
correlation functions obtained by the multiscale integration would not coincide
with the correlations which can be exactly computed from Luttinger's 
exact solution.

Very recently Benfatto and Mastropietro [BM1] proposed a new proof
of the vanishing of the Beta function, completely
independent from any exact solution and based on a rigorous implementation
of Ward identities.
Ward identities play a crucial role in Quantum Field Theory and
Statistical Mechanics, as they allow to prove cancellations in a
non perturbative way. The advantage of reducing the analysis of
Ashkin--Teller to a fermionic model like that in \equ(1.7) is that
such model can be written as the sum of a term formally verifying many 
symmetries which were not verified by AT, \eg {\it total gauge invariance}
symmetry $\psi_{\xx,\o}^\pm \to e^{\pm i\a_\xx} \psi_{\xx,\o}^\pm$
and {\it chiral gauge invariance} $\psi_{\xx,\o}^\pm\to e^{\pm
i\a_{\xx,\o}}\psi_{\xx,\o}^\pm$; plus mass terms and higher order corrections
which are weighted by small constants. The first term has an associated beta
function that is vanishing, as it can be proved through Ward identities 
following from its gauge invariance\footnote{$^6$}{\nota At the formal level
the proof of the vanishing of the Beta function through Ward identities
is well--known since the 1970's [DL][DM]. However the 
original proof of this statement discarded in the analysis the presence 
of cutoffs, which necessarily break exact gauge invariance; the 
problem of establishing whether 
gauge invariance and formal Ward identities were
recovered in the limit of cutoff removal was not considered by the 
authors of the original proof. In [BM1] the authors first considered
this problem and they proved that actually the Ward identities found after 
the removal of the cutoff are {\it different} from the formal ones: 
this is the phenomenon of chiral anomaly, well-known in the context
of similar models used in Relativistic Quantum Field Theory, \eg the Schwinger
model [ZJ].}; the second term produces 
summable corrections to the Beta functions, which are specific of 
Ashkin--Teller. One says that the symmetries used to prove the vanishing
of the Beta function are {\it hidden} in the spin models, 
as they are not verifyied even at a formal level; however they are exactly
realized by a model that is ``close'', in an RG sense, to Ashkin--Teller.
\\

So, we use the argument of [BM1] together with a detailed analysis of the 
structure of the perturbative expansion
to prove that $\l_h$ stays small under the multiscale integration.
Once this is established, 
we show that $\s_h,\m_h,Z_h$, under the iterations,
evolve as: $\s_h\simeq\s_1 \g^{b_2\l h}$, $\m_h\simeq \m_1 \g^{-b_2\l h}$, 
$Z_h\simeq\g^{-b_1\l^2h}$, with $b_1, b_2$ explicitely computable
in terms of a convergent power series.

We then perform the iterative integration described above up to a
scale $h^*_1$ such that $(|\s_{h^*_1}|+|\m_{h^*_1}|)\g^{-h^*_1}=O(1)$.
For scales lower than $h^*_1$ we return to the description in terms
of the original Majorana fermions
$\psi^{(1, \le h^*_1)}$, $\psi^{(2, \le h^*_1)}$
associated with the two Ising subsystems.
One of the two fields (say $\psi^{(1,\le h^*_1)}$) 
is massive on scale $h^*_1$
(so that the Ising subsystem with $j=1$ is ``far from criticality'' 
on the same scale); then we can integrate the massive Majorana
field $\psi^{(1,\le h^*_1)}$ without any further 
multiscale analysis, obtaining an effective theory
of a single Majorana field with mass $|\s_{h^*_1}|-|\m_{h^*_1}|$, which 
can be arbitrarly small; this is equivalent
to say that on scale $h^*_1$
we have an effective description of the system as a single 
perturbed Ising model
with {\it anomalous} parameters near criticality.
The integration of the scales $\le h^*_1$ is performed again by
a multiscale decomposition similar to the one just described; an
important feature is however that there are no more quartic
marginal terms, because the anticommutativity
of Grassmann variables forbids local quartic monomials
of a single Majorana fermion. This greatly
simplifies the analysis of the flow of the effective coupling constants,
which is convergent, as it follows just by dimensional estimates.
Criticality is found when the effective mass on scale $-\io$ is vanishing;
the values of $t,u$ for which this happens
are found by solving a non trivial implicit function problem.

Technically it is an interesting
feature of this problem that there are two regimes in which the
system must be described in terms of different fields:
a first one in which the natural variables are Dirac Grassmann variables,
and a second one in which they are Majorana; the scale $h^*_1$
separating the two
regimes is dynamically generated by the iterations. 
In the first regime
the two entangled Ising subsystems are 
undistinguishable,
the natural description is in terms of Dirac variables and
the effective interaction is marginal; in the integration
of such scales nonuniversal indexes appear and hidden Ward
identities must be used to control the flow of the effective
coupling constants. In the second region
the two Ising subsystems really look different, one appears to be 
(almost) at criticality and the other far from criticality on 
the same scale; the 
parameters of the two subsystems are deeply changed 
(in an anomalous way) by the previous integration;
in this region the effective interaction is irrelevant.\\
\\
\sub(1.7) {\bf Summary.}
\\
In Chap.2 we get the exact solution of Ising by rewriting 
the partition function as a Grassmann functional integral. This will
be the starting point for the subsequent perturbative construction.

In Chap.3 we describe the Grassmann formulation of a class 
of interacting Ising models in two dimensions, to which 
the multiscale method we will subsequently describe applies.
This class includes the Ashkin-Teller model and the 8V model (and 
models obtained as perturbations of both).
The general Grassmann formulation we describe is studied in detail
for the Ashkin--Teller model and for the latter we also give an alternative
Grassmann formulation that will be convenient in the following. 

In Chap.4 we describe how to integrate out the ultraviolet degrees of freedom
and we compute the effective action for the infrared part of the problem.
We also 
study in detail the symmetry properties of our model, and we classify 
the terms that can possibly appear in the theory by symmetry reasons.

In Chap.5 we describe the multiscale analysis in the first 
regime of scales, where the system is described in terms of Dirac fields.
In particular, this Chapter includes the 
definition of localization and a detailed analysis of the 
dimensional improvements that must be used to control 
the size of some contributions that, even if appearently marginal,
can be shown to be effectively irrelevant.  

In Chap.6 we study the flow of the running coupling constants, 
using the bounds previously derived in Chap. 5 and 
the vanishing of the Beta function. 

In Chap.7 we describe the multiscale analysis in the second regime of scales,
where the system is described in terms of a single Majorana field. 
We solve the equation for the scale $h^*_1$ dividing the first and the 
second regime and the equation for the critical temperatures. 

In Chap.8 we describe the expansion for the energy--energy correlation 
functions and we complete the proof of the main Theorem. 

In the remaining Appendices we collect a number of technical 
lemmas needed for the proof of the main Theorem. In particular 
in Appendix A6 we reproduce
the proof of the vanishing of the Beta function, 
following [BM1].\\

So, let's start.

\pagina
\setcap{2. The Ising model exact solution.}
\capindex{2}{The Ising model exact solution.}
\vskip1.truecm
\section(2,The Ising model exact solution.)
\capindex{2}{The Ising model exact solution.}

In this section we want to describe the Ising model exact solution, in a 
way that will be convenient for the subsequent perturbative analysis 
of the Ashkin--Teller model. We shall mainly follow the work of Samuel [S].

The Ising model partition function on a square lattice $\L_M\subset\ZZZ^2$, 
where $M$ is the side of the square, is defined as:
$$\Xi_I=\sum_{\s_{\L_M}}e^{\b J\sum_{<i,j>}\s_i\s_j}\;,\Eq(2.1)$$
where $<i,j>$ are nearest neighbor sites and $\b$ is the inverse 
temperature. We first consider open boundary conditions and,
after that, the more complicated case of periodic boundary conditions.
\\
\\
\sub(2.1) {\bf The multipolygon representation.}\\
It is well known that the partition function \equ(2.1) is equivalent 
to the partition function of a gas of multipolygons with hard core. This
representation was originally introduced to study the geometry of the 
microscopic configurations in the hot phase, and can be obtained 
as follows. 

One first rewrite the sum appearing at the exponent in \equ(2.1)
as $\sum_b\tilde\s_b$, where $\sum_b$ is the sum over the bonds linking 
nearest neighbor sites of $\L_M$
and $\tilde\s_b$ is the product of the spin variables over the 
two extremes of $b$.
If we expand the exponential in power series we find: 
$$\Xi_I=\sum_{\s_{\L_M}}\prod_b\left(\cosh\b J+\tilde\s_b\sinh\b J\right)=
(\cosh\b J)^{B}\sum_{\s_{\L_M}}\prod_b\left(1+\tilde\s_b\tanh\b J\right)
\Eq(2.2)$$
where $B$ is the number of bonds of $\L_M$. Developing the product, 
we are led to a sum of terms of the type:
$$(\tanh\b J)^k\tilde\s_{b_1}\cdots\tilde\s_{b_k}\Eq(2.3)$$
and we can conveniently describe them through the geometric set of lines 
$b_1,\ldots,b_k$. If we perform the summation over the configurations
$\s_{\L_M}$, many terms of the form \equ(2.3) give vanishing contribution.
The only terms which survive are those in which the vertices of the geometric
figure $b_1\cup b_2\cup\cdots\cup b_k$ belong to an even number of $b_j$'s.
These terms are those such that $\tilde\s_{b_1}\cdots\tilde\s_{b_k}\= 1$
and we shall call these geometric figures {\it multipolygons}. 
Let $P_k(\L_M)$ be the number of multipolygons with $k$ sides on the 
sublattice $\L_M$.
Then the partition function \equ(2.1) is easily rewritten as:
$$\Xi_I=(\cosh\b J)^{B} 2^{M^2} \sum_{k\ge 0}P_k(\L_M)(\tanh\b J)^{k}\;.
\Eq(2.4)$$
If open boundary conditions are assumed, only  
multipolygons {\it not} winding up the lattice are allowed. 
In the case of periodic 
boundary conditions the representation is the same, but the polygons
are allowed to wind up the lattice.\\
\\
\sub(2.2){\bf The Grassmann integration rules.}
\\
In this section we introduce some basic definitions about Grassmann 
integration. We will need them to reinterpret \equ(2.4) as a Grassmann
functional integral.

Let us consider a finite dimensional {\it Grassman algebra},
which is a set of anticommuting {\it Grassman variables}
$\{\psi^+_\a,\psi^-_\a\}$, with
$\a$ an index belonging to some finite set $A$. This means that
$$ \{ \psi_{\a}^{\s},\psi_{\a'}^{\s'}\} \=
\psi_{\a}^{\s},\psi_{\a'}^{\s'} +
\psi_{\a}^{\s},\psi_{\a'}^{\s'} = 0 \; , \qquad
\forall \a,\a'\in A \;,\quad \forall \s,\s'=\pm \; ; \Eq(2.5) $$
in particular $(\psi^{\s}_{\a})^2=0$
$\forall \a\in A$ and $\forall \s=\pm$.

Let us introduce another set of Grassman variables
$\{\der\psi^+_\a,\der\psi^-_\a\}$, $\a\in A$, anticommuting
with $\psi^+_\a,\psi^-_\a$, and an operation
({\it Grassman integration}) defined by
$$ \int \psi_\a^\s \der\psi_\a^\s=1 \; , \qquad
\int \der\psi_\a^\s =0 \; , \qquad
a\in A \; , \quad \s=\pm 1 \; . \Eq(2.6) $$
If $F(\psi)$ is a polynomial in
$\psi^+_{\a},\psi^-_{\a}$, $\a\in A$, the operation
$$\int \prod_{\a\in A} \der\psi^+_\a \der\psi^-_\a F(\psi) \Eq(2.7) $$
is simply defined by iteratively applying \equ(2.6)
and taking into account the anticommutation rules \equ(2.5).
It is easy to check that for all $\a\in A$ and $C\in \CCC$
$$ {\int \der \psi^+_{\a} d\psi^-_{\a} e^{-\psi_{\a}^+
C \psi^-_{\a}}\psi_{\a}^- \psi^+_{\a}
\over
\int \der \psi^+_{\a} d\psi^-_{\a} e^{-\psi_{\a}^+ C \psi^-_{\a}}}
= C^{-1} \; ; \Eq(2.8) $$
in fact $e^{-\psi_{\a}^+ C \psi^-_{\a}}=
1-\psi_{\a}^+ C \psi^-_{\a}$ and by \equ(2.6)
$$ \int \der \psi^+_{\a} \der\psi^-_{\a}
e^{-\psi_{\a}^+ C \psi^-_{\a}}= C \; , \Eq(2.9) $$
while 
$$\int \der\psi^+_{\a} d\psi^-_{\a} e^{-\psi_{\a}^+
C \psi^-_{\a}} \psi_{\a}^-\psi^+_{\a} = 1 \; . \Eq(2.10) $$
If one considers Grassmann variables whose quadratic action is not diagonal, 
one finds the generalizations of the above formulas, \eg
$$ { \int \prod_{\a\in A}\der\psi^+_{\a} \der\psi^-_{\a}
\,
e^{-\sum_{i,j\in A} \psi_i^+ M_{ij} \psi^-_j} \psi_{\a'}^- \psi^+_{\b'}
\over
\int \prod_{\a\in A}\der\psi^+_{\a} \der\psi^-_{\a}\, 
e^{-\sum_{i,j\in A} \psi_i^+ M_{ij} \psi^-_j}}
= [M^{-1}]_{\a'\b'} \; , \Eq(2.11) $$
with $M$ an $|A|\times |A|$ complex matrix.
Again \equ(2.11) can be easily verified
by using \equ(2.6) and the anticommutation rules \equ(2.5),
which also allow us to write
$$ \int \prod_{\a\in A}\der\psi^+_{\a} \der\psi^-_{\a} 
\, e^{-\sum_{ij\in A} \psi_i^+ M_{ij} \psi^-_j} \= \det M \Eq(2.12) $$
and
$$ \int \prod_{\a\in A}\der\psi^+_{\a} \der\psi^-_{\a}
\,
e^{-\sum_{ij\in A} \psi_i^+ M_{ij} \psi^-_j}
\psi_{\a'}^- \psi^+_{\b'}= M'_{\a'\b'} \; , \Eq(2.13) $$
if $M'_{\a'\b'}$ is the minor complementary to the entry
$M_{\a'\b'}$.

The above formulae closely remind us the Gaussian integrals:
note however that there is no need that $M$ is real or positive defined
(but of course they have to be invertible).

For the moment this is all we need for the Grassmann formulation of the Ising
model. More algebraic properties of the Grassmann integration can be found 
in Appendix A1.
\\
\\
\sub(2.3){\bf The Grassmann representation of the 2d Ising model
with open boundary conditions.}
\\
In order to represent the sum over multipolygons in \equ(2.4) as a Grassmann 
integral, we first associate to each site $\xx\in\L_M$, a set of four 
Grassmann variables, $\lis H_\xx,H_\xx,\lis V_\xx,V_\xx$, that
must be thought as associated to four new sites drawn very near to $\xx$ 
and to its right, left, up side, down side respectively, see Fig 3. We shall
denote these sites by $R_\xx, L_\xx, U_\xx, D_\xx$ respectively.


\midinsert
\*
\insertplotttt{170pt}{120pt}{
\ins{47pt}{57pt}{$\lis H_\xx$} 
\ins{2pt}{55pt}{$H_\xx$} 
\ins{27pt}{76pt}{$\lis V_\xx$} 
\ins{27pt}{36pt}{$V_\xx$} 
\ins{197pt}{57pt}{$\lis H_\yy$} 
\ins{152pt}{55pt}{$H_\yy$} 
\ins{177pt}{76pt}{$\lis V_\yy$} 
\ins{177pt}{36pt}{$V_\yy$} 
}{fig3}
\vskip.3truecm
\line{\vtop{\line{\hskip1.3truecm\vbox{\advance\hsize by -2.0 truecm
\0{\css Fig. 3.}
{\ottorm  The four Grassmann fields associated to the sites
$\xx$ and $\yy$.
} \hfill} }}}
\*
\endinsert

If $t\defin\tanh \b J$, we consider the action 
$$\eqalign{S(t)&=
t\sum_{\xx\in\L_M} \left[\lis H_{\xx} H_{\xx+\hat e_1}+
\lis V_{\xx} V_{\xx+\hat e_0}\right]
+\sum_{\xx\in\L_M}\left[\lis H_{\xx} H_{\xx}+
\lis V_{\xx}V_{\xx}+\lis V_{\xx} \lis H_{\xx}+
V_{\xx} \lis H_{\xx}+H_{\xx} \lis V_{\xx}+
V_{\xx} H_{\xx}\right]\;,\cr}\Eq(2.14)$$
where $\hat e_1, \hat e_0$ are the coordinate versors in the 
horizontal and vertical directions, respectively. Open boundary
conditions are assumed.
We claim that the following identity holds:
$${\Xi_I\over 2^{M^2}(\cosh\b J)^B}=(-1)^{M^2}
\int\prod_{\xx\in\L_M}\der\lis H_\xx
\der H_\xx\der\lis V_\xx\der V_\xx e^{S(t)}\Eq(2.15)$$
where $\Xi_I$ in the l.h.s. is calculated using open boundary conditions.
The proof of \equ(2.15) will occupy the rest of this section.

In order to prove \equ(2.15) we expand the exponential in the r.h.s., we
integrate term by term the Grassmann variables, 
and we get a summation over terms that we want to put in correspondence
with the terms in the summation over mutipolygons of \equ(2.4). We can do
as follows. We represent every quadratic term in \equ(2.14) with a line
connecting the two sites corresponding to the two Grassmann fields. 
Correspondingly, we represent every term obtained by the contraction 
of the Grassmann variables (that is the contraction of a 
suitable product of the quadratic terms appearing in $S(t)$) with the union 
of the lines representing the contracted monomials. The figure one 
obtaines (call it a dimer) 
resembles a multipolygon, and exactly coincide with 
a multipolygon if one shrinks the sites $R_\xx, L_\xx, U_\xx, D_\xx$ 
to let them coincide with $\xx$. 

This graphical construction allows to put in correspondence each 
dimer with a unique multipolygon. 
We then have to show that the total weight 
of the dimer corresponding to the same multipolygon
$\g$ is exactly $(-1)^{M^2} t^{|\g|}$, where $(-1)^{M^2}$ is the 
same factor appearing in the r.h.s. of \equ(2.15) (note that $M^2$
is the number of sites of $\L_M$) and, if $|\g|$ is the length of 
$\g$, $t^{|\g|}$ is the weight \equ(2.4) assigns to $\g$.

We first note that the correspondence between dimers
and multipolygons is not one to one, because
an empty site $\xx$ in the multipolygon representation corresponds 
to three different contractions of Grassmann fields, that is
either to $\int \der\lis H_\xx
\der H_\xx\der\lis V_\xx\der V_\xx\ \lis H_\xx H_\xx \lis V_\xx V_\xx$, 
or to $\int \der\lis H_\xx
\der H_\xx\der\lis V_\xx\der V_\xx\ V_\xx \lis H_\xx H_\xx \lis V_\xx $, 
or to $\int \der\lis H_\xx
\der H_\xx\der\lis V_\xx\der V_\xx\  V_\xx H_\xx \lis V_\xx \lis H_\xx$.
The total contribution of these three contractions is:
$$\int \der\lis H_\xx
\der H_\xx\der\lis V_\xx\der V_\xx\ (\lis H_\xx H_\xx \lis V_\xx V_\xx
+V_\xx \lis H_\xx H_\xx \lis V_\xx+V_\xx H_\xx \lis V_\xx \lis H_\xx)=
1-1-1=-1\;,\Eq(2.16)$$
as wanted.

It is easy to realize that, unless for the above ambiguity, the 
correspondence between dimers and multipolygons is unique. And, 
since each side of a dimer
is weighted by a factor $t$ and each empty site is weighted by $(-1)$,
the weights of the corresponding figures are the same, at least 
{\it in absolute value}. From now on we shall extract from the weight
of $\g$ the contribution of the empty sites together 
with the trivial factor $t^{|\g|}$ (that is we redefine
the weight of $\g$ by dividing it by $(-1)^{M^2-n_\g}t^{|\g|}$, where $n_\g$
is the number of sites belonging to $\g$, possibly different 
from $|\g|$, if $\g$ has self intersections). 

We are then left with proving that the weight of a dimer
(as just redefined) is exactly $(-1)^{n_\g}$; in this way the sign of every
configuration of 
dimers together with the minus signs of the empty sites,
\equ(2.16), would reproduce exactly the factor $(-1)^{M^2}$ in \equ(2.15).

We start with considering the simplest dimer,
that is the square with unit side. 
Let us denote its corner sites with
$(0,0)\=\xx_1$, $(1,0)\=\xx_2$, $(1,1)\=\xx_3$, 
$(0,1)\=\xx_4$ and let us prove that its weight is
$(-1)^{4}=1$. 
The explicit expression of its weight in terms 
of Grassmann integrals, as generated by the expansion of the exponent in 
\equ(2.15) is:
$$\eqalign{&\int \prod_{i=1}^4 \der\lis H_{\xx_i}
\der H_{\xx_i}\der\lis V_{\xx_i}\der V_{\xx_i} \cdot\cr
&\Big[
\lis H_{\xx_1} H_{\xx_2}\cdot 
V_{\xx_2} \lis H_{\xx_2}\cdot\lis V_{\xx_2}V_{\xx_3}\cdot
\lis V_{\xx_3}\lis H_{\xx_3}\cdot
(-H_{\xx_3}\lis H_{\xx_4}\big)\cdot H_{\xx_4}\lis V_{\xx_4}
\cdot(-V_{\xx_4}\lis V_{\xx_1}\big)\cdot V_{\xx_1}H_{\xx_1}
\Big]\cr}\Eq(2.17)$$
In the previous equation, we wrote the different binomials 
corresponding to the segments of the dimer
following the anticlockwise order, starting from 
$\lis H_{\xx_1}$.
We associated a sign to each binomial, $+$ if its fields are written
in the same order as they appear in \equ(2.14), and $-$
otherwise.

By collecting the minus signs and by permutating the position of 
$\lis H_{\xx_1}$ from the first to the last position, 
we find that \equ(2.17) is equal to 
$$\eqalign{&-\int \prod_{i=1}^4 \der\lis H_{\xx_i}
\der H_{\xx_i}\der\lis V_{\xx_i}\der V_{\xx_i} \cdot 
\Big[
H_{\xx_2}
V_{\xx_2} \lis H_{\xx_2}\lis V_{\xx_2}\cdot V_{\xx_3}
\lis V_{\xx_3}\lis H_{\xx_3}
H_{\xx_3}\cdot \lis H_{\xx_4}H_{\xx_4}\lis V_{\xx_4}
V_{\xx_4}\cdot \lis V_{\xx_1}V_{\xx_1}H_{\xx_1}\lis H_{\xx_1} 
\Big]\cr}\Eq(2.18)$$
where now we wrote separated from a dot the contributions corresponding 
to the same site. The explicit computation of \equ(2.18) gives 
$-[(-1)(-1)(+1)(-1)]=+1$, as desired.

Let us now consider a generic dimer $\g$ 
not winding up the lattice and without self intersections, 
and let us prove by induction that its weight is $(-1)^{n_\g}$.
We will then assume that the dimers with number of sites 
$k\le n_\g$
have weights $(-1)^{k}$. The first step from which the induction starts 
is the case $k=4$, that we have just considered. 

Let us consider the smallest rectangle $R$ containing $\g$.
Necessarely, each side of $R$ has non empty intersection with $\g$.
Let us enumerate the corners of $\g$ which are also extremes of straight
segments belonging to the sides of $R$, starting from the 
leftmost among the lowest of these points (possibly coinciding
with the lower left corner of $R$) and proceeding in anticlockwise 
order; call $\xx_j$ the site with label $j$. 
Note that two consecutive indeces $j,j+1$ could represent the same
site $\xx_j\=\xx_{j+1}\in\L_M$; in that case $\xx$ would be a corner of $R$. 
Call $2N$ the cardinality of the set
of the enumerated points (it is even by construction)
and let us identify the label
$2N+1$ with the label $1$. 

Let us denote with the symbol $(2j-1\to 2j)$, $j=1,\ldots,N$, the product of
Grassmann fields corresponding to the straight line connecting the point 
$2j-1$
with $2j$ (not including the fields located in $2j-1$ and in $2j$),
written in the anticlockwise order and with the sign induced by the 
expansion of the exponential in \equ(2.15). That is, if the 
two fields belonging to a binomial appearing in \equ(2.14), written 
following the anticlockwise order, are in the same order as they appear 
in \equ(2.14), we will assign a $+$ sign to the second of those two fields
(of course, second w.r.t. the anticlockwise order);
otherwise a $-$ sign. As an example, 
if $2j-1$ and $2j$ are two points on the upper horizontal side of $R$,
$(2j-1\to 2j)$ would be equal to 
$$(-\lis H_{\xx_{2j-1}-\hat e_1})\lis V_{\xx_{2j-1}-\hat e_1} V_{\xx_{2j-1}-\hat e_1}
H_{\xx_{2j-1}-\hat e_1}\cdots \cdots
(-\lis H_{\xx_{2j}+\hat e_1})\lis V_{\xx_{2j}+\hat e_1} V_{\xx_{2j}+\hat e_1}
H_{\xx_{2j}+\hat e_1}\Eq(2.18a)$$
With a small abuse of notation, in the following  
we shall also denote with the symbol
$(2j-1\to 2j)$ the straight line connecting $2j-1$ with $2j$ on the polygon
(\ie the geometric object, not only the algebraic one).   

Moreover, let us denote with the symbol $[2j\to 2j+1]$, $j=1,\ldots,N$, 
the product of
Grassmann fields corresponding to the {\it non} straight line connecting 
the point $2j$ with $2j+1$ (including the fields located in $2j$ and in $2j+1$)
in the order induced by the choice of proceeding in anticlockwise order
and with the sign induced by the expansion of the exponential in \equ(2.15).
With a small abuse of notation we shall also denote with the same symbol 
$[2j\to 2j+1]$
the corresponding line connecting $2j$ with $2j+1$ on the polygon $\g$. 
The sites $2j$ and $2j+1$ could either coincide
(in that case $2j$ is a corner of $R$) or,
if they do not, they could
belong to the same side of $R$ or to different
adjacent sides of $R$. Let us denote with $\g_j$ the 
union of $[2j\to 2j+1]$ with the shortest path on $R$ connecting 
$2j$ with $2j+1$. The key remark is that $n_{\g_j}<n_\g$ so that, 
by the inductive hypothesis, the weight of $\g_j$ is $(-1)^{n_{\g_j}}$.

With these notations and remarks, let us calculate the weight
of $\g$. We write the weight in terms
of a Grassmann integral as follows:
$$-\int \prod_{\xx\in\g}\der\lis H_\xx\der H_\xx\der\lis V_\xx\der 
V_\xx\quad (1\to 2)[2\to 3]\cdots (2N\!\!-\!\!1\to 2N)[2N\to 1]\Eq(2.18z)$$
The minus sign in front of the integral, appearing for the same reason
why it appears in \equ(2.18), is due to the permutation of 
the field $\lis H_{\xx_1}$ from the first position 
(that is the one one gets by expanding the exponential in \equ(2.15),
writing the Grassmann binomials starting from site 1 and proceeding 
in anticlockwise order) to the last one (that is the position it appears 
into the product $[2N\to 1]$). 

By a simple explicit calculation, 
it is straightforward to verify that the integral of the ``straight line''
$(2j-1\to 2j)$ gives a contribution $(-1)^{\ell_{2j-1}-1}$, where $\ell_{2j-1}$
is the length of the segment $(2j-1\to 2j)$ (note that $\ell_{2j-1}-1$ is the number
of sites belonging to $(2j-1\to 2j)$, excluding the extremes).
We are left with computing the integral of the ``non straight line''
$[2j\to 2j+1]$. 
We must distinguish 12 different cases, which we shall now study in detail.
\\
\\
\01) $j$ and $j+1$ are distinct and they belong to
the low side of $R$. In this case 
$$[j\to j+1]=\int H_{\xx_j}\cdot
V_{\xx_j}\lis H_{\xx_j}\cdot\big\{\lis V_{\xx_j}\cdots
(-\lis V_{\xx_{j+1}})\big\}\cdot
V_{\xx_{j+1}}H_{\xx_{j+1}}\cdot\lis H_{\xx_{j+1}}\;,\Eq(2.19)$$
as it follows from the rules explained above. We did not explicitely
write neither the integration elements (those appearing in 
the r.h.s. of \equ(2.15)) nor the fields corresponding to the sites 
between the site $\xx_j$ and the site $\xx_{j+1}$; note however that
the number of fields between braces is necessarely even.
In order
to compute \equ(2.19) we use the inductive hypothesis, telling
us that the weight of $\g_j$ is $(-1)^{n_{\g_j}}$, that is, explicitely:
$$(-1)^{D_j+d_j}=\int V_{\xx_j}H_{\xx_j}\cdot
\lis H_{\xx_j}(j\to j+1)H_{\xx_{j+1}}\cdot V_{\xx_{j+1}}\lis H_{\xx_{j+1}}
\cdot\big\{\lis V_{\xx_j}\cdots
(-\lis V_{\xx_{j+1}})\big\}$$
In the last equation we called $D_j$
the length of the non straight line $[j\to j+1]$ (note that $D_j+1$ 
is the number
of sites belonging to $[j\to j+1]$, including both extremes),
we denoted by the symbol 
$(j\to j+1)$ the product of
Grassmanian fields corresponding to the straight line on $R$ connecting
$\xx_j$ with $\xx_{j+1}$ and by $d_j$ its length (note that $d_j-1$ 
is the number of sites belonging to $(j\to j+1)$, excluding both extremes). 
By performing 
the integration over the fields in $(j\to j+1)$, we find:
$$ \eqalign{(-1)^{D_j+1}&=\int V_{\xx_j}H_{\xx_j}
\lis H_{\xx_j}H_{\xx_{j+1}}V_{\xx_{j+1}}\lis H_{\xx_{j+1}}
\big\{\lis V_{\xx_j}\cdots
(-\lis V_{\xx_{j+1}})\big\}=\cr
&=\int V_{\xx_j}H_{\xx_j}
\lis H_{\xx_j}
\big\{\lis V_{\xx_j}\cdots
(-\lis V_{\xx_{j+1}})\big\}H_{\xx_{j+1}}V_{\xx_{j+1}}\lis H_{\xx_{j+1}}\cr}$$
and the last line is clearly equal to the r.h.s. of \equ(2.19).\\
\\
\02) $j$ and $j+1$ coincide with 
the low right corner of $R$. In this case 
$$[j\to j+1]=\int H_{\xx_j}\cdot
V_{\xx_j}\lis H_{\xx_j}\cdot\lis V_{\xx_j}=-1\,.\Eq(2.20)$$
\03) $j$ and $j+1$ are distinct and they belong to
the low and the rights sides of $R$, respectively. In this case 
$$[j\to j+1]=\int H_{\xx_j}\cdot
V_{\xx_j}\lis H_{\xx_j}\cdot\big\{\lis V_{\xx_j}\cdots
H_{\xx_{j+1}}\big\}\cdot
V_{\xx_{j+1}}\lis H_{\xx_{j+1}}\cdot\lis V_{\xx_{j+1}}\;.\Eq(2.21)$$
Calling $\V0$ the lower right corner of $R$,
the inductive hypothesis tells us that:
$$(-1)^{D_j+d_j}=\int V_{\xx_j}H_{\xx_j}\cdot
\lis H_{\xx_j}(j\to \V0)H_\V0\cdot V_\V0\lis H_\V0\cdot \lis V_\V0
(\V0\to j+1)V_{\xx_{j+1}}\cdot\lis V_{\xx_{j+1}}\lis H_{\xx_{j+1}}
\cdot\big\{\lis V_{\xx_j}\cdots
H_{\xx_{j+1}}\big\}\;.$$
In the last equation we called $d_j$ the length of the shortest path 
on $R$ connecting $j$ with $j+1$ that is the sum of the lengths of 
$(j\to \V0)$ and $(\V0\to j+1)$.
By performing 
the integration over the fields in $(j\to \V0)$, in $\V0$ and in 
$(\V0\to j+1)$ we find:
$$ \eqalign{(-1)^{D_j+1}&=\int V_{\xx_j}H_{\xx_j}
\lis H_{\xx_j}V_{\xx_{j+1}}\lis V_{\xx_{j+1}}\lis H_{\xx_{j+1}}
\big\{\lis V_{\xx_j}\cdots
H_{\xx_{j+1}}\big\}=\cr
&=\int V_{\xx_j}H_{\xx_j}
\lis H_{\xx_j}
\big\{\lis V_{\xx_j}\cdots
H_{\xx_{j+1}}\big\}V_{\xx_{j+1}}\lis V_{\xx_{j+1}}\lis H_{\xx_{j+1}}\cr}$$
and the last line is clearly equal to the r.h.s. of \equ(2.21).
\\
\\
\04) $j$ and $j+1$ are distinct and they belong to
the right side of $R$. In this case 
$$[j\to j+1]=\int V_{\xx_j}\cdot
\lis V_{\xx_j}\lis H_{\xx_j}\cdot\big\{H_{\xx_j}\cdots
H_{\xx_{j+1}}\big\}\cdot
V_{\xx_{j+1}}\lis H_{\xx_{j+1}}\cdot\lis V_{\xx_{j+1}}\;.\Eq(2.22)$$
The inductive hypothesis tells us that:
$$(-1)^{D_j+d_j}=\int V_{\xx_j}\lis H_{\xx_j}\cdot
\lis V_{\xx_j}(j\to j+1)V_{\xx_{j+1}}\cdot \lis V_{\xx_{j+1}}
\lis H_{\xx_{j+1}}\cdot\big\{H_{\xx_j}\cdots
H_{\xx_{j+1}}\big\}\;.$$
By performing 
the integration over the fields in $(j\to j+1)$ we find:
$$ \eqalign{(-1)^{D_j+1}&=\int V_{\xx_j}\lis H_{\xx_j}
\lis V_{\xx_j}V_{\xx_{j+1}}\lis V_{\xx_{j+1}}\lis H_{\xx_{j+1}}
\big\{H_{\xx_j}\cdots
H_{\xx_{j+1}}\big\}=\cr
&=\int V_{\xx_j}\lis H_{\xx_j}
\lis V_{\xx_j}
\big\{H_{\xx_j}\cdots
H_{\xx_{j+1}}\big\}V_{\xx_{j+1}}\lis V_{\xx_{j+1}}\lis H_{\xx_{j+1}}\cr}$$
and the last line is clearly equal to the r.h.s. of \equ(2.22).
\\
\\
\05) $j$ and $j+1$ coincide with 
the upper right corner of $R$. In this case 
$$[j\to j+1]=\int V_{\xx_j}\cdot
\lis V_{\xx_j}\lis H_{\xx_j}\cdot H_{\xx_j}=-1\,.\Eq(2.23)$$
\06) $j$ and $j+1$ are distinct and they belong to
the right and upper sides of $R$, respectively. In this case 
$$[j\to j+1]=\int V_{\xx_j}\cdot
\lis V_{\xx_j}\lis H_{\xx_j}\cdot\big\{H_{\xx_j}\cdots
V_{\xx_{j+1}}\big\}\cdot
\lis V_{\xx_{j+1}}\lis H_{\xx_{j+1}}\cdot H_{\xx_{j+1}}\;.\Eq(2.24)$$
Calling $\V0$ the upper right corner of $R$,
the inductive hypothesis tells us that:
$$(-1)^{D_j+d_j}=\int V_{\xx_j}\lis H_{\xx_j}\cdot
\lis V_{\xx_j}(j\to \V0)V_\V0\cdot \lis V_\V0\lis H_\V0\cdot H_\V0
(\V0\to j+1)(-\lis H_{\xx_{j+1}})\cdot H_{\xx_{j+1}}\lis V_{\xx_{j+1}}
\cdot\big\{H_{\xx_j}\cdots
V_{\xx_{j+1}}\big\}\;.$$
By performing 
the integration over the fields in $(j\to \V0)$, in $\V0$ and in 
$(\V0\to j+1)$ we find:
$$ \eqalign{(-1)^{D_j+1}&=\int V_{\xx_j}\lis H_{\xx_j}
\lis V_{\xx_j}(-\lis H_{\xx_{j+1}})H_{\xx_{j+1}}\lis V_{\xx_{j+1}}
\big\{H_{\xx_j}\cdots
V_{\xx_{j+1}}\big\}=\cr
&=\int V_{\xx_j}\lis H_{\xx_j}
\lis V_{\xx_j}
\big\{H_{\xx_j}\cdots
V_{\xx_{j+1}}\big\}(-\lis H_{\xx_{j+1}})H_{\xx_{j+1}}\lis V_{\xx_{j+1}}\cr}$$
and the last line is clearly equal to the r.h.s. of \equ(2.24).
\\
\\
\07) $j$ and $j+1$ are distinct and they belong to
the upper side of $R$. In this case 
$$[j\to j+1]=\int (-\lis H_{\xx_j})\cdot
H_{\xx_j}\lis V_{\xx_j}\cdot\big\{V_{\xx_j}\cdots
V_{\xx_{j+1}}\big\}\cdot
\lis V_{\xx_{j+1}}\lis H_{\xx_{j+1}}\cdot H_{\xx_{j+1}}\;.\Eq(2.25)$$
The inductive hypothesis tells us that:
$$(-1)^{D_j+d_j}=\int \lis V_{\xx_j}\lis H_{\xx_j}\cdot
H_{\xx_j}(j\to j+1)(-\lis H_{\xx_{j+1}})\cdot H_{\xx_{j+1}}
\lis V_{\xx_{j+1}}\cdot\big\{V_{\xx_j}\cdots
V_{\xx_{j+1}}\big\}\;.$$
By performing 
the integration over the fields in $(j\to j+1)$ we find:
$$ \eqalign{(-1)^{D_j+1}&=\int \lis V_{\xx_j}\lis H_{\xx_j}
H_{\xx_j}(-\lis H_{\xx_{j+1}})H_{\xx_{j+1}}\lis V_{\xx_{j+1}}
\big\{V_{\xx_j}\cdots
V_{\xx_{j+1}}\big\}=\cr
&=\int \lis V_{\xx_j}\lis H_{\xx_j}
H_{\xx_j}
\big\{V_{\xx_j}\cdots
V_{\xx_{j+1}}\big\}(-\lis H_{\xx_{j+1}})H_{\xx_{j+1}}\lis V_{\xx_{j+1}}\cr}$$
and the last line is clearly equal to the r.h.s. of \equ(2.25).
\\
\\
\08) $j$ and $j+1$ coincide with 
the upper left corner of $R$. In this case 
$$[j\to j+1]=\int (-\lis H_{\xx_j})\cdot
H_{\xx_j}\lis V_{\xx_j}\cdot V_{\xx_j}=-1\,.\Eq(2.26)$$
\09) $j$ and $j+1$ are distinct and they belong to
the upper and left sides of $R$, respectively. In this case 
$$[j\to j+1]=\int (-\lis H_{\xx_j})\cdot
H_{\xx_j}\lis V_{\xx_j}\cdot\big\{V_{\xx_j}\cdots
(-\lis H_{\xx_{j+1}})\big\}\cdot
H_{\xx_{j+1}}\lis V_{\xx_{j+1}}\cdot V_{\xx_{j+1}}\;.\Eq(2.27)$$
Calling $\V0$ the upper left corner of $R$,
the inductive hypothesis tells us that:
$$(-1)^{D_j+d_j}=\int \lis V_{\xx_j}\lis H_{\xx_j}\cdot
H_{\xx_j}(j\to \V0)(-\lis H_\V0)\cdot H_\V0\lis V_\V0\cdot V_\V0
(\V0\to j+1)(-\lis V_{\xx_{j+1}})\cdot V_{\xx_{j+1}}H_{\xx_{j+1}}
\cdot\big\{V_{\xx_j}\cdots
(-\lis H_{\xx_{j+1}})\big\}\;.$$
By performing 
the integration over the fields in $(j\to \V0)$, in $\V0$ and in 
$(\V0\to j+1)$ we find:
$$ \eqalign{(-1)^{D_j+1}&=\int \lis V_{\xx_j}\lis H_{\xx_j}
H_{\xx_j}(-\lis V_{\xx_{j+1}})V_{\xx_{j+1}}H_{\xx_{j+1}}
\big\{V_{\xx_j}\cdots
(-\lis H_{\xx_{j+1}})\big\}=\cr
&=\int \lis V_{\xx_j}\lis H_{\xx_j}H_{\xx_j}
\big\{V_{\xx_j}\cdots
(-\lis H_{\xx_{j+1}})\big\}
(-\lis V_{\xx_{j+1}})V_{\xx_{j+1}}H_{\xx_{j+1}}\cr}$$
and the last line is clearly equal to the r.h.s. of \equ(2.27).
\\
\\
\010) $j$ and $j+1$ are distinct and they belong to
the left side of $R$. In this case 
$$[j\to j+1]=\int (-\lis V_{\xx_j})\cdot
V_{\xx_j}H_{\xx_j}\cdot\big\{\lis H_{\xx_j}\cdots
(-\lis H_{\xx_{j+1}})\big\}\cdot
H_{\xx_{j+1}}\lis V_{\xx_{j+1}}\cdot V_{\xx_{j+1}}\;.\Eq(2.28)$$
The inductive hypothesis tells us that:
$$(-1)^{D_j+d_j}=\int H_{\xx_j}\lis V_{\xx_j}\cdot
V_{\xx_j}(j\to j+1)(-\lis V_{\xx_{j+1}})\cdot V_{\xx_{j+1}}
H_{\xx_{j+1}}\cdot\big\{\lis H_{\xx_j}\cdots
(-\lis H_{\xx_{j+1}})\big\}\;.$$
By performing 
the integration over the fields in $(j\to j+1)$ we find:
$$ \eqalign{(-1)^{D_j+1}&=\int H_{\xx_j}\lis V_{\xx_j}
V_{\xx_j}(-\lis V_{\xx_{j+1}})V_{\xx_{j+1}}H_{\xx_{j+1}}
\big\{\lis H_{\xx_j}\cdots
(-\lis H_{\xx_{j+1}})\big\}=\cr
&=\int H_{\xx_j}\lis V_{\xx_j}
V_{\xx_j}\big\{\lis H_{\xx_j}\cdots
(-\lis H_{\xx_{j+1}})\big\}
(-\lis V_{\xx_{j+1}})V_{\xx_{j+1}}H_{\xx_{j+1}}\cr}$$
and the last line is clearly equal to the r.h.s. of \equ(2.28).
\\
\\
\011) $j$ and $j+1$ coincide with 
the lower left corner of $R$. In this case it is necessarely
$j\=2N$ and $j+1\=1$ and we have:
$$[2N\to 1]=\int (-\lis V_{\xx_1})\cdot
V_{\xx_1}H_{\xx_1}\cdot \lis H_{\xx_1}=+1\,.\Eq(2.29)$$
Note that this time the result is $+1$. This ``wrong'' sign 
exactly compensates the minus sign appearing in the 
r.h.s. of \equ(2.18z).
\\
\\
\012) $j$ and $j+1$ are distinct and they belong to
the left and lower sides of $R$, respectively. In this case 
it is necessarely $j\=2N$ and $j+1\=1$ and we have
$$[2N\to 1]=\int (-\lis V_{\xx_{2N}})\cdot
V_{\xx_{2N}}H_{\xx_{2N}}\cdot\big\{\lis H_{\xx_{2N}}\cdots
(-\lis V_{\xx_{1}})\big\}\cdot
V_{\xx_{1}}H_{\xx_{1}}\cdot \lis H_{\xx_{1}}\;.\Eq(2.30)$$
Calling $\V0$ the lower left corner of $R$,
the inductive hypothesis tells us that:
$$(-1)^{D_{N}+d_N}=\int H_{\xx_{2N}}\lis V_{\xx_{2N}}\cdot
V_{\xx_{2N}}({2N}\to \V0)(-\lis V_\V0)\cdot V_\V0 H_\V0\cdot \lis H_\V0
(\V0\to 1)H_{\xx_{1}}\cdot V_{\xx_{1}}\lis H_{\xx_{1}}
\cdot\big\{\lis H_{\xx_{2N}}\cdots
(-\lis V_{\xx_{1}})\big\}\;.$$
By performing 
the integration over the fields in $({2N}\to \V0)$, in $\V0$ and in 
$(\V0\to 1)$ we find:
$$ \eqalign{(-1)^{D_N}&=\int H_{\xx_{2N}}\lis V_{\xx_{2N}}
V_{\xx_{2N}}H_{\xx_{1}}V_{\xx_{1}}\lis H_{\xx_1}
\big\{\lis H_{\xx_{2N}}\cdots
(-\lis V_{\xx_{1}})\big\}=\cr
&=\int H_{\xx_{2N}}\lis V_{\xx_{2N}}
V_{\xx_{2N}}
\big\{\lis H_{\xx_{2N}}\cdots
(-\lis V_{\xx_{1}})\big\}H_{\xx_{1}}V_{\xx_{1}}\lis H_{\xx_1}\cr}$$
and the last line is clearly equal to the r.h.s. of \equ(2.27). It follows
that $[{2N}\to 1]=-(-1)^{D_N+1}$, consistently with the result in item (11) above.
Also in this case, the appearently ``wrong'' sign 
exactly compensates the minus sign appearing in the 
r.h.s. of \equ(2.18z).
\\
\\
Combining the results of previous items, we can simply say that
the integration of $(2j-1\to 2j)$ contributes 
to the weight of $\g$
with $(-1)^{\ell_{2j-1}-1}$; the integration of
$[2j\to 2j+1]$, with $j<N$, contributes 
with $(-1)^{L_{2j}+1}$ (here we defined $L_{2j}$
to be the length of $[2j\to 2j+1]$), while $[2N\to 1]$ with $(-1)^{L_{2N}}$. Substituting
these results into \equ(2.18z), we find that the weight of $\g$ 
is equal to $(-1)^{n_\g}$, as desired. \\

The above discussion concludes the proof in the case of polygons
without self intersections. Let us call
{\it simple} a polygon without self intersections.
If $\g$ is not simple, calling
$\n_\g$ the number of its self intersections, 
we can easily
prove that its weight is equal to $(-1)^{\n_\g}$
times the product of the weights of a number of simple polygons,
defined as follows. We draw with two colors, white and black, both the 
disconnetted interiors of the polygon and its exterior, call them 
$A_1,\ldots, A_n$ and $A_0$ respectively. The drawing is done in such a 
way that $A_0$ is white and two adjacent sets $A_i$ and $A_j$, $0\le i<j\le n$, 
have different colors 
(we call $A_i$ and $A_j$ adjacent if their boundaries have a 
common side). Then we consider
the set $\PP$ of simple polygons obtained as the boundaries of the black
sets, thought as completely disconnetted one from the other. 
The ``disconnection'' of the boundaries of the black regions (which 
originally could touch each other through the corners)
is realized by the elementary disconnetion of 
the intersection elements described in Fig.4.


\midinsert
\*
\insertplotttt{170pt}{260pt}{
\ins{-100pt}{185pt}{$a)$}
\ins{-50pt}{210pt}{$W$} 
\ins{0pt}{210pt}{$B$} 
\ins{0pt}{160pt}{$W$} 
\ins{-50pt}{160pt}{$B$} 
\ins{150pt}{210pt}{$W$} 
\ins{215pt}{225pt}{$B$} 
\ins{200pt}{160pt}{$W$} 
\ins{135pt}{145pt}{$B$}
\ins{-60pt}{185pt}{$1$} 
\ins{15pt}{185pt}{$3$}
\ins{-23pt}{225pt}{$4$} 
\ins{-23pt}{145pt}{$2$} 
\ins{-18pt}{180pt}{$0$}
\ins{125pt}{170pt}{$1$} 
\ins{230pt}{200pt}{$3$}
\ins{192pt}{240pt}{$4$} 
\ins{162pt}{130pt}{$2$} 
\ins{167pt}{165pt}{$0$} 
\ins{197pt}{195pt}{$0_1$}
\ins{-100pt}{40pt}{$b)$}
\ins{-50pt}{65pt}{$B$} 
\ins{0pt}{65pt}{$W$} 
\ins{0pt}{15pt}{$B$} 
\ins{-50pt}{15pt}{$W$} 
\ins{135pt}{80pt}{$B$} 
\ins{200pt}{65pt}{$W$} 
\ins{215pt}{0pt}{$B$} 
\ins{150pt}{15pt}{$W$} 
\ins{-60pt}{40pt}{$1$} 
\ins{15pt}{40pt}{$3$}
\ins{-23pt}{80pt}{$4$} 
\ins{-23pt}{0pt}{$2$} 
\ins{-18pt}{35pt}{$0$}
\ins{125pt}{55pt}{$1$} 
\ins{230pt}{25pt}{$3$}
\ins{162pt}{95pt}{$4$} 
\ins{192pt}{-15pt}{$2$} 
\ins{157pt}{62pt}{$0$} 
\ins{197pt}{20pt}{$0_1$}
}{fig4}
\vskip1.3truecm
\line{\vtop{\line{\hskip1.3truecm\vbox{\advance\hsize by -2.0 truecm
\0{\css Fig. 4.}
{\ottorm  The two elementary operations of disconnecting an intersection.
The labels ${\scriptstyle W}$ and ${\scriptstyle B}$ 
mean that the corresponding regions must 
be coloured white and black respectively. Note that the operation
of disconnecting an intersection involves the doubling of the site
${\scriptstyle 0}$ at the center of the intersection: 
in the figure we call ${\scriptstyle 0}$ and ${\scriptstyle 0_1}$
its two copies after the disconnection.
} \hfill} }}}
\*
\endinsert
 
We claim that the weight of $\g$ is $(-1)^{\n_\g}\prod_{\g'\in\PP}
(-1)^{n_{\g'}}$, which is the desired result 
(recall that $\PP$ is the set of polygons obtained
as boundaries of the black sets, {\it after} the disconnection described
in Fig.4). Note that the factor $(-1)^{\n_\g}$ in front of the 
product of the weights of the disconnected simple polygons 
is due to the doubling of the centers of the intersections,
implied by our definition of disconnection, see footnote to Fig. 4.

In order to prove the claim we explicitely write the contribution from the 
intersection in both cases $(a)$ and $(b)$ of Fig. 4, and we show that it 
is equal to the contribution of the two corner elements on the r.h.s.
of Fig. 4, unless for a minus sign, to be associated to the new
site $0_1$.

The contribution of the left hand side of case $(a)$ in Fig. 4 is:
$$\int \der\lis H_{\xx_0}\der H_{\xx_0}\der\lis V_{\xx_0}\der V_{\xx_0}
\Big[\lis H_{\xx_1} H_{\xx_0} \cdot \lis H_{\xx_0} H_{\xx_3}\cdot
\lis V_{\xx_2} V_{\xx_0} \cdot \lis V_{\xx_0} V_{\xx_4}\Big]\;.\Eq(2.31)$$
Multiplying \equ(2.31) by 
$$-\int\der\lis H_{\xx_{0_1}}\der H_{\xx_{0_1}}\der\lis V_{\xx_{0_1}}
\der V_{\xx_{0_1}}[\lis V_{\xx_{0_1}}\lis H_{\xx_{0_1}}\cdot
V_{\xx_{0_1}}H_{\xx_{0_1}}]=+1\;,\Eq(2.32)$$
we see that it can be equivalently rewritten as
$$\eqalign{&-
\int \big(\der\lis H_{\xx_0}\der H_{\xx_0}\der\lis V_{\xx_0}\der V_{\xx_0}
\big)\big(\der\lis H_{\xx_{0_1}}\der H_{\xx_{0_1}}\der\lis V_{\xx_{0_1}}
\der V_{\xx_{0_1}}\big)\cdot\cr
&\qquad\qquad\cdot
\Big[\lis H_{\xx_1} H_{\xx_0} \cdot\lis V_{\xx_{0_1}} \lis H_{\xx_{0_1}}
\cdot \lis V_{\xx_2} V_{\xx_0}\Big]\cdot\Big[
\lis H_{\xx_0} H_{\xx_3}\cdot V_{\xx_{0_1}}H_{\xx_{0_1}}\cdot
 \lis V_{\xx_0} V_{\xx_4}\Big]\;.\cr}\Eq(2.33)$$
Exchanging the names of the fields $\lis V_{\xx_0}\otto\lis V_{\xx_{0_1}}$
and $\lis H_{\xx_0}\otto\lis H_{\xx_{0_1}}$, we easily recognize that
\equ(2.33) is equal to $(-1)$ times 
the contribution of the r.h.s. of case $(a)$ in Fig. 4. The minus sign
compensate the fact that after the doubling the new polygon has a site 
more than the original one. 

The argument can be repeated in case $(b)$, so that the proof
of the claim is complete. 
\\
\\
This concludes the proof of \equ(2.15) in the case of open boundary 
conditions (\ie in the case where polygons winding 
up over the lattice are not allowed).
\\
\\
\sub(2.4){\bf The Grassmann representation of the 2d Ising model
with periodic boundary conditions.}
\\
In the case periodic boundary conditions are assumed, 
the representation in terms of multipolygons is the same, except
for the fact that also 
polygons winding up over the lattice are allowed. In order
to construct a Grassmann representation for
the multipolygon expansion of Ising with p.b.c., let us start 
with considering the following expression:
$$\int\prod_{\xx\in\L_M}\der\lis H_\xx
\der H_\xx\der\lis V_\xx\der V_\xx e^{S_{\e,\e'}(t)}\;,\Eq(2.340)$$
where $\e,\e'=\pm$ and $S_{\e,\e'}(t)$ is defined by \equ(2.14), but
with different boundary conditions, \ie
$$\eqalign{
&\lis H_{\xx+M\hat e_0}=\e\lis H_{\xx}
\virg\lis H_{\xx+M\hat e_1}=\e'\lis H_{\xx}\cr
&H_{\xx+M\hat e_0}=\e H_{\xx}\virg
H_{\xx+M\hat e_1}=\e' H_{\xx}\cr}\virg\e,\e'=\pm\;,\Eq(2.34)$$
where we recall that $M$ is the side of the lattice $\L_M$.
Identical definitions are set for the variables $V,{\lis
V}$. We shall say that $\lis H, H,\lis V,V$ 
satisfy $\e$--periodic
($\e'$--periodic) boundary conditions in vertical (horizontal) 
direction. Note that, unless for a sign and for the replacement $S(t)\to
S_{\e,\e'}(t)$, \equ(2.340) is the same as the r.h.s. of \equ(2.15).

Clearly, by expanding the exponential in \equ(2.340) and by integrating 
the Grassmann fields as described in previous
section, we get a summation over dimers very similar to the 
one seen above. In particular the weights assigned to the closed
polygons not winding up the lattice are exactly the same as those calculated
in previous section. In this case, however, also Grassmann polygons 
winding up the lattice are allowed. Let us calculate the weight 
that \equ(2.340) assigns to these polygons (as above we define the weight 
by descarding the ``trivial'' factors $t^{|\g|}$ and $(-1)^{M^2-n_\g}$).

As an example, let us first calculate the contribution from the simplest
polygon $\g$ winding up the lattice, the horizontal straight line winding once
in horizontal direction. Its weight is given by:
$$\int \lis V_{\V0} V_{\V0}\cdot \lis H_{\V0} H_{\hat e_1}\cdot
  \lis V_{\hat e_1} V_{\hat e_1}\cdot \lis H_{\hat e_1} 
H_{2 \hat e_1}\cdots \lis H_{(M-1)\hat e_1} 
H_{M\hat e_1}\;.\Eq(2.35)$$
Now, using \equ(2.34) we can rewrite $\lis H_{M\hat e_1}$ as $\e' H_\V0$.
Also, permutating the field $H_\V0$ from the last position to the third one,
we see that \equ(2.35) is equal to:
$$\eqalign{&(-\e')\int \lis V_{\V0} V_{\V0} H_\V0\lis H_{\V0}\cdot 
  \lis V_{\hat e_1} V_{\hat e_1}H_{\hat e_1}\lis H_{\hat e_1} 
\cdots \lis V_{(M-1)\hat e_1} V_{(M-1)\hat e_1}H_{(M-1)\hat e_1}\lis H_{
(M-1)\hat e_1}=\cr
&\qquad\qquad=(-\e')(-1)^{M}=(-\e')(-1)^{n_\g}\;,\cr}\Eq(2.36)$$
where, in the last identity, we used that the length of the straight
polygon $\g$ is exactly $M$. 
Repeating the lengthy construction of previous section, 
it can be (straightforwardly) proven that a generic polygon $\g$ 
winding up once in horizontal direction has a weight (as assigned by 
\equ(2.340)) equal to $(-\e')(-1)^{n_\g}$. Analogously 
a polygon $\g$ 
winding up once in horizontal direction has a weight (as assigned by 
\equ(2.340)) equal to $(-\e)(-1)^{n_\g}$. 

Let us now consider the simplest polygon $\g$ winding up $h$ 
times in horizontal direction and $v$ times in vertical direction, that is the
union of $h$ distinct horizontal lines and $v$ distinct vertical lines 
each of them winding once over the lattice in horizontal or 
vertical direction, respectively. Repeating the same simple calculation 
of \equ(2.35)--\equ(2.36), we easily see that the
weight assigned by \equ(2.340) to $\g$ is $(-\e')^h(-\e)^v (-1)^{M(h+v)}$.
Note that $\g$ has $(-1)^{h\cdot v}$ self intersections,
so that $n_\g=M(h+v)-h\cdot v$ and the weight can be rewritten as 
$(-\e')^h(-\e)^v (-1)^{h\cdot v}(-1)^{n_\g}$.
Again, repeating the lengthy construction of previous section, 
it can be (straightforwardly) proven that a generic polygon $\g$ 
winding up $h$ times in horizontal direction and $v$ times
in vertical direction has a weight (as assigned by 
\equ(2.340)) equal to $(-\e')^h(-\e)^v (-1)^{h\cdot v}(-1)^{n_\g}$.
\\

Since the weight assigned to a generic polygon 
is the one just computed, which is in general different from $(-1)^{n_\g}$,
it is clear that there exists {\it no choice} of 
$\e,\e'=\pm 1$ such that 
\equ(2.340) is equal to
$(-1)^{M^2}(2\cosh^2\b J)^{-M^2}$ times $\Xi_I$, where now $\Xi_I$
is the Ising model partition function in the volume $\L_M$ with periodic 
boundary conditions. 
However it is easy to realize that 
$(-1)^{M^2}\Xi_I(2\cosh^2\b J)^{- M^2}$ is equal to a suitable linear
combination of the expressions in \equ(2.340), with different choices of 
$\e,\e'=\pm 1$: it holds that
$$(-1)^{M^2}{\Xi_I\over (2\cosh^2\b J)^{M^2}}={1\over 2}\sum_{\e,\e'=\pm 1}
\int\prod_{\xx\in\L_M}\der\lis H_\xx
\der H_\xx\der\lis V_\xx\der V_\xx (-1)^{\d_{(\e,\e')}}e^{S_{\e,\e'}(t)}\;,
\Eq(2.37)$$
where $\d_{+,-}=\d_{-,+}=\d_{-,-}=0$ and $\d_{+,+}=1$. In order
to verify the last identity it is sufficient to verify
that the weight assigned from the r.h.s. of \equ(2.37) to each polygon 
$\g$ is exactly $(-1)^{n_\g}$. If $\g$ winds up
the lattice $h$ times in horizontal direction and $v$ times in vertical 
direction, from the calculation above it follows that the weight assigned 
to $\g$ by the r.h.s. of \equ(2.37) is:
$$\eqalign{&{1\over 2}\sum_{\e,\e'=\pm 1}(-1)^{\d_{(\e,\e')}}
(-\e')^h(-\e)^v (-1)^{h\cdot v}(-1)^{n_\g}=\cr
&={1\over 2}(-1)^{n_\g}\Big[
(-1)^{h+v+hv+\d_{+,+}}+
(-1)^{v+hv+\d_{+,-}}+
(-1)^{h+hv+\d_{-,+}}+
(-1)^{hv+\d_{-,-}}\Big]\cr}
\Eq(2.38)$$
The expression between square brackets on the last line is equal to
$(-1)^{hv}[-(-1)^{h+v}+(-1)^{v}+(-1)^{h}+1]$. Now, if 
$h$ and $v$ are both even, this is equal to $(+1)[-1+1+1+1]=2$;
if $h$ is even and $v$ is odd (or viceversa), it is equal to $(+1)[+1-1+1+1]
=2$;
if they are both odd, it is equal to $(-1)[-1-1-1+1]=2$. That is, 
\equ(2.38) is identically equal to $(-1)^{n_\g}$, as wanted, and 
\equ(2.37) is proven.
\\
\\
\sub(2.5){\bf The Ising model's free energy}
\\
From the Grassmann representation of the Ising model partition
function, it is easy to derive the well--known expression
for the Ising's free energy. Even if in the following
we won't need it, we reproduce here the calculation, for completeness.

The unitary transformation of the Grassmann fields diagonalizing
the action $S_{\e,\e'}(t)$ is the following:
$$\eqalign{&
H_\xx={1\over |\L_M|^{1/2}}\sum_{\kk\in\DD_{\e,\e'}}\hat H_\kk e^{-i\kk\xx}
\;,\qquad\qquad
\lis H_\xx={1\over |\L_M|^{1/2}}\sum_{\kk\in\DD_{\e,\e'}}\hat{\lis H}_\kk 
e^{-i\kk\xx}
\;,\cr
&V_\xx={1\over |\L_M|^{1/2}}\sum_{\kk\in\DD_{\e,\e'}}\hat V_\kk e^{-i\kk\xx}
\;,\qquad\qquad
\lis V_\xx={1\over |\L_M|^{1/2}}\sum_{\kk\in\DD_{\e,\e'}}\hat{\lis V}_\kk 
e^{-i\kk\xx}
\;,\cr}\Eq(2.39)$$
where $\kk=(k,k_0)$ and $D_{\e,\e'}$ is the set of $\kk$'s such that
$$k={2\pi n_1\over M}+{(\e'-1)\pi\over M}\quad
k_0={2\pi n_0\over M}+{(\e-1)\pi\over M}
\Eq(2.40)$$
with $-[M/2]\le n_0\le [(M-1)/2]$, $-[M/2]\le n_1\le [(M-1)/2]$,
$n_0,n_1\in\ZZZ$.
In terms of the new fields $\hat H_\kk,\hat{\lis H}_\kk,
\hat V_\kk,\hat{\lis V}_\kk$, the action $S_{\e,\e'}(t)$ can be written
as:
$$S_{\e,\e'}(t)=\sum_{\kk\in\DD_{\e,\e'}}\Big[
t \hat{\lis H}_\kk\hat H_{-\kk}e^{i k}+t \hat{\lis V}_\kk
\hat V_{-\kk}e^{i k_0}+ \hat{\lis H}_\kk\hat H_{-\kk}+\hat{\lis V}_\kk
\hat V_{-\kk}+\hat{\lis V}_\kk
\hat{\lis H}_{-\kk}+\hat{V}_\kk
\hat{\lis H}_{-\kk}+\hat{H}_\kk
\hat{\lis V}_{-\kk}+\hat{V}_\kk
\hat{H}_{-\kk}\Big]\Eq(2.41)$$
Let us say that $\kk>0$ if its first component $k_0$ is $>0$.
Then we can rewrite \equ(2.41) as:
$$\eqalign{&\sum_{\kk>0}\Big[
t \hat{\lis H}_\kk\hat H_{-\kk}e^{i k}
-t \hat H_{\kk}\hat{\lis H}_{-\kk} e^{-i k}
+t \hat{\lis V}_\kk\hat V_{-\kk}e^{i k_0}
-t \hat{V}_\kk\hat{\lis V}_{-\kk}e^{-i k_0}
+ \hat{\lis H}_\kk\hat H_{-\kk}
-\hat{H}_\kk\hat{\lis H}_{-\kk}
+\hat{\lis V}_\kk\hat V_{-\kk}
-\hat{V}_\kk\hat{\lis V}_{-\kk}+\cr
&\qquad+\hat{\lis V}_\kk\hat{\lis H}_{-\kk}
-\hat{\lis H}_{\kk}\hat{\lis V}_{-\kk}
+\hat{V}_\kk\hat{\lis H}_{-\kk}
-\hat{\lis H}_{\kk}\hat{V}_{-\kk}
+\hat{H}_\kk\hat{\lis V}_{-\kk}
-\hat{\lis V}_{\kk}\hat{H}_{-\kk}
+\hat{V}_\kk\hat{H}_{-\kk}
-\hat{H}_{\kk}\hat{V}_{-\kk}\Big]\=\cr
&\=\sum_{\kk>0}\Psi_\kk^T M_\kk \Psi_{-\kk}\;,\cr}
\Eq(2.42)$$
where $\Psi_\kk^T\defin
\big(\hat{\lis H}_\kk,\hat{H}_\kk,\hat{\lis V}_\kk,\hat{V}_\kk\big)$
and the matrix $M_\kk$ is defined as:
$$M_\kk\defin\left( \matrix{0 & 1+ t e^{i k} & -1 & -1\cr
-(1+ t e^{-i k}) & 0 & 1 & -1 \cr
1 & -1 & 0 & 1+te^{i k_0}\cr
1 & 1 & -(1+te^{-ik_0}) & 0\cr}
\right)\;.\Eq(2.43)$$
Then, unless for a sign, 
$$\int \prod_{\xx\in\L_M}\der\lis H_\xx
\der H_\xx\der\lis V_\xx\der V_\xx e^{S_{\e,\e'}(t)}=
\prod_{\kk>0}\Big[\int \der\hat{\lis H}_\kk\der\hat{\lis H}_{-\kk}
\der \hat{H}_\kk\der \hat{H}_{-\kk}\der\hat{\lis V}_\kk\der\hat{\lis V}_{-\kk}
\der \hat{V}_\kk \der \hat{V}_{-\kk}\cdot e^{\Psi_\kk^T M_\kk \Psi_{-\kk}} 
\Big]\;,\Eq(2.44)$$
and, using \equ(2.12), we see that the r.h.s. of \equ(2.44) is equal 
$\prod_{\kk>0}\det M_\kk$.
The explicit computation of $\det M_\kk$ leads to:
$$\eqalign{\det M_\kk&=\Big[1+t^2+2t\cos k\Big]\Big[1+t^2+2t\cos k_0\Big]
-4t(\cos k+\cos k_0)-4t^2\cos k\cos k_0=\cr
&=(1+t^2)^2-2t(1-t^2)(\cos k+\cos k_0)
\;.\cr}\Eq(2.46)$$
Now, using \equ(2.37), we find that
$$\eqalign{&-\b f_{Ising}\defin\lim_{M\to\io}{1\over M^2}\log \Xi_I=\cr
&=\log(2\cosh^2\b J)+
{1\over 2}\int_{-\p}^\p {d k\over 2\p}\int_{-\p}^\p {d k_0\over 2\p}
\log\{(1+t^2)^2-2t(1-t^2)(\cos k+\cos k_0)\}=\cr
&={1\over 2}\int_{-\p}^\p {d k\over 2\p}\int_{-\p}^\p {d k_0\over 2\p}
\log\Big\{4\big[\cosh^22\b J-\sinh 2\b J(\cos k+\cos k_0)\big]\Big\}
\;,\cr}\Eq(2.47)$$
that is the celebrated Onsager's result. Note that the argument
of the logarithm in the last expression is always $\ge 0$ and it 
vanishes iff $\sinh 2\b J=1$, that is the equation for the
critical temperature. In the following we shall also write this condition
in the equivalent form $\tanh\b J=\sqrt 2-1$.

\pagina
\setcap{3. The Grassmann formulation of Ashkin--Teller.}
\capindex{3}{The Grassmann formulation of Ashkin--Teller.}
\vskip1.truecm
\section(3,The Grassmann formulation of Ashkin--Teller.)
\capindex{3}{The Grassmann formulation of Ashkin--Teller.}

In this section, using the Grassmann representation 
of the Ising model, derived in previous Chapter, we will
derive the Grassmann representation for a class of interacting 
spin models, defined as a pair of Ising models coupled by
suitable multi--spin 
interactions.

We will first derive the general representation for a wide class 
of models, to be defined in next section, including the 
Ashkin--Teller model defined in \equ(1.1), the four states
Potts model, the 8V model and the next 
to nearest neighbor Ising model. Then we will focus on AT, 
and we will perform more algebraic manipulations to get to a final 
representation that will be convenient for the following 
multiscale integration, necessary to construct a convergent
expansion for some correlation functions, as explained in the Introduction.

The reason why we choose to focus on AT is for definiteness and 
for avoiding too cumbersome abstract expressions, that necessarely would turn 
out in trying to describe our method in a too general settling. However
it will be clear that the same method we will apply
to the study of the AT model could equally well be applied 
to 8V (in a suitable range of parameters), to Ising perturbed with 
a small non nearest neighbor interaction or 
to linear combinations of the above models. Note that, even if 
the four states Potts model can be represented by a Grassmann functional 
integral as proved below, the subsequent multiscale analysis we will apply 
to AT would {\it not} work for Potts. This is because Potts
is equivalent to a system of strongly interacting fermions, while
our perturbative methods are applicable only in the range of
weak coupling.
\\
\\
\sub(3.1){\bf The Grassmann representation for a pair of Ising models
with multi spin interactions.}
\\
Let us start with considering a pair of nearest neighbor
Ising models with periodic boundary conditions, 
labeled by $j=1,2$, with couplings allowed
to depend on the bonds $b\in\L_M^*$ (here $\L_M^*$ is the dual of $\L_M$,
that is the set of bonds linking the nearest neighbor sites of $\L_M$):
$$H_I^{(j)}\{J^{(j)}_b\}=-\sum_{b\in\L_M^*}J^{(j)}_b
\tilde\s_b^{(j)}
\;,\Eq(3.1)$$
where the bond spin $\tilde\s_b^{(j)}$ 
was defined in \sec(2.1) above. Repeating
the construction of previous Chapter, one finds that  
the partition function of the model \equ(3.1) can be written as:
$$\eqalign{\Xi_I^{(j)}&=\sum_{\s_{\L_M}^{(j)}}e^{-\b H_I^{(j)}
\{J^{(j)}_b\}}=\cr
&=
(-1)^{M^2}2^{M^2}\Big[\prod_{b\in\L_M^*}\cosh\b J_b^{(j)}\Big]
{1\over
2}\sum_{\e,\e'=\pm} \int \prod_{\xx\in\L_M} dH^{(j)}_\xx
d\lis H^{(j)}_\xx d V^{(j)}_\xx d\lis V^{(j)}_\xx
(-1)^{\d_{\g}} e^{S^{(j)}_\g\{t^{(j)}_b\}}\;,\cr}
\Eq(3.2)$$
where $\g= (\e,\e')$ labels the boundary conditions of
the Grassmann fields, $\d_\g$ was
defined after \equ(2.37) and
$$\eqalign{
&S^{(j)}_{\g}\{t^{(j)}_b\}=\sum_{\xx\in\L_M} \left[\tanh( J^{(j)}_{
\xx,\xx+\hat e_1})\lis H^{(j)}_{\xx} H^{(j)}_{\xx+\hat e_1}+
\tanh( J^{(j)}_{\xx,\xx+\hat e_0})
\lis V^{(j)}_{\xx} V^{(j)}_{\xx+\hat e_0}\right]+\cr
&+\sum_{\xx\in\L_M}\left[\lis H^{(j)}_{\xx} H^{(j)}_{\xx}+
\lis V^{(j)}_{\xx}
V^{(j)}_{\xx}+\lis V^{(j)}_{\xx} \lis H^{(j)}_{\xx}+
V^{(j)}_{\xx} \lis H^{(j)}_{\xx}+
H^{(j)}_{\xx} \lis V^{(j)}_{\xx}+
V^{(j)}_{\xx} H^{(j)}_{\xx}\right]\;.\cr}\Eq(3.3)$$
Let us now consider a multi spin interaction 
$V(\s^{(1)},\s^{(2)})$ between 
the two layers, linear combination of interactions of the form:
$$\eqalign{V_{\II}&=-\sum_{\xx\in\L_M}
(\s_{\xx+\zz_1}^{(i_1)}\s_{\xx+\zz_1+\hat e_{j_1}}^{(i_1)})\cdot
(\s_{\xx+\zz_2}^{(i_2)}\s_{\xx+\zz_2+\hat e_{j_2}}^{(i_2)})\cdots 
(\s_{\xx+\zz_k}^{(i_k)}\s_{\xx+\zz_k+\hat e_{j_k}}^{(i_k)})\=\cr
&\=-\sum_{\xx\in\L_M}\Big[\prod_{(b,i)\in\II}
\tilde\s_{b+\xx}^{(i)}\Big]\;,\cr}\Eq(3.4)$$
where $\II=\{(b_p,i_p)\}_{p=1}^k$ is a set
of indeces (with $b_p\in\L_M^*$ and $i_p=1,2$) and by $b+\xx$
we denote the bond obtained by rigidly translating $b$ of a vector $\xx$.  
Periodic boundary conditions are assumed.
Note that the interaction of Ashkin--Teller in \equ(1.1)
can be written as $\l V_{\II_1}+\l V_{\II_2}$, 
where, if we define $b_0$ to be the bond connecting $(0,0)$ with $(0,1)$
and $b_1$ that connecting $(0,0)$ with $(1,0)$,
$\II_1=\{(b_0,1),(b_0,2)\}$ and $\II_2=\{(b_1,1),(b_1,2)\}$.

The key feature of the interaction \equ(3.4) is 
to be a product of bond interactions appearing either in 
$H_I^{(1)}\{J^{(1)}_b\}$ or in $H_I^{(2)}\{J^{(2)}_b\}$, so that
the partition function associated to the Hamiltonian
$H_I^{(1)}\{J^{(1)}_b\}+H_I^{(2)}\{J^{(2)}_b\}+
\l_1 V_{\II_1}+\cdots+\l_nV_{\II_n}$
can be expressed a suitable derivative of $\Xi^{(1)}_I\Xi^{(2)}_I$ with
respect to the couplings $J_b^{(j)}$. In fact:
$$\eqalign{\Xi&=\sum_{\s^{(1)}_{\L_M},\s^{(2)}_{\L_M}}\exp\Big\{-\b (H_I^{(1)}
\{J^{(1)}_b\}+H_I^{(2)}\{J^{(2)}_b\}+
\l_1 V_{\II_1}+\cdots+\l_nV_{\II_n})\Big\}=\cr
&=\sum_{\s^{(1)}_{\L_M},\s^{(2)}_{\L_M}}\exp\Big\{{\b\sum_b(J_b^{(1)}
\tilde\s^{(1)}_b+J_b^{(2)}
\tilde\s^{(2)}_b)+\b\sum_{q=1}^n\l_q\sum_\xx\Big[\prod_{(b,j)\in\II_q}
\tilde\s^{(i)}_{b+\xx}\Big]\Big\}}\;.\cr}\Eq(3.5)$$
Defining $\hat\l_q\defin\tanh\b\l_q$, the last expression can be rewritten as:
$$\eqalign{&\Big[\prod_{q=1}^n\cosh\b\l_q\Big]^{M^2}
\sum_{\s^{(1)}_{\L_M},\s^{(2)}_{\L_M}}e^{\b\sum_b(J_b^{(1)}
\tilde\s^{(1)}_b+J_b^{(2)}
\tilde\s^{(2)}_b)}\prod_{\xx\in\L_M}\prod_{q=1}^n\left(1+\hat\l_q
\Big[\prod_{(b,j)\in\II_q}
\tilde\s^{(i)}_{b+\xx}\Big]\right)=\cr
=&\Big[\prod_{q=1}^n\cosh\b\l_q\Big]^{M^2}
\prod_{\xx\in\L_M}\prod_{q=1}^n\left(1+\hat\l_q
\Big[\prod_{(b,j)\in\II_q}\b^{-1}{\dpr\over\dpr 
J^{(i)}_{b+\xx}}
\Big]\right)\Xi^{(1)}\{J_b^{(1)}\}\Xi^{(1)}\{J_b^{(1)}\}
\;.\cr}\Eq(3.6)$$
Substuting into the r.h.s. of \equ(3.6) the representation \equ(3.2),
we find that $\Xi$ can be expressed as the sum of 16 Grassmann partition
functions, differing for the boundary conditions and labeled by
$\g_1=(\e_1,\e_1')$, $\g_2=(\e_2,\e_2')$:
$$\Xi=\Big[4\prod_{q=1}^n\cosh\b\l_q\Big]^{M^2}{1\over 4}\sum_{\g_1,\g_2}
(-1)^{\d_{\g_1}+\d_{\g_2}}\Xi^{\g_1,\g_2}\;,\Eq(3.7)$$
with $\Xi^{\g_1,\g_2}$ given by:
$$\eqalign{&\prod_{\xx\in\L_M}\prod_{q=1}^n\Big(1+\hat\l_q
\Big[\prod_{(b,j)\in\II_q}\b^{-1}{\dpr\over\dpr 
J^{(i)}_{b+\xx}}
\Big]\Big)\cdot\Biggl\{\cr
&\cdot\Big[\prod_{b\in\L_M^*}
\prod_{j=1}^2\cosh\b J_b^{(j)}\Big]
\int \Big[\prod_{\xx\in\L_M} \prod_{j=1}^2 dH^{(j)}_\xx
d\lis H^{(j)}_\xx d V^{(j)}_\xx d\lis V^{(j)}_\xx\Big]
e^{S^{(1)}_\g\{t^{(1)}_b\}+S^{(2)}_\g\{t^{(2)}_b\}}\Biggr\}\;.\cr}\Eq(3.8)$$
We now want to explicitely write the effect of the derivatives 
in the last expression and rewrite \equ(3.8) as the Grassmann
integral over an exponential of a (non quadratic) action.
Let us first note that the effect of a singol derivative $\b^{-1}\dpr/\dpr J^{(i)
}_{b_0}$
over $\big[\prod_{b,j}\cosh\b J_b^{(j)}\big]e^{S^{(1)}_\g+S^{(2)}_\g}$ is:
$$\b^{-1}{\dpr\over\dpr 
J^{(i)}_{b_0}}\Big\{
\big[\prod_{b,j}\cosh\b J_b^{(j)}\big]e^{S^{(1)}_\g+S^{(2)}_\g}\Big\}=
\big[\prod_{b,j}\cosh\b J_b^{(j)}\big]e^{S^{(1)}_\g+S^{(2)}_\g}\left(
t_{b_0}^{(j)}+s_{b_0}^{(j)}D_{b_0}^{(j)}\right)\;,\Eq(3.9)$$
where we introduced the definitions $t_b^{(j)}\defin\tanh\b J_b^{(j)}$,
$s_b^{(j)}\defin 1/\cosh^2\b J_b^{(j)}$ and $D_b^{(j)}$
is a Grassmann binomial such that, 
if $b=(\xx,\xx+\hat e_0)$,
$D_b^{(j)}\defin \lis V^{(j)}_\xx V^{(j)}_{\xx+\hat e_0}$ while, if 
$b=(\xx,\xx+\hat e_1)$, $D_b^{(j)}\defin \lis H^{(j)}_\xx 
H^{(j)}_{\xx+\hat e_1}$.

Using \equ(3.9) we see that \equ(3.8) can be rewritten as:
$$\eqalign{&\Big[\prod_{b\in\L_M^*}
\prod_{j=1}^2\cosh\b J_b^{(j)}\Big]
\int \Big[\prod_{\xx\in\L_M} \prod_{j=1}^2 dH^{(j)}_\xx
d\lis H^{(j)}_\xx d V^{(j)}_\xx d\lis V^{(j)}_\xx\Big]\cdot\cr
&\qquad\qquad\qquad\cdot
e^{S^{(1)}_\g\{t^{(1)}_b\}+S^{(2)}_\g\{t^{(2)}_b\}}
\prod_{\xx\in\L_M}\prod_{q=1}^n\Big(1+\hat\l_q
\Big[\prod_{(b,j)\in\II_q}\left(
t_{b+\xx}^{(j)}+s_{b+\xx}^{(j)}
D_{b+\xx}^{(j)}\right)
\Big]\Big)\;.\cr}\Eq(3.10)$$
Let us denote with $\ii$ the elements of $\II_q$. and, if $\ii=(b,j)$,
define $\l_\xx(\ii)\defin s_{b+\xx}^{(j)}/t_{b+\xx}^{(j)}$ and 
$D_\xx(\ii)\defin D_{b+\xx}^{(j)}$. Let us also assign an ordering
to the elements of $\II$ and let us write $\ii_1<\ii_2$ 
if $\ii_1$ precedes $\ii_2$ w.r.t. this ordering.
With these definitions we can rewrite the last product in \equ(3.10) as:
$$\eqalign{&\prod_{\xx\in\L_M}\prod_{q=1}^n
\Biggl\{1+\hat\l_q\Big(\prod_{(b,j)\in\II_q}t_{b+\xx}^{(j)}\Big)
\Big[1+\sum_{\ii_1\in\II_q}\l_\xx(\ii_1)D_\xx(\ii_1)+
\sum_{\ii_1<\ii_2}\l_\xx(\ii_1)D_\xx(\ii_1)\l_\xx(\ii_2)D_\xx(\ii_2)+
\cdots\cr
&\qquad\qquad\cdots+
\sum_{\ii_1<\ii_2<\cdots\ii_{|\II_q|}}\l_\xx(\ii_1)D_\xx(\ii_1)
\cdots\l_\xx(\ii_{|\II_q|})D_\xx(\ii_{|\II_q|})\Big]\Biggr\}\cr}\Eq(3.11)$$
and, calling $T_\xx(\II_q)\defin\hat\l_q
\Big(\prod_{(b,j)\in\II_q}t_{b+\xx}^{(j)}\Big)\cdot\left[1+\hat\l_q
\Big(\prod_{(b,j)\in\II_q}t_{b+\xx}^{(j)}\Big)\right]^{-1}$ we still
can rewrite the last expression as:
$$\eqalign{&\prod_{\xx\in\L_M}\prod_{q=1}^n
\Biggl\{\Big(1+\hat\l_q
\prod_{(b,j)\in\II_q}t_{b+\xx}^{(j)}\Big)\cdot
\Big[1+\sum_{\ii_1\in\II_q}T_\xx(\II_q)\l_\xx(\ii_1)D_\xx(\ii_1)+\cr
&+\sum_{\ii_1<\ii_2}T_\xx(\II_q)
\l_\xx(\ii_1)D_\xx(\ii_1)\l_\xx(\ii_2)D_\xx(\ii_2)+
\cdots+
\sum_{\ii_1<\ii_2<\cdots\ii_{|\II_q|}}T_\xx(\II_q)\l_\xx(\ii_1)D_\xx(\ii_1)
\cdots\l_\xx(\ii_{|\II_q|})D_\xx(\ii_{|\II_q|})\Big]\Biggr\}=\cr
&=\prod_{\xx\in\L_M}\prod_{q=1}^n\Biggl\{\Big(1+\hat\l_q
\prod_{(b,j)\in\II_q}t_{b+\xx}^{(j)}\Big)
\exp\Big\{
\sum_{\ii_1\in\II_q}\tilde\l_\xx^{(q)}(\ii_1)D_\xx(\ii_1)+
\sum_{\ii_1<\ii_2}\tilde\l^{(q)}_\xx(\ii_1,\ii_2)D_\xx(\ii_1)D_\xx(\ii_2)+
\cdots+\cr
&\qquad\qquad\qquad\cdots+
\sum_{\ii_1<\ii_2<\cdots\ii_{|\II_q|}}\tilde\l^{(q)}_\xx
(\ii_1,\ldots,\ii_{|\II_q|})
D_\xx(\ii_1)\cdots D_\xx(\ii_{|\II_q|})
\Big\}\Biggr\}\,.\cr}\Eq(3.12)$$
In the last expression $\tilde\l^{(q)}_\xx(\ii_1)=T_\xx(\II_q)\l_\xx(\ii_1)$
and the couplings $\tilde\l^{(q)}_\xx(\ii_1,\ldots,\ii_k)$,
$2\le k\le |\II_q|$, are defined by the following recursive relations:
$$T_\xx(\II_q)\l_\xx(\ii_1)\cdots\l_\xx(\ii_k)=
\sum_{p=1}^{k}\ \ \sum_{\JJ_1\cup\JJ_2\cup\cdots\cup\JJ_p=(\ii_1,\ldots,\ii_k)}
\tilde\l^{(q)}_\xx(\JJ_1)\cdots\tilde\l^{(q)}_\xx(\JJ_p)\;,\Eq(3.13)$$
where $\JJ_r=(\jj_1^{(r)},\ldots,\jj_{|\JJ_r|}^{(r)})$ are ordered 
(\ie $\jj_1^{(r)}<\ldots<\jj_{|\JJ_r|}^{(r)}$) subsets of $(\ii_1,\ldots,
\ii_k)$, such that $|\JJ_1|+\cdots+|\JJ_p|=k$.  

Substituting \equ(3.12) into \equ(3.10) we finally find:
$$ \eqalign{&\Xi^{\g_1,\g_2}=\Big[\prod_{b\in\L_M^*}
\prod_{j=1}^2\cosh\b J_b^{(j)}\Big]\Big[
\prod_{\xx\in\L_M}\prod_{q=1}^n\Big(1+\hat\l_q
\prod_{(b,j)\in\II_q}t_{b+\xx}^{(j)}\Big)\Big]\cdot\cr
&\qquad\qquad\cdot
\int \Big[\prod_{\xx\in\L_M} \prod_{j=1}^2 dH^{(j)}_\xx
d\lis H^{(j)}_\xx d V^{(j)}_\xx d\lis V^{(j)}_\xx\Big]
e^{S^{(1)}_\g\{t^{(1)}_b\}+S^{(2)}_\g\{t^{(2)}_b\}+V_\l}
\;;\cr
& V_\l\defin \sum_\xx \sum_q \sum_{k=1}^{|\II_q|}
\ \ \sum_{\ii_1<\ii_2<\cdots\ii_{k}}\tilde\l^{(q)}_\xx(\ii_1,\ldots,\ii_{k})
D_\xx(\ii_1)\cdots D_\xx(\ii_{k})\;.\cr}\Eq(3.14)$$
This concludes the derivation of the Grassmann representation
for pairs of Ising models coupled by an interaction 
$V(\s^{(1)},\s^{(2)})$, linear combination of interactions of the form 
\equ(3.4). 
\\
\\
\sub(3.2){\bf The Grassmann representation for the Ashkin--Teller model.}
\\
We now specialize to the case of the Ashkin--Teller model.
We first write the explicit form of $V_\l$ in \equ(3.14) for AT. 
As already discussed in previous section, the AT model corresponds to
an interaction of the form $\l V_{\II_1}+\l V_{\II_2}$, 
with $\II_1=\{(b_0,1),(b_0,2)\}$ and $\II_2=\{(b_1,1),(b_1,2)\}$,
$b_0$ being the bond connecting $(0,0)$ with $(0,1)$
and $b_1$ the one connecting $(0,0)$ with $(1,0)$. 
We shall assume the sets $\II_q$, $q=1,2$, ordered
so that $(b_{q-1},1)<(b_{q-1},2)$.
We are interested in writing the explicit expressions in the 
case $t_b^{(j)}\=t^{(j)}$ and $s_b^{(j)}\=s^{(j)}$ independent of $b$ 
(but in general depending on the lattice $j=1,2$). 

The definitions of $\l_q$, $T_\xx$ and $\l_\xx$ introduced in last section
become in this case:
$$\hat\l_1=\hat\l_2=\tanh\b\l\=\hat\l\;,\qquad
T_\xx(\II_q)={\hat\l t^{(1)}t^{(2)}\over 1+\hat\l t^{(1)}t^{(2)}}\;,\qquad
\l_\xx(b_{q-1},j)={s^{(j)}\over t^{(j)}}\;.\Eq(3.15)$$
Then \equ(3.13) can be rewritten as:
$$\eqalign{&
{\hat\l t^{(1)}t^{(2)}\over 1+\hat\l t^{(1)}t^{(2)}}\l_\xx(b_{q-1},1)
\l_\xx(b_{q-1},2)=\tilde\l_\xx^{(q)}\Big((b_{q-1},1),(b_{q-1},2)\Big)+
\tilde\l_\xx(b_{q-1},1)\tilde\l_\xx(b_{q-1},2)\;,\cr
&\tilde\l_\xx(b_{q-1},j)=
{\hat\l t^{(1)}t^{(2)}\over 1+\hat\l t^{(1)}t^{(2)}}{s^{(j)}\over t^{(j)}}
\;,\cr}\Eq(3.16)$$
implying:
$$\eqalign{&\tilde\l_\xx(b_{q-1},1)=
{\hat\l s^{(1)}t^{(2)}\over 1+\hat\l t^{(1)}t^{(2)}}\=\l^{(1)}\;,\qquad
\tilde\l_\xx(b_{q-1},2)=
{\hat\l s^{(2)}t^{(1)}\over 1+\hat\l t^{(1)}t^{(2)}}\=\l^{(2)}\,,\cr
&\tilde\l_\xx^{(q)}\Big((b_{q-1},1),(b_{q-1},2)\Big)=
{\hat\l s^{(1)}s^{(2)}\over \big(1+\hat\l t^{(1)}t^{(2)}\big)^2}\=
\widetilde\l\;.\cr}\Eq(3.17)$$
With these definitions $V_\l$ in \equ(3.14) can be written as:
$$\eqalign{&V_\l=\sum_{\xx\in\L_M}
\Biggl\{\left[\l^{(1)}\lis H^{(1)}_{\xx} H^{(1)}_{\xx+\hat e_1}+
\l^{(2)}\lis H^{(2)}_{\xx}
H^{(2)}_{\xx+\hat e_1}+\widetilde\l\lis
H^{(1)}_{\xx} H^{(1)}_{\xx+\hat e_1}
\lis H^{(2)}_{\xx} H^{(2)}_{\xx+\hat e_1}\right]+\cr
&\qquad\qquad+\left[\l^{(1)}\lis V^{(1)}_{\xx} V^{(1)}_{\xx+\hat e_0}+
\l^{(2)}\lis V^{(2)}_{\xx}
V^{(2)}_{\xx+\hat e_0}+\widetilde\l\lis
V^{(1)}_{\xx} V^{(1)}_{\xx+\hat e_0}
\lis V^{(2)}_{\xx} V^{(2)}_{\xx+\hat e_0}\right]\Biggr\}\cr
}\Eq(3.18)$$
and the first of \equ(3.14) becomes:
$$\eqalign{&\Xi^{\g_1,\g_2}_{AT}=\Big[(1+\hat\l t^{(1)}t^{(2)})\cosh\b J^{(1)}
\cosh\b J^{(2)}\Big]^{2M^2}\cdot\cr
&\qquad\qquad\cdot
\int \Big[\prod_{\xx\in\L_M} \prod_{j=1}^2 dH^{(j)}_\xx
d\lis H^{(j)}_\xx d V^{(j)}_\xx d\lis V^{(j)}_\xx\Big]
e^{S^{(1)}_\g(t^{(1)}_\l)+S^{(2)}_\g(t^{(2)}_\l)+\tilde\l \VV}\;,\cr}
\Eq(3.19)$$
where $S^{(j)}(t)$ was defined in previous Chapter, see \equ(2.340);
moreover $t^{(j)}_\l\defin t^{(j)}+\l^{(j)}$ and
$$\VV=\sum_{\xx\in\L_M} 
\left(\lis H^{(1)}_{\xx} H^{(1)}_{\xx+\hat e_1}
\lis H^{(2)}_{\xx} H^{(2)}_{\xx+\hat e_1}+ \lis V^{(1)}_\xx V^{(1)}_{
\xx+\hat e_0} \lis V^{(2)}_{\xx} V^{(2)}_{\xx+\hat e_0}\right)
\;.\Eq(3.20)$$
\sub(3.3) Starting from \equ(3.19) we will now make more 
algebraic manipulations
by introducing new Grassmann fields, linear combinations of the fields 
$\lis H,H,\lis V,V$. This will be convenient in order to set
up the Renormalization Group scheme we will use to study
in detail the specific heat of AT. The aim is to rewrite the 
formal action appearing at the exponent in 
\equ(3.19) as the formal action of a perturbed
{\it massive Luttinger model}, the latter being a model 
for which Renormalization Group technique is already
well developed [BM][GM].

We shall consider for simplicity the partition function 
$\Xi_{AT}^{-}\defin \Xi_{AT}^{(-,-),(-,-)}$, \ie the partition function
in which all Grassmannian variables verify antiperiodic boundary
conditions. The other fifteen partition functions in
the analogue of \equ(3.7) admit similar expressions.  
In fact we shall see in Chap. 7 and Appendix A9 that, 
if $(\l, t, u)$ {\it does not belong} 
to the {\it critical surface}\footnote{${}^1$}{\nota
The critical surface is a suitable 2--dimensional subset of 
$[-\e,\e]\times D\times [-{|D|\over
2},{|D|\over 2}]$, that is of the 3--dim set in the parameters space where we
are interested to study the AT model, 
see the assumptions in the main Theorem in the Introduction;
we will explicitely determine the critical surface in Chap. 7 below and 
we will prove that it can be parametrized as $(\l,t_c^{\pm}(\l,u),u)$,
with $t_c^{\pm}(\l,u)$ given by \equ(1.3).}
the partition function $\Xi_{AT}^{\g_1,\g_2}$ divided by 
$\Xi_I^{(1)\g_1}\Xi_I^{(2)\g_2}$
is exponentially insensitive to boundary conditions as $M\to\io$.

It is convenient to perform the following change of
variables [ID], $j=1,2$
$$\eqalign{
&\lis H_{\l,\xx}^{(j)}+i H_{\l,\xx}^{(j)}=e^{i{\pi\over 4}}
\psi_\xx^{(j)}-e^{i{\pi\over 4}}\chi_\xx^{(j)}\qquad
\lis H_{\l,\xx}^{(j)}-i H_{\l,\xx}^{(j)}=
e^{-i{\pi\over 4}}\lis\psi_\xx^{(j)}-
e^{-i{\pi\over 4}}\lis\chi_\xx^{(j)}\cr
&\lis V_{\l,\xx}^{(j)}+i V_{\l,\xx}^{(j)}=
\psi_\xx^{(j)}+\chi_\xx^{(j)}\qquad
\lis V_{\l,\xx}^{(j)}-i V_{\l,\xx}^{(j)}=\lis\psi_\xx^{(j)}+
\lis\chi_\xx^{(j)}\;.\cr}\Eq(3.21)$$
The effect of this change of variables is the following one.
If $S^{(j)}(t_\l^{(j)})\defin\sum_\xx S^{(j)}_{\xx}$, 
after the change of variables \equ(2.19) we get:
$$S^{(j)}_{\xx}=S^{(j,\psi)}_{\xx}+S^{(j,\chi)}_{\xx}
+Q_{\xx}^{(j)}\Eq(3.22)$$
where
$$\eqalign{&S^{(j,\psi)}_{\xx}={t_\l^{(j)}\over 4}\left[
\psi^{(j)}_\xx(\partial_1-i\partial_0)\psi^{(j)}_\xx+
\lis\psi_\xx^{(j)}(\partial_1+i\partial_0)\lis\psi^{(j)}_\xx\right]
+\cr &+{t_\l^{(j)}\over 4}
\left[-i\lis\psi^{(j)}_\xx(\partial_1\psi^{(j)}_x+
\partial_0\psi^{(j)}_\xx)+i\psi_\xx^{(j)}
(\partial_1\lis\psi_\xx^{(j)}+\partial_0\lis\psi_\xx^{(j)})\right]
+i\left(\sqrt 2-1-t_\l^{(j)}\right)\lis\psi_\xx^{(j)}
\psi_\xx^{(j)}\cr}\Eq(3.23))$$
with 
$$\partial_1\psi_\xx^{(j)}=\psi_{\xx+\hat e_1}^{(j)}-
\psi_\xx^{(j)}\qquad\partial_0\psi_\xx^{(j)}=\psi_{\xx+\hat e_0}^{
(\a)}-\psi_\xx^{(j)}\;.\Eq(3.24)$$
Moreover
$$\eqalign{& S^{(j,\chi)}_{\xx}={t_\l^{(j)}\over 4}
\left[\chi^{(j)}_\xx(\partial_1-i\partial_0)\chi^{(j)}_\xx+
\lis\chi_\xx^{(j)}(\partial_1+i\partial_0)\lis\chi^{(j)}_\xx\right]+\cr
&+{t_\l^{(j)}\over 4}
\left[-i\lis\chi^{(j)}_\xx(\partial_1\chi^{(j)}_x+
\partial_0\chi^{(j)}_\xx)+
i\chi_\xx^{(j)}
(\partial_1\lis\chi_\xx^{(j)}+\partial_0\lis\chi_\xx^{(j)})\right]
-i\left(\sqrt{2}+1+t_\l^{(j)}\right)
\lis\chi_\xx^{(j)}\chi_\xx^{(j)}\cr}\Eq(3.25)$$
and finally
$$\eqalign{
&Q^{(j)}_{\xx}={t_\l^{(j)}\over 4}
\Bigl[-\psi^{(j)}_\xx(\partial_1\chi^{(j)}_\xx+
i\partial_0\chi^{(j)}_\xx)
-\lis\psi^{(j)}_\xx(\partial_1\lis\chi^{(j)}_\xx-
i\partial_0\lis\chi^{(j)}_\xx)-\cr
&-\chi^{(j)}_\xx(\partial_1\psi^{(j)}_\xx+i\partial_0\psi^{(j)}_\xx)-
\lis\chi^{(j)}_\xx(\partial_1\lis\psi^{(j)}_\xx-i\partial_0
\lis\psi^{(j)}_\xx)
+i\lis\psi^{(j)}_\xx(\partial_1\chi^{(j)}_\xx-
\partial_0\chi^{(j)}_\xx)+\cr
&+i\psi^{(j)}_\xx
(-\partial_1\lis\chi^{(j)}_\xx+\partial_0\lis\chi^{(j)}_\xx)
+i\lis\chi^{(j)}_\xx(\partial_1\psi^{(j)}_\xx-\partial_0\psi^{(j)}_\xx)
+i\chi^{(j)}_\xx(-\partial_1\lis\psi^{(j)}_\xx+
\partial_0\lis\psi^{(j)}_\xx)\Bigr]\;.\cr}\Eq(3.26)$$
Formally $S^{j,\psi}$ and  $S^{j,\chi}$
are the actions of a pair of {\it Majorana}
$d=2$ fermions on a lattice with masses $\sqrt{2}-1-t^{(j)}_\l$,
and $\sqrt{2}+1+t^{(j)}_\l$, respectively; 
note that, since $-c|\l|\le t^{(j)}_\l\le 1+c|\l|$, for some $c>0$, the mass
of the $\chi$ field is always $O(1)$. On the contrary the mass of the $\psi$ 
field can be arbitrarily small; in the free case ($\l=0$) the condition 
for the theory to be massless is equivalent to the condition $t=\sqrt2-1$, that is the 
Ising's criticality condition (see the end of Chap.3); this is consistent
with the well--known property that the Ising's correlation functions decay 
as power laws if and only if we are at criticality.

It is convenient to pass from Majorana to {\it Dirac}
fermions via the change of variables 
$$\psi^\mp_{1,\xx}= {1\over\sqrt{2}} (\psi^{(1)}_\xx\pm i
\psi^{(2)}_\xx), \qquad \psi^\mp_{-1,\xx}=
{1\over\sqrt{2}}(
\lis\psi^{(1)}_\xx\pm i\lis\psi^{(2)}_\xx)\;,\Eq(2.13)$$
$$\chi^\mp_{1,\xx}= {1\over\sqrt{2}} (\chi^{(1)}_\xx\pm i
\chi^{(2)}_\xx), \qquad \chi^\mp_{-1,\xx}=
{1\over\sqrt{2}}(
\lis\chi^{(1)}_\xx\pm i\lis\chi^{(2)}_\xx)\Eq(3.27)$$
and, if $\a=\pm$, $\o=\pm 1$, 
we define $\hat\phi^\a_{\o,\kk}\defin\sum_\xx e^{-i\a\kk\xx}\phi^\a_{\o,\xx}$,
with $\phi$ denoting either $\psi$ or $\c$.\\

Let us introduce some more definitions. Let
$$t_\l\defin{t_\l^{(1)}+t_\l^{(2)}\over 2}\virg u_\l\defin 
{t_\l^{(1)}-t_\l^{(2)}\over 2}\Eq(3.28)$$
and note that $t_\l,u_\l$ as functions of $t,u$ are given by 
$$t_\l=t{1+\hat\l\over 1+\hat\l(t^2-u^2)}\virg u_\l=u{1-
\hat\l\over 1+\hat\l(t^2-u^2)}\;.\Eq(3.28a)$$
Furthermore, let 
$$Q(\psi_,\c)\defin\sum_{\xx,j}Q^{(j)}_\xx\virg 
V(\psi,\c)\defin\VV
\;,\Eq(3.30)$$
where $Q(\psi,\c)$ and $V(\psi,\c)$ must be thought as functions 
of $\psi^{\pm}$ and $\c^\pm$.
With the above definitions and using \equ(2.13), \equ(2.26)
it is straightforward algebra to verify that $\Xi^-_{AT}$
can be rewritten as:
$$\Xi_{AT}^-=e^{-E M^2}\int P(d\psi)
P(d\chi)e^{Q(\psi,\c)+\tilde\l V(\psi,\c)}\;,\Eq(3.31)$$
where $E$ is a suitable constant (we won't need its explicit
value) and $P(d\phi)$, $\phi=\psi,\c$, is:
$$\eqalign{
&P(d\phi)=\NN^{-1}_\phi\prod_{\kk\in D_{-,-}} \prod_{\o=\pm 1}
d\phi^{+}_{\kk,\o}
d\phi^{-}_{\kk,\o}
\exp\Bigl\{-{t_\l\over 4M^2}\sum_{\kk\in D_{-,-}}
{\bf\Phi}_\kk^{+,T} A_\phi(\kk){\bf\Phi}_\kk\Bigr\}\;,\cr
&A_\phi(\kk)=
\left( \matrix{i\sin k+\sin k_0 & -i \s_\phi(\kk)
& -{\m\over 2}(i\sin k+\sin k_0) & i \m(\kk)\cr
i \s_\phi(\kk) & i\sin k-\sin k_0& -i \m(\kk)& 
-{\m\over 2}(i\sin k-\sin k_0) \cr
-{\m\over 2}(i\sin k+\sin k_0)& 
i \m(\kk) & i\sin k+\sin k_0 & -i \s_\phi(\kk)\cr
-i \m(\kk)& -{\m\over 2}(i\sin k-\sin k_0)&i \s_\phi(\kk)&i\sin k-\sin k_0\cr}
\right)\;,\cr}\Eq(3.32)$$
where
$${\bf\Phi^{+,T}}_\kk=(\hat\phi^+_{1,\kk},
\hat\phi^+_{-1,\kk},\hat\phi^-_{1,-\kk},
\hat\phi^-_{-1,-\kk})\virg
{\bf\Phi^{T}}_\kk=(\hat\phi^-_{1,\kk},\hat\phi^-_{-1,\kk},
\hat\phi^+_{1,-\kk},\hat\phi^+_{-1,-\kk})
\;,\Eq(3.33)$$
$\NN_\phi$ is chosen in such a way that $\int P(d\phi)=1$ and
$$\eqalign{&\s_\phi(\kk)=2\Bigl(1+{\pm\sqrt 2+1 \over 
t_\l}\Bigr)+\cos k_0+\cos k-2\virg \mu(\kk)=-(u_\l/t_\l)(\cos k+\cos k_0)
\;.\cr}\Eq(3.34)$$
In the first of \equ(2.34) the $-$ ($+$) sign corresponds to $\phi=\psi$ 
($\phi=\c$). The parameter $\m$ in \equ(3.32) is given by 
$\m\defin\m({\bf 0})$.

It is convenient to split the $\sqrt2-1$ appearing in the definition of 
$\s_{\psi}(\kk)$ as:
$$\sqrt2-1=(\sqrt2-1+{\n\over 2})-{\n\over 2}\defin t_\psi-{\n\over 2}
\;,\Eq(nu)$$
where $\n$ is a parameter 
to be properly chosen later as a function of $\l$,
in such a way that the average location of the critical points
will be given by $t_\l=t_\psi$; in other words 
$\n$ has the role of a {\it counterterm} fixing 
the middle point of the critical temperatures. The splitting \equ(nu)
induces the following splitting of $P(d\psi)$:
$$P(d\psi)=P_\s(d\psi)e^{-\n F_\n(\psi)}\virg
F_\n(\psi)\defin {1\over 2M^2}\sum_{\kk,\o}
(-i\o)\hat\psi^+_{\o,\kk}\hat\psi^-_{-\o,\kk}\;,\Eq(ppsi)$$
where $P_\s(d\psi)$
is given by \equ(2.27) with $\phi=\psi$ and 
$\s\defin 2(1-t_\psi/t_\l)$ replacing $\s_\psi({\bf 0})$.\\

The final expression we found will be the starting point for the 
multiscale analysis of the partition function and of the correlation function,
which will occupy us in the following Chapters.

\pagina
\setcap{4. The ultraviolet integration.}
\capindex{4}{The ultraviolet integration.}
\vskip1.truecm
\section(4,The ultraviolet integration.)
\capindex{4}{The ultraviolet integration.}

Starting from this Chapter and up to Chapter 7, we will 
construct the expansion for the free energy $f$ of the Ashkin--Teller model,
see \equ(cv), and we will prove that $f$ is well
defined and analytic in $\l,t,u$ for any $t\not =t_c^\pm$, see \equ(1.3). 

It will soon be clear that a naive perturbative expansion in $\widetilde\l$
of the Grassmann functional integral in \equ(3.31) would give us
poor bounds for the partition function. This is because the propagator
of the $\psi$ fields introduced in last Chapter has a mass that is {\it
vanishing} at $t_\l=\sqrt 2-1+{\n\over 2}\pm u$, that is in correspondence of the 
``bare'' critical points. This produces infrared divergences
in the integrals defining the $n$--th order contribution
to the free energy, as obtained by this naive perturbative
expansion. It is then necessary to find out an iterative resummation 
rule, giving sense to the perturbation series. The iterative construction
we will develop is inspired to the multiscale analysis of 
Grassmann functional integrals similar to \equ(3.31), as those 
appearing in the context of non relativistic spinless fermions
in 1+1 dimensions or of 1--dim quantum spin chains [BGPS][GM][BM];
in all these problems the partition function can be written as
the integral of an exponential of a fermionic action, of the form 
of a Luttinger model action plus a perturbation, containing both a quadratic 
and a quartic term. In our case, looking at \equ(3.31) and \equ(3.32),
the Luttinger model part of the action corresponds to the diagonal elements 
of $A_{\phi}(\kk)$ plus the {\it local part} of $\tilde\l V$; 
the quadratic corrections to the non diagonal terms
of $A_{\phi}(\kk)$; the quartic corrections to the {\it non local part} 
of $\tilde\l V$.
The difference between our problem
and those already studied in the literature consists in the 
form of this perturbation; more precisely, in the form of the
quadratic corrections, which can be {\it relevant} or {\it marginal}
in a Renormalization Group sense, see next Chapter. These 
terms generate new {\it effective coupling constants}, 
whose size must be controlled throughout the Renormalization 
Group iterations. 
Moreover, our problem, formulated as a problem of 1--dim fermions,
does not have many natural symmetries that usually are present 
in a fermionic theory, such as gauge symmetry, conservation
of the particle number and of the quasi--particle number. A priori,
this could be a reason why other relevant or marginal terms, not
originaly present in the action \equ(3.31), 
could be generated by the iterative 
construction. We will use a number of hidden symmetries,
induced by the symmetries of the original spin model, 
to guarantee that these terms are not generated; by ``hidden'' here 
we mean that these symmetries, very natural in the original
spin language, are not appearent in the fermionic one.\\

In this Chapter we describe the integration of the ultraviolet
degrees of freedom, that is of the massive fields in \equ(3.31) (\ie the
$\c$ fields). This will be the first step
of our iterative construction. The subsequent steps
will be for various aspects technically very similar to the ultraviolet one, 
which we will now present in all details.
We will introduce and describe many of the technical
tools we will use throughout the work, such as the Pfaffian
expansion, the Gram--Hadamard bounds and the symmetry relations for
the fermionic fields.\\
\\
\sub(4.1){\bf The effective interaction on scale $1$}.
\\
The propagators $<\phi_{\xx,\o}^\s\phi_{\yy,\o'}^{\s'}>$ 
of the fermionic integration $P(d\phi)$, defined in \equ(3.32), verify
the following bound, for some $A,\k>0$:
$$|<\phi_{\xx,\o}^\s\phi_{\yy,\o'}^{\s'}>|\le A e^{-\k \bar m_\phi
|\xx-\yy|}\;,\Eq(4.1)$$
where $\bar m_\phi$ is the minimum between
$|m^{(1)}_{\phi}|$ and $|m^{(2)}_{\phi}|$ and
$$m^{(1)}_{\phi}\defin
2(t_\l^{(1)}-t_\phi)/t_\l\;,\qquad m^{(1)}_{\phi}\defin
2(t_\l^{(1)}-t_\phi)/t_\l\;,\Eq(4.1a)$$
where $\phi=\psi,\c$, $t_\psi$ was defined in \equ(nu) and $t_\c\defin
-\sqrt 2-1$. Note that both $m^{(1)}_{\c}$
and $m^{(2)}_{\c}$ are $O(1)$.
This suggests to integrate first the $\chi$ variables.\\

Aim of the present and of the subsequent sections is to perform the integration
of the $\c$ variables and, after that, to rewrite \equ(3.31)
in the form
$$\Xi^-_{AT}=e^{-M^2 E_1}\int P_{Z_1,\s_1,\m_1,C_1}(d\psi)
e^{-\VV^{(1)}(\sqrt{Z_1}\psi)}\;,
\qquad \VV^{(1)}(0)=0\;,\Eq(4.2)$$
where $C_1(\kk)\=1$, $Z_1=t_\psi$, $\s_1={\s/(1-{\s\over 2})}$,
$\m_1={\m/(1-{\s\over 2})}$
and $P_{Z_1,\s_1,\m_1,C_1}(d\psi)$ is the exponential 
of a quadratic form:
$$\eqalign{& P_{Z_1,\s_1,\m_1,C_1}(d\psi)=\NN_1^{-1}
\prod_{\kk\in D_{-,-}}^{\o=\pm 1}
d\psi^{+}_{\o,\kk}
d\psi^{-}_{\o,\kk}
\exp\Bigl[-{1 \over 4 M^2}\sum_{\kk\in D_{-,-}}
Z_1 C_1(\kk)
\Psi_\kk^{+,T} A_\psi^{(1)}(\kk)\Psi_\kk\Bigr]\;,\cr
&A_\psi^{(1)}(\kk)=\left( \matrix{M^{(1)}(\kk)& N^{(1)}(\kk)\cr
N^{(1)}(\kk)&M^{(1)}(\kk)\cr}\right)\cr
&M^{(1)}(\kk)=\left(\matrix{i \sin k+\sin k_0+a^+_1(\kk)& 
-i\left(\s_1+c_1(\kk)\right)\cr
i\left(\s_1+c_1(\kk)\right) & i \sin k-\sin k_0+a^-_1(\kk) 
\cr}\right)\cr
&N^{(1)}(\kk)=\left(\matrix{b^+_1(\kk)&i\left(\m_1+d_1(\kk)\right)\cr
-i\left(\mu_1+d_1(\kk)\right)& b^-_1(\kk)\cr}
\right)\;,\cr}\Eq(4.3)$$
where $\NN_1$ is chosen 
in such a way that $\int P_{Z_1,\s_1,\m_1,C_1}(d \psi)=1$.
We shall call $\VV^{(1)}$ the {\it effective interaction} on scale $1$;
it can be expressed as a sum
of monomials in $\psi$ of arbitrary order:
$$\VV^{(1)}(\psi)=\sum_{n=1}^\io
\sum_{\kk_1,\ldots,\kk_{2n}\atop\aa,\oo}
\prod_{i=1}^{2n}\hat\psi^{\a_i(\le 1)}_{\o_i,\kk_i}
\widehat W_{2n,\aa,\oo}^{(1)}(\kk_1,\ldots,\kk_{2n-1})
\d(\sum_{i=1}^{2n}\a_i\kk_i)\Eq(4.4)$$
where $\aa=(\a_1,\ldots,\a_{2n})$, $\oo=(\o_1,\ldots,\o_{2n})$,
$\a_i=\pm$, $\o_i=\pm 1$ 
and $\d(\kk)=\sum_{\nn\in\zzz^2}\d_{\kk,2\p\nn}$. The constant $E_1$ in 
\equ(4.2), the functions $a^\pm_1,b^\pm_1,c_1,d_1$ in \equ(4.3)
and the kernels $\widehat W_{2n,\aa,\oo}^{(1)}$
in \equ(4.4) satisfy natural dimensional bounds and a number
of symmetry relations, which will be described and proved below, 
where we will also show in detail how to get to \equ(4.2). 
At the end of the Chapter we will collect the results in Theorem 4.1.

Note that from now on we will consider all functions
appearing in the theory as functions of $\l,\s_1,\m_1$ (of course
$t$ and $u$ can be analytically and elementarily expressed in terms 
of $\l,\s_1,\m_1$). We shall also assume $|\s_1|,|\m_1|$ bounded 
by some $O(1)$ constant.
Note that if $t\pm u$ belong
to a sufficiently small interval $D$ centered around $\sqrt2-1$, 
as assumed in the hypothesis of the Main Theorem in the Introduction, 
then of course 
$|\s_1|,|\m_1|\le c_1$ for a suitable constant $c_1$ (for instance, 
if $D=[{3(\sqrt2-1)\over 4},{5(\sqrt2-1)\over 4}]$, that is
a possible choice for the interval $D$,
we find $|\s_1|\le 1+O(\e)$ and $|\m_1|\le 2+O(\e)$).\\
\\
\sub(4.2){\bf The integration of the $\c$ fields}.
\\
We start with considering \equ(3.31), with $P(d\psi)$ rewritten as in 
\equ(ppsi), and we define:
$$e^{-\tilde E_1 M^2-Q^{(1)}(\psi)-\VV^{(1)}(\psi)}
\defin\int P(d\c)e^{Q(\psi,\c)-\n F_\s(\psi)+
\tilde\l V(\psi,\c)}\;,\Eq(4.5)$$
where $\widetilde E_1$ is a constant, $Q^{(1)}$ is quadratic in $\psi$ and 
$O(1)$ w.r.t. $\l,\n$ and $\VV^{(1)}$ is at least quadratic in $\psi$
and $O(\l,\n)$.
$Q^{(1)}$ will contribute to the free measure $P_{Z_1,\s_1,\m_1,C_1}$.

We calculate $\VV^{(1)}$ in terms of {\it truncated expectations} (see Appendix A1), 
defined as:
$$\EE^T_\chi(X;n)={\partial^n\over\partial\a^n}\log\int P(d\chi)
e^{\a X(\chi)}|_{\a=0}\;,\Eq(4.5a)$$
where $P(d\c)$ is defined in \equ(3.32) and the associated propagator is 
given by
$$\eqalign{&g^\c_{(-,\o),(+,\o)}(\xx-\yy)\defin<\c^-_{\xx,\o}\c^+_{\yy,\o}>=
g^{\c(1)}_\o(\xx-\yy)+g^{\c(2)}_\o(\xx-\yy)\cr
&g^\c_{(-,\o),(+,-\o)}(\xx-\yy)\defin<\c^-_{\xx,\o}\c^+_{\yy,-\o}>=
g^{\c(1)}_{\o,-\o}(\xx-\yy)+g^{\c(2)}_{\o,-\o}(\xx-\yy)\cr
&g^\c_{(\a,\o),(\a,-\o)}(\xx-\yy)\defin<\c^\a_{\xx,\o}\c^\a_{\yy,-\o}>=
g^{\c(1)}_{\o,-\o}(\xx-\yy)-g^{\c(2)}_{\o,-\o}(\xx-\yy)\cr
&g^\c_{(\a,\o),(\a,\o)}(\xx-\yy)\defin<\c^\a_{\xx,\o}\c^\a_{\yy,\o}>=
g^{\c(1)}_{\o}(\xx-\yy)-g^{\c(2)}_{\o}(\xx-\yy)
\;,\cr}\Eq(4.6)$$
where, for $j=1,2$,
$$\eqalign{&g^{\c(j)}_{\o}(\xx-\yy)={2\over t_\l}
{1\over M^2}
\sum_\kk e^{-i\kk(\xx-\yy)} {\z_j\left(-i\sin k+\o\sin k_0\right)
\over \z_j^2\left(\sin^2 k+
\sin^2 k_0\right)+(m^{(j)}_{\c,\kk})^2}\cr
&g^{\c(j)}_{\o,-\o}(\xx-\yy)={2\over t_\l}
{1\over M^2}
\sum_\kk e^{-i\kk(\xx-\yy)} {-i\o m^{(j)}_{\c,\kk}\over \z_j^2
\left(\sin^2 k+\sin^2 k_0\right)+(m^{(j)}_{\c,\kk})^2}\;,\cr}\Eq(4.7)$$
with $m^{(j)}_{\c,\kk}=\s_\c(\kk)+(-1)^j\mu(\kk)$ and $\z_j=1+(-1)^j
(\m/2)$. Calling $m^{(j)}_{\c}\defin m^{(j)}_{\c,{\bf 0}}$ 
one can easily verify that $m^{(j)}_{\c}$ is given by \equ(4.1a)
and the propagators are bounded as in \equ(4.1), for some $\k>0$.
The similar equations and bounds for the $\psi$ propagators
are proven in the same way.\\

Calling
$$-\lis \VV(\psi,\c)=Q(\psi,\c)-\n F_\s(\psi)
+\tilde\l V(\psi,\c)\;,\Eq(4.8)$$
and using the rules in Appendix A1,
we obtain
$$M^2 \widetilde E_1+Q^{(1)}(\psi)+
\VV^{(1)}(\psi)=-\log \int P(d\chi)e^{-\lis \VV(\psi,\c)}
=\sum_{n=0}^\io{(-1)^{n+1}\over n!}\EE^T_\chi(\lis\VV;n)\;.
\Eq(4.9)$$
We label each one of the monomials in $\lis\VV$ 
by an index $v_i$, so that
each monomial in $\lis\VV$ can be written as
$$\sum_{\xx_{v_i}} K_{v_i}(\xx_{v_i})
\prod_{f\in \tilde P_{v_i}}
\psi^{\a(f)}_{\o(f),\xx(f)}
\prod_{f\in P_{v_i}}\chi^{\a(f)}_{\o(f),\xx(f)}\;,\Eq(4.10)$$
where $\xx_{v_i}$ is the total set of coordinates
associated to $v_i$, $K_{v_i}(\xx_{v_i})$ is a bounded compact
support function and $P_{v_i}$ and $\widetilde P_{v_i}$
are the set of indices labelling the $\chi$ or $\psi$-fields
in the monomial $v_i$; the labels $\a(f),\o(f),\xx(f)$ assume
values in the sets $\{\pm\},\{\pm 1\}$ and $\L_M$ respectively. 
We can write
$$\eqalign{&\VV^{(1)}(\psi)=\sum_{\tilde P_{v_0}\not=0}
\VV^{(1)}(\widetilde P_{v_0})\virg 
\VV^{(1)}(\widetilde P_{v_0})= \sum_{\xx_{v_0}}
\big[\prod_{f\in\tilde P_{v_0}}\psi_{\o(f),\xx(f)}^{\a(f)}\big]
K_{\tilde P_{v_0}}(\xx_{v_0})\cr
& K_{\tilde P_{v_0}}(\xx_{v_0})=\sum_{s=1}^\io
{1\over s!} \sum_{i_1,\ldots,v_s}^*
\EE^T_{\chi}\left(
\widetilde\chi(P_{v_1}),\ldots,
\widetilde\chi(P_{v_s})\right)
\prod_{i=1}^s K_{v_i}(\xx_{v_i})\;,\cr}\Eq(4.11)$$
where $\widetilde\chi(P_{v_i})=\prod_{f\in P_{v_i}}
\chi^{\a(f)}_{\xx(f),\o(f)}$
and the $*$ on the sum means that we are excluding the case
$v_1,\ldots,v_s$ all come from $Q(\psi,\c)$ (such terms
will contribute, by definition, to $Q^{(1)}(\psi)$). 
Furthermore $\sum_{i_1,\ldots,i_s}\le c^s$, for some constant $c$, 
$\widetilde P_{v_0}=\bigcup_i \widetilde P_{v_i}$ and $\xx_{v_0}=\bigcup_i
\xx_{v_i}$.

We use now a generalization of a well known expression for $\EE^T_\chi$ [Le],
proven in Appendix A2:
$$\EE^T_{\chi}(\widetilde\chi(P_{v_1}),\ldots,\widetilde\chi(P_{v_n}))=
\sum_{T}\a_T\prod_{\ell\in T}
g_\c(f^1_\ell,f^2_\ell)
\int dP_{T}(\tt) \Pf G^{T}(\tt)\Eq(4.12)$$
where:\\
\0a) $T$ is a set of lines forming an {\it anchored tree} between
the cluster of points $P_{v_1},\ldots,P_{v_s}$ \ie $T$ is a set
of lines which becomes a tree if one identifies all the points
in the same clusters;\\
\0b) $\a_T$ is a sign (irrelevant for the subsequent bounds);\\
\0c) given $\ell\in T$, let $f^1_\ell, f^2_\ell$ the field labels associated
to the points connected by $\ell$; $g_\c(f^1_\ell,f^2_\ell)$ is defined as:
$$g_\c(f^1_\ell,f^2_\ell)\defin g^\c_{\underline a(f^1_\ell),\underline 
a(f^2_\ell)}(\xx(f^1_\ell)-\xx(f^2_\ell))\virg \underline a(f)=(\a(f),\o(f))
\;;\Eq(4.13)$$
\0d) $\tt=\{t_{i,i'}\in [0,1], 1\le i,i' \le s\}$, $dP_{T}(\tt)$
is a probability measure with support on a set of $\tt$ such that
$t_{i,i'}=\uu_i\cdot\uu_{i'}$ for some family of vectors $\uu_i\in \RRR^n$ of
unit norm;\\
\0e) if $2n=\sum_{i=1}^s|P_{v_i}|$, then 
$G^{T}(\tt)$ is a $(2n-2s+2)\times (2n-2s+2)$ antisymmetrix matrix, whose
elements are given by $G^{T}_{f,f'}=t_{i(f),i(f')}g_\c(f,f')$, where:
$f, f'\not\in F_T$ and $F_T\defin\cup_{\ell\in T}\{f^1_\ell,f^2_\ell\}$;
$i(f)$ is s.t. $f\in P_{i(f)}$;\\
\0f) $\Pf G^T$ is the {\it Pfaffian} of $G^T$; given an antisymmetrix matrix
$A_{ij}=-A_{ji}$, $i,j=1,\ldots,2k$, its Pfaffian is defined as
$$\eqalign{\Pf A&={1\over 2^k k!}\sum_{\p}(-1)^\p A_{\p(1)\p(2)}\cdots 
A_{\p(2k-1)\p(2k)}\cr
&=\int d\psi_1\cdots d\psi_{2k}e^{-{1\over 2}\sum_{i,j}\psi_i 
A_{ij}\psi_j }\;,\cr}\Eq(4.14)$$
where in the first line $\p$ is a permutation of $\{1,\ldots,2k\}$ and 
$(-1)^\p$ is its parity
while, in the second line, $\psi_1,\ldots,\psi_{2k}$ are Grassmanian variables.
A well known property is that $(\Pf A)^2=\det A$.
\\

If $s=1$ the sum over $T$ is empty, but we can still
use the above equation by interpreting the r.h.s.
as $1$ if $P_{v_1}$ is empty, and $\Pf G^T(P_{v_1})$ otherwise.\\

In order to bound $\Pf G^T$ we first use $|\Pf G^T|=\sqrt{|\det G^T|}$
and then, in order to bound the determinant, 
the {\it Gram-Hadamard inequality}, proven in Appendix A3, stating
that, if $M$ is a square matrix with elements $M_{ij}$ of the form
$M_{ij}=<A_i,B_j>$, where $A_i$, $B_j$ are vectors in a Hilbert space with
scalar product $<\cdot,\cdot>$, then
$$|\det M|\le \prod_i ||A_i||\cdot ||B_i||\;.\Eq(4.15)$$
where $||\cdot||$ is the norm induced by the scalar product.
 
Let $\HH=\RRR^n\otimes \HH_0$, where $\HH_0$ is the Hilbert space of complex
four dimensional vectors $F(\kk)=(F_1(\kk),\ldots,$
$\ldots,F_4(\kk))$, $F_i(\kk)$
being a function on the set $\DD_{M}$, with scalar product
$$<F,G>=\sum_{i=1}^4 {1\over M^2}\sum_{\kk} F^*_i(\kk) 
G_i(\kk)\;.\Eq(4.16)$$
It is easy to verify that
$$G_{f,f'}=t_{i(f),i(f')} g_\c(f,f')=
<\uu_{i(f)}\otimes A_{f},
\uu_{i(f')}\otimes B_{f'}>\;,\Eq(4.17)$$
where $\uu_i\in \RRR^n$, $i=1,\ldots,n$, are vectors such that
$t_{i,i'}=\uu_i\cdot\uu_{i'}$, and, if 
$\hat g^\c_{\underline a,\underline a'}(\kk)$ is the Fourier transform of
$g^\c_{\underline a,\underline a'}(\xx-\yy)$, 
$A_f(\kk)$ and $B_{f'}(\kk)$ are given by
$$\eqalign{& A_f(\kk)=e^{-i\kk\xx(f)}\Big(\hat g^\c_{
\underline a(f),(-,1)}(\kk),
\hat g^\c_{\underline a(f),(-,-1)}(\kk),
\hat g^\c_{\underline a(f),(+,1)}(\kk),
\hat g^\c_{\underline a(f),(+,-1)}(\kk)\Big)\;,\cr
&B_{f'}(\kk)=e^{-i\kk\xx(f')}\cases{(1,0,0,0),& if $\underline a(f')=(-,1)$,\cr
(0,1,0,0),& if $\underline a(f')=(-,-1)$,\cr
(0,0,1,0),& if $\underline a(f')=(+,1)$,\cr
(0,0,0,1),& if $\underline a(f')=(+,-1)$,\cr}}
\Eq(4.18)$$
Note that $||A_f||\le C$, for some $C=O(1)$, and $||B_{f'}||=4$.
Hence we have proved that
$$|\Pf G^{T}|=\sqrt{|\det G^T|}\le c^s\;,\Eq(4.19)$$
for some $c=O(1)$ (we used that $2n\le 4s$).
Finally we get
$$\sum_{\xx_{v_0}}|K_{\tilde P_{v_0}}(\xx_{v_0})|
\le \sum_{s=1}^\io{c^s\over s!}
\sum_{v_1,\ldots,v_s}
\sum_{\xx_{v_1},\ldots,\xx_{v_s}}\sum_{T}\prod_{\ell\in T}
|g_\c(f^1_\ell,f^2_\ell)|
\prod_{i=1}^s |K_i(\xx_{v_i})|\Eq(4.20)$$
where we have used that $\int d P_{T}(\tt)=1$.
The number of addenda in $\sum_T$
is bounded by $s!c^s$. Finally $T$ and the $\bigcup_i \xx_{v_i}$
form a tree connecting all points, so that, using that 
the propagators decay exponentially on scale $O(1)$ 
and that the interactions are short ranged, we find
that, if $|\n|\le c|\l|$,
$$\sum_{v_1,\ldots,v_s}\sum_{\xx_{v_1},\ldots,\xx_{v_s}}\sum_{T}\prod_{
\ell\in T}
|g_\c(f^1_\ell,f^2_\ell)|
\prod_{i=1}^s |K_i(\xx_{v_i})|\le c^s s! |\l|^m M^2\;,\Eq(4.21)$$
where $m$ is the number of 
couplings $O(\l,\n)$ ($m\ge 1$ by construction).

Note that if $v_i$ only come from 
$-\lis \VV(\psi,\chi)-Q(\psi,\c)$,
then $m=s$. Let us consider now the case in which
there are $n_0$ end-points associated to $Q(\psi,\chi)$,
which have $O(1)$ coupling. In this case
$n_0\le|\widetilde P_{v_0}|$. In fact in $Q(\psi,\chi)$
there are only terms of the form $\psi_\xx\chi_{\xx'}$, where $\xx'$
is either $\xx$ or $\xx\pm\hat e_0$ or $\xx\pm\hat e_1$, 
so at most the number of them is equal to the number of $\psi$ fields.
If we call $n_\l\le m$
the number of vertices quartic in the fields it is clear that
$n_\l\ge\max\{1,|\widetilde P_{v_0}|/2-1\}$. Hence
$$\sum_{\xx_{v_0}}|K_{\tilde P_{v_0}}(\xx_{v_0})|
\le M^2 \sum_{n_0=0}^{|\tilde P_{v_0}|}c^{n_0}\sum_{m=1}^\io c^m|\l|^{m/2}
|\l|^{\max\{1/2,|\tilde P_{v_0}|/4-1/2\}}\Eq(4.22)$$
The last bound implies that the kernels $\widehat W_{2n,\aa,\oo}^{(1)}$ 
in \equ(4.4), which are the Fourier transforms of 
$K_{\tilde P_{v_0}}(\xx_{v_0})$, see \equ(4.11), can be bounded as:
$$|\widehat W_{2n,\aa,\oo}^{(1)}(\kk_1,\ldots,\kk_{2n-1})|
\le M^2 C^n|\l|^{\max\{1,n/2\}}\;;
\Eq(4.23)$$

We now turn to the construction of $P_{Z_1,\s_1,\m_1,C_1}$. 
We define:
$$e^{-t_1 M^2} P_{Z_1,\s_1,\m_1,C_1}(d\psi)
\defin P_\s(d\psi)e^{-Q^{(1)}(\psi)}
\;,\Eq(4.24)$$
where $t_1$ is chosen in such a way $\int P_{Z_1,\s_1,\m_1,C_1}(d\psi)=1$.
From definition \equ(4.24), \equ(4.2) follows, with 
$E_1=\widetilde E_1+t_1$ ($\widetilde E_1$ was defined in \equ(4.5))
and $\VV^{(1)}(\psi)$ constructed above.

Let us now study in more detail the structure of $P_{Z_1,\s_1,\m_1,C_1}
(d\psi)$. In order to write it  
as an exponential of a quadratic form, it is sufficient to calculate
the correlations
$$\eqalign{<\psi^{\a_1}_{\o_1,\kk}
\psi^{\a_2}_{\o_2,-\a_1\a_2\kk}>_1&\defin
\int P_{Z_1,\s_1,\m_1,C_1}(d\psi)\psi^{\a_1}_{\o_1,\kk}
\psi^{\a_2}_{\o_2,-\a_1\a_2\kk}=\cr
&=e^{-t_1M^2}
\int P_\s(d\psi)P(d\c)e^{Q(\c,\psi)}\psi^{\a_1}_{\o_1,\kk}
\psi^{\a_2}_{\o_2,-\a_1\a_2\kk}\;.\cr}\Eq(4.25)$$
It is easy to realize that the measure
$\sim P_\s(d\psi)P(d\c)e^{Q(\c,\psi)}$ 
factorizes into the product of two
measures generated by the fields $\psi^{(j)}_{\o,\xx}$, $j=1,2$,
defined by $\psi^\a_{\o,\xx}=(\psi^{(1)}_{\o,\xx}+i(-1)^\a\psi^{(2)}_{
\o,\xx})/\sqrt2$.
In fact, using this change of variables, one finds that 
$$ P_\s(d\psi)P(d\c)e^{Q(\c,\psi)}=\prod_{j=1,2}\lis P^{(j)}
(d\psi^{(j)},d\c^{(j)})\;,\Eq(4.26)$$
with
$$\lis P^{(j)}(d\psi^{(j)},d\c^{(j)})\defin
{1\over \NN^{(j)}}\prod_{\xx}d\psi_\xx^{(j)}d
\lis\psi_\xx^{(j)}d\c_\xx^{(j)}d
\lis\c_\xx^{(j)}e^{\sum_\xx[S^{(j,\psi)}_{\n,\xx}+ 
S^{(j,\c)}_{\xx}+
Q_\xx^{(j)}]}
\;,\Eq(4.27)$$
with $j=1,2$, $S^{(j,\c)}_\xx, Q_\xx^{(j)}$ defined 
as in \equ(3.25), \equ(3.26) and 
$$S^{(j,\psi)}_{\n,\xx}=
S^{(j,\psi)}_{\xx}+i(t_\psi-\sqrt2 +1)\lis\psi^{(j)}_\xx\psi^{(j)}_\xx
\;,\Eq(4.28)$$
where $S^{(j,\psi)}_{\xx}$ is defined as in \equ(3.23).
Substituting these expressions in \equ(4.27), we find
that, if $\x^{(j),T}_\kk\defin$ $\defin(\psi_\kk^{(j)}, \lis\psi_\kk^{(j)}, 
\c_\kk^{(j)}, \lis\c_\kk^{(j)})$,
$$\eqalign{&\lis P^{(j)}(d\psi^{(j)},d\c^{(j)})=
{1\over \NN^{(j)}}\exp\{-{t_\l^{(j)}\over 4M^2}\sum_{\kk}
\x^{(j),T}_\kk C^{(j)}_\kk \x^{(j)}_{-\kk}\}\cr
& C^{(j)}_\kk\defin\left(\matrix{-i\sin k-\sin k_0& -i m^{(j)}_{\psi,\kk}
& i\sin k-\sin k_0& i(\cos k-\cos k_0)\cr
i m^{(j)}_{\psi,\kk}&-i\sin k+\sin k_0&-i(\cos k-\cos k_0)&i\sin k+
\sin k_0\cr
i\sin k-\sin k_0& i(\cos k-\cos k_0)&-i\sin k-\sin k_0& -i m^{(j)}_{\c,\kk}\cr
-i(\cos k-\cos k_0)&i\sin k+\sin k_0&i m^{(j)}_{\c,\kk}&-i\sin k+\sin k_0\cr}
\right)\;.\cr}\Eq(4.29)$$
A lengthy but straightforward calculation shows that 
the determinant $B^{(j)}(\kk)\defin \det C^{(j)}_\kk$ is equal to
$$B^{(j)}(\kk)={16\over (t_\l^{(j)})^4}\big\{
2t_\l^{(j)}[1-(t_\l^{(j)})^2](2-\cos k-\cos k_0)+(t_\l^{(j)}-t_\psi)^2
(t_\l^{(j)}-t_\c)^2\big\}\Eq(4.30)$$
Using, for $l,m=1,\ldots,4$, the algebraic identity 
$${1\over \NN^{(j)}}\int\big[\prod_{\kk,i}(d\x^{(j)}_\kk)_i\big] 
(\x_{-\kk'}^{(j)})_l(\x_{\kk'}^{(j)})_m\exp\{
-{t_\l^{(j)}\over 4M^2}\sum_{\kk}\x_\kk^{(j),T}
C^{(j)}_\kk\x^{(j)}_{-\kk}\}
={4M^2\over t_\l^{(j)}}(C^{(j)}_{\kk'})^{-1}_{lm}\;,\Eq(4.31)$$
we find:
$$\eqalign{
&<\psi^{(j)}_{-\kk}\psi^{(j)}_{\kk}>_1={4M^2\over t_\l^{(j)}}
{c_{1,1}^{(j)}(\kk)\over B^{(j)}(\kk)}\virg 
<\lis\psi^{(j)}_{-\kk}\psi^{(j)}_{\kk}>_1={4M^2\over t_\l^{(j)}}
{c_{-1,1}^{(j)}(\kk)\over B^{(j)}(\kk)}\virg\cr
&<\lis\psi^{(j)}_{-\kk}\lis\psi^{(j)}_{\kk}>_1={4M^2\over t_\l^{(j)}}
{c_{-1,-1}^{(j)}(\kk)\over B^{(j)}(\kk)}\;,\cr}\Eq(4.32)$$
where, if $\o=\pm 1$, recalling that $t_\psi=\sqrt 2 -1+\n/2$ and $t_\c=
-\sqrt 2 - 1$,
$$\eqalign{c_{\o,\o}^{(j)}(\kk)\defin&
{4\over (t_\l^{(j)})^2}\big\{2t_\l^{(j)}t_\c(-i\sin k\cos k_0+\o 
\sin k_0\cos k)+[(t_\l^{(j)})^2+t_\c^2](i\sin k-\o\sin k_0)\big\}\cr
c^{(j)}_{\o,-\o}(\kk)\defin& -i\o{4\over (t_\l^{(j)})^2}\big\{
-t_\l^{(j)}(3t_\c+t_\psi)\cos k\cos k_0+[(t_\l^{(j)})^2+2t_\c t_\psi+
t_\c^2](\cos k+\cos k_0)-\cr
&\qquad\qquad -\big(t_\l^{(j)}(t_\psi+t_\c)+2{t_\psi t_\c^2\over 
t_\l^{(j)}}\big)\big\}\;.\cr}\Eq(4.33)$$
$P_{Z_1,\s_1,\m_1,C_1}(d\psi)$ can now be written in terms of these
correlations, as 
$$P_{Z_1,\s_1,\m_1,C_1}(d\psi)=
P^{(1)}(d\psi^{(1)})P^{(2)}(d\psi^{(2)})\;,\Eq(4.33a)$$
with
$$\eqalign{&
P^{(j)}(d\psi^{(j)})\defin{1\over N_j}\prod_\kk d\psi^{(j)}_\kk 
d\lis\psi^{(j)}_\kk\cdot\cr
&\cdot\exp\Bigl\{-{t_\l^{(j)}B^{(j)}(\kk)\over 4M^2\det c^{(j)}_\kk}
(\psi^{(j)}_\kk, \lis\psi^{(j)}_{\kk})\left(\matrix{c^{(j)}_{-1,-1}(\kk)
&-c^{(j)}_{1,-1}(\kk)\cr-c^{(j)}_{-1,1}(\kk)&c^{(j)}_{1,1}(\kk)\cr}\right)
\left(\matrix{\psi^{(j)}_{-\kk}\cr \lis\psi^{(j)}_{-\kk}\cr}\right)
\Bigr\}\;,\cr}\Eq(4.34)$$
where $\det c^{(j)}_\kk=c^{(j)}_{1,1}(\kk)c^{(j)}_{-1,-1}(\kk)-
c^{(j)}_{1,-1}(\kk)c^{(j)}_{-1,1}(\kk)$.
If we now use the identity $t_\l^{(j)}=t_\psi(2+(-1)^j\m)/(2-\s)$
and rewrite
the measure $P^{(1)}(d\psi^{(1)})P^{(2)}(d\psi^{(2)})$
in terms of $\psi^{\pm}_{\o,\kk}$ we find:
$$P_{Z_1,\s_1,\m_1,C_1}(d\psi)=
{1\over \NN^{(1)}}\prod_{\kk,\o} d\psi^{+}_{\o,\kk}d\psi^{-}_{\o,\kk}
\exp\{-{Z_1 C_1(\kk)
\over 4M^2}\Psi^{+,T}_\kk A_\psi^{(1)}\Psi^-_\kk\}\;,\Eq(4.35)$$
with $C_1(\kk)$, $Z_1$, $\s_1$ and $\m_1$
defined as after \equ(4.2), and
$A^{(1)}_\psi(\kk)$ as in \equ(4.3), with
$$\eqalign{& M^{(1)}(\kk)={2\over 2-\s}\left(\matrix{-c^+_{-1,-1}(\kk)&
c^+_{-1,1}(\kk)\cr
c^+_{1,-1}(\kk) & -c^+_{1,1}(\kk)\cr}\right)\virg
N^{(1)}(\kk)={2\over 2-\s}\left(\matrix{-c^-_{-1,-1}(\kk)&
c^-_{-1,1}(\kk)\cr
c^-_{1,-1}(\kk) & -c^-_{1,1}(\kk)\cr}\right)\;,\cr}
\Eq(4.36)$$
where $c^\a_{\o_1,\o_2}(\kk)\defin 
[(1-\m/2)B^{(1)}(\kk)c^{(1)}_{\o_1,\o_2}(\kk)/\det c^{(1)}_\kk+\a(1+\m/2)
B^{(2)}(\kk)c^{(2)}_{\o_1,\o_2}(\kk)/\det c^{(2)}_\kk]/2$.
It is easy to verify that $A_\psi^{(1)}(\kk)$ can be 
written in the same form as \equ(4.3).
In fact, computing the functions in \equ(4.36), one finds that, for $\kk$,
$\s_1$ and $\m_1$ small,
$$\eqalign{&M^{(1)}(\kk)=\left(\matrix{(i\sin k+\sin k_0)\big(1+O(\s_1)
\big)+O(\kk^3) &
-i\s_1+O(\kk^2)\cr
i\s_1+O(\kk^2)
&(i\sin k-\sin k_0)\big(1+O(\s_1)\big)+O(\kk^3)\cr}\right)\cr
&N^{(1)}(\kk)=\left(\matrix{(i\sin k+\sin k_0)O(\m_1)+O(\kk^3)&
i\m_1+O(\m_1\kk^2)\cr
-i\m_1+O(\m_1\kk^2) & 
(i\sin k-\sin k_0)O(\m_1)+O(\kk^3)\cr}\right)\;,\cr}
\Eq(4.37)$$
where the higher order terms in $\kk$, $\s_1$ and $\m_1$ 
contribute to the corrections $a_1^\pm(\kk)$, $b_1^{\pm}(\kk)$,
$c_1(\kk)$ and $d_1(\kk)$. 
\\
\\
\sub(4.3){\bf Symmetry properties.}
\\
In this section we identify some symmetries of 
model \equ(3.19) and, using these symmetry properties, 
we prove that the quadratic and quartic terms in $\VV^{(1)}$ 
and the corrections $a_1^\pm(\kk)$, $b_1^{\pm}(\kk)$,
$c_1(\kk)$ and $d_1(\kk)$ appearing in \equ(4.3)
have 
a special structure, described in Theorem 4.1 below.\\

We start with noting that the formal action appearing in \equ(3.19) (see also
\equ(2.14), \equ(2.340) and \equ(3.20) for an explicit form
of the different contributions appearing in \equ(3.19))
is invariant under the following transformations.\\  

\01) {\it Parity}:
$$H_\xx^{(j)}\to\lis H_{-\xx}^{(j)}\;,\qquad \lis H_\xx^{(j)}\to 
-H_{-\xx}^{(j)}\;,\qquad V_\xx^{(j)}\to\lis V_{-\xx}^{(j)}\;,\qquad 
\lis V_\xx^{(j)}\to -V_{-\xx}^{(j)}\;.
\Eq(b1)$$
In terms of the variables $\hat\psi^{\a}_{\o,\kk}$, this transformation
is equivalent to $\hat\psi^\a_{\o,\kk}\to i\o\hat\psi^\a_{\o,-\kk}$ 
(the same for $\c$) and we shall call it {\it parity}.\\

\02) {\it Complex conjugation}:
$$\psi_\xx^{(j)}\to \lis\psi_\xx^{(j)}\;,\qquad
\lis\psi_\xx^{(j)}\to\psi_\xx^{(j)}\;,\qquad
\chi_\xx^{(j)}\to \lis\chi_\xx^{(j)}\;,\qquad
\lis\chi_\xx^{(j)}\to\chi_\xx^{(j)}\;,\qquad c\to c^*\;,\Eq(b2)$$
where $c$ is a generic constant appearing in the formal action and
$c^*$ is its complex conjugate. In terms of the variables 
$\hat\psi^{\a}_{\o,\kk}$, this transformation is equivalent to  
$\hat\psi^\a_{\o,\kk}\to\hat\psi^{-\a}_{-\o,\kk}$ (the same for $\c$), 
$c\to c^*$ and we shall 
call it {\it complex conjugation}.\\

\03) {\it Hole-particle}:
$$\eqalign{&H_\xx^{(j)}\to(-1)^{j+1}H_{\xx}^{(j)}\;,
\qquad \lis H_\xx^{(j)}\to 
(-1)^{j+1}\lis H_{\xx}^{(j)}\;,\cr
&V_\xx^{(j)}\to(-1)^{j+1} V_{\xx}^{(j)}\;,\qquad 
\lis V_\xx^{(j)}\to (-1)^{j+1}\lis V_{\xx}^{(j)}\;.\cr}
\Eq(b3)$$
This transformation is equivalent to $\hat\psi^\a_{\o,\kk}\to\hat
\psi^{-\a}_{\o,-\kk}$ (the same for $\c$) and we shall 
call it {\it hole-particle}.\\

\04) {\it Rotation}:
$$\eqalign{&H_{x,x_0}^{(j)}\to i\lis V_{-x_0,-x}^{(j)}\;,
\qquad \lis H_{x,x_0}^{(j)}\to i V_{-x_0,-x}^{(j)}\;,\cr
&V_{x,x_0}^{(j)}\to i\lis 
H_{-x_0,-x}^{(j)}\;,\qquad \lis V_{x,x_0}^{(j)}\to i H_{-x_0,-x}^{(j)}\;.
\cr}\Eq(b4)$$
This transformation is equivalent to
$$\hat\psi^\a_{\o,(k,k_0)}\to-\o e^{-i\o\p/4}\hat
\psi^{\a}_{-\o,(-k_0,-k)}\virg\hat\c^\a_{\o,(k,k_0)}\to\o e^{-i\o\p/4}\hat
\c^{\a}_{-\o,(-k_0,-k)}\Eq(b5)$$ 
and we shall call it {\it rotation}.\\

\05) {\it Reflection}:
$$\eqalign{&H_{x,x_0}^{(j)}\to i\lis H_{-x,x_0}^{(j)}\;,
\qquad \lis H_{x,x_0}^{(j)}\to i H_{-x,x_0}^{(j)}\;,\cr
&V_{x,x_0}^{(j)}\to -iV_{-x,x_0}^{(j)}\;,\qquad \lis V_{x,x_0}^{(j)}\to 
i \lis V_{-x,x_0}^{(j)}\;.
\cr}\Eq(b4a)$$
This transformation is equivalent to $\hat\psi^\a_{\o,(k,k_0)}\to i\hat
\psi^{\a}_{-\o,(-k,k_0)}$
(the same for $\c$) and we shall call it {\it reflection}.\\

\06) {\it The $(1)\otto (2)$ symmetry}.
$$\eqalign{&H_{\xx}^{(1)}\otto H_{\xx}^{(2)}\virg
\lis H_{\xx}^{(1)}\otto\lis H_{\xx}^{(2)}\;,\cr
&V_{\xx}^{(1)}\otto V_{\xx}^{(2)}\virg
\lis V_{\xx}^{(1)}\otto\lis V_{\xx}^{(2)}\virg u\to -u\;.
\cr}\Eq(b4otto)$$
This transformation is equivalent to
$\hat\psi^\a_{\o,\kk}\to-i\a\hat\psi^{-\a}_{\o,-\kk}$ (the same for $\c$)
together with $u\to -u$ and we shall call it {\it 
$(1)\otto(2)$ symmetry}.\\

It is easy to verify that the quadratic forms $P(d\c)$, $P(d\psi)$ 
and $P_{Z_1,\s_1,\m_1,C_1}(d\psi)$
are separately invariant
under the symmetries above. Then the effective action $\VV^{(1)}(\psi)$
is still invariant under the same symmetries.
Using the invariance of $\VV^{(1)}$ under transformations (1)--(6), we now
study in detail the structure of its quadratic and quartic terms.\\

\0{\it Quartic term.} Let us consider in \equ(4.4) the term with
$2n=4$, $\a_1=\a_2=-\a_3=-\a_4=+$, $\o_1=-\o_2=\o_3=-\o_4=1$;
for simplicity of notation, let us denote it with 
$\sum_{\kk_i}W(\kk_1,\kk_2,\kk_3,\kk_4)
\hat\psi^+_{1,\kk_1}\hat\psi^+_{-1,\kk_2}
\hat\psi^-_{-1,\kk_3}\hat\psi^-_{1,\kk_4}$ $\d(\kk_1+\kk_2-\kk_3-\kk_4)$.
Under complex conjugation it becomes equal to
$\sum_{\kk_i}W^*(\kk_1,\kk_2,\kk_3,\kk_4)
\hat\psi^-_{-1,\kk_1}\hat\psi^-_{1,\kk_2}
\hat\psi^+_{1,\kk_3}\hat\psi^+_{-1,\kk_4}$ $\d(\kk_3+\kk_4-\kk_1-\kk_2)$,
so that $W(\kk_1,\kk_2,\kk_3,\kk_4)=W^*(\kk_3,\kk_4,\kk_1,\kk_2)$.

Then, defining $L_1=W(\bk++,\bk++,\bk++,\bk++)$, where
$\bk++=(\p/M,\p/M)$,
and $l_1=\PP_0 L_1\defin L_1\big|_{\s_1=\m_1=0}$,
we see that $L_1$ and $l_1$ are
real. From the explicit computation of the 
lower order term we find $l_1=\widetilde\l/Z_1^2+O(\l^2)$.\\

\0{\it Quadratic terms.}
We distinguish 4 cases (items (a)--(d) below).\\
{\it \0a)} Let us consider in \equ(4.4) the term with
$2n=2$, $\a_1=-\a_2=+$ and $\o_1=-\o_2=\o$; let us denote it with 
$\sum_{\o,\kk}W_\o(\kk;\m_1)\hat\psi^+_{\o,\kk}
\hat\psi^-_{-\o,\kk}$.
Under parity it becomes 
$$\sum_{\o,\kk}W_\o(\kk;\m_1)
(i\o)\hat\psi^+_{\o,-\kk}(-i\o)\hat\psi^-_{-\o,-\kk}=
\sum_{\o,\kk}W_\o(-\kk;\m_1)\hat\psi^+_{\o,\kk}\hat\psi^-_{-\o,\kk}
\;,\Eq(B1)$$
so that $W_\o(\kk;\m_1)$ is even in $\kk$.\\
Under complex conjugation it becomes 
$$\sum_{\o,\kk}W_\o(\kk;\m_1)^*
\hat\psi^-_{-\o,\kk}\hat\psi^+_{\o,\kk}=-\sum_{\o,\kk}W_\o(\kk;\m_1)^*
\hat\psi^+_{\o,\kk}\hat\psi^-_{-\o,\kk}\;,\Eq(B2)$$
so that $W_\o(\kk;\m_1)$ is purely imaginary.\\
Under hole-particle it becomes 
$$\sum_{\o,\kk}W_\o(\kk;\m_1)\hat\psi^-_{\o,-\kk}
\hat\psi^+_{-\o,-\kk}=-\sum_{\o,\kk}W_{-\o}(\kk;\m_1)\hat\psi^+_{\o,\kk}
\hat\psi^-_{-\o,\kk}\;,\Eq(B3)$$
so that $W_\o(\kk;\m_1)$ is odd in $\o$.\\
Under $(1)\otto(2)$ it becomes:
$$\sum_{\o,\kk}W_\o(\kk;-\m_1)(-i)
\hat\psi^-_{-\o,-\kk}(i)\hat\psi^+_{\o,-\kk}=
\sum_{\o,\kk}W_\o(\kk;-\m_1)\hat\psi^+_{\o,\kk}\hat\psi^-_{-\o,\kk}\;,
\Eq(B4)$$
so that $W_\o(\kk;\m_1)$ is even in $\m_1$. 
Let us define $S_1=i\o/2\sum_{\h,\h'=\pm 1}W_\o(\bk\h{\h'})$,
where $\bk\h{\h'}=(\h\p/M,\h'\p/M)$, and $\g n_1=\PP_0 S_1$,
$s_1=\PP_1 S_1=\s_1\dpr_{\s_1}S_1\big|_{\s_1=\m_1=0}+
\m_1\dpr_{\m_1}S_1\big|_{\s_1=\m_1=0}$. From the 
previous discussion we see that
$S_1, s_1$ and $n_1$ are real and $s_1$ is independent
of $\m_1$. From the computation of the lower order terms
we find $s_1=O(\l\s_1)$ and $\g n_1=\n/Z_1+c^\n_1\l+O(\l^2)$,
for some constant $c^\n_1$ independent of $\l$. 
Note that, since $W_\o(\kk;\m_1)$ is even in $\kk$ 
(so that in particular no linear terms in $\kk$ appear) 
in real space no terms of the form
$\psi^+_{\o,\xx}\dpr\psi^-_{-\o,\xx}$ can appear.\\
\\
{\it \0b)} Let us consider in \equ(4.4) the term with 
$2n=2$, $\a_1=\a_2=\a$ and $\o_1=-\o_2=\o$ and let us denote it with 
$\sum_{\o,\a,\kk}W_\o^\a(\kk;\m_1)
\hat\psi^\a_{\o,\kk}
\hat\psi^\a_{-\o,-\kk}$. We proceed as in item (a) and, 
by using parity, we see that $W_\o^\a(\kk;\m_1)$ is even in $\kk$ 
and odd in $\o$.\\ 
By using complex conjugation, we see that 
$W_\o^\a(\kk;\m_1)=-W^{-\a}_{\o}(\kk;\m_1)^*$.\\
By using hole-particle, we see that 
$W_\o^\a(\kk;\m_1)$ is even in $\a$ and $W_\o^\a(\kk;\m_1)=-
W^{-\a}_{\o}(\kk;\m_1)^*$ implies that $W_\o^\a(\kk;\m_1)$ 
is purely imaginary.\\
By using $(1)\otto(2)$ we see that 
$W_\o^\a(\kk;\m_1)$ is odd in $\m_1$.\\ 

If we define 
$M_1=-i\o/2\sum_{\h,\h'}W_\o^\a(\bk\h{\h'};\m_1)$ 
and $m_1=\PP_1 M_1$, from the previous properties follows
that $M_1$ and $m_1$ are real, $m_1$ is independent of $\s_1$
and, from the computation of its lower order, $m_1=O(\l\m_1)$.
Note that, since $W_\o^\a(\kk;\m_1)$ is even in $\kk$ 
(so that in particular no linear terms in $\kk$ appear) 
in real space no terms of the form
$\psi^\a_{\o,\xx}\dpr\psi^\a_{-\o,\xx}$ can appear.\\
\\
{\it \0c)} Let us consider in \equ(4.4) the term with $2n=2$, 
$\a_1=-\a_2=+$, $\o_1=\o_2=\o$ and let us denote it with
$\sum_{\o,\kk}W_\o(\kk;\m_1)\hat\psi^+_{\o,\kk}
\hat\psi^-_{\o,\kk}$. By using parity we see that
$W_\o(\kk;\m_1)$ is odd in $\kk$.\\
By using reflection we see that $W_\o(k,k_0;\m_1)=W_{-\o}(k,-k_0;\m_1)$.\\
By using complex conjugation we see that
$W_\o(k,k_0;\m_1)=W^*_{\o}(-k,k_0;\m_1)$.\\
By using rotation we find $W_\o(k,k_0;\m_1)=-i\o W_\o(k_0,-k;\m_1)$.\\
By using $(1)\otto(2)$ we see that
$W_{\o}(\kk;-\m_1)$ is even in $\m_1$.\\ 

We now define 
$$G_1(\kk)={1\over 4}\sum_{\h,\h'}W_\o(\bk\h{\h'};\m_1)(\h
{\sin k\over \sin\p/M}+\h'{\sin k_0\over \sin\p/M})\;.$$ 
We can rewrite
$G_1(\kk)=a_\o\sin k+b_\o\sin k_0$, with
$$\eqalign{&
a_\o={1\over 2\sin{\p\over M}}\left[W_\o({\p\over M},{\p\over M};\m_1)+
W_\o({\p\over M},-{\p\over M};\m_1)\right]\cr
&b_\o={1\over 2\sin{\p\over M}}\left[W_\o({\p\over M},{\p\over M};\m_1)-
W_\o({\p\over M},-{\p\over M};\m_1)\right]\;.\cr}\Eq(4.36B)$$
From the properties of $W_\o(\kk;\m_1)$ discussed above, we get:
$$\eqalign{&W_\o({\p\over M},{\p\over M};\m_1)=W_{-\o}({\p\over M},
-{\p\over M};\m_1)
=-W_\o^*({\p\over M},-{\p\over M};\m_1)=-i\o W_\o({\p\over M},
-{\p\over M};\m_1)\cr
&W_\o({\p\over M},-{\p\over M};\m_1)=W_{-\o}({\p\over M},{\p\over M};\m_1)
=-W_\o^*({\p\over M},{\p\over M};\m_1)=i\o W_\o({\p\over M},{\p\over M};\m_1)
\cr}\Eq(4.37B)$$
so that
$$\eqalign{&a_\o=a_{-\o}=-a^*_\o=i\o b_\o\defin ia\cr
&b_\o=-b_{-\o}=b^*_\o=-i\o a_\o\defin \o b=-i\o i a\cr}\Eq(4.38B)$$
with $a=b$ real and independent of $\o$. As a consequence,
$G_1(\kk)=G_1(i\sin k+\o\sin k_0)$ for some real constant $G_1$. If $z_1
\defin\PP_0 G_1$ and we compute the lowest order contribution to $z_1$,
we find $z_1=O(\l^2)$.\\
\\
{\it \0d)} Let us consider in \equ(4.4) the term with
$2n=2$, $\a_1=\a_2=\a$, $\o_1=\o_2=\o$ and let us denote it with
$\sum_{\a,\o,\kk}W_\o^\a(\kk;\m_1)\hat\psi^\a_{\o,\kk}\hat\psi^\a_{\o,-\kk}$.
Repeating the proof in item c) we see that 
$W_\o^\a(\kk;\m_1)$ is odd in $\kk$ and in $\m_1$ and, if we define
$$F_1(\kk)={1\over 4}\sum_{\h,\h'}W_\o^\a(\bk\h{\h'};\m_1)(\h
{\sin k\over \sin\p/M}+\h'{\sin k_0\over \sin\p/M})\;,$$ 
we can rewrite $F_1(\kk)=F_1(i\sin k+\o\sin k_0)$. Since 
$W_\o^\a(\kk;\m_1)$ is odd in $\m_1$, we find $F_1=O(\l\m_1)$.\\

This conlcudes the study of the properties of the kernels
of $\VV^{(1)}$ we shall need in the following. Repeating the proof
above it can also seen that the corrections $a_1^\pm(\kk)$, $b_1^{\pm}(\kk)$,
appearing in \equ(4.3), are analytic odd functions of $\kk$, while $c_1(\kk)$
and $d_1(\kk)$ are real and even; the explicit computation of the
lower order terms in the Taylor expansion in $\kk$ shows that, 
in a neighborhood of $\kk= {\bf 0}$,  
$a_1^\pm(\kk)=O(\s_1\kk)
+O(\kk^3)$, $b_1^{\pm}(\kk)=O(\m_1\kk)+O(\kk^3)$,
$c_1(\kk)=O(\kk^2)$ and $d_1(\kk)=O(\m_1\kk^2)$.\\
\\
The result of the previous discussion can be collected in the
following Theorem.\\
\\
{\cs Theorem 4.1} {\it Assume that $|\s_1|,|\m_1|\le c_1$
for some constant $c_1>0$.
There exist a constant
$\e$ such that,
if $|\l|,|\n|\le \e$, 
then $\Xi^-_{AT}$ can be written as in \equ(4.2), \equ(4.3), \equ(4.4),
where:\\ 
\01) $E_1$ is an $O(1)$ constant;\\
\02) $a^\pm_1(\kk), b^\pm_1(\kk)$ are analytic odd
functions of $\kk$ and $c_1(\kk),d_1(\kk)$ real analytic  
even functions of $\kk$; in a neighborhood of $\kk= {\bf 0}$,  
$a_1^\pm(\kk)=O(\s_1\kk)
+O(\kk^3)$, $b_1^{\pm}(\kk)=O(\m_1\kk)+O(\kk^3)$,
$c_1(\kk)=O(\kk^2)$ and $d_1(\kk)=O(\m_1\kk^2)$;\\ 
\03) the determinant 
$|\det A^{(1)}_\psi(\kk)|$ can be bounded above and below by two positive
constants times $\big[(\s_1-\m_1)^2+|c(\kk)|\big]\big[(\s_1+\m_1)^2+|c(\kk)|
\big]$ 
and $c(\kk)=\cos k_0+\cos k-2$;\\ 
\04) $\widehat W_{2n,\aa,\oo}^{(1)}$ are analytic functions of
$\kk_i,\l,\n,\s_1,\m_1$, $i=1,\ldots,2n$ and, for some constant $C$,
$$|\widehat W_{2n,\aa,\oo}^{(1)}(\kk_1,\ldots,\kk_{2n-1})|
\le M^2 C^n|\l|^{\max\{1,n/2\}}\;;
\Eq(4.40)$$
\04--a) the terms in \equ(4.4) with $n=2$ can be written as
$$\eqalign{
&L_1\sum_{\kk_1,\ldots,\kk_4} \hat\psi^{+}_{1,\kk_1}
\hat\psi^{+}_{-1,\kk_2}\hat\psi^{-}_{-1,\kk_3}\hat\psi^{-}_{1,
\kk_4}\d(\kk_1+\kk_2-\kk_3-\kk_4)+
\cr &+\sum_{\kk_1,\ldots,\kk_4}\sum_{\aa,\oo}
\widetilde W_{4,
\underline\a,\underline\o}(\kk_1,\kk_2,\kk_3)
\hat\psi^{\a_1}_{\o_1,\kk_1}\hat\psi^{\a_2}_{\o_2,\kk_2}
\hat\psi^{\a_3}_{\o_3,\kk_3}\hat\psi^{\a_4}_{\o_4,\kk_4}
\d(\sum_{i=1}^4\a_i\kk_i)\;,\cr}\Eq(4.41)$$
where $L_1$ is real and $\widetilde W_{4,
\underline\a,\underline\o}(\kk_1,\kk_2,\kk_3)$ vanishes at
$\kk_1=\kk_2=\kk_3=\left({\p\over M},{\p\over
M}\right)$;\\
\04--b)
the term in \equ(4.4) with $n=1$ can be written as:
$$\eqalign{
&{1\over 4}\sum_{\o,\a=\pm}\
\sum_\kk \Bigl[S_1(-i\o)\hat\psi^{+}_{\o,\kk}\hat
\psi^{-}_{-\o,\kk}+ M_1(i\o)\hat\psi^{\a}_{\o,\kk}\hat\psi^{\a}_{
-\o,-\kk}+F_1(i\sin k+\o\sin k_0)\hat\psi^{\a}_{\o,\kk}\hat\psi^{\a
}_{\o,-\kk}+\cr
&+G_1(i\sin k+\o\sin k_0)\hat\psi^{+}_{\o,\kk}\hat\psi^{-
}_{\o,\kk}\Bigr]
+\sum_{\kk}\sum_{\aa,\oo}\widetilde
W_{2,\underline\a,\underline\o}(\kk)
\hat\psi^{\a_1}_{\o_1,\kk}
\hat\psi_{\o_2,-\a_1\a_2\kk}^{\a_{2}}\cr}\Eq(4.42)$$
where: $\widetilde
W_{2,\underline\a,\underline\o}(\kk)$ is $O(\kk^2)$ in a neighborhood 
of $\kk={\bf 0}$; $S_1, M_1, F_1,G_1$ are real analytic functions of
$\l,\s_1,\m_1,\nu$ s.t. $F_1=O(\l\m_1)$ and
$$\eqalign{&L_1=l_1+O(\l\s_1)+O(\l\m_1)\virg
S_1=s_1+\g n_1+O(\l\s_1^2)+O(\l\m_1^2)\cr
&M_1=m_1+O(\l\m_1\s_1)+O(\l\m_1^3)\virg
G_1=z_1+O(\l\s_1)+O(\l\m_1)\cr}\Eq(4.43)$$
with $s_1=\s_1 f_1$, $m_1=\m_1 f_2$ and
$l_1, n_1, f_1, f_2, z_1$ independent of $\s_1,\m_1$;
moreover $l_1=\widetilde\l/Z_1^2+O(\l^2)$, 
$f_1,f_2=O(\l)$, $\g n_1=\n/Z_1+c^\n_1\l+O(\l^2)$, for some $c^\n_1$ 
independent of $\l$, and $z_1=O(\l^2)$.}\\
\\
{\it Remark.} 
The meaning of Theorem 2.1 is that after the integration of the $\chi$
fields we are left with a fermionic integration similar
to \equ(3.32) up to corrections which are at least $O(\kk^2)$,
and an effective interaction containing terms with any number
of fields. 
{\it A priori} many 
bilinear terms with kernel $O(1)$ or $O(\kk)$ with respect to
$\kk$ near $\kk={\bf 0}$ 
could be generated by the $\chi$--integration
besides the ones originally present in \equ(2.27); however
{\it symmetry considerations restrict drastically the number
of possible bilinear terms} $O(1)$ or $O(\kk)$. 
Only one new term of the form
$\sum_\kk(i\sin k+\o\sin k_0)\hat\psi^{\a}_{\o,\kk}\hat\psi^{\a}_{\o,-\kk}$
appears, which is ``dimensionally'' {\it marginal}
in a RG sense; however it is weighted by a constant 
$O(\l\m_1)$ and this will improve its 
``dimension'', so that it will result to be {\it irrelevant},
see next Chapter.\\

\pagina
\setcap{5. Renormalization Group for light fermions. The anomalous regime.}
\capindex{5}{Renormalization Group for light fermions. The anomalous regime.}
\vskip1.truecm
\section(5,Renormalization Group for light fermions. The anomalous regime.)
\capindex{5}{Renormalization Group for light fermions. The anomalous regime.}

In this Chapter we begin to describe the iterative integration scheme 
we shall follow in order to compute
the Grassmann functional integral in \equ(4.2). Each step of the 
iteration will resemble for many technical aspects the ultraviolet 
step described
in the previous Chapter. We first split the light field $\psi$ in a sum
of independent Grassmann fields $\sum_h\psi_h$
with masses smaller and smaller, labeled by a {\it scale index} $h\le 1$.
Then we begin to integrate step by step each of them, starting from that
with the biggest mass. After each integration step we rewrite 
the partition function in a way similar to the r.h.s. of \equ(4.2),
with new effective parameters $Z_h,\s_h,\m_h$ and a new effective interaction
$\VV^{(h)}$ replacing $Z_h,\s_h,\m_h$ and $\VV^{(h)}$ respectively. 
As a consequence, a new fundamental problem must be faced:
the size of these parameters and of the new effective interaction must
be controlled, and in particular it must be proven that the 
weight of the local quartic term in $\VV^{(h)}$ remains small under the 
iterations. This is not trivial at all, and in fact one of the major
difficulties of the problem is in finding a suitable definition 
of the new parameters after each integration step. It will in fact
become clear that there is some arbitrariness in their definition
and the choice must be done with care, so that the flow of the effective 
coupling constants can be controlled.

In the present Chapter we will first describe the iterative procedure,
including the definition of {\it localization}, crucial for the 
definition of the effective coupling constants. In the present Chapter
we shall describe only the regime in which the effective parameters
$\s_h,\m_h$ are small; we shall call this regime the {\it anomalous} one,
because $\s_h,\m_h$ grow exponentially in this regime, with
an exponent that is a non trivial function of $\l$. 
We then describe the 
result of the iteration in this regime, that is the bounds the kernels
of the effective interaction satisfy at each step, {\it under the assumption
that the size of the effective local quartic term remain small}. 
This key property (also called {\it vanishing of the Beta
function}, for reasons that will become clear later) will be proven in 
next Chapter. The subsequent regime (in which $\s_h,\m_h$
are of the same order of the mass of the field) must be studied 
with a different iterative procedure, and will be done in Chapter 8.\\

\sub(5.1) {\bf Multiscale analysis}.
\\
From the bound on $\det A^{(1)}_\psi(\kk)$ described in Theorem 4.1, 
we see that the $\psi$ fields have a mass 
given by $\min\{|\s_1-\m_1|,|\s_1+\m_1|\}$, which can be arbitrarly small; 
their integration in the
infrared region (small $\kk$) needs a multiscale analysis.
We introduce a {\sl scaling parameter} $\g>1$ which will be used to
define a geometrically growing sequence of length scales
$1,\g,\g^2,\ldots$, \ie of geometrically decreasing momentum
scales $\g^h,\,h=0,-1,-2,\ldots$
Correspondingly we introduce 
$C^\io$ compact support functions $f_h(\kk)$ $h\le 1$, 
with the following properties: if $|\kk|\defin\sqrt{\sin^2 k+\sin^2 k_0}$, 
when $h\le 0$, $f_h(\kk) = 0$ for $|\kk| <\g^{h-2}$ or $|\kk|
>\g^{h}$, and $f_h(\kk)= 1$, if $|\kk| =\g^{h-1}$; 
$f_1(\kk)=0$ for $|\kk|\le\g^{-1}$ and $f_1(\kk)=1$ for $|\kk|\ge 1$;
furthermore:
$$1=\sum_{h=h_{M} }^1 f_h(\kk)\virg {\rm where:}\qquad h_M=\min
\{h:\g^{h}> \sqrt2\sin{\p\over M}\}
\;,\Eq(5.2)$$
and $\sqrt2\sin(\p/ M)$ is the smallest momentum allowed by
the antiperiodic boundary conditions, \ie it is equal to
$\min_{\kk\in D_{-,-}}|\kk|$.

The purpose is to perform the integration of \equ(2.38) over the 
fermion fields in an iterative way.
After each iteration we shall be left with a ``simpler''
Grassmannian integration to perform: if
$h=1,0,-1,\ldots,h_M$, we shall write
$$\Xi^-_{AT}=\int P_{Z_h,\s_h,\m_h,C_h}(d\psi^{(\le h)}) \, e^{-\VV^{(h)}
(\sqrt{Z_h}\psi^{(\le h)})-M^2 E_h}\;,\quad \VV^{(h)}(0)=0\;,\Eq(5.3)$$
where the quantities $Z_h$,
$\s_h$, $\m_h$, $C_h$, $P_{Z_h,\s_h,\m_h,C_h}(d\psi^{(\le h)})$, 
$\VV^{(h)}$ and $E_h$ have to be defined recursively and the result of
the last iteration will be $\Xi_{AT}^-=e^{-M^2 E_{-1+h_M}}$, 
\ie the value of the partition function. 
$P_{Z_h,\s_h,\m_h,C_h}(d\psi^{(\le h)})$ is defined as
$$\eqalign{& P_{Z_h,\s_h,\m_h,C_h}(d\psi^{(\le h)})=\cr
&=\NN_h^{-1}\prod_{\kk\in D_{-,-}} \prod_{\o=\pm 1}
d\psi^{+(\le h)}_{\kk,\o}d\psi^{-(\le h)}_{\kk,\o}
\exp\Bigl[-{1\over 4 M^2}\sum_{\kk\in D_{-,-}\atop C_h^{-1}(\kk)>0}
Z_h C_h(\kk)
{\Psi_\kk^{+(\le h),T}} A_\psi^{(h)}(\kk)\Psi_\kk^{(\le h)}\Bigr]\;,\cr
&A_\psi^{(h)}(\kk)=\left( \matrix{M^{(h)}(\kk)& N^{(h)}(\kk)\cr
N^{(h)}(\kk)&M^{(h)}(\kk)\cr}\right)\cr
&M^{(h)}(\kk)=\left(\matrix{i \sin k+\sin k_0+a^+_h(\kk)& 
-i\left(\s_h(\kk)+c_h(\kk)\right)\cr
i\left(\s_h(\kk)+c_h(\kk)\right) & i \sin k-\sin k_0+a^-_h(\kk) 
\cr}\right)\cr
&N^{(h)}(\kk)=\left(\matrix{b^+_h(\kk)&i\left(\m_h(\kk)+d_h(\kk)\right)\cr
-i\left(\mu_h(\kk)+d_h(\kk)\right)& b^-_h(\kk)\cr}
\right)\;,\cr}\Eq(5.4)$$
and
$$\Psi^{+(\le h),T}_\kk=(\hat\psi^{+(\le h)}_{1,\kk},
\hat\psi^{+(\le h)}_{-1,\kk},\hat\psi^{-(\le h)}_{1,-\kk},
\hat\psi^{-(\le h)}_{-1,-\kk})
\quad\Psi^{(\le h),T}_\kk=(\hat\psi^{-(\le h)}_{1,\kk},
\hat\psi^{-(\le h)}_{-1,\kk},
\hat\psi^{+(\le h)}_{1,-\kk},\hat\psi^{+(\le h)}_{-1,-\kk})
\;,\Eq(5.5)$$
$\NN_h$ is such that $\int P_{Z_h,\s_h,\m_h,C_h}(d\psi^{(\le h)})=1$,
$C_h(\kk)^{-1}=\sum_{j=h_{M}}^hf_j(\kk)$. Moreover
$$\eqalign{\VV^{(h)}(\psi)&= \sum_{n=1}^\io{1\over M^{2n}}
\sum_{\kk_1,\ldots,\kk_{2n-1},\atop \aa,\oo}
\prod_{i=1}^{2n}\hat\psi^{\a_i(\le h)}_{\o_i,\kk_i}
\widehat W_{2n,\aa,\oo}^{(h)}
(\kk_1,\ldots,\kk_{2n-1})\d(\sum_{i=1}^{2n}\a_i\kk_i)\defin\cr
&\defin
\sum_{n=1}^\io
\sum_{\xx_1,\ldots,\xx_{2n},\atop \ss,\underline j,\oo,\aa}
\prod_{i=1}^{2n}\partial^{\s_i}_{j_i}\psi^{\a_i(\le h)}_{\o_i,\xx_i}
W_{2n,\ss,\underline j,\aa,\oo}^{(h)}(\xx_1,\ldots,\xx_{2n})\;,\cr}\Eq(5.6)$$
where in the last line $j_i=0,1$, $\s_i\ge 0$ and 
$\dpr_j$ is the forward discrete derivative in the $\hat e_j$ direction.

Note that the field $\psi^{(\le h)}$, whose propagator is given by the 
inverse of $Z_h C_h(\kk)A^{(h)}_\psi$, 
has the same support of $C_h^{-1}(\kk)$, that is 
on a strip of width $\g^h$ around the singularity $\kk={\bf 0}$. The field
$\psi^{(\le 1)}$ coincides with the field $\psi$ of previous section, so that
\equ(4.2) is the same as \equ(5.3) with $h=1$.  

It is crucial for the following to think 
$\widehat W_{2n,\aa,\oo}^{(h)}$, $h\le 1$, as functions of the 
variables $\s_k(\kk),\m_k(\kk)$,
$k=h, h+1,\ldots,0,1$, $\kk\in D_{-,-}$. The iterative construction below
will inductively imply that the dependence on these variables 
is well defined (note that for $h=1$
we can think the kernels of $\VV^{(1)}$ as functions of $\s_1,\m_1$, see 
Theorem 4.1). 

\\
\sub(5.2){\bf The localization operator.}
\\
We now begin to describe the iterative construction leading to \equ(5.4).
The first step consits in defining a {\it localization} operator $\LL$ 
acting on the kernels of $\VV^{(h)}$, in terms of which we shall rewrite
$\VV^{(h)}=\LL\VV^{(h)}+\RR\VV^{(h)}$, where $\RR=1-\LL$. The iterative 
integration procedure will use such splitting, see \sec(5.3) below.
 
$\LL$ will be non zero only if 
acting on a kernel $\widehat W_{2n,\aa,\oo}^{(h)}$ with $n=1,2$. In this case  
$\LL$ will be the combination of four different operators:
$\LL_j$, $j=0,1$, 
whose effect on a function of $\kk$ 
will be essentially to extract the term of order $j$
from its Taylor series in $\kk$; and $\PP_j$, $j=0,1$, 
whose effect on a functional of the sequence
$\s_h(\kk),\m_h(\kk),\ldots,\s_1,\m_1$ will be essentially 
to extract the term of order $j$ from its power series 
in $\s_h(\kk),\m_h(\kk),\ldots,\s_1,\m_1$.

The action of $\LL_j$, $j=0,1$, on the kernels 
$\widehat W_{2n,\aa,\oo}^{(h)}(\kk_1,\ldots,\kk_{2n})$
is defined as follows.\\
\\
\01) If $n=1$, 
$$\eqalign{&\LL_0\widehat W_{2,\aa,\oo}^{(h)}(\kk,\a_1\a_2\kk)=\fra14 
\sum_{\h,\h'=\pm 1}\widehat W_{2,\aa,\oo}
^{(h)}(\bk\h{\h'},\a_1\a_2\bk\h{\h'})
\cr
&\LL_1\widehat W_{2,\aa,\oo}^{(h)}(\kk,\a_1\a_2\kk)=\fra14 
\sum_{\h,\h'=\pm 1}\widehat W_{2,\aa,\oo}
^{(h)}(\bk\h{\h'},\a_1\a_2\bk\h{\h'})
\big[\h {\sin k\over \sin{\p\over M}}  +
\h'{\sin k_0\over \sin{\p\over M}}\big]\;,\cr}\Eq(5.7)$$
where $\bk\h{\h'} = \left(\h{\p\over M},\h'{\p\over
M}\right)$ are the smallest momenta allowed by
the antiperiodic boundary conditions.\\
\\
\02) If $n=2$, $\LL_1\widehat W_{4,\aa,\oo}^{(h)}=0$ and
$$\LL_0 \widehat W_{4,\aa,\oo}^{(h)}(\kk_1,\kk_2,\kk_3,\kk_4)\defin
\widehat W_{4,\aa,\oo}^{(h)}(\bk++,\bk++,\bk++,\bk++)\;.\Eq(5.8)$$
\03) If $n>2$, $\LL_0\widehat W_{2n,\aa,\oo}=\LL_1\widehat W_{2n,\aa,\oo}
=0$.\\

The action of $\PP_j$, $j=0,1$, on the kernels $\widehat W_{2n,\aa,\oo}$,
thought as functionals of the sequence $\s_h(\kk),\m_h(\kk),\ldots$
$\ldots,\s_1,\m_1$
is defined as follows. 
$$\eqalign{&
\PP_0 \widehat W_{2n,\aa,\oo}\defin
\widehat W_{2n,\aa,\oo}\Big|_{\ss^{(h)}=\underline\m^{(h)}=0}\cr
&\PP_1 \widehat W_{2n,\aa,\oo}\defin\sum_{k\ge h,\kk}\Big[
\s_{k}(\kk){\dpr \widehat W_{2n,\aa,\oo}\over \dpr\s_k(\kk)}
\Big|_{\ss^{(h)}=\underline\m^{(h)}=0}+
\m_{k}(\kk){\dpr \widehat W_{2n,\aa,\oo}\over \dpr\m_k(\kk)}
\Big|_{\ss^{(h)}=\underline\m^{(h)}=0}\Big]\;.\cr}\Eq(5.9)$$
Given $\LL_j,\PP_j$, $j=0,1$ as above, we define the action of $\LL$
on the kernels $\widehat W_{2n,\aa,\oo}$ as follows.\\
\\
\01) If $n=1$, then
$$\LL \widehat W_{2,\aa,\oo}\defin\cases{
\LL_0(\PP_0+\PP_1)
\widehat W_{2,\aa,\oo} & if $\o_1+\o_2=0$ and $\a_1+\a_2=0$,\cr
\LL_0\PP_1
\widehat W_{2,\aa,\oo} & if $\o_1+\o_2=0$ and $\a_1+\a_2\not=0$,\cr
\LL_1\PP_0\widehat W_{2,\aa,\oo} & if $\o_1+\o_2\not=0$ and 
$\a_1+\a_2=0$,\cr
0 & if $\o_1+\o_2\not=0$ and $\a_1+\a_2\not=0$.}$$
\02) If $n=2$, then
$\LL \widehat W_{4,\aa,\oo}\defin \LL_0\PP_0\widehat W_{4,\aa,\oo}$.\\
\03) If $n>2$, then $\LL \widehat W_{2n,\aa,\oo}=0$.\\ 

Finally, the effect of $\LL$ on $\VV^{(h)}$ is, by definition,
to replace on the r.h.s. of \equ(4.6) $\widehat W_{2n,\aa,\oo}$ with 
$\LL\widehat W_{2n,\aa,\oo}$. Note that $\LL^2\VV^{(h)}=\LL\VV^{(h)}$.

Using the previous definitions we get the following result. 
We use the notation $\ss^{(h)}=\{\s_k(\kk)\}^{k=h,\ldots,1}_{
\kk\in D_{-,-}}$ and $\underline\m^{(h)}=\{\m_k(\kk)\}^{k=h,\ldots,1}_{
\kk\in D_{-,-}}$.
\\
\\
{\cs Lemma 5.1.} {\it Let the action of $\LL$ on $\VV^{(h)}$ 
be defined as above. Then
$$\LL\VV^{(h)}(\psi^{(\le h)})=(s_h+\g^h n_h) F_\s^{(\le h)}
+m_h F^{(\le h)}_{\m}+l_h
F_\l^{(\le h)} +z_{h} F_\z^{(\le h)}
\;,\Eq(5.10) $$
where $s_h,n_h,m_h,l_h$ and $z_h$ are real constants and:
$s_h$ is linear in $\ss^{(h)}$ and independent of $\underline\m^{(h)}$; 
$m_h$ is linear in $\underline\m^{(h)}$ and independent of $\ss^{(h)}$;
$n_h,l_h,z_h$ are independent of $\ss^{(h)},\underline\m^{(h)}$;
moreover, 
if $D_h\defin D_{-,-}\cap\{\kk:C_h^{-1}(\kk)>0\}$,
$$\eqalign{
F_\s^{(\le h)}(\psi^{(\le h)})&=
{1\over 2 M^2}\sum_{\kk\in D_h}\sum_{\o=\pm 1} (-i\o)
\widehat\psi^{+(\le h)}_{\o,\kk}
\widehat\psi^{-(\le h)}_{-\o,\kk}\ \ \defin\ \ 
{1\over M^2}\sum_{\kk\in D_h}
\widehat F_\s^{(\le h)}(\kk)\;,\cr
F_\m^{(\le h)}(\psi^{(\le h)})&=
{1\over 4M^2}\sum_{\kk\in D_h}\sum_{\a,\o=\pm 1} i\o
\widehat\psi^{\a(\le h)}_{\o,\kk}
\widehat\psi^{\a(\le h)}_{-\o,-\kk}\ \ \defin\ \ 
{1\over M^2}\sum_{\kk\in D_h}
\widehat F_\m^{(\le h)}(\kk)\;,\cr
F_\l^{(\le h)}(\psi^{(\le h)})&={1\over M^8}\sum_{\kk_1,\ldots,\kk_4
\in D_h}
\widehat\psi^{+(\le h)}_{1,\kk_1}
\widehat\psi^{+(\le h)}_{-1,\kk_2} \widehat\psi^{-(\le h)}_{-1,\kk_3}
\widehat\psi^{-(\le h)}_{1,\kk_4}\d(\kk_1+\kk_2-\kk_3-\kk_4)\;\cr
F_{\z}^{(\le h)}(\psi^{(\le h)})&={1\over 2M^2}
\sum_{\kk\in D_h}\sum_{\o=\pm 1}
(i\sin k+\o\sin k_0)\widehat\psi^{+(\le h)}_{\o,\kk}
\widehat\psi^{-(\le h)}_{\o,\kk}\ \ \defin\ \ 
{1\over M^2}\sum_{\kk\in D_h}
\widehat F_\z^{(\le h)}(\kk)\;.\cr}\Eq(5.11)$$
where $\d(\kk)=M^2\sum_{{\bf n}\in\zzz^2}\d_{\kk,2\p{\bf n}}$.}\\
\\
{\it Remark.}
The application of $\LL$ to the kernels of the effective potential generates
the sum in \equ(5.10), \ie a linear combination of the Grassmannian
monomials in \equ(5.11) which, in the renormalization group language,
are called ``{\it relevant}'' (the first two) or ``{\it marginal}''
operators (the two others).\\
\\
{\cs Proof of Lemma 5.1} Lemma 5.1 can be proven repeating the 
discussion in \sec(4.3) above. Note in fact that the result of \sec(4.3),
as presented in Theorem 4.1, can be reformulated by saying that
$$\LL\VV^{(1)}(\psi)=(s_1+\g n_1) F_\s^{(\le 1)}
+m_1 F^{(\le 1)}_{\m}+l_1
F_\l^{(\le 1)} +z_1 F_\z^{(\le 1)}
\;,\Eq(5.BB.10) $$
where $s_1,n_1,m_1,l_1$ and $z_1$ are real constants and:
$s_1$ is linear in $\s_1$ and independent of $\m_1$; 
$m_1$ is linear in $\m_1$ and independent of $\s_1$;
$n_1,l_1,z_1$ are independent of $\s_1,\m_1$.

It is now sufficient to note that the symmetries (1)--(6) discussed 
in \sec(4.3) are preserved by the iterative 
integration procedure: in fact it is easy to verify that
$\LL\VV^{(h)}$, $\RR\VV^{(h)}$ and $P_{Z_{h-1},\s_{h-1},\m_{h-1},\widetilde 
f_h}(d\psi^{(h)})$ are, step by step, separately
invariant under the transformations (1)--(6). Then the same proof
leading to \equ(5.BB.10) leads to \equ(5.10) (it is sufficient to replace
any scale label $=1$ with $h$).\\
\\
We now consider the operator $\RR\defin 1-\LL$. The following result
holds. We use the notation
$\RR_1=1-\LL_0$, $\RR_2=1-\LL_0-\LL_1$, $\SS_1=1-\PP_0$, $\SS_2=1-\PP_0-
\PP_1$.\\
\\
{\cs Lemma 5.2.} {\it The action of $\RR$ on $\widehat W_{2n,\aa,\oo}$
for $n=1,2$ is the following.\\
\01) If $n=1$, then
$$\RR \widehat W_{2,\aa,\oo}=\cases{
[\SS_2+\RR_2(\PP_0+\PP_1)]
\widehat W_{2,\aa,\oo} & if $\o_1+\o_2=0$,\cr 
[\RR_1\SS_1+\RR_2\PP_0]
\widehat W_{2,\aa,\oo} & if 
$\o_1+\o_2\not=0$ and $\a_1+\a_2=0$,\cr
\RR_1\SS_1\widehat W_{2,\aa,\oo} & if 
$\o_1+\o_2\not=0$ and $\a_1+\a_2\not=0$,\cr}$$
\02) If $n=2$, then
$\RR \widehat W_{4,\aa,\oo}= [\SS_1+\RR_1\PP_0]
\widehat W_{4,\aa,\oo}$.}\\
\\
{\it Remark.}
The effect of $\RR_j$, $j=1,2$ on $\widehat W_{2n,\aa,\oo}^{(h)}$ 
consists in  
extracting the rest of a Taylor series in $\kk$ of order $j$. 
The effect of $\SS_j$, $j=1,2$ on $\widehat W_{2n,\aa,\oo}^{(h)}$ 
consists in extracting the rest of a
power series in $(\ss^{(h)},\underline\m^{(h)})$ 
of order $j$. The definitions are given in such a way that 
$\RR\widehat W_{2n,\aa,\oo}$ is at least quadratic in $\kk,\ss^{(h)},
\underline\m^{(h)}$ if $n=1$ and at least linear in $\kk,\ss^{(h)},
\underline\m^{(h)}$ when $n=2$. This will give dimensional gain
factors in the bounds for $\RR\widehat W_{2n,\aa,\oo}^{(h)}$ w.r.t. 
the bounds for $\widehat W_{2n,\aa,\oo}^{(h)}$, $n=1,2$, as we shall see in 
details in \sec(5.5).\\
\\
{\cs Proof of Lemma 5.2} 
It is sufficient to note that the symmetry 
properties discussed in \sec(4.3) imply that: $\LL_1 W_{2,\aa,\oo}=0$ if 
$\o_1+\o_2=0$; $\LL_0 W_{2,\aa,\oo}=0$ if 
$\o_1+\o_2\not=0$; $\PP_0 W_{2,\aa,\oo}=0$ if 
$\a_1+\a_2\not=0$; and use the definitions of $\RR_i$, $\SS_i$, $i=1,2$.
\\
\\
\sub(5.3){\bf Renormalization.}
\\
Once that the above definitions are given we can describe our
integration procedure for $h\le 0$.
\*
We start from \equ(5.3) and we rewrite it as
$$\int P_{Z_h,\s_h,\m_h,C_h}(d\psi^{(\le h)}) \, e^{-\LL\VV^{(h)}
(\sqrt{Z_h}\psi^{(\le h)})-\RR\VV^{(h)}
(\sqrt{Z_h}\psi^{(\le h)}) -M^2 E_h}\;,\Eq(5.3a)$$
with $\LL\VV^{(h)}$ as in \equ(5.10). 
Then we include the quadratic part of  
$\LL\VV^{(h)}$ (except the term proportional to $n_h$)
in the fermionic integration, so obtaining
$$\int P_{\hat Z_{h-1},\s_{h-1},\m_{h-1},C_h}(d\psi^{(\le h)}) \, e^{-l_h 
F_\l(\sqrt{Z_h}\psi^{(\le h)})-\g^h n_h F_\s(\sqrt{Z_h}\psi^{(\le h)})-
\RR\VV^{(h)}
(\sqrt{Z_h}\psi^{(\le h)}) -M^2 E_h}\;,\Eq(5.3aa)$$
where $\widehat Z_{h-1}(\kk)\defin Z_h (1+z_h C_h^{-1}(\kk))$ and
$$\eqalign{
&\s_{h-1}(\kk)\defin{Z_h\over\widehat Z_{h-1}(\kk)}
(\s_h(\kk)+s_hC_h^{-1}(\kk))\virg
\m_{h-1}(\kk)\defin {Z_h\over \widehat Z_{h-1}(\kk)}
(\m_h(\kk)+m_h C_h^{-1}(\kk))\cr
&a^\o_{h-1}(\kk)\defin{Z_h\over \widehat Z_{h-1}(\kk)}
a^\o_h(\kk)\virg
b^\o_{h-1}(\kk)\defin {Z_h\over \widehat Z_{h-1}(\kk)}
b^\o_h(\kk)\cr
&c_{h-1}(\kk)\defin{Z_h\over\widehat Z_{h-1}(\kk)}c_h(\kk)\virg
d_{h-1}(\kk)\defin{Z_h\over\widehat Z_{h-1}(\kk)}d_h(\kk)\;.\cr}\Eq(5.17a)$$
The integration in \equ(5.3aa) differs from the one in \equ(5.3)
and \equ(5.3a): $P_{\hat Z_{h-1},
\s_{h-1},\m_{h-1},C_h}$ is defined by \equ(5.4)
with $Z_h$ and 
$A^{(h)}_\psi$ replaced by $\widehat Z_{h-1}(\kk)$ and 
$A^{(h-1)}_\psi$.

\*
Now we can perform the integration of the $\psi^{(h)}$ field.
It is convenient to rescale the fields:
$${\widehat\VV}^{(h)}(\sqrt{Z_{h-1}}\psi^{(\le h)})\ \defin\ 
\l_h F_\l(\sqrt{Z_{h-1}}\psi^{(\le h)})+
\g^h\n_h F_\s(\sqrt{Z_{h-1}}\psi^{(\le h)})
+\RR\VV^{(h)}(\sqrt{Z_h}\psi^{(\le h)})
\;,\Eq(5.18)$$
where 
$$\l_h=\big({Z_h\over Z_{h-1}}\big)^2l_h\;,\qquad\n_h={Z_h\over Z_{h-1}}n_h
\;,\Eq(5.18z)$$
and $\RR\VV^{(h)}=(1-\LL)\VV^{(h)}$ is the irrelevant part of $\VV^{(h)}$, and
rewrite \equ(5.3aa) as
$$e^{-M^2(t_h+E_h)}\int
P_{Z_{h-1},\s_{h-1},\m_{h-1},C_{h-1}}(d\psi^{(\le h-1)}) \, 
\int P_{Z_{h-1},\s_{h-1},\m_{h-1},
{\widetilde f}^{-1}_h}
(d\psi^{(h)})\, e^{-\widehat
\VV^{(h)}(\sqrt{Z_{h-1}}\psi^{(\le h)})}
\Eq(5.20)$$
where we used the decomposition $\psi^{(\le h)}=\psi^{(\le h-1)}+\psi^{(h)}$
(and $\psi^{(\le h-1)},\psi^{(h)}$ are independent) and 
${\widetilde f}_h(\kk)$ is defined by the relation $C_h^{-1}(\kk)\widehat 
Z_{h-1}^{-1}(\kk)=C_{h-1}^{-1}(\kk)Z_{h-1}^{-1}+\widetilde f_h(\kk)
Z_{h-1}^{-1}$, namely:
$$\widetilde f_h(\kk)\ \defin\ Z_{h-1}\Bigl[
{C_h^{-1}(\kk)\over \widehat Z_{h-1}
(\kk)}-{C_{h-1}^{-1}(\kk)\over Z_{h-1}}\Bigr]=f_h(\kk)\Bigl[
1+ {z_hf_{h+1}(\kk)\over
1+z_hf_h(\kk)}
\Bigr]\;.\Eq(5.21)$$
Note that $\widetilde f_h(\kk)$ has the same support as $f_h(\kk)$.
Moreover 
$P_{Z_{h-1},\s_{h-1},\m_{h-1},{\widetilde f}^{-1}_h}
(d\psi^{(h)})$ is defined in the same way as 
$P_{\widehat Z_{h-1},\s_{h-1},\m_{h-1},C_h}
(d\psi^{(h)})$, with $\widehat Z_{h-1}(\kk)$ resp. $C_h$
replaced by $Z_{h-1}$ resp. $\widetilde f^{-1}_h$. 
The {\it single scale} propagator is
$$\int P_{Z_{h-1},\s_{h-1},\m_{h-1},\widetilde f_h^{-1}}(d\psi^{(h)})\,
\psi^{\a(h)}_{\xx,\o}\psi^{\a'(h)}_{\yy,\o'} ={1\over Z_{h-1}}
g^{(h)}_{\underline a,\underline a'}(\xx-\yy)\virg 
\underline a=(\a,\o)\virg \underline a'=(\a',\o')\;,\Eq(5.210a)$$
where 
$$g^{(h)}_{\underline a,\underline a'}(\xx-\yy)={1\over 2M^2}
\sum_{\kk}e^{i\a\a'\kk(\xx-\yy)}
\widetilde f_h(\kk)[A_\psi^{(h-1)}(\kk)]^{-1}_{j(\underline a),
j'(\underline a')}\Eq(5.210b)$$
with $j(-,1)=j'(+,1)=1$, $j(-,-1)=j'(+,-1)=2$, $j(+,1)=j'(-,1)=3$ and
$j(+,-1)=j'(-,-1)=4$. One finds that 
$g^{(h)}_{\underline a,\underline a'}(\xx)=g_{\o,\o'}^{(1,h)}(\xx)-\a\a'
g_{\o,\o'}^{(2,h)}(\xx)$, where 
$g_{\o,\o'}^{(j,h)}(\xx)$, $j=1,2$ are defined in Appendix A4.

The long distance behaviour of the propagator is given by the following 
Lemma, proved in Appendix A4.\\
\\
{\cs Lemma 5.3.} {\it Let $\s_h\defin\s_h({\bf 0})$ and $\m_h\defin
\m_h({\bf 0})$ and assume $|\l|\le \e_1$ for a small constant $\e_1$. 
Suppose that for $h>\bar h$
$$|z_h|\le {1\over 2}\virg |s_h|\le {1\over 2}|\s_h|\virg 
|m_h|\le {1\over 2}|\m_h|\;,\Eq(5.40yz)$$
that there exists $c$ s.t. 
$$e^{-c|\l|}\le \Big|{\s_h\over\s_{h-1}}\Big|\le e^{c|\l|} 
\virg
e^{-c|\l|}\le \Big|{\m_h\over\m_{h-1}}\Big|\le e^{c|\l|}\virg
e^{-c|\l|^2}\le \Big|{Z_h\over Z_{h-1}}\Big|\le e^{c|\l|^2}\;,\Eq(5.40z)$$
and that, for some constant $C_1$, 
$${|\s_{\bar h}|\over\g^{\bar h}}\le C_1\virg
{|\m_{\bar h}|\over\g^{\bar h}}\le C_1\;;\Eq(5.40a)$$
then, for all $h\ge \bar h$, 
given the positive integers $N, n_0,n_1$ and putting $n=n_0+n_1$,
there exists a constant $C_{N,n}$ s.t. 
$$|\dpr^{n_0}_{x_0}\dpr^{n_1}_{x}g_{\underline a,\underline a'}^{(h)}
(\xx-\yy)|\le C_{N,n}{\g^{(1+n)h}\over 1+(\g^{h}
|\dd(\xx-\yy)|)^N}\virg{\it where}\quad
\dd(\xx)={M\over\p}\big(\sin{\p x\over M},\sin{\p x_0\over M})\;.
\Eq(5.400)$$ 
Furthermore, if $\PP_0$, $\PP_1$ are 
defined as in \equ(5.9) and $\SS_1$, $\SS_2$ are defined as in
Lemma 5.2, 
we have that $\PP_jg^{(h)}_{\underline a,
\underline a'}$, $j=0,1$ and $\SS_jg^{(h)}_{\underline a,
\underline a'}$, $j=1,2$, satisfy the same bound \equ(5.400),
times a factor $\big({|\s_{h}|+|\m_{h}|\over \g^{h}}\big)^j$.
The bounds for $\PP_0g^{(h)}_{\underline a,
\underline a'}$ and $\PP_1g^{(h)}_{\underline a,
\underline a'}$ hold even without hypothesis \equ(5.40a).}\\ 

After the integration of the field on scale $h$ we are left with an 
integral involving the fields 
$\psi^{(\le h-1)}$ and the new effective interaction
$\VV^{(h-1)}$, defined as 
$$e^{-\VV^{(h-1)}(\sqrt{Z_{h-1}}\psi^{(\le h-1)})-\tilde E_h M^2}=
\int P_{Z_{h-1},\s_{h-1},\m_{h-1},\widetilde f_h}(d\psi^{(h)})
e^{-\widehat \VV^{(h)}(\sqrt{Z_{h-1}}\psi^{(\le h)})}\;.\Eq(5.20a)$$
It is easy to see that $\VV^{(h-1)}$ is of the form \equ(5.6) and that 
$E_{h-1}=E_h+t_h+\tilde E_h$. It is sufficient to use the well known identity
$$M^2\tilde E_h+\VV^{(h-1)}(\sqrt{Z_{h-1}}\psi^{(\le h-1)})=
\sum_{n\ge 1}{1\over n!}(-1)^{n+1}\EE^T_h(\widehat\VV^{(h)}(\sqrt{Z_{h-1}}
\psi^{(\le h)});n)\;,\Eq(5.20az)$$
where $\EE^T_h(X(\psi^{(h)});n)$ 
is the truncated expectation of order $n$ w.r.t. the 
propagator $Z_{h-1}^{-1}g^{(h)}_{\underline a,\underline a'}$, defined
as
$$\EE^T_h(X(\psi^{(h)});n)={\dpr\over\dpr\l^n}\log\int P_{Z_{h-1},
\s_{h-1},\m_{h-1},\widetilde f_h}(d\psi^{(h)})e^{\l X(\psi^{(h)})}\Big|_{\l=
0}\;.\Eq(eth)$$

Note that the above procedure allow us to write the 
{\it running coupling constants} $\vec v_{h-1}=(\l_{h-1},\n_{h-1})$, $h\le 1$, 
in terms of $\vec v_k$, $h\le k\le 1$, namely
$$\vec v_{h-1}=\b_h(\vec v_h,\ldots,\vec v_1)\;,\Eq(beta)$$
where $\b_h$ is the so--called {\it Beta function}.
\\
\\
\sub(5.4){\bf Analiticity of the effective potential}
\\
We have expressed the effective potential $\VV^{(h)}$
in terms of the {\it running coupling constants} $\l_k,\n_k$, $k\ge h$,
and of the {\it renormalization constants} $Z_k,\m_k(\kk),\s_k(\kk)$, $k\ge h$.

In next section we will prove the following result.\\
\\
{\cs Theorem 5.1.} {\it Let $\s_h\defin\s_h({\bf 0})$ and $\m_h\defin
\m_h({\bf 0})$ and assume
$|\l|\le \e_1$ for a small constant $\e_1$. 
Suppose that for $h> \bar h$ the hypothesis \equ(5.40yz),
\equ(5.40z) and \equ(5.40a) hold.
If, for some constant $c$,
$$\max_{h> \bar h}\{|\l_h|,|\n_h|\}\le c|\l|\;,\Eq(5.40)$$
then there exists $C>0$ s.t. the kernels in \equ(5.6) satisfy
$$\int d\xx_1\cdots d\xx_{2n}|W^{(\bar h)}_{2n,\ss,\underline j,\aa,\oo}
(\xx_1,\ldots,\xx_{2n})|
\le M^2 \g^{-\bar h D_k(n)} \,(C\,|\l|)^{max(1,n-1)}\Eq(5.45)$$
where $D_k(n)=-2+n+k$
and $k=\sum_{i=1}^{2n}\s_i$.

Moreover $|\tilde E_{\bar h+1}|+|t_{\bar h+1}|\le c|\l|\g^{2\bar h}$ and
the kernels of $\LL\VV^{(\bar h)}$ satisfy
$$|s_{\bar h}|\le C|\l||\s_{\bar h}|\virg
|m_{\bar h}|\le C|\l||\m_{\bar h}|\Eq(5.45y)$$
and
$$|n_{\bar h}|\le C|\l|\virg
|z_{\bar h}|\le C|\l|^2\virg|l_{\bar h}|\le C|\l|^2\;.\Eq(5.45yz)$$
The bounds \equ(5.45y) holds even if \equ(5.40a) does not hold.
The bounds \equ(5.45yz) holds even if \equ(5.40a) and the 
first two of \equ(5.40z) do not hold.}\\
 
{\it Remarks.}\\
1) The above result immediately implies analyticity of the effective potential
of scale $h$
in the running coupling constants $\l_k,\n_k$, $k\ge h$, under the assumptions 
\equ(5.40yz), \equ(5.40z), \equ(5.40a) and \equ(5.40).\\
2) The assumptions \equ(5.40z) and \equ(5.40) 
will be proved in next Chapter, solving the 
{\it flow equations} for $\vec v_h=(\l_h,\n_h)$ and $Z_h,\s_h,\m_h$,
given by $\vec v_{h-1}=\b_h(\vec v_h,\ldots,\vec v_1)$, $Z_{h-1}=Z_h(1+z_h)$
and \equ(5.17a). They will be proved to 
be true up to $h=-\io$.\\ 
%
%
\\
\sub(5.5){\bf Proof of Theorem 5.1.}
\\
It is possible to write $\VV^{(h)}$  \equ(5.6)
in terms of {\it Gallavotti--Nicolo' trees.} The detailed 
derivation of this representation can be found in the 
reviews papers [G1][GM] and in my diploma thesis [G]. We do
not repeat here the details, we only give the basic definitions,
in order to make the subsequent discussion self consistent.\\

\midinsert
\*
\insertplot{300pt}{150pt}%
{\ins{30pt}{85pt}{$r$}\ins{50pt}{85pt}{$v_0$}\ins{130pt}{100pt}{$v$}%
\ins{35pt}{-2pt}{$h$}\ins{55pt}{-2pt}{$h+1$}\ins{135pt}{-2pt}{$h_v$}%
\ins{215pt}{-2pt}{$0$}\ins{235pt}{-2pt}{$+1$}\ins{255pt}{-2pt}{$+2$}}%
{fig51}{}
\vskip.7truecm
\line{\vtop{\line{\hskip6.3truecm\vbox{\advance\hsize by -8.0 truecm
\0{\css Fig. 5.}
{\ottorm  A tree with its scale labels.}
} \hfill} }}
\*
\endinsert


Let us introduce the following definitions and notations.

\0 1) Let us consider the family of all trees which can be constructed
by joining a point $r$, the {\it root}, with an ordered set of $n\ge 1$
points, the {\it endpoints} of the {\it unlabeled tree},
so that $r$ is not a branching point. $n$ will be called the
{\it order} of the unlabeled tree and the branching points will be called
the {\it non trivial vertices}.
Two unlabeled trees are identified if they can be superposed by a suitable
continuous deformation, so that the endpoints with the same index coincide.
Then the number of unlabeled trees with $n$ end-points
is bounded by $4^n$.

\0 2) We associate a label $h\le 0$ with the root and we denote $\TT_{h,n}$
the corresponding set of labeled trees with $n$ endpoints. Moreover, 
we introduce
a family of vertical lines, labeled
by an integer taking values in
$[h,2]$, and we represent any
tree $\t\in\TT_{h,n}$ so that, if $v$ is an
vendpoint or a non trivial vertex, it is contained in a vertical line with
index $h_v>h$, to be called the {\it scale} of $v$, while the root is on the
line with index $h$. There is the constraint that, if $v$ is an endpoint,
$h_v>h+1$; if there is only one end-point its scale must
be equal to $h+2$,
for $h\le 0$.
Moreover, there is only one vertex immediately following
the root, which will be denoted $v_0$ and can not be an endpoint;
its scale is $h+1$.

\0 3) With each endpoint $v$ of scale $h_v=+2$ we associate one of the
contributions to $\VV^{(1)}$ given by \equ(4.4);
with each endpoint $v$ of
scale $h_v\le 1$ one of the terms in
$\LL \VV^{(h_v-1)}$ defined in \equ(5.10).
Moreover, we impose the constraint that, if $v$ is an endpoint and
$h_v\le 1$,
$h_v=h_{v'}+1$, if $v'$ is the non trivial vertex immediately preceding $v$.


\0 4) We introduce a {\it field label} $f$ to distinguish the field variables
appearing in the terms associated with the endpoints as in item 3);
the set of field labels associated with the endpoint $v$ will be called $I_v$.
Analogously, if $v$ is not an endpoint, we shall
call $I_v$ the set of field labels associated with the endpoints following
the vertex $v$; $\xx(f)$, $\s(f)$ and $\o(f)$ will denote the space-time
point, the $\s$ index and the $\o$ index, respectively, of the
field variable with label $f$.

\0 5) We associate with any vertex $v$ of the tree a subset $P_v$ of $I_v$,
the {\it external fields} of $v$. These subsets must satisfy various
constraints. First of all, if $v$ is not an endpoint and $v_1,\ldots,v_{s_v}$
are the $s_v$ vertices immediately following it, then $P_v \subset \cup_i
P_{v_i}$; if $v$ is an endpoint, $P_v=I_v$. We shall denote $Q_{v_i}$ the
intersection of $P_v$ and $P_{v_i}$; this definition implies that $P_v=\cup_i
Q_{v_i}$. The subsets $P_{v_i}\bs Q_{v_i}$, whose union will be made, by
definition, of the {\it internal fields} of $v$, have to be non empty, if
$s_v>1$, that is if $v$ is a non trivial vertex.
Given $\t\in\TT_{j,n}$, there are many possible choices of the subsets $P_v$,
$v\in\t$, compatible with the previous constraints; let us call $\bP$ one of
this choices. 
Given $\bP$, we consider the family $\cal G_\bP$ of all
connected Feynman graphs, such that, for any $v\in\t$, the internal fields of
$v$ are paired by propagators of scale $h_v$, so that the following condition
is satisfied: for any $v\in\t$, the subgraph built by the propagators
associated with all vertices $v'\ge v$ is connected. The sets $P_v$ have, in
this picture, the role of the external legs of the subgraph associated with$v$.
The graphs belonging to $\cal G_\bP$ will be called {\it compatible with
$\bP$} and we shall denote $\PP_\t$ the family of all choices of $\bP$ such
that $\cal G_\bP$ is not empty.

\06) we associate with any vertex $v$ an index $\r_v\in\{s,p\}$
and correspondingly an operator $\RR_{\r_v}$, where 
$\RR_s$ or $\RR_p$ are defined as 
$$\RR_s\defin\cases{\SS_2 & if $n=1$ and $\o_1+\o_2=0$,\cr
\RR_1\SS_1 & if $n=1$ and $\o_1+\o_2\not=0$,\cr
\SS_1 & if $n=2$,\cr
1 & if $n>2$;\cr}\Eq(rs)$$
and 
$$\RR_p\defin\cases{\RR_2(\PP_0+\PP_1) & if $n=1$ and $\o_1+\o_2=0$,\cr
\RR_2\PP_0 & if $n=1$, $\o_1+\o_2\not=0$
and $\a_1+\a_2=0$,\cr
0 & if $n=1$, $\o_1+\o_2\not=0$
and $\a_1+\a_2\not=0$,\cr
\RR_1\PP_0 & if $n=2$,\cr
0 & if $n>2$.\cr}\Eq(rp)$$
Note that $\RR_s+\RR_p=\RR$, see Lemma 5.2. 

\*
The effective potential can be written in the following way:
$$\VV^{(h)}(\sqrt{Z_h}\psi^{(\le h)}) + M^2 \tilde E_{h+1}=
\sum_{n=1}^\io\sum_{\t\in\TT_{h,n}}
\VV^{(h)}(\t,\sqrt{Z_h}\psi^{(\le h)});,\Eq(5.46)$$
where, if $v_0$ is the first vertex of $\t$ and $\t_1,\ldots,\t_s$ 
are the subtrees of $\t$ with root $v_0$,\\
$\VV^{(h)}(\t,\sqrt{Z_h}\psi^{(\le h)})$ is defined inductively by the relation
$$\eqalign{
&\qquad \VV^{(h)}(\t,\sqrt{Z_h}\psi^{(\le h)})=\cr
&{(-1)^{s+1}\over s!} \EE^T_{h+1}[\bar
V^{(h+1)}(\t_1,\sqrt{Z_{h}}\psi^{(\le h+1)});\ldots; \bar
V^{(h+1)}(\t_{s},\sqrt{Z_{h}}\psi^{(\le h+1)})]\;,\cr}\Eq(5.47)$$
and $\bar V^{(h+1)}(\t_i,\sqrt{Z_{h}}\psi^{(\le h+1)})$:
 
\0 a) is equal to $\RR_{\r_{v_i}}\widehat
\VV^{(h+1)}(\t_i,\sqrt{Z_{h}}\psi^{(\le h+1)})$ if
the subtree $\t_i$ with first vertex $v_i$
is not trivial (see \equ(5.18) for the definition of
$\widehat \VV^{(h)}$);
 
\0 b) if $\t_i$ is trivial and $h\le -1$, it
is equal to one of the terms in $\LL\widehat\VV^{(h+1)}$,
see \equ(5.18), or,
if $h=0$, to one of the terms contributing to 
$\widehat\VV^{(1)}(\sqrt{Z_1}\psi^{\le 1})$.\\
\\
\sub(5.a.0) The explicit 
expression for the kernels of $\VV^{(h)}$ can be found from 
\equ(5.46) and \equ(5.47) by writing the truncated expectations 
of monomials of $\psi$ fields using the analogue of \equ(4.12): if 
$\widetilde\psi(P_{v_i})=\prod_{f\in P_{v_i}}\psi^{\a(f)(h_v)}_{\xx(f),
\o(f)}$, the following identity holds:
$$\EE^T_{h_v}(\widetilde\psi(P_{v_1}),\ldots,\widetilde\psi(P_{v_s}))=
\Big({1\over Z_{h_v-1}}\Big)^n\sum_{T_v}\a_{T_v}\prod_{\ell\in T_v}
g^{(h_v)}(f^1_\ell,f^2_\ell)
\int dP_{T_v}(\tt) \Pf G^{T_v}(\tt)\Eq(5.48)$$
where $g^{(h)}(f,f')=g_{\underline a(f),\underline a(f')}(\xx(f)-\xx(f'))$
and the other symbols in \equ(5.48) have the same meaning as those in 
\equ(4.12).

Using iteratively \equ(5.48) we can express the kernels
of $\VV^{(h)}$ as sums of products of propagators of the fields (the ones
associated to the anchored trees $T_v$) and Pfaffians of matrices $G^{T_v}$.
%
\\
\\
\sub(5.a.1) {\it If the $\RR$ operator were not applied to 
the vertices $v\in\t$} then the result of the iteration would lead to the 
following relation:
$$\VV^*_h(\t,\sqrt{Z_h}\psi^{(\le h)})=\sqrt{Z_h}^{|P_{v_0}|}\sum_{
\bP\in\PP_\t}\sum_{T\in\bT}\int d\xx_{v_0}W^*_{\t,\bP,\bT}
(\xx_{v_0})\Big\{\prod_{f\in P_{v_0}}
\psi^{\a(f)(\le h)}_{
\xx(f),\o(f)}\Big\}\;,\Eq(5.49)$$
where $\xx_{v_0}$ is the set of integration variables
asociated to $\t$ and $T=\bigcup_v T_v$; 
$W^*_{\t,\bP,\bT}$ is given by
$$\eqalign{&W^*_{\t,\bP,\bT}(\xx_{v_0})=
\Big[\prod_{v\,\hbox{\ottorm not e.p.}}
\Big({Z_{h_v}\over Z_{h_v-1}}\Big)^{|P_v|\over 2}\Big]
\Big[\prod_{i=1}^n K^{h_i}_{v^*_i}(\xx_{v^*_i})\Big]
\Big\{\prod_{v\,\hbox{\ottorm not e.p.}}{1\over s_v!} \int
dP_{T_v}(\tt_v) \;\cdot\cr
&\cdot\; \Pf G^{h_v,T_v}(\tt_v)
\Big[\prod_{l\in T_v} g^{(h_v)}(f^1_l,f^2_l)\Big]\Big\}\;,\cr}
\Eq(5.50)$$
where: $e.p.$ is an abbreviation of ``end points'';
$v_1^*,\ldots, v_n^*$ are the endpoints of $\t$, $h_i\=h_{v_i^*}$
and $K^{h_v}_v(\xx_v)$ are the corresponding kernels (equal to 
$\l_{h_v-1}\d(\xx_v)$ or $\n_{h_v-1}\d(\xx_v)$ if $v$ is an endpoint 
of type $\l$ or $\n$ on scale $h_v\le 1$; or equal to one of the kernels
of $\VV^{(1)}$ if $h_v=2$).

Bounding \equ(5.50) 
using \equ(5.400) and the Gram--Hadamard inequality,
see Appendix A3, we would find:
$$\int d\xx_{v_0} |W^*_{\t,\bP,T}(\xx_{v_0})|\le
C^n M^2|\l|^n \g^{-h(-2+|P_{v_0}|/2 )}
\prod_{v\,\hbox{\ottorm not e.p.}} \left\{ {1\over s_v!}
\Big({Z_{h_v}\over Z_{h_v-1}}\Big)^{|P_v|\over 2}
\g^{-[-2+{|P_v|\over 2}]}\right\}\;.\Eq(5.51)$$
We call $D_v=-2+{|P_v|\over 2}$ the {\it dimension}
of $v$, depending on the number of the external
fields of $v$. If $D_v<0$ for any $v$ one can
sum over $\t,\bP,T$ obtaining convergence
for $\l$ small enough; however 
$D_v\le 0$ when there are two or four external lines.
We will take now into account the effect of the $\RR$ operator
and we will see how the bound \equ(5.51) is improved.\\
\\
\sub(5.a.1b) 
The effect of application of $\PP_j$ and $\SS_j$ 
is to replace a kernel $W^{(h)}_{2n,\ss,\underline j,\aa,\oo}$
with $\PP_j W^{(h)}_{2n,\ss,\underline j,\aa,\oo}$ and 
$\SS_j W^{(h)}_{2n,\ss,\underline j,\aa,\oo}$. 
If inductively, starting from the end--points,
we write the kernels $W^{(h)}_{2n,\ss,\underline j,\aa,\oo}$ 
in a form similar to 
\equ(5.50), we easily realize that, eventually, 
$\PP_j$ or $\SS_j$ will act on some 
propagator of an anchored tree or on some Pfaffian $\Pf G^{T_v}$, 
for some $v$.
It is easy to realize that $\PP_j$ and $\SS_j$, when applied to Pfaffians,
do not break the Pfaffian structure. In fact the effect 
of $\PP_j$ on the Pfaffian of an antisymmetric matrix $G$ 
with elements $G_{f,f'}$, $f,f'\in J$, $|J|=2k$,
is the following (the proof is trivial):
$$\PP_0\Pf G=\Pf G^0\virg
\PP_1\Pf G=
{1\over 2 }\sum_{f_1,f_2\in J}\PP_1G_{f_1,f_2}
(-1)^\p\Pf G_1^0\;,\Eq(5.52)$$
where $G^0$ is the matrix with elements $\PP_0 G_{f,f'}$, $f,f'\in J$;
$G_1^0$ is the matrix with elements $\PP_0 G_{f,f'}$, $f,f'\in J_1\defin
J\setminus
\{f_1\cup f_2\}$ and $(-1)^\p$ is the sign of the permutation
leading from the ordering $J$ of the labels $f$ in the l.h.s. to the 
ordering $f_1,f_2,J_1$ in the r.h.s. The effect of 
$\SS_j$ is the following, see Appendix A5 for a proof:
$$\SS_1\Pf G={1\over 2\cdot k!}\sum_{f_1,f_2\in J}\SS_1G_{f_1,f_2}
\sum_{J_1\cup J_2=
J\setminus\cup_i f_i}^*(-1)^\p k_1!\, k_2!\,
\Pf G_1^0\,\Pf G_2\;,\Eq(5.53)$$
where: the $*$ on the sum means that $J_1\cap J_2=\emptyset$; $|J_i|=2k_i$, 
$i=1,2$; $(-1)^\p$ is the sign of the permutation leading 
from the ordering $J$ of the fields labels on the l.h.s. to the ordering 
$f_1,f_2,J_1,J_2$ on the r.h.s.;
$G_1^0$ is the matrix with elements $\PP_0 G_{f,f'}$, $f,f'\in J_1$;
$G_2$ is the matrix with elements $G_{f,f'}$, $f,f'\in J_2$.
The effect of $\SS_2$ on $\Pf G^T$ is given by a formula 
similar to \equ(5.53).
Note that the
number of terms in the sums appearing in
\equ(5.52), \equ(5.53) (and in the analogous equation for $\SS_2\Pf G^T$),
is bounded by $c^k$ for some constant $c$.\\
\\
\sub(5.a.2) It is possible to show 
that the $\RR_j$ operators   
produce derivatives applied to the propagators of the anchored 
trees and on the elements of $G^{T_v}$; and a product of ``zeros''
of the form $d_{j}^{b}(\xx(f^1_\ell)-\xx(f^2_\ell))$, $j=0,1$, $b=0,1,2$, 
associated to the lines $\ell\in T_v$.
This is a well known result, and a very detailed discussion
can be found in \S 3 of [BM]. By such analysis, and using
\equ(5.52),\equ(5.53),
we get the following expression for 
$\RR\VV^{(h)}(\t,\sqrt{Z_h}\psi^{(\le h)})$:
$$\eqalign{&\RR\VV^{(h)}(\t,\sqrt{Z_h}\psi^{(\le h)})=\cr
&\qquad=\sqrt{Z_h}^{|P_{v_0}|}\sum_{
\bP\in\PP_\t}\sum_{T\in\bT}\sum_{\b\in B_T}\int d\xx_{v_0}W_{\t,\bP,\bT,\b}
(\xx_{v_0})\Big\{\prod_{f\in P_{v_0}}
\hat\dpr^{q_\b(f)}_{j_\b(f)}\psi^{\a(f)(\le h)}_{
\xx_\b(f),\o(f)}\Big\}\;,\cr}\Eq(5.54)$$
where: $B_T$ is a set of indeces which allows to distinguish the
different terms produced by the non trivial $\RR$ operations;
$\xx_\b(f)$ is a coordinate obtained by interpolating two points in $
\xx_{v_0}$, in a suitable way depending on $\b$; $q_\b(f)$ is a nonnegative 
integer $\le 2$; $j_\b(f)=0,1$ and $\hat\dpr^q_j$ is a suitable differential 
operator, dimensionally equivalent to $\dpr^q_j$ (see [BM] for a precise 
definition); $W_{\t,\bP,\bT,\b}$ is given by:
$$\eqalign{W_{\t,\bP,\bT,\b}(\xx_{v_0})&=
\Big[\prod_{v\,\hbox{\ottorm not e.p.}}
\Big({Z_{h_v}\over Z_{h_v-1}}\Big)^{|P_v|\over 2}\Big]
\Big[\prod_{i=1}^n d^{b_\b(v_i^*)}_{j_\b(v_i^*)}(\xx^i_\b,
\yy^i_{\b})\PP_{I_\b(v^*_i)}^{C_\b(v^*_i)}
\SS_{i_\b(v^*_i)}^{c_\b(v^*_i)}K^{h_i}_{v^*_i}(\xx_{v^*_i})\Big]\cdot\cr
&\cdot\Big\{\prod_{v\,\hbox{\ottorm not e.p.}}{1\over s_v!} \int
dP_{T_v}(\tt_v)\PP_{I_\b(v)}^{C_\b(v)}
\SS_{i_\b(v)}^{c_\b(v)}\Pf G^{h_v,T_v}_\b(\tt_v)\cdot\cr
&\cdot\Big[\prod_{l\in T_v}\hat\dpr^{q_\b(f^1_l)}_{j_\b(f^1_l)}
\hat\dpr^{q_\b(f^2_l)}_{j_\b(f^2_l)}[d^{b_\b(l)}_{j_\b(l)}(\xx_l,\yy_l) 
\PP_{I_\b(l)}^{C_\b(l)}
\SS_{i_\b(l)}^{c_\b(l)}g^{(h_v)}(f^1_l,f^2_l)]\Big]\Big\}\;,\cr}
\Eq(5.55)$$
where: $v^*_1,\ldots,v^*_n$ are the endpoints of $\t$; 
$b_\b(v)$, $b_\b(l)$, $q_\b(f^1_l)$ and $q_\b(f^2_l)$ are
nonnegative integers $\le 2$; $j_\b(v)$, $j_\b(f^1_l)$, $j_\b(f^2_l)$
and $j_\b(l)$ can be 
$0$ or $1$; $i_\b(v)$ and $i_\b(l)$ can be $1$ or $2$;
$I_\b(v)$ and $I_\b(l)$ can be $0$ or $1$; $C_\b(v)$, $c_\b(v)$, $C_\b(l)$
and $c_\b(l)$ can be $0,1$ and $\max\{C_\b(v)+c_\b(v),C_\b(l)+ c_\b(l\})\le 1$;
$G^{h_v,T_v}_\b(\tt_v)$ is obtained from $G^{h_v,T_v}(\tt_v)$
by substituting the element $t_{i(f),i(f')}g^{(h_v)}(f,f')$ with 
$t_{i(f),i(f')}\hat\dpr^{q_\b(f)}_{j_\b(f)}
\hat\dpr^{q_\b(f')}_{j_\b(f')}g^{(h_v)}(f,f')$.\\

It would be very difficult to give a precise description
of the various contributions of the sum over $B_T$, but fortunately
we only need to know some very general properties, which easily
follows from the construction in \sec(5.1)--\sec(5.3).\\
\\
1) There is a constant $C$ such that, $\forall T\in {\bf T}_\t$,
$|B_T|\le C^n$;
for any $\b\in B_T$, the following inequality is satisfied
$$\Big[\prod_{f\in \cup_v P_v} \g^{h(f) q_\b(f)} \Big]
\Big[\prod_{l\in T} \g^{-h(l) b_\b(l)} \Big]\le
\prod_{v\,\hbox{\ottorm not e.p.}} \g^{-z(P_v)}\;,\Eq(5.56)$$
where: $h(f)=h_{v_0}-1$ if $f\in P_{v_0}$, otherwise it is the scale of
the vertex where the field with label $f$ is contracted;
$h(l)=h_v$, if $l\in T_v$ and
$$z(P_v)=\cases{
1 & if $|P_v|=4$ and $\r_v=p\;,$ \cr
2 & if $|P_v|=2$ and $\r_v=p\;,$ \cr
1 & if $|P_v|=2$, $\r_v=s$ and $\sum_{f\in P_v}\o(f)\not=0\;,$ \cr
0 & otherwise.\cr}\Eq(5.57)$$
2) If we define
$$\prod_{v\in\t}\Big[\Big({|\s_{h_v}|+|\m_{h_v}|\over 
\g^{h_v}}\Big)^{c_\b(v)i_\b(v)}
\prod_{\ell\in T_v}\Big({|\s_{h_v}|+|\m_{h_v}|\over 
\g^{h_v}}\Big)^{c_\b(\ell)i_\b(\ell)}\Big]
\defin \prod_{v\in V_\b}\Big({|\s_{h_v}|+|\m_{h_v}|\over 
\g^{h_v}}\Big)^{i(v,\b)}\;.\Eq(5.58)$$
the indeces $i(v,\b)$ satisfy, for any $B_T$, the following property:
$$\sum_{w\ge v}i(v,\b)\ge z'(P_v)\;,\Eq(5.59)$$
where
$$z'(P_v)=\cases{
1 & if $|P_v|=4$ and $\r_v=s\;,$ \cr
2 & if $|P_v|=2$ and $\r_v=s$ and $\sum_{f\in P_v}\o(f)=0\;,$ \cr
1 & if $|P_v|=2$, $\r_v=s$ and $\sum_{f\in P_v}\o(f)\not=0\;,$ \cr
0 & otherwise.\cr}\Eq(5.60)$$

\sub(5.a.6) 
We can bound 
any $|\PP_{I_\b(v)}^{C_\b(v)}\SS_{i_\b(v)}^{c_\b(v)}
\Pf G^{h_v,T_v}_\b|$ in \equ(5.55), with $C_\b(v)+c_\b(v)=0,1$, 
by using \equ(5.52), \equ(5.53) and 
Gram inequality, as illustrated in previous Chapter for the case of the 
integration of the $\c$ fields. Using that the elements of $G$ are all
propagators on scale $h_v$, dimensionally bounded as in Lemma 5.3, we find:
$$\eqalign{&
|\PP_{I_\b(v)}^{C_\b(v)}\SS_{i_\b(v)}^{c_\b(v)}
\Pf G^{h_v,T_v}_\b|\le C^{\sum_{i=1}^{s_v}|P_{v_i}|-|P_v|-
2(s_v-1)}\cdot\cr
&\qquad\cdot
\g^{{h_v\over 2}\left(\sum_{i=1}^{s_v}|P_{v_i}|-|P_v|-2(s_v-1)\right)}
\Big[\prod_{f\in J_v}
\g^{h_v q_\b(f)}\Big]\Big({|\s_{h_v}|+|\m_{h_v}|\over 
\g^{h_v}}\Big)^{c_\b(v)i_\b(v)+C_\b(v)I_\b(v)}\;,\cr}\Eq(5.61)$$
where $J_v=\cup_{i=1}^{s_v}P_{v_i}\setminus Q_{v_i}$.
We will bound
the factors $\Big({|\s_{h_v}|+|\m_{h_v}|\over 
\g^{h_v}}\Big)^{C_\b(v)I_\b(v)}$ 
using \equ(5.40a) times a constant.\\

If we call 
$$\eqalign{
J_{\t,\bP,T,\b}&=\int d\xx_{v_0}
\Big| \Big[ \prod_{i=1}^n d_{j_\b(v^*_i)}^{b_\b(v^*_i)}(\xx^i_\b,\yy^i_\b)
\PP_{I_\b(v^*_i)}^{C_\b(v^*_i)}
\SS_{i_\b(v^*_i)}^{c_\b(v^*_i)}K^{h_i}_{v^*_i}(\xx_{v^*_i})\Big]\cdot\cr
&\cdot \Big\{ \prod_{v\,\hbox{\ottorm not e.p.}} {1\over s_v!}
\Big[\prod_{l\in T_v} \hat\partial^{q_\b(f^1_l)}_{j_\b(f^1_l)}
\hat\partial^{q_\b(f^2_l)}_{j_\b(f^2_l)} [d^{b_\b(l)}_{j_\b(l)}(\xx_l,\yy_l)
\PP_{I_\b(l)}^{C_\b(l)}\SS_{i_\b(l)}^{c_\b(l)}
g^{(h_v)}(f^1_l,f^2_l)]\Big]\Big\}\Big|\;,\cr}
\Eq(5.62)$$
we have, under the hypothesis \equ(5.40),
$$\eqalign{&
J_{\t,\bP,T,\a}\le C^n M^2 |\l|^n\Big[\prod_{i=1}^n
\Big({|\s_{h^*_i}|+|\m_{h^*_i}|\over 
\g^{h^*_i}}\Big)^{c_\b(v^*_i)i_\b(v^*_i)}\Big]\cdot\cr
&\cdot\Big\{\prod_{v\,\hbox{\ottorm not e.p.}}{1\over s_v!} C^{2(s_v-1)}
\g^{h_v n_\n(v)}\g^{-h_v\sum_{l\in T_v}b_\b(l)}
\g^{-h_v\sum_{i=1}^n b_\b(v_i^*)}
\g^{-h_v(s_v-1)}\cdot\cr
&\cdot\g^{h_v\sum_{l\in T_v}\left[q_\b(f^1_l)+q_\b(f^2_l)\right]}
\Big\}\Big[\prod_{\ell\in T}\Big({|\s_{h_v}|+|\m_{h_v}|\over 
\g^{h_v}}\Big)^{c_\b(\ell)i_\b(\ell)}
\Big]\;,\cr} \Eq(5.63)$$ 
where $n_\n(v)$ is the number
of vertices of type $\n$ with scale $h_v+1$. 

Now, substituting \equ(5.61), \equ(5.63) into \equ(5.55), using
\equ(5.56), we find that:
$$\eqalign{
&\int d\xx_{v_0} |W_{\t,\bP,T,\b}(\xx_{v_0})|\le
C^n M^2|\l|^n \g^{-h D_k(|P_{v_0}|)}
\prod_{v\in V_\b}\Big({|\s_{h_v}|+|\m_{h_v}|\over\g^{h_v}}\Big)^{i(v,\b)}
\;\cdot\cr
&\cdot\; \prod_{v\,\hbox{\ottorm not e.p.}} \left\{ {1\over s_v!}
C^{\sum_{i=1}^{s_v}|P_{v_i}|-|P_v|}
\Big({Z_{h_v}\over Z_{h_v-1}}\Big)^{|P_v|\over 2}
\g^{-[-2+{|P_v|\over 2}+z(P_v)]}\right\}\;,\cr}\Eq(5.64)$$
where, if $k=\sum_{f\in P_{v_0}}q_\b(f)$, $D_k(p)=-2+p+k$ and we 
have used \equ(5.58).
Note that, given $v\in \t$ and $\t\in \TT_{h,n}$ and
using \equ(5.40a) together with the first two of \equ(5.40z), 
$$\eqalign{&{|\sigma_{h_v}|\over \g^{h_v}}={|\sigma_h|\over \g^h}
{|\sigma_{h_v}|\over |\sigma_h|}
\g^{h-h_v} \le {|\sigma_h|\over \g^h} \g^{(h-h_v)(1-c|\l|)}\le
C_1\g^{(h-h_{\bar v})(1-c|\l|)}\cr
&{|\m_{h_v}|\over \g^{h_v}}={|\m_h|\over \g^h}
{|\m_{h_v}|\over |\m_h|}
\g^{h-h_v} \le {|\m_h|\over \g^h} \g^{(h-h_v)(1-c|\l|)}\le
C_1\g^{(h-h_v)(1-c|\l|)}\cr}\Eq(5.65)$$
Moreover  
the indeces $i(v,\b)$ satisfy, for any $B_T$, 
\equ(5.60) so that, using \equ(5.65) and \equ(5.59), we find
$$\prod_{v\in V_\b}\Big({|\s_{h_v}|+|\m_{h_v}|\over 
\g^{h_v}}\Big)^{i(v,\b)}\le C_1^n \prod_{v\, {\rm not}\, {\rm e.p.}}
\g^{-(1-c|\l|)z'(P_v)}\;.\Eq(5.66)$$
%
%
%
%
%
Substituting \equ(5.65) into \equ(5.64) and using \equ(5.59), we find:
$$\eqalign{
&\int d\xx_{v_0} |W_{\t,\bP,T,\b}(\xx_{v_0})|\le
C^n M^2|\l|^n \g^{-h D_k(|P_{v_0}|)}
\;\cdot\cr
&\cdot\; \prod_{v\,\hbox{\ottorm not e.p.}} \left\{ {1\over s_v!}
C^{\sum_{i=1}^{s_v}|P_{v_i}|-|P_v|}
\Big({Z_{h_v}\over Z_{h_v-1}}\Big)^{|P_v|\over 2}
\g^{-[-2+{|P_v|\over 2}+z(P_v)+(1-c|\l|)z'(P_v)]}\right\}\;.\cr}\Eq(5.67)$$
and it holds: 
$$-2+{|P_v|\over 2}+z(P_v)+(1-c|\l|)z'(P_v)\ge {|P_v|\over 6}\;.
\Eq(5.68)$$
Then
\equ(5.45) in Theorem 5.1 
follows from the previous bounds and the remark that
$$\sum_{\t\in\TT_{h,n}}
\sum_{\bP\in\PP_\t}\sum_{T\in\bT}\sum_{\b\in B_T}
\prod_{v}{1\over s_v!}\g^{-{|P_v|\over 6}}\le
c^n\;,\Eq(5.69)$$
for some constant $c$, see [BM][GM] or [G] for further details.\\

The bound on $\tilde E_h$, $t_h$, \equ(5.45y) and \equ(5.45yz) follow 
from a similar analysis. 
The remarks following \equ(5.45y) and \equ(5.45yz) follow from noticing 
that in the expansion for $\LL\VV^{(h)}$ appear only propagators of type 
$\PP_0 g^{(h_v)}_{\underline a,\underline a'}$ or
$\PP_1 g^{(h_v)}_{\underline a,\underline a'}$ (in order to bound 
these propagators we do not need \equ(5.40a), see 
the last statement in Lemma 5.3). 
Furthermore, by construction $l_h,n_h$ and $z_h$ are 
independent of $\s_k,\m_k$, so that, in order to prove \equ(5.45yz) 
we do not even need the first two inequalities in \equ(5.40z).
\qed
\\

\sub(short) The sum over all the trees with root scale $h$
and  with at least a $v$ with $h_v=k$ is $O(|\l|\g^{{1\over 2}(h-k)})$;
this follows from the fact that
the bound \equ(5.69) holds, for some $c=O(1)$, 
even if $\g^{-|P_v|/6}$ is replaced
by $\g^{-\k |P_v|}$, for any constant $\k>0$ independent of $\l$;
and that $D_v$, instead of using \equ(5.68), can also 
be bounded as $D_v\ge 1/2 +|P_v|/12$. This property 
is called {\it short memory property.}

\pagina
\setcap{6. The flow of the running coupling constants.}
\capindex{6}{The flow of the running coupling constants.}
\vskip1.truecm
\section(6,The flow of the running coupling constants.)
\capindex{6}{The flow of the running coupling constants.}

The convergence of the expansion for the effective potential
is proved by Theorem 5.1 under the hypothesis that
the running coupling constants are small, see \equ(5.40),
and that the bounds \equ(5.40yz), \equ(5.40z) and \equ(5.40a) 
are satisfied. We now want to show that, choosing $\l$ small enough and 
$\n$ as a suitable function of $\l$, such hypothesis are indeed verified.
In the present Chapter, we will  
prove these hypotheses under the assumption that the Luttinger
model Beta function is vanishing; we will do more, and we
will find an explicit solution for the flow equation of $Z_h,\s_h,\m_h$,
satisfying in particular the bounds \equ(5.40yz), \equ(5.40z) 
and satisfying \equ(5.40a)
for any scale $\bar h\ge h^*_1$, where $h^*_1$ is a scale we will
explicitely choose in the present Chapter (it is the scale
dividing the anomalous regime from the non anomalous one).
The proof of the vanishing of the Beta function
will be done in Appendix A6, following the recent work [BM1].
The proof of Appendix A6 will be based on the implementation 
in our constructive formalism of some non perturbative identities
between Schwinger functions, that is of two different approximate
Ward identities 
for the two and four legs Schwinger functions respectively,
of the Dyson equation, and of some correction identities, expressing the 
corrections to the formal Ward identities in terms of two or four legs 
Schwinger functions.
It worths to stress that these non perturbative
identities are derived by making use of (chiral) gauge invariance,
that is {\it not} satisfied by the Ashkin--Teller model. 
However, since there is a model near to AT in a Renormalization Group sense
(we shall call it the {\it reference model}) satisfying these symmetries, 
the cancellations appearing in the perturbation theory of the 
reference model 
also imply cancellations for AT itself. We can say that some {\it hidden}
symmetries of Ashkin--Teller allow us to control the flow
of its running coupling constants. Note that here the word ``hidden''
has a different (and much deeper) meaning than in the introducion of 
Chapter 4.
\\
\\
\sub(6.1){\bf The flow equations}
\\ 
We will first solve the flow equations for the 
renormalization constants (following 
from \equ(5.17a) and preceding line):
$${Z_{h-1}\over Z_h} = 1+ z_h\virg
{\s_{h-1}\over \s_h} = 1+{s_h/\s_h-z_h\over 1+z_h}\virg
{\m_{h-1}\over \m_h} = 1+{m_h/\m_h-z_h\over 1+z_h}\;,\Eq(6.4z)$$
together with those for the running coupling constants \equ(beta):
$$\eqalign{&\l_{h-1}=\l_h+\b^h_\l(\l_h,\n_h;\ldots;\l_1,\n_1)\cr
&\n_{h-1}=\g\n_h+\b^h_\n(\l_h,\n_h;\ldots;\l_1,\n_1)\;.\cr}\Eq(6.1)$$
The functions $\b^h_\l, \b^h_\n$ are called
the $\l$ and $\n$ components of the Beta function, see the comment after 
\equ(eth), and, by construction, are
{\it independent} of $\s_k,\m_k$, so that their convergence 
follow just from \equ(5.40) and the last of \equ(5.40z), \ie 
without assuming \equ(5.40a), see Theorem 5.1.
While for a general kernel we will
apply Theorem 5.1 just up to a finite scale $h^*_1$ (in order
to insure the validity of \equ(5.40a) with $\bar h=h^*_1$), 
we will inductively study the flow generated by 
\equ(6.1) up to scale $-\io$, and we shall prove that it is bounded
for all scales. The main result on the flows of $\l_h$ and $\n_h$,
proven in next section, is the following.\\
\\
{\cs Theorem 6.1.} {\it If $\l$ is small enough, there exists an 
analytic function $\n^*(\l)$ independent of $t,u$
such that the running coupling constants 
$\{\l_h,\n_h\}_{h\le 1}$ with $\n_1=\n^*(\l)$
verify
$|\n_h|\le c|\l|\g^{(\th/2)h}$ and $|\l_h|\le c|\l|$.
Moreover the kernels $z_h,s_h$ and $m_h$ satisfy
\equ(5.40yz) and the solutions of the flow equations \equ(6.4z) 
satisfy \equ(5.40z).}\\
\\
\sub(6.2){\bf Proof of Theorem 6.1.}
\\
We consider the space $\MMM_\th$ of sequences $\un=\{\n_h\}_{h\le 1}$
such that $|\n_h|\le c|\l|\g^{(\th/2)h}$; we shall think $\MMM_\th$
as a Banach space with norm $||\cdot||_\th$, where $||\un||_\th\defin
\sup_{k\le 1}|\n_k|\g^{-(\th/2)k}$. We will proceed as follows: we 
first show that, for any sequence $\un\in\MMM_\th$, 
the flow equation for $\n_h$, the hypothesis 
\equ(5.40yz), \equ(5.40z) and the property $|\l_h(\un)|\le c|\l|$
are verified, uniformly in $\un$. Then we fix $\un\in\MMM_\th$
via an exponentially convergent iterative procedure, in such a way that
the flow equation for $\n_h$ is satisfied.\\
\\
Given $\un\in\MMM_\th$,
let us suppose inductively that \equ(5.40yz), \equ(5.40z) and that,
for $k>\bar h+1$,
$$|\l_{k-1}(\un)-\l_k(\un)|\le c_0|\l|^{2}
\g^{(\th/2)k}\;,\Eq(6.5.7)$$
for some $c_0>0$. Note that \equ(6.5.7) is certainly true for $h=1$ 
(in that case the r.h.s. of \equ(6.5.7)
is just the bound on $\b^1_\l$).
Note also that \equ(6.5.7) implies  
that $|\l_k|\le c|\l|$, for any $k> \bar h$.

Using \equ(5.45y), the second of
\equ(5.45yz) and \equ(6.4z) we find that \equ(5.40yz), \equ(5.40z)
are true with $\bar h$ replaced by $\bar h-1$. 

We now consider the equation $\l_{h-1}=\l_h+\b^h_\l(\l_h,\n_h;\ldots;
\l_1,\n_1)$, $h>\bar h$. 
The function $\b^h_\l$ can be expressed as a convergent
sum over tree diagrams, as described in \sec(5.5); note that
it depends on $(\l_h,\n_h;\ldots;
\l_1,\n_1)$ directly through the end--points of the trees and indirectly
through the factors $Z_h/Z_{h-1}$.

We can write $\PP_0 g^{(h)}_{(+,\o),(-,\o)}(\xx-\yy)=
g^{(h)}_{L,\o}(\xx-\yy)+
r^{(h)}_\o(\xx-\yy)$, where 
$$g^{(h)}_{L,\o}(\xx-\yy)\defin {4\over M^2}\sum_{\kk}e^{-i\kk(\xx-\yy)}
\widetilde f_h(\kk){1\over ik+\o k_0}\Eq(6.gL)$$
and $r^{(h)}_\o$ is the rest, satisfying the same bound as 
$g^{(h)}_{(+,\o),(-,\o)}$, times a factor $\g^h$. This decomposition
induces the following decomposition for $\b^h_\l$:
$$\eqalign{&\b^h_\l(\l_h,\n_h;\ldots;\l_1,\n_1)=\cr
&\qquad=\b_{\l,L}^h
(\l_h,\ldots,\l_h)+
\sum_{k=h+1}^1 
D^{h,k}_\l + r^h_\l(\l_h,\ldots,\l_1)+\sum_{k\ge h}\n_k\tilde\b_\l^{h,k}
(\l_k,\n_k;\ldots;\l_1,\n_1)\;,\cr}\Eq(6.5.1aaa)$$
with 
$$\eqalign{&|\b^h_{\l,L}|\le c|\l|^2 \g^{\th h}\;,\qquad
|D^{h,k}_\l|\le c |\l| \g^{\th(h-k)} |\l_k-\l_h|\;,\qquad\cr
&|r^h_\l|\le c|\l|^2\g^{(\th/2) h}\;,\qquad |\tilde\b_\l^{h,k}|\le
c|\l|\g^{\th(h-k)}\;.\cr}\Eq(6.5.1bbb)$$
The first two terms in \equ(6.5.1aaa)
$\b^h_{\l,L}$ collect the contributions
obtained by posing $r^{(k)}_\o=0$, $k\ge h$ and substituting
the discrete $\d$ function defined after \equ(5.11) with 
$M^2\d_{\kk,{\bf 0}}$.  
The first of \equ(6.5.1bbb) is called 
the {\it vanishing of the Luttinger
model Beta function} property, and it is a crucial and non trivial property
of interacting fermionic systems in $d=1$. It will
be proved in Appendix A6.

Using the decomposition \equ(6.5.1aaa) and the bounds \equ(6.5.1bbb)
we prove the following bounds for $\l_{\bar h}(\un)$, $\un\in\MMM_\th$:
$$|\l_{\bar h}(\un)-\l_1(\un)|\le c_0|\l|^2
\virg
|\l_{\bar h}(\un)-\l_{\bar h+1}(\un)|\le c_0|\l|^2\g^{(\th/2) \bar h}
\;,\Eq(6.5.1cc)$$
for some $c_0>0$.
Moreover, given $\un,\un'\in\MMM_\th$, we show that:
$$|\l_{\bar h}(\un)-\l_{\bar h}(\un')|\le c|\l|
||\un-\un'||_0\;,\Eq(6.5.1d)$$
where $||\un-\un'||_0\defin\sup_{h\le 1}|\n_h-\n_h'|$.\\
\\
{\cs Proof of \equ(6.5.1cc).} 
We decompose $\l_{\bar h}-\l_{\bar h+1}=\b^{\bar h+1}_\l$ 
as in \equ(6.5.1aaa). Using the bounds
\equ(6.5.1bbb) and the inductive hypothesis \equ(6.5.7), we find:
$$\eqalign{|\l_{\bar h}(\un)-\l_{\bar h+1}(\un)| 
&\le c|\l|^2\g^{\th (\bar h+1)}+
\sum_{k\ge \bar h+2}c|\l|\g^{\th(\bar h+1-k)}\sum_{k'=\bar h+2}^k 
c_0|\l|^{2}\g^{(\th/ 2)k'}+\cr
&+c|\l|^2\g^{(\th/2)(\bar h+1)}+\sum_{k\ge \bar h+1}
c^2|\l|^2\g^{(\th/2)k}\g^{(\th(\bar h+1-k))}
\;,\cr}
\Eq(6.5.8)$$
which, for $c_0$ big enough, immediately implies the second of
\equ(6.5.1cc) with $h\to h-1$; from this bound and the 
hypothesis \equ(6.5.7) follows the first of \equ(6.5.1cc).\qed\\
\\
{\cs Proof of \equ(6.5.1d).} If we take two sequences 
$\un,\un'\in\MMM_\th$, we
easily find that the beta function for $\l_{\bar h}(\un)-\l_{\bar h}(\un')$
can be represented by a tree expansion similar to the one for $\b^h_\l$,
with the property that the trees giving a non vanishing contribution 
have necessarily one end--point on scale $k\ge h$ associated to a 
coupling constant $\l_k(\un)-\l_k(\un')$ or $\n_k-\n_k'$. Then we find:
$$\l_{\bar h}(\un)-\l_{\bar h}(\un')
=\l_{1}(\un)-\l_{1}(\un')+\sum_{\bar h+1\le k\le 1}
[\b^k_\l(\l_k(\un),\n_k;
\ldots;\l_1,\n_1)-\b^k_\l(\l_k(\un'),\n_k';
\ldots;\l_1,\n_1')]
\;.\Eq(6.5.8aaa)$$
Note that $|\l_1(\un)-\l_1(\un')|\le c_0|\l||\n_1-\n_1'|$,
because $\l_1=\l/Z_1^2+O(\l^2/Z_1^4)$
and $Z_1=\sqrt2-1+\n/2$. 
If we inductively suppose that,
for any $k> \bar h$, $|\l_k(\un)-\l_k(\un')|\le 2c_0|\l|
||\un-\un'||_0$, we
find, by using the decomposition
\equ(6.5.1aaa):
$$|\l_{\bar h}(\un)-\l_{\bar h}(\un')|\le c_0|\l||\n_1-\n_1'|+
c|\l|\sum_{k\ge \bar h+1}\g^{(\th/2)k}\sum_{k'\ge k}
\g^{\th(k-k')}\Big[2c_0|\l|\,||\un-\un'||_0+
|\n_k-\n'_k|\Big]\;.\Eq(6.5.8aab)$$
Choosing $c_0$ big enough, \equ(6.5.1d) follows.\qed\\

We are now left with fixing the sequence $\un$ in such a way that 
the flow equation for $\n$ is satisfied.
Since we want to fix $\un$ in such a way that $\n_{-\io}=0$, we must have:
$$\n_1=-\sum_{k=-\io}^1\g^{k-2}\b^k_\n(\l_k,\n_k;\ldots;\l_1,\n_1)\;.
\Eq(6.5.8a)$$
If we manage to fix $\n_1$ as in \equ(6.5.8a), we also get:
$$\n_h=-\sum_{k\le h}\g^{k-h-1}\b^k_\n(\l_k,\n_k;\ldots;\l_1,\n_1)\;.
\Eq(6.5.8b)$$
We look for a fixed point 
of the operator $\bT:\MMM_\th\to\MMM_\th$ defined as:
$$(\bT\un)_h=-\sum_{k\le h}\g^{k-h-1}\b^k_\n(\l_k(\un),\n_k;\ldots;\l_1,\n_1)
\,.\Eq(6.5.8c)$$
where $\l_k(\un)$ is the solution of the first line of \equ(6.1), 
obtained as a function of the {\it parameter} $\un$, as described above.

If we find a fixed point $\un^*$ of \equ(6.5.8c), 
the first two lines in \equ(6.1) 
will be simultaneously solved by $\ul(\un^*)$ and $\un^*$ respectively, 
and the solution will have the desired smallness properties for $\l_h$ and 
$\n_h$.

First note that, if $|\l|$ is sufficiently small, then $\bT$ leaves 
$\MMM_\th$ invariant: in fact, as a consequence of parity
cancellations, the $\n$--component
of the Beta function satisfies:
$$
\b^h_\n(\l_h,\n_h;\ldots;\l_1,\n_1)=
\b^h_{\n,1}(\l_h;\ldots;\l_1)+\sum_{k}\n_k
\tilde\b^{h,k}_\n(\l_h,\n_h;\ldots;\l_1,\n_1)\Eq(6.gg)$$
where, if $c_1, c_2$ are suitable constants
$$|\b^h_{\n,1}|\le c_1 |\l|\g^{\th h}\;\quad
|\tilde\b^{h,k}_\n|\le c_2|\l|\g^{\th(h-k)}\;.\Eq(6.kmn)$$
by using \equ(6.gg)
and choosing $c=2 c_1$ we obtain
$$|(\bT \n)_h|\le \sum_{k\le h} 2 c_1|\l|\g^{(\th/2) k}\g^{k-h}\le 
c|\l|\g^{(\th/2)h}\;,\Eq(6.5.8d)$$
Furthermore, using \equ(6.gg) and \equ(6.5.1d), we find that 
$\bT$ is a contraction on $\MMM_\th$:
$$\eqalign{&|(\bT \n)_h-(\bT\un')_h|\le \sum_{k\le h}
\g^{k-h-1}|\b^k_\n(\l_k(\un),\n_k;
\ldots;\l_1,\n_1)-\b^k_\n(\l_k(\un'),\n_k';
\ldots;\l_1,\n_1')|\le\cr
&\le c\sum_{k\le h}\g^{k-h-1}
\left[\g^{\th k}\sum_{k'=k}^1 |\l_{k'}(\un)-
\l_{k'}(\un')|+\sum_{k'=k}^1\g^{\th(k-k')}|\l||\n_{k'}-\n_{k'}'|
\right]\le \cr
&\le c'\sum_{k\le h}\g^{k-h-1}\Big[|k|\g^{\th k}|\l|\,||\un-\un'||_0
+\sum_{k'=k}^1\g^{\th(k-k')}
|\l|\g^{(\th/2)k'}||\un-\un'||_\th
\le\cr
&\le c''|\l|\g^{(\th/2)h}
||\un-\un'||_\th\;.\cr}\Eq(6.5.8e)$$
hence $||(\bT \n)-(\bT\un')||_\th\le c''|\l|||\un-\un'||_\th$.
Then, a unique fixed point $\un^*$ for $\bT$ exists on $\MMM_\th$. 
Proof of 
Theorem 6.1 is concluded by noticing that $\bT$ 
is analytic (in fact 
$\b^h_\n$ and $\ul$ are analytic in $\un$ in the domain $\MMM_\th$).\qed\\
\\
\sub(6.3){\bf The flow of the renormalization constants.}
\\
Once that $\n_1$ is conveniently
chosen as in Theorem 6.1, one can study in more detail the flows of the 
renormalization constants. We will now prove the following.
\\
\\
{\cs Lemma 6.1.} {\it If $\l$ is small enough and $\n_1$ is 
chosen as in Theorem 
6.1, the solution of \equ(6.4z) can be written as:
$$Z_h=\g^{\h_z(h-1)+F^h_\z}\virg \mu_h=\m_1\g^{\h_\m(h-1)+F^h_\m}\virg
\s_h=\s_1\g^{\h_\s(h-1)+F^h_\s}\Eq(6.lem4.1)$$
where $\h_z,\h_\m,\h_z$ and $F^h_\z,F^h_\m,F^h_\s$ are $O(\l)$
functions, independent of $\s_1,\m_1$. 

Moreover $\h_\s-\h_\m=-b\l+O(|\l|^2)$, $b>0$.}\\
\\
{\cs Proof of Lemma 6.1} 
From now on we shall think $\l_h$ and $\n_h$ fixed, with $\n_1$ conveniently
chosen as above ($\n_1=\n_1^*(\l)$). Then we have $|\l_h|\le c|\l|$ 
and $|\n_h|\le c|\l|\g^{(\th/2)h}$, for some $c,\th>0$. Having fixed 
$\n_1$ as a convenient function 
of $\l$, we can also think $\l_h$ and $\n_h$ as functions of $\l$.\\

\0{\it The flow of $Z_h$.}
The flow of $Z_h$ is given by the first of \equ(6.4z) with $z_h$
independent of $\s_k,\m_k$, $k\ge h$. By Theorem 3.1
we have that $|z_h|\le c|\l|^2$, uniformly in $h$. Again, as for $\l_h$ and 
$\n_h$, we can formally study this equation up to $h=-\io$.
We define $\g^{-\h_z}
\defin\lim_{h\to-\io} 1+z_h$, so that
$$\eqalign{&
\log_\g Z_h=\sum_{k\ge h+1}\log_\g (1+z_k)=\h_z(h-1)+\sum_{k\ge h+1}
r_\z^k\virg
r^k_\z\defin\log_\g\big(1+{z_k-z_{-\io}\over 1+z_{-\io}}
\big)\;.\cr}\Eq(6.5.9)$$
Using the fact that $z_{k-1}-z_k$ is necessarily proportional
to $\l_{k-1}-\l_k$ or to $\n_{k-1}-\n_k$ and that $\l_{k-1}-\l_k$
is bounded as in \equ(6.5.7), we easily find:
$|r^k_\z|\le c\sum_{k'\le k}|z_{k'-1}-z_{k'}|\le c'|\l|^2\g^{(\th/2)k}$. 
So, if $F^h_\z\defin \sum_{k\ge h+1}
r_\z^k$ and 
$F^1_\z=0$, then $F^h_\z=O(\l)$ and $Z_h=\g^{\h_z(h-1)+F^h_\z}$.
Clearly, by definition, $\h_z$ and $F^h_\z$ only depend on $\l_k$, $\n_k$, 
$k\le 1$, so they are independent of $t$ and $u$.\\

\0{\it The flow of $\m_h$.}
The flow of $\m_h$ is given by the last of \equ(6.4z). 
One can easily show inductively 
that $\m_k(\kk)/\m_h$, $k\ge h$, is independent of $\m_1$, so that
one can think that $\m_{h-1}/\m_h$ is just a function of $\l_h$, $\n_h$.
Then, again we can study the flow equation for $\m_h$ up to $h\to -\io$.
We define $\g^{-\h_\m}\defin \lim_{h\to-\io} 1+(m_h/\m_h-z_h)/(1+z_h)$,
so that, proceeding as for $Z_h$, we see that 
$$\mu_h=\m_1\g^{\h_\m(h-1)+F^h_\m}\;,\Eq(6.mu)$$ 
for a suitable $F^h_\m=O(\l)$.
Of course $\h_\m$ and $F^h_\m$ are independent of $t$ and $u$.\\

\0{\it The flow of $\s_h$.}
The flow of $\s_h$ can be studied as the one of $\m_h$. If 
we define $\g^{-\h_\s}\defin\lim_{h\to-\io}1+(s_h/\s_h-z_h)/(1+z_h)$, we find
that 
$$\s_h=\s_1\g^{\h_\s(h-1)+F^h_\s}\;,\Eq(6.sigma)$$ 
for a suitable $F^h_\s=O(\l)$.
Again, $\h_\s$ and $F^h_\s$ are independent of $t,u$.\\

We are left with proving that $\h_\s-\h_\m\not=0$. It is sufficient to note 
that, by direct computation of the lowest order terms, 
for some $\th>0$, \equ(6.4z) can
be written as:
$$\eqalign{&
z_h=b_1 \l_h^2+O(|\l|^2\g^{\th h})+O(|\l|^3)\virg b_1>0\cr
&{s_h/ \s_h}=-b_2\l_h+O(|\l|\g^{\th h})+O(|\l|^2)\virg b_2>0\cr
&{m_h/ \m_h}=b_2\l_h+O(|\l|\g^{\th h})+O(|\l|^2)\virg b_2>0\;,\cr}
\Eq(6.5.5z)$$
where $b_1,b_2$ are constants independent of $\l$ and $h$.
Using \equ(6.5.5z) and the definitions of $\h_\m$ and $\h_\s$ we
find: $\h_\s-\h_\m=(2b_2/\log\g)\l+O(\l^2)$.\qed\\
\\
\sub(6.100){\bf The scale $h^*_1$}
\\
The integration described in Chapter 5 is iterated
until a scale $h^*_1$ defined in the following way: 
$$h^*_1\defin\cases{\min\big\{1,
\big[\log_\g|\s_1|^{1\over 1-\h_\s}\big]\big\} 
& if $|\s_1|^{1\over 1-\h_\s}
>2|\m_1|^{1\over 1-\h_\m}$,\cr 
\min\big\{1,\big[\log_\g|u|^{1\over 1-\h_\m}\big]\big\} & if 
$|\s_1|^{1\over 1-\h_\s}\le 2|\m_1|^{1\over 1-\h_\m}$.\cr}\Eq(6.hhh)$$

From \equ(6.hhh) it follows that 
$$C_2\g^{h^*_1}\le |\s_{h^*_1}|+|\m_{h^*_1}|
\le C_1\g^{h^*_1}\;,\Eq(6.5.20)$$ 
with $C_1,C_2$ independent of $\l,\m_1,\s_1$. 

This is obvious in the case $h^*_1=1$. If $h^*_1<1$ and 
$|\s_1|^{1\over 1-\h_\s}
>2|\m_1|^{1\over 1-\h_\m}$, then $\g^{h^*_1-1}=c_\s
|\s_1|^{1\over 1-\h_\s}$, with $1\le c_\s<\g$, 
so that, using the third of \equ(6.lem4.1), we see that
$C_2\g^{h^*_1}\le |\s_{h^*_1}|\le C_1'g^{h^*_1}$, for some $C_1',C_2=O(1)$. 
Furthermore, using also the second of \equ(6.lem4.1), we find
$${|\m_{h^*_1}|\over|\s_{h^*_1}|}=c_\s^{\h_\m-\h_\s}
|\m_1||\s_1|^{-{1-\h_\m\over 1-\h_\s}}
\g^{F^{h^*_1}_\m-F^{h^*_1}_\s}<1\Eq(6.1/3)$$ 
and \equ(6.5.20) follows.

If $h^*_1<1$ and $|\s_1|^{1\over 1-\h_\s}\le 2|\m_1|^{1\over 1-\h_\m}$, then
$\g^{h^*_1-1}=c_u
|u|^{1\over 1-\h_\m}$, with $1\le c_\m<\g$, 
so that, using the second of \equ(6.lem4.1) and $|\m_1|=O(|u|)$, we see that
$C_2\g^{h^*_1}\le |\m_{h^*_1}|\le C_1'\g^{h^*_1}$. Furthermore, 
using the third \equ(6.lem4.1), we find
$${|\s_{h^*_1}|\over|\m_{h^*_1}|}=c_u^{\h_\s-\h_\m}
|\s_1||u|^{-{1-\h_\s\over 1-\h_\m}}
\g^{F^{h^*_1}_\s-F^{h^*_1}_\m}<C_1''\;,\Eq(6.1/0)$$ 
for some $C_1''=O(1)$, and \equ(6.5.20) again follows.\\
\\
{\it Remark.} The specific value of $h^*_1$ is not crucial:
if we change $h^*_1$ in $h^*_1+n$, $n\in\ZZZ$, 
the constants $C_1,C_2$ in \equ(6.5.20) 
are replaced by different $O(1)$ constants
and the estimates below
are not qualitatively modified. Of course, the specific values 
of $C_1,C_2$ (then, the specific value of $h^*_1$) can affect 
the convergence radius of the pertubative series 
in $\l$. The optimal value of $h^*_1$ should be chosen by maximizing the
corresponding convergence radius. Since here we are not interested in 
optimal estimates, we find the choice in \equ(6.hhh) convenient.

Note also that $h^*_1$ is a non analytic function of $(\l,t,u)$ 
(in particular for small $u$ we have $\g^{h^*_1}\sim |u|^{1+O(\l)}$). 
As a consequence, the asymptotic expression for the specific heat
near the critical points (that we shall obtain in next section)
will contain non analytic functions of $u$ (in fact it will contain
terms depending on $h^*_1$). 
However, as remarked after the Main Theorem in Chapter 1, this does not
imply that $C_v$ is non analytic: it is clear that in this case
the non analyticity 
is introduced ``by hands'' by our specific choice of $h^*_1$.
\\

From the results of Theorem 6.1 and Lemma 6.1, together with \equ(6.hhh) and 
\equ(6.5.20), it follows that the assumptions of Theorem 5.1
are satisfied for any $\bar h\ge h^*_1$. The integration of the scales 
$\le h^*_1$ must be performed in a different way, as will be discussed in 
next Chapter.

\pagina
\setcap{7. Renormalization Group for light fermions. The non anomalous regime.}
\capindex{7}{Renormalization Group for light fermions. 
The non anomalous regime.}
\vskip1.truecm
\section(7, Renormalization Group for light fermions. 
The non anomalous regime.)
\capindex{7}{Renormalization Group for light fermions. 
The non anomalous regime.}

In the preceding Chapters, we have explained how to integrate
the $\psi$ fields up to the scale $h^*_1$, defined in the last section 
of previous Chapter. We managed to prove that, up to that scale,
the running coupling constants can be bounded as in \equ(5.40z), \equ(5.40a)
and \equ(5.40), so that the iterative construction is inductively well 
defined, and the kernels of the effective potentials can be bounded, 
step by step, as stated by Theorem 5.1. Once we reach the scale $h^*_1$,
the bound \equ(5.40a) stops to be true and the bounds leading to 
Theorem 5.1 fail. In particular the crucial bound \equ(5.65) stops to be true. 
As a consequence, from this scale on, we have to proceed via a different
iterative procedure. The idea is to use conditions \equ(5.40), which
hold with an equal sign on scale $\bar h=h^*_1$, to prove that 
the $\psi^{(\le h^*_1)}$ field can be written, by a rotation 
which is essentially the inverse of \equ(2.13), as a sum of two fields
$\psi^{(1,\le h^*_1)}$, $\psi^{(2,\le h^*_1)}$, one of 
whom is massive on scale $h^*_1$ (\ie with mass $O(\g^{h^*_1})$).
It is then easy to show that one can integrate in one step (\ie without
any further multiscale integration) the massive field, so that 
one is left with an effective theory involving only
the (nearly) massless field.

The two fields correspond to the variables associated to the two original
Ising layers. We can then say that on scale $h^*_1$ the theory is effectively
described by a theory of two interacting Ising layers, with 
(anomalously) renormalized parameters. On scale $h^*_1$ one of the two layers
(the one corresponding to the massive field) is well far from criticality
and the corresponding variables can be easily integrated out; we are
left with the theory of a single perturbed Ising model with 
renormalized parameters. The multiscale integration for the latter
will be much easier than that described above, and in particular it will
not involve any anomalous flow of the effective renormalization constants.\\

In the present Chapter we will first describe the integration
of the massive field and the iterative integration of left over massless 
field. A corollary of the construction will be 
the analyticity of the free energy for temperatures different from 
the critical ones. Finally we will derive and solve the equation for the 
critical temperatures, leading to \equ(1.3).\\
\\
\sub(7.6.aaa){\bf Integration of the $\psi^{(1)}$ fields}
\\
If $h^*_1$ is fixed as in \sec(6.100), we can use Theorem 5.1 up to
the scale $\bar h=h^*_1+1$.  

Once that all the scales $> h^*_1$ are integrated out,
it is more convenient to describe the system in terms of
the fields $\psi^{(1)}_{\o},\psi^{(2)}_{\o}$, $\o=\pm 1$,  
defined through the following
change of variables:
$$\hat\psi^{\a(\le h^*_1)}_{\o,\kk}= {1\over\sqrt{2}} (\hat\psi^{(1,
\le h^*_1)}_{\o,-\a\kk}-i\a \hat\psi^{(2,\le h^*_1)}_{\o,-\a\kk})\;,
\quad \psi^{(j)}_{\o,\xx}={1\over M^2}\sum_{\kk}e^{-i\kk\xx}\hat\psi^{
(j)}_{\o,\kk}\;.\Eq(7.4.21b)$$
If we perform this change of variables, we find 
$P_{Z_{h^*_1},\s_{h^*_1},\m_{h^*_1},C_{h^*_1}}
=
\prod_{j=1}^2 P^{(j)}_{
Z_{h^*_1},m_{h^*_1}^{(j)},C_{h^*_1}}$ where, if 
we define $\Psi^{(j,\le h^*_1),T}_\kk\defin(\psi^{(j,\le h^*_1)}_{1,\kk},\ 
\psi^{(j,\le h^*_1)}_{-1,\kk})$,
$$\eqalign{& P^{(j)}_{Z_{h^*_1},m_{h^*_1}^{(j)},C_{h^*_1}}
(d\psi^{(j,\le h^*_1)})\defin\cr
&\defin{1\over N_{h^*_1}^{(j)}}\prod_{\kk,\o} d\psi^{(j,\le h^*_1)}_{\o,\kk}
\exp\Big\{-{Z_{h^*_1}\over 4M^2}\sum_{\kk\in
D_{h^*_1}} 
C_{h^*_1}(\kk)\Psi^{(j,\le h^*_1),T}_{\kk}
A^{(h^*_1)}_j(\kk)\Psi^{(j,\le h^*_1)}_{-\kk}
\Big\}
\cr
&A^{(h^*_1)}_j(\kk)\defin\pmatrix
{(-i\sin k-\sin k_0)+a^{+(j)}_{h^*_1}(\kk)&
-i \big(m_{h^*_1}^{(j)}(\kk)+c_{h^*_1}^{(j)}(\kk)\big)\cr i
\big(m_{h^*_1}^{(j)}(\kk)+c_{h^*_1}^{(j)}(\kk)\big)&
(-i\sin k+\sin k_0)+a^{-(j)}_{h^*_1}(\kk)\cr}\cr}\Eq(7.4.22)$$
and $a^{\o(j)}_{h^*_1}$, $m_{h^*_1}^{(j)}$, $c_{h^*_1}^{(j)}$
are given by \equ(A4.24y) with $h=h^*+1$.

The propagators $g^{(j,\le h^*_1)}_{\o_1,\o_2}$ 
associated with the fermionic integration \equ(7.4.22) 
are given by \equ(A4.24) with $h=h^*_1+1$.
Note that, by \equ(6.5.20), $\max\{|m_{h^*_1}^{(1)}|,|m_{h^*_1}^{(2)}|\}=
|\s_{h^*_1}|+|\m_{h^*_1}|=O(\g^{h^*_1})$ (see \equ(A4.24y) for the definition
of $m_{h^*_1}^{(1)}$, $m_{h^*_1}^{(2)}$).
From now on, for definiteness we shall suppose that 
$\max\{|m_{h^*_1}^{(1)}|,|m_{h^*_1}^{(2)}|\}\=|m_{h^*_1}^{(1)}|$. 
Then, it is easy to realize that the propagator
$g_{\o_1,\o_2}^{(1,\le h^*_1)}$ is bounded as follows.
$$|\dpr^{n_0}_{x_0}\dpr^{n_1}_{x}g_{\o_1,\o_2}^{(1,\le h^*_1)}
(\xx)|\le C_{N,n}{\g^{(1+n)h^*_1}\over 1+(\g^{h^*_1}
|\dd(\xx)|)^N}\virg n=n_0+n_1\;,\Eq(7.4.4*)$$
namely $g_{\o_1,\o_2}^{(1,\le h^*_1)}$ satisfies the same bound as
the single scale propagator on scale $h=h^*_1$. This suggests to 
integrate out $\psi^{(1,\le h^*_1)}$, without any
other scale decomposition. We find the following result.\\
\\
{\cs Lemma 7.1} {\it If $|\l|\le \e_1$, 
$|\s_1|,|\m_1|\le c_1$ ($c_1,\e_1$ being the same as in Theorem 4.1)
and $\n_1$ is fixed as in Theorem 6.1,
we can rewrite the partition function as
$$\Xi_{AT}^-=\int P^{(2)}_{Z_{h^*_1}, \widehat m^{(2)}_{h^*_1}, C_{h^*_1}}
(d\psi^{(2,\le h^*_1)})e^{-\lis \VV^{(h^*_1)}({\sqrt Z_{h^*_1}}\psi^{(2,
\le h^*_1)})-M^2\lis E_{h^*_1}}
\;,\Eq(7.zh*)$$
where: $\widehat m^{(2)}_{h^*_1}(\kk)=m^{(2)}_{h^*_1}(\kk)-
\g^{h^*_1}\p_{h^*_1}C_{h^*_1}^{-1}(\kk)$,
with $\p_{h^*_1}$ a free parameter, s.t. $|\p_{h^*_1}|\le c|\l|$; 
$|\lis E_{h^*_1}-E_{h^*_1}|\le c|\l|\g^{2h^*_1}$;
and
$$\eqalign{\lis\VV^{(h^*_1)}(\psi^{(2)})-
\g^{h^*_1}\p_{h^*_1}F_\s^{(2,\le h^*_1)}(
\psi^{(2\le h^*_1)})
&=\sum_{n=1}^{\io}\sum_{\oo}\prod_{i=1}^{2n}
\hat\psi^{(2)}_{\o_i,\kk_i}
\lis W^{(h^*_1)}_{2n,\oo}(\kk_1,\ldots,\kk_{2n-1})
\d(\sum_{i=1}^{2n}\kk_i)=\cr
&=\sum_{n=1}^{\io}
\sum_{\ss,\underline j,\oo}\prod_{i=1}^{2n}
\dpr^{\s_i}_{j_i}\psi^{(2)}_{\o_i,\xx_i}
\lis W^{(h^*_1)}_{2n,\ss,\underline j,\oo}(\xx_1,\ldots,\xx_{2n})
\;,\cr}\Eq(7.4.50)$$
with $F_\s^{(2,\le h)}$ given by the first of
\equ(5.11) with $\hat\psi^{(2,\le h)}_{\o,\kk}\hat\psi^{(2,\le h)}_{\o',-\kk}$
replacing $\hat\psi^{+(\le h)}_{\o,\kk}\hat\psi^{-(\le h)}_{\o',\kk}$; and
$\lis W^{(h^*_1)}_{2n,\ss,\underline j,\oo}$ satisfying the same 
bound \equ(5.45) as $W^{(\bar h)}_{2n,\ss,\underline j,\aa,\oo}$ 
with $\bar h=h^*_1$.}\\

In order to prove the Lemma it 
is sufficient to consider \equ(5.3) with $h=h^*_1$
and rewrite 
$P_{Z_{h^*_1},\s_{h^*_1},\m_{h^*_1},C_{h^*_1}}$ as
the product $\prod_{j=1}^2 P^{(j)}_{
Z_{h^*_1},m_{h^*_1}^{(j)},C_{h^*_1}}$. Then the integration
over the $\psi^{(1,\le h^*_1)}$ field is done as the integration
of the $\chi$'s in Chapter 4, recalling the bound \equ(7.4.4*). 

Finally we rewrite $m^{(2)}_{h^*_1}(\kk)$
as $\widehat m^{(2)}_{h^*_1}(\kk)+\g^{h^*_1}\p_{h^*_1}C_{h^*_1}^{-1}
(\kk)$, where 
$\p_{h^*_1}$ is a parameter to be suitably fixed below as a function 
of $\l,\s_1,\m_1$.\\

\sub(7.6.2aaa){\bf The localization operator}
\\
The integration of the r.h.s. of \equ(7.zh*)
is done in an iterative way similar to the one described
in Chapter 5.
If $h=h^*_1,h^*_1-1,\ldots$, we shall write:
$$\Xi_{AT}^-=\int P^{(2)}_{Z_{h}, \widehat m^{(2)}_{h}, C_{h}}
(d\psi^{(2,\le h)})e^{-\lis\VV^{(h)}
(\sqrt{Z_h}\psi^{(2,\le h)})-M^2E_h}\;,\Eq(7.veff)$$
where $\lis\VV^{(h)}$ is given by an expansion similar to \equ(5.50), with
$h$ replacing $h^*_1$ and $Z_{h}, \widehat m^{(2)}_{h}$
are defined recursively in the following way. We first
introduce a {\it localization operator} $\LL$. As in \sec(5.2), 
we define $\LL$ as a combination of four operators $\LL_j$ 
and $\lis\PP_j$, $j=0,1$. $\LL_j$ are defined as in \equ(5.7) and \equ(5.8),
while $\lis\PP_0$ and $\lis\PP_1$,
in analogy with \equ(5.9), are defined as the operators extracting from a 
functional of $\widehat m^{(2)}_h(\kk)$, $h\le h^*_1$, the contributions
independent and linear in $\widehat m^{(2)}_h(\kk)$.
Note that inductively the kernels $\lis W_{2n,\oo}^{(h)}$ can be thought
as functionals of $\widehat m_k(\kk)$, $h\le k\le h^*_1$.
Given $\LL_j,\lis\PP_j$, $j=0,1$ as above, we define the action of $\LL$
on the kernels $\lis W_{2n,\oo}^{(h)}$ as follows.\\
\\
\01) If $n=1$, then
$$\LL \lis W_{2,\oo}^{(h)}\defin\cases{
\LL_0(\lis\PP_0+\lis\PP_1)
\lis W_{2,\oo}^{(h)} & if $\o_1+\o_2=0$,\cr
\LL_1\lis\PP_0\lis W_{2,\oo}^{(h)} & if $\o_1+\o_2\not=0$.}$$
\02) If $n>2$, then $\LL \lis W_{2n,\oo}^{(h)}=0$.\\ 

It is easy to prove the analogue of Lemma 5.1: 
$$\LL\lis\VV^{(h)}=(s_h+\g^h p_h)F_\s^{(2,\le h)}+z_h 
F_\z^{(2,\le h)}\;,\Eq(7.4.54)$$
where $s_h,p_h$ and $z_h$ are real constants and: $s_h$ is linear
in $\widehat m_k^{(2)}(\kk)$, $h\le k\le h^*_1$; $p_h$ and $z_h$ are
independent of $\widehat m_k^{(2)}(\kk)$. Furthermore 
$F_\s^{(2,\le h)}$ and $F_\z^{(2,\le h)}$ are given by the first and the last 
of
\equ(5.11) with $\hat\psi^{(2,\le h)}_{\o,\kk}\hat\psi^{(2,\le h)}_{\o',-\kk}$
replacing $\hat\psi^{+(\le h)}_{\o,\kk}\hat\psi^{-(\le h)}_{\o',\kk}$.\\
\\
{\it Remark.} Note that the action of $\LL$ on the quartic 
terms is trivial. The reason of such a choice is that in the present case
no quartic local term can appear, because of Pauli principle:
$\psi^{(2,h)}_{1,\xx}\psi^{(2,h)}_{1,\xx}\psi^{(2,h)}_{-1,\xx}
\psi^{(2,h)}_{-1,\xx}\=0$,
so that $\LL_0\lis W_{4,\oo}=0$.\\

Using the symmetry properties exposed in \sec(4.3), we can 
prove the analogue of Lemma 5.2: if $n=1$, then
$$\RR \lis W_{2,\oo}=\cases{
[\lis\SS_2+\RR_2(\lis\PP_0+\lis\PP_1)]
\lis W_{2,\oo} & if $\o_1+\o_2=0$,\cr 
[\RR_1\lis\SS_1+\RR_2\lis\PP_0]
\lis W_{2,\aa,\oo} & if 
$\o_1+\o_2\not=0$,\cr}\Eq(7.rr*)$$
where $\lis\SS_1=1-\lis\PP_0$ and $\lis\SS_2=1-\lis\PP_0-\lis\PP_1$;
if $n=2$, then
$\lis W_{4,\oo}= \RR_1\lis W_{4,\oo}$.\\

\sub(7.6.3aaa){\bf Renormalization for $h\le h^*_1$}
\\
If $\LL$ and $\RR=1-\LL$ are defined as in previous subsection, we can rewrite
\equ(7.veff) as:
$$\int P^{(2)}_{Z_{h}, \widehat m^{(2)}_{h}, C_{h}}
(d\psi^{(2,\le h)})e^{-\LL\lis\VV^{(h)}(\sqrt{Z_h}\psi^{(2,\le h)})
-\RR\lis\VV^{(h)}(\sqrt{Z_h}\psi^{(2,\le h)})-M^2E_h}\;.\Eq(7.veff1)$$
Furthermore, using \equ(7.4.54) and defining:
$$\widehat Z_{h-1}(\kk)\defin
Z_h(1+C_h^{-1}(\kk) z_h)\virg
\widehat m_{h-1}^{(2)}(\kk)\defin{Z_h\over \widehat Z_{h-1}(\kk)}
\left(\widehat m_h^{(2)}(\kk)+C_h^{-1}(\kk) s_h\right)\;,\Eq(7.flow)$$
we see that \equ(7.veff1) is equal to
$$\int P^{(2)}_{\widehat Z_{h-1},\widehat m^{(2)}_{h-1}, C_{h}}
(d\psi^{(2,\le h)})e^{-\g^h p_h F_\s^{(2,\le h)}(\sqrt{Z_{h}}
\psi^{(2),\le h})-
\RR\lis\VV^{h}(\sqrt{Z_{h}}\psi^{(2),\le h})-M^2 (E_h+t_h)}\Eq(7.4.58*)$$
Again, we rescale the potential: 
$$\widetilde\VV^{(h)}(\sqrt{Z_{h-1}}
\psi^{(\le h)})\defin \g^h\p_h F_\s^{(2,\le h)}(\sqrt{Z_{h-1}}
\psi^{(2,\le h)})+
\RR\lis\VV^{h}(\sqrt{Z_{h}}\psi^{(2,\le h)})\;,\Eq(7.recsale)$$ 
where $Z_{h-1}=\widehat Z_{h-1}({\bf 0})$ and 
$\p_h=( Z_h/ Z_{h-1})p_h$; we
define $\widetilde f_h^{-1}$ as in \equ(5.21), we perform
the single scale integration and we define the new effective potential as
$$e^{-\lis\VV^{(h-1)}(\sqrt{Z_{h-1}}\psi^{(2,\le h-1)})-M^2 
\tilde E_{h}}\defin \int 
P^{(2)}_{Z_{h-1},\widehat m^{(2)}_{h-1}, \widetilde f^{-1}_h}
(d\psi^{(2,h)})
e^{-\widetilde\VV^{h}(\sqrt{Z_{h}}\psi^{(2,\le h)})}
\;.\Eq(7.4.58)$$
Finally we pose $E_{h-1}=E_h+t_h+\tilde E_h$.
Note that the above procedure allow us to write the 
$\p_h$ in terms of $\p_k$, $h\le k\le h^*_1$, namely
$\p_{h-1}=\g^h\p_h+\b^h_\p(\p_h,\ldots,\p_{h^*_1})$,
where $\b^h_\p$ is the {\it Beta function}.

Proceeding as in \sec(4) we can inductively show that $\lis\VV^{(h)}$ has 
the structure of \equ(7.4.50), with $h$ replacing $h^*_1$ and that 
the kernels of $\lis\VV^{(h)}$ are bounded as follows.\\
\\
{\cs Lemma 7.2.} {\it Let the hypothesis of Lemma 5.1 be satisfied
and suppose that, for $\bar h< h\le  h^*_1$ and some constants $c,\th>0$ 
$$e^{-c|\l|}\le{\widehat m^{(2)}_h\over \widehat m^{(2)}_{h-1}}\le 
e^{c|\l|}\virg
e^{-c|\l|^2}\le{Z_h\over Z_{h-1}}\le e^{c|\l|^2}
\virg |\p_h|\le c|\l|\virg |\widehat m^{(2)}_{\bar h}|\le 
\g^{\bar h}\;.\Eq(7.lem5.2)$$
Then the partition 
function can be rewritten as in \equ(7.veff)
and there exists $C>0$ s.t. the kernels of  
$\lis\VV^{(h)}$ satisfy:
$$\int d\xx_1\cdots d\xx_{2n}|\lis W^{(\bar h)}_{2n,\ss,\underline j,\oo}
(\xx_1,\ldots,\xx_{2n})|
\le M^2 \g^{-\bar h D_k(n)} \,(C\,|\l|)^{max(1,n-1)}\Eq(7.5.45*)$$
where $D_k(n)=-2+n+k$ and $k=\sum_{i=1}^{2n}\s_i$.
Finally $|E_{\bar h+1}|+|t_{\bar h+1}|\le c|\l|\g^{2\bar h}$.}\\

The proof of Lemma 7.2 is essentially identical to the proof
of Theorem 5.1 and we do not repeat it here.\\

It is possible to fix $\p_{h^*_1}$
so that the first three assumptions
in \equ(7.lem5.2) are valid for any $h\le h^*_1$. More precisely,
the following result holds, see Appendix A8 for the proof.\\
\\
{\cs Lemma 7.3.} {\it If $|\l|\le \e_1$, 
$|\s_1|,|\m_1|\le c_1$ and $\n_1$ is fixed as in Theorem 4.1,
there exists  
$\p^*_{h^*_1}(\l,\s_1,\m_1)$ such that, if we fix $\p_{h^*_1}=
\p^*_{h^*_1}(\l,\s_1,\m_1)$, for $h\le h^*_1$ we have: 
$$|\p_h|\le c|\l|\g^{(\th/2)(h-h^*_1)}\virg
\widehat m^{(2)}_h=\widehat m^{(2)}_{h^*_1}\g^{ F^h_m}\virg 
Z_h=Z_{h^*_1}\g^{\lis F^h_\z}\;,\Eq(7.4.58b)$$
where $F^h_m$ and $\lis F^h_\z$ are $O(\l)$. Moreover: 
$$\Big|\p^*_{h^*_1}(\l,\s_1,\m_1)-\p^*_{h^*_1}(\l,\s_1',\m_1')\Big|
\le c|\l|\left(\g^{(\h_\s-1)h^*_1}|\s_1-\s_1'|+
\g^{(\h_\m-1)h^*_1}|\m_1-\m_1'|\right)\;.\Eq(7.ph)$$}

\sub(7.6.101){\bf The integration of the scales $\le h^*_2$}
\\
In order to insure that the last assumption in \equ(7.lem5.2) holds, 
we iterate the preceding construction up to the scale
$h^*_2$ defined as the scale s.t. $|\widehat m_k^{(2)}|\le\g^{k-1}$ for any
$h^{*}_2\le k\le h^{*}_1$ and 
$|\widehat m_{h^{*}_2-1}^{(2)}|> \g^{h^*_2-2}$.

Once we have integrated all the fields $\psi^{(>h^*_2)}$, we can integrate
$\psi^{(2,\le h^*_2)}$ without any further multiscale 
decomposition. Note in fact that by definition the propagator
satisfies the same bound \equ(7.4.4*) with $h^*_2$ replacing $h^*_1$.
Then, if we define
$$e^{-M^2\tilde E_{\le h^*_2}}\defin \int P_{Z_{h^*_2-1},
\widehat m^{(2)}_{h^*_2-1},C_{h^*_2}}e^{-\widetilde \VV^{(h^*_2)}(\sqrt{Z_{
h^*_2-1}}\psi^{(2,\le h^*_2)})}\;,\Eq(7.e*2)$$
we find that $|\tilde E_{\le h^*_2}|
\le c|\l|\g^{2h^*_2}$ (the proof is a repetition
of the estimates on the single scale integration).

Combining this bound with the results of Theorem 5.1, Lemma 7.1, Lemma 7.2
and Lemma 7.3,
together with the results of Chapter 5, we finally find
that the free energy associated to $\Xi_{AT}^-$ is given by the following
{\it finite} sum, uniformly convergent with the size of $\L_M$:
$$\lim_{M\to\io}
{1\over M^2}\log \Xi_{AT}^-=E_{\le h^*_2}+(\lis E_{h^*_1}-E_{h^*_1})
+\sum_{h=h^*_2+1}^1 (\tilde E_h+t_h)\;,\Eq(7.oh)$$
where $E_{\le h^*_2}=\lim_{M\to\io}\tilde E_{\le h^*_2}$ and it is easy to
see that $E_{\le h^*_2}$, for any finite $h^*_2$, exists and satisfies 
the same bound of $\tilde E_{h^*_2}$.\\

\sub(7.an){\bf Keeping $h^*_2$ finite.}
\\
From the discussion of previous subsection, it follows that, for
any finite $h^*_2$, \equ(7.oh) is an analytic function of $\l,t,u$,
for $|\l|$ sufficiently small, uniformly in $h^*_2$
(this is an elementary consequence of Vitali's convergence theorem). 
Moreover, in Appendix A9 it is 
proved that,
for any $\g^{h^*_2}>0$, the limit \equ(7.oh) coincides with 
$\lim_{M\to\io}1/M^2 \log \Xi_{AT}^{\g_1,\g_2}$ for any choice 
$\g_1,\g_2$ of boundary conditions; hence this limit coincides
with $-2\log(2\cosh\b\l)$ plus 
the free energy in \equ(cv). We can state the result 
as follows.\\
\\
{\cs Lemma 7.4.} {\it There exists $\e_1>0$ such that, if $|\l|\le\e_1$ 
and $t\pm u\in D$ (the same as in the Main Theorem in Chapter 1), 
the free energy $f$ defined in \equ(cv) 
is real analytic in $\l,t,u$, except
possibly for the choices of $\l,t,u$ such that $\g^{h^*_2}=0$.}\\

We shall see in next Chapter that the specific heat is logarithmically 
divergent as $\g^{h^*_2}\to 0$. So the 
critical point is really given by the condition $\g^{h^*_2}=0$.
We shall explicitely solve the equation for the critical point
in next subsection. \\
\\
\sub(7.6.1){\bf The critical points}.
\\
In the present subsection we check that, if $t\pm u\in D$, $D$ being a suitable
interval centered around $\sqrt2-1$, see Main Theorem in Chapter 1, 
there are precisely two critical points, of the form \equ(1.3). 
More precisely, keeping in mind that the equation 
for the critical point is simply $\g^{h^*_2}=0$ (see the end of 
previous subsection), we prove the following.\\
\\
{\cs Lemma 7.5.} {\it Let 
$|\l|\le\e_1$, $t\pm u\in D$ and $\p_{h^*_1}$
be fixed as in Lemma 7.3.  
Then $\g^{h^*_2}=0$ only if $(\l,t,u)=
(\l,t_c^\pm(\l,u),u)$, where $t_c^\pm(\l,u)$ is given by \equ(1.3).}\\
\\
\proof
From the definition of $h^*_2$ given above, see \sec(7.6.101),
it follows that $h^*_2$ satisfies the following equation:
$$\g^{h^*_2-1}=c_m\g^{F^{h^*_2}_m}\Big|
|\s_{h^*_1}|-|\m_{h^*_1}|-\a_\s\g^{h^*_1}\p_{h^*_1}\Big|\;,
\Eq(7.4.61a)$$
for some $1\le c_m<\g$ and $\a_\s=\sign\s_1$. 
Then, the equation $\g^{h^*_2}=0$ can be rewritten as: 
$$|\s_{h^*_1}|-|\mu_{h^*_1}|-\a_\s \g^{h^*_1}\p_{h^*_1}=0\;.
\Eq(7.4.61aa)$$
First note that the result of Lemma 7.5 is trivial when $h^*_1=1$.
If $h^*_1<1$, \equ(7.4.61aa)
cannot be solved when $|\s_1|^{1\over 1-\h_\s}>2
|\m_1|^{1\over 1-\h_\m}$. In fact, 
$$\eqalign{&|\s_1|\g^{\h_\s(h^*_1-1)+F^{h^*_1}_\s}-|\m_1|\g^{\h_\m(h^*_1-1)+
F^{h^*_1}_\m}
-\a_\s\g^{h^*_1}\p_{h^*_1}=\cr
&=|\s_1|^{1+{\h_\s\over 1-\h_\s}}c_1-
\Big(|\m_1||\s_1|^{-{1-\h_\m\over 1-\h_\s}}\Big)|\s_1|^{{1-\h_\m\over 1-\h_\s}
-{\h_\m\over 1-\h_\s}}c_1'-\a_\s\g^{h^*_1}\p_{h^*_1}\ge {\g^{h^*_1-1}
\over 3\g}\;,\cr}\Eq(7.4.61aaa)$$
where $c_1,c_1'$ are constants $=1+O(\l)$,
$\p_{h^*_1}=O(\l)$ and $\g^{h^*_1-1}=c_\s
|\s_1|^{1\over 1-\h_\s}$,
with $1\le c_\s< \g$. Now, if $|\m_1|>0$, the r.h.s. of \equ(7.4.61aaa)
equation is strictly positive.\\

So, let us consider the case $h^*_1<1$ and $|\s_1|^{1\over 1-\h_\s}\le 2
|\m_1|^{1\over 1-\h_\m}$ 
(s.t. $\g^{h^*_1}=c_u\log_\g|u|^{1\over 1-\h_\m}$, with $1\le c_u\le \g$). 
In this case
\equ(7.4.61aa) can be easily solved to find:
$$|\s_1|=|\m_1||u|^{\h_\m-\h_\s\over 1-\h_\m}c_u^{\h_\m-\h_\s}\g^{F^{h^*_1}_\m
-F^{h^*_1}_\s}+|u|^{1-\h_\s\over 1-\h_\m}
c_u^{1-\h_\s}\a_\s\g^{1-F^{h^*_1}_\s}\p_{h^*_1}\;.
\Eq(7.6.102)$$
Note that $c_u^{\h_\m-\h_\s}\g^{F^{h^*_1}_\m
-F^{h^*_1}_\s}=1+O(\l)$ is just a function of $u$,
(it does not depend on $t$), because
of our definition of $h^*_1$. 
Moreover $\p_{h^*_1}$ is a smooth
function of $t$: if we call $\p_{h^*_1}(t,u)$ resp. 
$\p_{h^*_1}(t',u)$
the correction corresponding 
to the initial data $\s_1(t,u),\m_1(t,u)$ resp. $\s_1(t',u),\m_1(t',u)$, 
we have
$$|\p_{h^*_1}(t,u)-\p_{h^*_1}(t',u)|\le c|\l|
|u|^{\h_\s-1\over 1-\h_\m}|t-t'|\;,\Eq(7.6.106)$$
where we used \equ(7.ph) and the bounds $|\s_1-\s_1'|\le c|t-t'|$ and
$|\m_1-\m_1'|\le c|u||t-t'|$, following from 
the definitions of $(\s_1,\m_1)$ in terms of $(\s,\m)$ and
of $(t,u)$, see Chapter 4.

Using the same definitions we also realize that \equ(7.6.102) 
can be rewritten as
$$t=\Big[\sqrt2-1+{\n(\l)\over 2}\pm|u|^{1+\h}\Big(1+\l f(t,u)\Big)
\Big]{1+\hat\l(t^2-u^2)\over 1+\hat\l}\;,\Eq(7.6.105)$$
where 
$$1+\h\defin{1-\h_\s\over 1-\h_\m}\;,\Eq(7.eta)$$
and the crucial property is that $\h=-b\l+O(\l^2)$, $b>0$, see Lemma 6.1
and \equ(6.5.5z). We also recall that both $\h$ and $\n$ are
functions of $\l$ and are independent of $t,u$. 
Moreover $f(t,u)$ is a suitable bounded function s.t. 
$|f(t,u)-f(t',u)|\le c|u|^{-(1+\h)}|t-t'|$, as it follows from the Lipshitz 
property of $\p_{h^*_1}$ \equ(7.6.106). 
The r.h.s. of \equ(7.6.105) is
Lipshitz in $t$ with constant $O(\l)$, so that \equ(7.6.105) can be inverted 
w.r.t. $t$ by contractions and, for both choices of the sign, 
we find a unique solution
$$t=t_c^\pm(\l,u)=\sqrt2-1+\n^*(\l)\pm|u|^{1+\h}\big(1+F^\pm(\l,u)\big)\;,
\Eq(7.6.103)$$
with $|F^\pm(\l,u)|\le c\big|\l|$, for some $c$.\qed\\

\sub(7.6.2){\bf Computation of $h^*_2$.}
\\
Let us now solve \equ(7.4.61a) in the general case of 
$\g^{h^*_2}\ge0$. 
Calling $\e\defin\g^{h^*_2-h^*_1-F^{h^*_2}_m}/c_m$, we find:
$$\eqalign{\e&=\left||\s_1|\g^{(\h_\s-1)(h^*_1-1)+F^{h^*_1}_\s}-|\m_1|
\g^{(\h_\m-1)(h^*_1-1)+F^{h^*_1}_\m}
-\a_\s\g\p_{h^*_1}\right|=\cr
&=\g^{(\h_\s-1)(h^*_1-1)+F^{h^*_1}_\s}\left||\s_1|-|\m_1|
\g^{(\h_\m-\h_\s)(h^*_1-1)+F^{h^*_1}_\m-F^{h^*_1}_\s}
-\a_\s\g^{1+(1-\h_\s)(h^*_1-1)-F^{h^*_1}_\s}\p_{h^*_1}\right|\;.
\cr}
\Eq(7.6.103)$$
If $|\s_1|^{1/(1-\h_\s)}\le 2|\m_1|^{1/(1-\h_\m)}$,
we use $\g^{h^*_1-1}=c_u|u|^{1/(1-\h_\m)}$ and,
from the second row of \equ(7.6.103), we find: $\e=C{\left||\s_1|-
|\s_{1,c}^{\a_\s}|\right||u|^{-(1+\h)}}$, where 
$\s_{1,c}^\pm=\s_1(\l,t_c^\pm,u)$
and $C=C(\l,t,u)$ is bounded above and below by $O(1)$ constants; defining
$\D$ as in \equ(1.6), we can rewrite:
$$\e=C{\left||\s_1|-
|\s_{1,c}^{\a_\s}|\right|\over|u|^{1+\h}}=C'{\left|\s_1^2-
(\s_{1,c}^{\a_\s})^2\right|\over \D|u|^{1+\h}}=C''{|t-t_c^+|\cdot|t-t_c^-|\over
\D^2}\;,\Eq(7.epsilon)$$
where $C'=C'(\l,t,u)$ and $C''=C''(\l,t,u)$ 
are bounded above and below by $O(1)$ constants.

In the opposite case ($|\s_1|^{1/(1-\h_s)}> 2|\m_1|^{1/(1-\h_\m)}$),
we use $\g^{h^*_1-1}=c_\s|\s_1|^{1/(1-\h_\s)}$ and, from the first row
of \equ(7.6.103), we find
$\e=\tilde C(1-{|\m_1||\s_1|^{-1/(1+\h)}}-\a_\s\g\p_{h^*_1})=\bar C$,
where $\tilde C$ and $\bar C$ are bounded above and below by $O(1)$ constants.
Since in this region of parameters $|t-t_c^\pm|\D^{-1}$ 
is also bounded above and below
by $O(1)$ constants, we can in both cases write 
$$\e=C_\e(\l,t,u) {|t-t_c^+|\cdot|t-t_c^-|\over
\D^2}\virg C_{1,\e}\le C_\e(\l,t,u) \le C_{2,\e}\Eq(7.eps2)$$
and $C_{j,\e}$, $j=1,2$, are suitable positive $O(1)$ constants.

\pagina
\setcap{8. The specific heat.}
\capindex{8}{The specific heat.}
\vskip1.truecm
\section(8, The specific heat.)
\capindex{8}{The specific heat.}

In this Chapter we describe the expansion for the energy--energy 
correlation functions, from which we can derive a convergent expansion
for the specific heat of the Ashkin--Teller model. We then compute
the leading order contributing to the specific heat, we derive
the expression \equ(1.4), so concluding the proof of the Main
Theorem in the Introduction.\\
\\ 
Consider the specific heat defined in \equ(cv).
The correlation function $<H^{AT}_{\xx}H^{AT}_{\yy}>_{\L_M,T}$
can be conveniently written as
$$<H^{AT}_{\xx}H^{AT}_{\yy}>_{\L,T}=
{\dpr^2\over \dpr\phi_\xx\dpr\phi_\yy}\log \Xi_{AT}(\phi)\Big|_{\phi=0}\virg
\Xi_{AT}(\phi)\defin\sum_{\s^{(1)},\s^{(2)}}e^{-\sum_{\xx\in\L}(1+\phi_\xx)
H^{AT}_\xx}\Eq(8.7.2)$$
where $\phi_\xx$ is a real commuting auxiliary field (with periodic boundary 
conditions).\\
\\
Repeating the construction of Chapter 3, we see that $\Xi_{AT}(\phi)$
admit a Grassmanian representation similar to the one of $\Xi_{AT}$,  
and in particular, if $\xx\not =\yy$:
$$\eqalign{&{\dpr^2\over \dpr\phi_\xx\dpr\phi_\yy}\log
\Xi_{AT}(\phi)\Big|_{\phi=0}={\dpr^2\over \dpr\phi_\xx\dpr\phi_\yy}
\log\sum_{\g_1,\g_2}(-1)^{\d_{\g_1}+\d_{\g_2}}
\widehat \Xi_{AT}^{\g_1,\g_2}(\phi)\Big|_{\phi=0}\cr
&\widehat 
\Xi_{AT}^{\g_1,\g_2}(\phi)=\int\prod_{\xx\in\L_M}^{j=1,2} dH^{(j)}_\xx d
\lis H^{(j)}_\xx d V^{(j)}_\xx
d\lis V^{(j)}_\xx\, e^{S^{(1)}_{\g_1}(t^{(1)})+S^{(2)}_{\g_2}(t^{(2)})+
V_\l+\BB(\phi)}\cr}\Eq(8.7.3)$$
where $\d_\g$, $S^{(j)}(t^{(j)})$ and $V_\l$ where defined in Chapter 3, 
the apex $\g_1,\g_2$ attached to $\widehat \Xi_{AT}$ refers to the
boundary conditions assigned to the Grassmanian fields
and finally $\BB(\phi)$ is defined as:
$$\eqalign{\BB(\phi)=
\sum_{\xx\in\L}&\phi_\xx\Big\{a^{(1)}\big(\lis H^{(1)}_{\xx} H^{(1)}_{
\xx+\hat e_1}+\lis V^{(1)}_{\xx} V^{(1)}_{\xx+\hat e_0}\big)+
a^{(2)}\big(\lis H^{(2)}_{\xx} H^{(2)}_{
\xx+\hat e_1}+\lis V^{(2)}_{\xx} V^{(2)}_{\xx+\hat e_0}\big)+\cr
&+\l \widetilde a
\big(\lis H^{(1)}_{\xx} H^{(1)}_{
\xx+\hat e_1}\lis H^{(2)}_{\xx} H^{(2)}_{\xx+\hat e_1}+
\lis V^{(1)}_{\xx} V^{(1)}_{
\xx+\hat e_0}\lis V^{(2)}_{\xx} V^{(2)}_{\xx+\hat e_0}\big)\Big\}\defin
\sum_{\xx\in\L}\phi_\xx A_\xx\;,\cr}
\Eq(8.7.4)$$
where $a^{(1)}$, $a^{(2)}$ and $\widetilde a$ are $O(1)$ constants, with
$a^{(1)}-a^{(2)}=O(u)$. Using \equ(8.7.3) and \equ(8.7.4) we can rewrite:
$$<H^{AT}_{\xx}H^{AT}_{\yy}>_{\L,T}={1\over 4}(\cosh J)^{2M^2}
\sum_{\g_1,\g_2}(-1)^{\d_{\g_1}+\d_{\g_2}}
{\Xi_{AT}^{\g_1,\g_2}\over \Xi_{AT}}<A_\xx A_\yy>_{\L_M,T}^{\g_1,\g_2}\;,
\Eq(8.axay)$$
where $<\cdot>_{\L_M,T}^{\g_1,\g_2}$ is the average w.r.t. the
boundary conditions $\g_1,\g_2$. Proceeding as in Appendix A9
one can show that, if $\g^{h^*_2}>0$, 
$<A_\xx A_\yy>_{\L_M,T}^{\g_1,\g_2}$ is exponentially
insensitive
to boundary conditions and $\sum_{\g_1,\g_2}(-1)^{\d_{\g_1}+\d_{\g_2}}
{\Xi_{AT}^{\g_1,\g_2}/ \Xi_{AT}}$ is an $O(1)$ constant. 
Then from now on we will
study only $\Xi_{AT}^-(\phi)\defin$
$\defin\widehat \Xi_{AT}^{(-,-),(-,-)}(\phi)$ and $<A_\xx A_\yy>_{\L_M,T}^{
(-,-),(-,-)}$.

Proceeding as in Chapter 4
we integrate out the $\c$ fields and we find:
$$\Xi_{AT}^-(\phi)=
\int P_{Z_1,\s_1,\m_1,C_1}(d\psi) e^{\VV^{(1)}+\BB^{(1)}}\;,\Eq(8.7.5)$$
where
$$\BB^{(1)}(\psi,\phi)=\sum_{m,n=1}^\io
\sum^{\underline\s,\underline j,
\underline\a,\oo}_{\xx_1\cdots\xx_m\atop\yy_1\cdots \yy_{2n}}
B^{(1)}_{m,2n;\underline\s,\underline j,
\underline\a,\oo}(\xx_1,\ldots,\xx_m;\yy_1,
\ldots,\yy_{2n})
\Big[\prod_{i=1}^m\phi_{\xx_i}\Big] \Big[\prod_{i=1}^{2n}
\partial^{\s_i}_{j_i}\psi^{\a_i}_{\yy_i,\o_i}\Big]\;.\Eq(8.6.6)$$
We proceed as for the partition function, namely as described in Chapter 5
above. We introduce
the scale decomposition described in Chapter 5
and we perform iteratively
the integration of the single scale fields, starting from the field of
scale $1$.
After the integration of the fields $\psi^{(1)},\ldots,\psi^{(h+1)}$, 
$h^*_1<h\le 0$, we are left with
$$\Xi_{AT}^-(\phi)=e^{-M^2 E_h+S^{(h+1)}(\phi)}\int P_{Z_h,\s_h, 
\m_h,C_h}(d\psi^{\le
h})e^{-\VV^{(h)}(\sqrt{Z_h}\psi^{(\le h)})+\BB^{(h)}
(\sqrt{Z_h}\psi^{(\le h)},\phi)}\;,\Eq(8.7.6)$$
where $P_{Z_h,\s_h,\m_h m_h,C_h}(d\psi^{(\le h)})$
and $\VV^{(h)}$ are the same as in Chapter 5, $S^{(h+1)}$ $(\phi)$ 
denotes the sum of the contributions dependent on $\phi$ but independent of
$\psi$, and finally $\BB^{(h)}(\psi^{(\le h)},\phi)$ denotes the
sum over all terms containing at least one $\phi$ field and two $\psi$
fields. $S^{(h+1)}$ and $\BB^{(h)}$ can be represented as
$$\eqalign{&S^{(h+1)}(\phi)=\sum_{m=1}^\io\sum_{\xx_1\cdots\xx_m}
S^{(h+1)}_m(\xx_1,\ldots,\xx_m)
\prod_{i=1}^m\phi_{\xx_i}\cr
&\BB^{(h)}(\psi^{(\le h)},\phi)=
\sum_{m,n=1}^\io
\sum_{\xx_1\cdots\xx_m\atop \yy_1\cdots\yy_{2n}}^{\ss,\underline j,\aa,\oo}
B^{(h)}_{m,2n;\ss,\underline j,\aa,\oo}(\xx_1,
\ldots,\xx_m;\yy_1,\ldots,\yy_{2n})
\Big[\prod_{i=1}^m\phi_{\xx_i}\Big] \Big[\prod_{i=1}^{2n}
\partial^{\s_i}\psi^{(\le h)\a_i}_{\yy_i,\o_i}\Big]\;.\cr}\Eq(8.7.7)$$
Since the field $\phi$ is equivalent, as regarding dimensional
bounds, to two $\psi$ fields (see Theorem 8.1 below for a more precise 
statement), the only terms in the 
expansion for $\BB^{(h)}$ 
which are not irrelevant are those with $m=n=1$, $\s_1=\s_2=0$ 
and they are marginal. 
Hence we extend the definition of the localization operator $\LL$,
so that its action on $\BB^{(h)}(\psi^{(\le
h)},\phi)$ is defined by its action on the kernels
$\widehat B^{(h)}_{m,2n;\aa,\oo}
(\qq_1,\ldots,\qq_m;\kk_1,\ldots,\kk_{2n})$:\\
\01) if $m=n=1$ and $\a_1+\a_2=\o_1+\o_2=0$, then
$\LL \widehat B^{(h)}_{1,2;\ss,\aa,\oo}(\qq_1;\kk_1,\kk_2)\defin
\PP_0\widehat B^{(h)}_{1,2;\aa,\oo}
(\kk_+;\kk_+,\kk_+)$,
where $\PP_0$ is defined as in \equ(5.9);\\
\02) in all other cases $\LL
\widehat B^{(h)}_{m,2n;\aa,\oo}=0$.
\*

Using the symmetry considerations of \sec(4.3) together with the remark that
$\phi_\xx$ is invariant under {\it Complex conjugation}, {\it 
Hole--particle} and $(1)\otto(2)$, while under {\it Parity} 
$\phi_\xx\to\phi_{-\xx}$ and under 
{\it Rotation} $\phi_{(x,x_0)}\to\phi_{(-x_0,
-x)}$,
we easily realize that $\LL\BB^{(h)}$ has necessarily the following form:
$$\LL\BB^{(h)}(\psi^{(\le h)},\phi)={\lis Z_h\over Z_h}
\sum_{\xx,\o}{(-i\o)\over 2}\phi_\xx\psi^{
(\le h)+}_{\o,\xx}\psi^{(\le h)-}_{-\o,\xx}\;,\Eq(8.7.10)$$
where $\lis Z_h$ is real and $\lis Z_1=a^{(1)}|_{\s=\m=0}\=
a^{(2)}|_{\s=\m=0}$.

Note that apriori a term $\sum_{\xx,\o,\a}
\phi_\xx\psi^{(\le h)\a}_{\o,\xx}\psi^{(\le h)\a}_{-\o,\xx}$ is allowed
by symmetry but, using $(1)\otto(2)$ symmetry, 
one sees that its kernel is proportional
to $\m_k$, $k\ge h$. So, with our definition of localization,
such term contributes to $\RR\BB^{(h)}$.\\

Now that the action of $\LL$ on $\BB$ is defined, 
we can describe the single scale integration, for $h> h^*_1$. 
The integral in the r.h.s. of \equ(8.7.6) can be rewritten as:
$$\eqalign{&e^{-M^2 t_h}\int P_{Z_{h-1},\s_{h-1}, 
\m_{h-1},C_{h-1}}(d\psi^{\le
h-1})\cdot\cr
&\qquad\qquad\cdot\int P_{Z_{h-1},\s_{h-1}, 
\m_{h-1},\widetilde f^{-1}_{h}}(d\psi^{(h)})
e^{-\widehat \VV^{(h)}(\sqrt{Z_{h-1}}\psi^{(\le h)})+\widehat \BB^{(h)}
(\sqrt{Z_{h-1}}\psi^{(\le h)},\phi)}\;,\cr}\Eq(8.7.11)$$
where $\widehat \VV^{(h)}$ was defined in \equ(5.18) and 
$$\widehat \BB^{(h)}
(\sqrt{Z_{h-1}}\psi^{(\le h)},\phi)\defin \BB^{(h)}
(\sqrt{Z_h}\psi^{(\le h)},\phi)\;.\Eq(8.7.12)$$
Finally we define 
$$\eqalign{&
e^{-\widetilde E_h M^2+\widetilde S^{(h)}(\phi)-\VV^{(h-1)}
(\sqrt{Z_{h-1}}\psi^{(\le h-1)})+\BB^{(h-1)}(\sqrt{Z_{h-1}}
\psi^{(\le h-1)},\phi)}\defin\cr
&\qquad\defin  \int P_{Z_{h-1},\s_{h-1}, 
\m_{h-1},\widetilde f^{-1}_{h}}(d\psi^{(h)})
e^{-\widehat \VV^{(h)}(\sqrt{Z_{h-1}}\psi^{(\le h)})+\widehat \BB^{(h)}
(\sqrt{Z_{h-1}}\psi^{(\le h)},\phi)}\;,\cr}\Eq(8.7.13)$$
and
$$E_{h-1}\defin E_h+t_h+\widetilde E_h\virg S^{(h)}(\phi)\defin S^{(h+1)}(\phi)
+\widetilde S^{(h)}(\phi)\;.\Eq(8.7.14)$$
With the definitions above, it is easy to verify that $\lis Z_{h-1}$
satisfies the equation $\lis Z_{h-1}=\lis Z_h(1+\lis z_h)$,
where $\lis z_h=\lis b\l_h+O(\l^2)$, for some $\lis b\not =0$. 
Then, for some $c>0$,
$\lis Z_1e^{-c|\l|h}\le \lis Z_h\le\lis Z_1 e^{c|\l|h}$. 
The analogous of Theorem 5.1 for the kernels of $\BB^{(h)}$ holds:\\
\\
{\cs Theorem 8.1.} {\it Suppose that the hypothesis of Lemma 7.1 are satisfied.
Then, for $h^*_1\le \bar h\le 1$ and a suitable constant $C$, 
the kernels of $\BB^{(h)}$ satisfy
$$\int d\xx_1\cdots d\xx_{2n}|B^{(\bar h)}_{2n,m;\ss,\underline j,\aa,\oo}
(\xx_1,\ldots,\xx_m;\yy_1,\ldots,\yy_{2n})|
\le M^2 \g^{-\bar h (D_k(n)+m)} \,(C\,|\l|)^{max(1,n-1)}\;,\Eq(8.7.17)$$
where $D_k(n)=-2+n+k$ and $k=\sum_{i=1}^{2n}\s_i$.}\\

Note that, consistently with our definition of localization,
the dimension of $B^{(h)}_{2,1;(0,0),(+,-),(\o,-\o)}$ is $D_0(1)+1=0$.

Again, proceeding as in Chapter 6, 
we can study the flow of $\lis Z_h$ up to $h=-\io$ and prove that
$\lis Z_h=\lis Z_1\g^{\lis\h(h-1)+F^h_{\bar z}}$,
where $\lis \h$ is a non trivial analytic function of $\l$ (its linear part is 
non vanishing) and $F^h_{\bar z}$ is a suitable
$O(\l)$ function (independent of $\s_1,\m_1$). 
We recall that $\lis Z_1=O(1)$.

We proceed as above up to the scale $h^*_1$. Once that the scale $h^*_1$
is reached we pass to the $\psi^{(1)},\psi^{(2)}$ variables,
we integrate out (say) the $\psi^{(1)}$ fields and we get
$$\int P^{(2)}_{Z_{h^*_1},\widehat m^{(2)}_{h^*_1},C_{h^*_1}}
(d\psi^{(2)(\le h^*_1)})e^{-\lis\VV^{(h^*_1)}
(\sqrt{Z_{h^*_1}}\psi^{(2,\le h^*_1)})+\lis
B^{(h^*_1)}(\sqrt{Z_{h^*_1}}
\psi^{(2,\le h^*_1)})}\;,\Eq(8.7.19)$$
with $\LL\lis B^{h^*_1}(\sqrt{Z_{h^*_1}}\psi^{(2),\le h^*_1})=
\lis Z_{h^*_1}\sum_\xx i\phi_{\xx}
\psi^{(2,\le h^*_1)}_{1,\xx}\psi^{(2,\le h^*_1)}_{-1,\xx}$.

The scales $h^*_2\le h\le h^*_1$
are integrated as in Chapter 7 and
one finds that the flow of $\lis Z_{h}$ in this regime is trivial, \ie
if $h^*_2\le h\le h^*_1$, $\lis Z_h=\lis Z_{h^*_1}\g^{F^h_{z}}$,
with $F^h_{z}=O(\l)$.

The result is that the correlation function $<H^{AT}_\xx H^{AT}_\yy>_{\L_M,T}$
is given by a convergent power series in $\l$, uniformly in $\L_M$.
Then, the leading behaviour of the specific heat is given by the sum
over $\xx$ and $\yy$ of the lowest order contributions to 
$<H^{AT}_\xx H^{AT}_\yy>_{\L_M,T}$, namely
by the diagrams in Fig.6. Absolute convergence of the power series
of $<H^{AT}_\xx H^{AT}_\yy>_{\L_M,T}$ implies that the rest is 
a small correction. 
\midinsert
\*
\insertplotbm{300pt}{50pt}%
{\ins{-3pt}{8pt}{$\sum_{h=h^*_2}^{h^*_1}$}
\ins{53pt}{-3pt}{$\xx$}
\ins{91pt}{-3pt}{$\yy$}
\ins{73pt}{25pt}{$h$}
\ins{73pt}{-15pt}{$h$}
\ins{130pt}{4pt}{$+$}
\ins{150pt}{8pt}{$\sum_{h=h^*_1}^1$}
\ins{215pt}{-3pt}{$\xx$}
\ins{253pt}{-3pt}{$\yy$}
\ins{233pt}{25pt}{$h$}
\ins{233pt}{-15pt}{$h$}
}%
{fig1}{}
\vskip1.5truecm
\line{\vtop{\line{\hskip1.3truecm\vbox{\advance\hsize by -2.0 truecm
\0{\css Fig. 6.}
{\ottorm The lowest order diagrams contributing to ${\scriptstyle 
<H^{AT}_\xx H^{AT}_\yy>_{\L_M,T}}$.
The wavy lines ending in the points labeled ${\scriptstyle \xx}$ and 
${\scriptstyle \yy}$ represent
the fields ${\scriptstyle \phi_\xx}$ and ${\scriptstyle \phi_\yy}$
respectively. 
The solid lines labeled by ${\scriptstyle h}$
and going from ${\scriptstyle \xx}$ to ${\scriptstyle \yy}$ 
represent the propagators ${\scriptstyle g^{(h)}(\xx-\yy)}$.
The sums are over the scale indeces and, even if not explicitly written, over
the indexes ${\scriptstyle \aa,\oo}$ 
(and the propagators depend on these indexes).}
\hfill} }}}
\*
\endinsert
The conclusion is that $C_v$, for $\l$ small and
$|t-\sqrt2+1|,|u|\le (\sqrt2-1)/4$, is given by:
$$\eqalign{&C_v={\b^2\over |\L|}\sum_{\xx,\yy\in\L_M}\sum_{\o_1,\o_2=\pm1}
\sum_{h,h'=h^*_2}^{1}{(Z_{h\vee h'}^{(1)})^2\over Z_{h-1} Z_{h'-1}}
\Bigg[G^{(h)}_{(+,\o_1),(+,\o_2)}(\xx-\yy)
G^{(h')}_{(-,-\o_2),(-,-\o_1)}(\yy-\xx)+\cr
&+
G^{(h)}_{(+,\o_1),(-,-\o_2)}(\xx-\yy)
G^{(h')}_{(-,-\o_1),(+,\o_2)}(\xx-\yy)\Bigg]+{1\over |\L|}\sum_{\xx,\yy\in\L_M}
\sum_{h^*_2}^1
\Big({\lis Z_h\over Z_h}\Big)^2\O_{\L_M}^{(h)}(\xx-\yy)\;,\cr}\Eq(8.wow)$$
where $h\vee h'=\max\{h,h'\}$ and $G^{(h)}_{(\a_1,\o_1),(\a_2,\o_2)}(\xx)$
must be interpreted as
$$G^{(h)}_{(\a_1\o_1),(\a_2,\o_2)}(\xx)=\cases{
g^{(h)}_{(\a_1\o_1),(\a_2,\o_2)}(\xx) & if $h>h^*_1$,\cr
g^{(1,\le h^*_1)}_{\o_1,\o_2}(\xx)+g^{(2,h^*_1)}_{\o_1,\o_2}(\xx)
& if $h=h^*_1$,\cr
g^{(2,h)}_{\o_1,\o_2}(\xx) & if $h^*_2<h<h^*_1$,\cr
g^{(2,\le h^*_2)}_{\o_1,\o_2}(\xx) & if $h=h^*_2$.\cr}$$
Moreover, if $N,n_0,n_1\ge 0$ and $n=n_0+n_1$, 
$|\dpr_{x}^{n_0}\dpr_{x_0}\O_{\L_M}^{(h)}(\xx)|\le C_{N,n}|\l|{\g^{(2+n)h}
\over 1+(\g^h|\dd(\xx)|)^N}$.
Now, calling $\h_c$ the exponent associated to $\lis Z_h/Z_h$, from \equ(8.wow)
we find:
$$\eqalign{&C_v=-C_1 \g^{2\h_c h^*_1}\log_\g\g^{h^*_1-h^*_2}
\big(1+\O_{h^*_1,h^*_2}^{(1)}(\l)\big)
+C_2{1-\g^{2\h_c(h^*_1-1)}\over 2\h_c}
\big(1+\O^{(2)}_{h^*_1}(\l)\big)\;,\cr}\Eq(8.7.22)$$
where $|\O^{(1)}_{h^*_1,h^*_2}(\l)|,|\O^{(2)}_{h^*_1}(\l)|
\le c|\l|$, for some $c$, and $C_1,C_2$ are functions of
$\l,t,u$, bounded above and below by $O(1)$ constants.
Note that, defining $\D$ as in the line following \equ(1.4), 
$\g^{(1-\h_\s)h^*_1}\D^{-1}$
is bounded above and below by $O(1)$ constants. Then, using \equ(7.eps2),
\equ(1.4) follows.
\\

\pagina
\setcap{9. Conclusions and open problems.}
\capindex{9}{Conclusions and open problems.}
\vskip1.truecm
\section(9, Conclusions and open problems.)
\capindex{9}{Conclusions and open problems.}

In the previous Chapters we described a constructive approach
to the study of the thermodynamic properties of weakly interacting 
spin systems in two dimensions, 
arbitrarely near to the critical temperature(s). 

The approach was based on an exact mapping of the interacting spin system
into an interacting system of $1+1$--dim non relativistic fermions.
It applies to a wide class of perturbations of Ising, including the 
Ashkin--Teller model, the 8 vertex model, the next to nearest neighbor
Ising model and linear combinations of these models. 

As an application of the method, we studied the free energy
and the specific heat for the anisotropic Ashkin--Teller model, which
is a well--known model of statistical mechanics, widely studied with
a number of different theoretical and empirical techniques. However 
exact results were lacking since long time: in the 1970's Baxter,
Kadanoff and others conjectured that (1) anisotropic AT has in general two
different critical temperetures (whose location was unknown) and (2) AT
belongs to the same universality class of Ising except at the isotropic point. 

Our calculation of the free energy and of the specific heat allowed
to (rigorously) prove for the first time the two conjectures above in 
the regime of weak coupling and
to derive detailed asymptotic expressions for the specific heat itself
and for the shape of the critical surface (\ie for the critical temperatures
as functions of the anisotropy parameter and of the coupling). 
The latter calculation revealed the existence of a previously unknown 
critical exponent, describing how the difference of the 
critical temperatures rescale, when we let the anisotropy
go to 0.\\

Important open problems are the following.\\
\\
\01) The study of the free energy and of the correlation functions
{\it directly at the critical point}, where it is expected that 
the correlation functions are, in the thermodynamic limit, 
homogeneous functions of the coordinates and, moreover,
{\it conformal invariant}. 
Even for the Ising model this is a widely
expected by still unproved conjecture. Technically we have to face 
the difficulty of dealing with a linear combination
of 16 Grassmann partition functions, differing for the boundary conditions;
up to now we are able to control this combination only outside 
the critical point (but arbitrarely near to it).\\
\\
\02) The study of more complicated correlation functions, such as the 
spin--spin correlation functions $<\s_\V0\s_\xx>$. 
These are difficult to study in the
Grassmann formulation, because  
they correspond to the average of an exponential 
of a relevant non traslationally invariant operator in the Grassmann 
fields. Such operator is concentrated along a path connecting the two
points $\V0,\xx$ and, moreover, is weighted by an order 1 constant!
Note that even in the free case (Ising) the calculation of the spin--spin 
correlation functions is very non trivial and is based on 
the analysis of a Toeplitz determinant, in the limit in 
which the size of the Toeplitz matrix diverging to infinity, through
an application of Szego's Theorem. 

A first step towards the understanding of such objects would be 
the calculation of averages of exponentials of simpler relevant 
non translational invariant operators, such as those appearing
in the study of large deviations for the magnetization or the particle 
number in a bounded region of $\ZZZ^2$.

\pagina
\setcap{appendix a1: Grassmann integration. Truncated expectations.}
\appindex{A1}{Grassmann integration. Truncated expectations.}
\vskip1.truecm

\appendix(A1, Grassmann integration. Truncated expectations.)

In the present Appendix we list some more properties of Grassmann
integration (the basic ones were introduced in \sec(2.2)). 
In particular we introduce the definition of truncated expectation,
and we describe a possible graphical interpretation for the truncated
expectations, the so--called {\it Feynman diagrams}.\\
\\
\asub(A1.0){\bf Truncated expectations and some more rules}.\\
Pursuing further the analogy with Gaussian integrals stressed in \sec(2.2),
we can consider a ``measure'' (a similar expression is
found replacing $g$ with a matrix, see \equ(4.24A) below)
$$ P(\der\psi)=\prod_{\a\in A} \der\psi^+_{\a} \der
\psi^-_{\a} g_{\a}
e^{-\sum_{\a\in A} \psi_{\a}^+ g_{\a}^{-1} \psi^-_{\a}}
\; ; \Eqa(4.10A) $$
by construction one has
$$ \int P(\der\psi) = 1 \; , \qquad
\int P(\der\psi) \, \psi^{-}_{\a} \psi^{+}_{\b}
= \d_{\a,\b} g_{\a} \; . \Eqa(4.11A) $$

In general $P(\der\psi)$ will be called a {\it Gaussian fermionic
integration measure} (or {\it Grassman integration measure} or,
as we shall do in the following, integration {\it tout court})
with covariance $g$: for any analytic function $F$
defined on the Grassman algebra we can write
$$ \int P(\der\psi) \, F(\psi) = \EE(F) \; . \Eqa(4.12A) $$
However note that $P(\der\psi)$ is not at all a real measure,
as it does not satisfy the necessary positivity conditions,
so that the terminology is only formal and the use of the
symbol $\EE$ (which stands for expectation value)
is meant only by analogy.

Given $p$ functions $X_1,\ldots,X_p$
defined on the Grassman algebra
and $p$ positive integer numbers $n_1,\ldots,n_p$,
the {\it truncated expectation} is defined as
$$ \EE^T(X_1,\ldots,X_p;n_1,\ldots,n_p)
= \left. {\dpr^{n_1+\ldots+n_p} \over
\dpr\l_{1}^{n_1}\ldots\dpr\l_{p}^{n_p} }
\log\int P(\der\psi) \, e^{\l_1 X_1(\psi) + \ldots +
\l_{p}X_p(\psi) } \right|_{\l=0} \; , \Eqa(4.13A) $$
where $\l=\{\l_1,\ldots,\l_p\}$.
It is easy to check that $\EE^T$ is a linear operation,
that is, formally,
$$ \EE^T(c_1X_1+\ldots+c_pX_p;n) =
\sum_{n_1+\ldots + n_p=n}
{n! \over n_1!\ldots n_p!}
c_1^{n_1}\ldots c_p^{n_p}
\EE^T(X_1,\ldots,X_p;n_1,\ldots,n_p) \; , \Eqa(4.14A) $$
so that the following relations immediately follow:
$$ \eqalign{
& (1) \qquad \EE^T (X;1) = \EE(X) \; , \cr
& (2) \qquad \EE^T (X;0) = 0 \; , \cr
& (3) \qquad \EE^T (X,\ldots,X;n_1,\ldots,n_p) =
\EE^T(X;n_1+\ldots+n_p) \; . \cr} \Eqa(4.15A) $$
Moreover one has
$$ \EE^{T} (X_1,\ldots,X_1,\ldots,X_p,\ldots,X_p;
1,\ldots,1,\ldots,1,\ldots,1) =
\EE^{T}(X_1,\ldots,X_p;n_1,\ldots,n_p) \; , \Eqa(4.15aA) $$
where, for any $j=1,\ldots,p$, in the l.h.s.
the function $X_j$ is repeated $n_j$ times and
1 is repeated $n_1+\ldots+n_p$ times.

We define also
$$ \EE^{T}(X_1,\ldots,X_p) \= \EE^{T}(X_1,\ldots,X_p;1,\ldots,1)
\; . \Eqa(4.15bA) $$
By \equ(4.15aA) we see that all truncated expectations
can be expressed in terms of \equ(4.15bA);
it is easy to see that \equ(4.15bA) is vanishing if
$X_j=0$ for at least one $j$.

The truncated expectation appears naturally considering
the integration of an exponential; in fact
as a particular case of \equ(4.13A) one has
$$ \EE^T(X;n) = \left. {\partial^n\over\partial\l^n}
\log\int P(\der\psi) \, e^{\l X(\psi)}
\right|_{\l=0} \; , \Eqa(4.16A) $$
so that
$$ \eqalign{
\log \int P(\der\psi) \, e^{X(\psi)} & = \left. \sum_{n=0}^{\io}
{1\over n!} {\partial^n\over\partial\l^n}\log\sum_{n=0}^\io
\int P(\der\psi)\, e^{\l X(\psi)} \right|_{\l=0} \cr
& = \sum_{n=0}^\io {1\over n!}\EE^T(X;n) \; . \cr} \Eqa(4.17A) $$

The following properties, immediate consequence of \equ(2.6) and
very similar to the properties of Gaussian integrations, follow.

\*

\0(1) {\it Wick rule.}
Given two sets of labels $\{\a_1,\ldots,\a_n\}$
and $\{\b_1\,\ldots,\b_m\}$ in $A$, one has
$$ \int P(\der\psi) \, \psi^{-}_{\a_1}...\psi^{-}_{\a_n}
\psi^{+}_{\b_1},\ldots,\psi^{+}_{\b_m} = \d_{n,m}
\sum_{\p} (-1)^{p_{\p}}
\prod_{i=1}^{n} \d_{\a_{i},\b_{\p(j)}}
g_{\a_{i}} \; , \Eqa(4.18A) $$
where the sum is over all the permutations
$\p=\{\p(1),\ldots,\p(n)\}$ of the indices $\{1,\ldots,n\}$
with parity $p_{\p}$ with respect to the fundamental permutation.

\*

\0(2) {\it Addition principle.} Given two integrations
$P(\der\psi_1)$ and $P(\der\psi_2)$,
with covariance $g_1$ and $g_2$ respectively,
then, for any function $F$ which can be written as
sum over monomials of Grassman variables, \ie
$F=F(\psi)$, with $\psi=\psi_1+\psi_2$, one has
$$ \int P(\der\psi_1)\int P(\der\psi_2)
F(\psi_1+\psi_2)=\int P(\der\psi) \, F(\psi) \; , \Eqa(4.19A) $$
where $P(\der\psi)$ has covariance $g\=g_1+g_2$.
It is sufficient to prove it for $F(\psi)=\psi^-\psi^+$, then
one uses the anticommutation rules \equ(2.5). One has
$$ \eqalign{
& \int P(\der\psi_1)\int P(d\psi_2)
\left( \psi_1^{-} +\psi_2^{-} \right)
\left(\psi_1^{+} + \psi_2^{+} \right) \cr
& \qquad \qquad =
\int P(\der\psi_1) \, \psi_1^{-} \psi_1^{+} \int P(d\psi_2)
+ \int P(d\psi_1) \int P(d\psi_2) \, \psi_2^{-} \psi_2^{+} = g_1+g_2
\; . \cr} \Eqa(4.20A) $$
where \equ(4.11A) has been used.

\*

\0(3) {\it Invariance of exponentials.}
From the definition of truncated expectations,
it follows that, if $\phi$ is an ``external field", \ie a
not integrated field, then
$$ \int P(\der\psi) \, e^{X(\psi+\phi)} =
\exp \left[ \sum_{n=0}^{\io} {1\over n!} \EE^T
\left( X(\cdot+\phi);n \right) \right]
\equiv e^{X'(\phi)} \; , \Eqa(4.21A) $$
which is a main technical point: \equ(4.21A) says that
integrating an exponential one still gets an exponential,
whose argument is expressed by the sum of truncated expectations.

\*

\0(4) {\it Change of integration.}
If $P_{g}(\der\psi)$ denotes the integration with
covariance $g$, then, for any analytic
function $F(\psi)$, one has
$$ {1\over \NN_{\n} }
\int P_{g}(\der\psi) \, e^{-\n \psi^+\psi^-} F(\psi)
= \int P_{\tilde g}(\der\psi) \, F(\psi) \; , \qquad
\tilde g^{-1} = g^{-1}+\n \; , \Eqa(4.22A) $$
where
$$ \NN_{\n} = {g^{-1} +\n \over g^{-1}} = 1+g\n =
\int P_{g} (\der\psi) \, e^{-\nu \psi^+\psi^-} \; . \Eqa(4.23A) $$
The proof is very easy from the definitions.
More generally one has that, if $A$ is a set of labels for the Grassmann
fields, if $M$ is an invertible
$|A|\times |A|$ matrix and $P_{M}(\der\psi)$ is given by
$$ P_{M}(\der\psi) = \int 
\big(\prod_{\a\in A}\der\psi^+_{\a} \der\psi^-_{\a}\big) 
\, \det M \,  e^{-\sum_{i,j\in A} \psi_i^+ M^{-1}_{ij} \psi^-_j}
\; , \Eqa(4.24A) $$
then
$$ {1 \over \NN_{N}}
\int P_{M}(\der \psi) \, e^{-\sum_{i,j\in A}
\psi_i^+ N^{-1}_{ij} \psi^-_j} F(\psi)
= \int P_{\tilde M}(\der \psi) \, F(\psi) \; , \Eqa(4.26aA) $$
where
$$ \tilde M^{-1} = M^{-1}+N^{-1} \Eqa(4.26bA) $$
and
$$ \NN_{N} = \det \left( \openone + N^{-1}M \right)
= {\det \left( M^{-1} + N^{-1} \right)
\over \det M^{-1} } = \int P_{M} (\der\psi) \,
e^{-\sum_{i,j\in A} \psi_{i}^+ N_{ij}^{-1}
\psi_{j}^{-} } \; . \Eqa(4.26cA) $$

\*

\asub(A1.1){\bf Graphical representation for truncated expectations.}
\\
Given a Grassman algebra as in \equ(2.5) and
an integration measure like \equ(4.10A) we define
the {\it simple expectation} as in \equ(4.12A). Then
$$ g_{\a} = \EE (\psi_{\a}^{-}\psi_{\a}^{+} ) \; . \Eqa(A1.1) $$
Given a monomial
$$ X(\psi) \= \tilde \psi_{B}
= \prod_{\a\in B} \psi_{\a}^{\s_{\a}} \; , \Eqa(A1.2) $$
where $B$ is a subset of $A$ and $\s_{\a}\in \{\pm \}$,
the expectation $\EE(\tilde\psi_{B})$ can be graphically
represented in the following way.

Represent the indices $\a\in B$ as points on the plane.
With each $\psi_{\a}^{+}$, $\a\in B$, we associate a line
exiting from $\a$, while with each $\psi_{\a}^{-}$,
$\a\in B$, we associate a line entering $\a$.
Let $\TT$ be the set of graphs obtained by contracting
such lines in all possible ways so that only
lines with opposite $\s_{\a}$ are contracted:
given $\a,\b\in B$, denote by $(\a\b)$ the line joining $\a$ and
$\b$ and by $\t$ an element of $\TT$, \ie a graph in $\TT$.

Then we can easily verify that
$$ \EE(\tilde\psi_{B}) = \sum_{\t\in\TT}
\prod_{(\a\b)\in \t} (-1)^{\p_{\t}}
g_{\a}\d_{\a,\b} \; , \Eqa(A1.2a) $$
which is the {\it Wick rule} stated in \sec(A1.0):
here $\p_{\t}$ is a sign which depends on the graph $\t$
(see \equ(4.18A)).

Then define the {\it truncated expecation}
$$ \EE^{T} \left( \tilde\psi_{B_{1}},\ldots,\tilde\psi_{B_{p}};
n_{1},\ldots,n_{p} \right) \; , \Eqa(A1.3) $$
with $B_{j}\subset A$ for any $j$, as in \equ(4.13A).

One easily check that, if $X_j$ are analytic functions
of the Grassman variables (each depending on an even number
of variables, for simplicity, so that
no change of sign intervenes in permuting the order
of the $X_{j}$), then
$$\eqalign{
& (1) \qquad \EE^{T}(X_{1},X_{2}) = \EE(X_{1}X_{2}) -
\EE(X_{1})\EE(X_{2})=\EE(X_{1}X_{2}) -
\EE^T(X_{1})\EE^T(X_{2}) \; , \cr
& (2) \qquad \EE^{T}(X_{1},X_{2},X_{3}) = 
\EE(X_{1}X_{2}X_{3}) -
\EE(X_{1}X_{2})\EE(X_{3})-
\EE(X_{1}X_{3})\EE(X_{2}) \cr
& \qquad \qquad - \EE(X_{2}X_{3})\EE(X_{1}) + 2
\EE(X_{1})\EE(X_{2})\EE(X_{3})=\EE(X_{1}X_{2}X_{3}) -\cr
& \qquad\qquad -\EE^T(X_{1}X_{2})\EE^T(X_{3})-
\EE^T(X_{1}X_{3})\EE^T(X_{2})- \EE^T(X_{2}X_{3})\EE^T(X_{1})\; , \cr
& (3) \qquad \EE^{T}(X_{1},X_{2},X_{3},X_{4}) = 
\EE(X_{1}X_{2}X_{3}X_{4}) -
\EE(X_{1}X_{2}X_{3})\EE(X_{4})-
\EE(X_{1}X_{2}X_{4})\EE(X_{3}) \cr
& \qquad \qquad -\EE(X_{1}X_{3}X_{4})\EE(X_{2})-
\EE(X_{2}X_{3}X_{4})\EE(X_{1}) \cr
& \qquad \qquad
- \EE(X_{1}X_{2})\EE(X_{3}X_{4})
- \EE(X_{1}X_{3})\EE(X_{2}X_{4})
- \EE(X_{1}X_{4})\EE(X_{2}X_{3}) \cr
& \qquad \qquad
+ 2\EE(X_{1}X_{2})\EE(X_{3})\EE(X_{4})
+ 2\EE(X_{1}X_{3})\EE(X_{2})\EE(X_{4})
+ 2\EE(X_{1}X_{4})\EE(X_{2})\EE(X_{3}) \cr
& \qquad \qquad
+ 2\EE(X_{2}X_{3})\EE(X_{1})\EE(X_{2})
+ 2\EE(X_{2}X_{4})\EE(X_{1})\EE(X_{3})
+ 2\EE(X_{3}X_{4})\EE(X_{1})\EE(X_{2})
\cr
& \qquad \qquad - 6\EE(X_{1})\EE(X_{2})\EE(X_{3})\EE(X_{4})=
\EE(X_{1}X_{2}X_{3}X_{4}) -
\EE^T(X_{1}X_{2}X_{3})\EE^T(X_{4})-\cr
& \qquad\qquad 
-\EE^T(X_{1}X_{2}X_{4})\EE^T(X_{3})
-\EE^T(X_{1}X_{3}X_{4})\EE^T(X_{2})
-\EE^T(X_{2}X_{3}X_{4})\EE^T(X_{1})-\cr
& \qquad\qquad
- \EE^T(X_{1}X_{2})\EE^T(X_{3}X_{4})
- \EE^T(X_{1}X_{3})\EE^T(X_{2}X_{4})
- \EE^T(X_{1}X_{4})\EE^T(X_{2}X_{3})
\; . \cr} \Eqa(A1.3a) $$
and so on. One can always write the truncated expectations
in terms of simple expectations and viceversa: it is easy to check that
in general one has
$$ \EE(X_{1}\ldots X_{s}) =
\sum_{p=1}^{s} \sum_{Y_{1},\ldots,Y_{p}}
\EE^T(X_{\p_1(1)},\ldots,X_{\p_{|Y_{1}|}(1)})\ldots 
\EE^T(X_{\p_1(p)},\ldots,X_{\p_{|Y_{p}|}(p)})\; , \Eqa(A1.3b) $$
where:\\
\0(1) the sum is over all the possible sets $Y_i,\quad i=1,\ldots, p$, which 
are unions of $|Y_i|$ sets $X_j$, such that $\cup_{j=1}^{s} X_{j} = 
\cup_{k=1}^{|Y_1|+\cdots+|Y_p|}Y_{k}$;\\
\0(2) $\{ \p_1(1),\ldots,\p_{|Y_{1}|}(1),\p_1(2),\ldots,\ldots, \p_{|Y_{p}|}(p)\}$
is a permutation of $\{ 1,\ldots, s\}$.\\
\equ(A1.3b) can be verified by induction, see Appendix A4 in [G].

We can now describe the rules to graphically represent the 
truncated expectations $\EE^T(\tilde\psi_{B_1},\ldots,\tilde\psi_{B_p})$
in \equ(A1.3). Draw in the plane $p$ boxes $G_1,\ldots,G_p$,
such that $G_i$ contains all points
representing the indices belonging to $B_{i}$; from each of the points $\a\in G_i$ 
draw the line corresponding to the field $\psi_\a^{\s_a}$ contained in 
the monomial $\tilde\psi_{B_i}$, with the direction consistent with $\s_\a$
(the line enters or exists $\a$ depending if $\s_\a$ is $-$ or $+$).
We call {\it clusters} such boxes
Then consider all possible graphs $\t$ obtained by
contracting as before all the lines emerging
from the points in such a way that no line is left
uncontracted and with the property that if
the clusters were considered as points then $\t$
would be connected. If we denote the lines as before we have
$$ \EE^{T} \left( \tilde\psi_{B_{1}},\ldots,\tilde\psi_{B_{p}};
n_{1},\ldots,n_{p} \right) = \sum_{\t\in\TT_{0}}
\prod_{(\a\b)\in\t} (-1)^{\p_{\t}} g_{\a} \d_{\a,\b} \; , \Eqa(A1.4) $$
where $\TT_0$ denotes the set of all graphs obtained following the
just given prescription; again $\p_{\t}$ is a sign depending on $\t$.

The reason why we have to sum only over the connected graphs follows 
from \equ(A1.3b), as it can be easily verified by induction.

\*

\pagina
\setcap{appendix a2: The Pfaffian expansion.}
\appindex{A2}{The Pfaffian expansion.}
\vskip1.truecm
\appendix(A2, The Pfaffian expansion.)

In this Appendix we prove \equ(4.12).

Given $s$ set of indices $P_{1},\ldots,P_{s}$,
consider the quantity $\EE^T(\tilde\phi(P_1),
\ldots,\tilde\phi(P_s))$,
with $\tilde\phi(P_i)=
\prod_{f\in P_i}\phi^{\a(f)}_{\xx(f),\o(f)}$ and $\phi=\c,\psi$.

Define
$$ {\cal D}\phi=\prod_{j=1}^{n}
\prod_{f\in P_{j}} \der\phi^{\a(f)}_{\xx(f),\o(f)} 
\quad \quad\left ( \phi,G\phi \right) =
\sum_{f,f'\in \cup_{i}P_i}
\phi^{\a(f)}_{\xx(f),\o(f)}G_{f,f'}\phi^{\a(f')}_{\xx(f'),\o(f')}
\Eqa(A2.9) $$
where, if 
$ 2n = \sum_{j=1}^{s} \left| P_{j} \right|$
then $G$ is the $2n\times 2n$ antisymmetric matrix with entries
$$ G_{f,f'}\defin <\phi^{\a(f)}_{\xx(f),\o(f)}
\phi^{\a(f')}_{\xx(f'),\o(f')}> \; . \Eqa(A2.11) $$
Then one has
$$ \EE \left( \prod_{j=1}^{s} \tilde \phi(P_{j}) \right)
= \Pf G = \int {\cal D}\phi \, \exp\left[
- {1\over 2}\left( \phi,G\phi \right) \right] \; . \Eqa(A2.12) $$

Setting $X\=\{1,\ldots,s\}$ and
$$ \overline V_{jj'}
={1\over 2} \sum_{f\in P_j} \sum_{f'\in P_{j'}}
\phi^{\a(f)}_{\xx(f),\o(f)}G_{f,f'}\phi^{\a(f')}_{\xx(f'),\o(f')} \; ,
\Eqa(A2.13) $$
we write
$$ V(X) = \sum_{j,j'\in X} \overline V_{jj'} = \sum_{j \le j'}
V_{jj'} \; , \Eqa(A2.14) $$
so defining the quantity $V_{jj'}$ as
$$ V_{jj'} = \cases{
\overline V_{jj'} \; , & if $j=j' \; , $ \cr
\overline V_{jj'} + \overline V_{j'j} \; , & if $j<j' \; .$
\cr} \Eqa(A2.14a) $$
Then \equ(A2.12) can be written, by the definition
of Grassman variables, as
$$ \EE \left( \prod_{j=1}^{s} \tilde \phi(P_{j}) \right) =
\int {\cal D}\phi \, e^{-V(X) } \; . \Eqa(A2.13b) $$
We now want to express the last expression in terms 
of the functions $W_{X}$, defined as follows:
$$ W_{X}(X_{1},\ldots,X_{r};t_{1},\ldots,t_{r}) =
\sum_{\ell} \prod_{k=1}^{r} t_{k}(\ell) \, V_{\ell} \; , \Eqa(2A4.25) $$
where:\\
\0(1) $X_{k}$ are subsets of $X$ with $|X_{k}|=k$,
inductively defined as:
$$ \cases{
X_{1} = \{1\} \; , & \cr
X_{k+1} \supset X_{k} \; , \cr} \Eqa(2A4.25a) $$\\
\0(2) $\ell=(jj')$ is a pair of elements $j,j'\in X$
and the sum in \equ(2A4.25) is over all the possible pairs $(jj')$,\\
\0(3) the functions $t_{k}(\ell)$ are defined as follows:
$$ t_{k}(\ell) = \cases{
t_{k} \; , & if $\ell \sim \dpr X_{k} \; , $ \cr
1 \; , & otherwise $\; ,$ \cr} \Eqa(2A4.27) $$
where $\ell\sim X_{k}$ means that $\ell=(jj')$ ``intersects the boundary'' of 
$X_{k}$, \ie connects a point in $P_{j}$, $j \in X_{k}$,
to a point in  $P_{j'}$, $j'\notin X_{k}$.
See Fig. A2.1.

\midinsert
\*
\insertplotttt{150pt}{90pt}{
\ins{31pt}{43pt}{$P_{1}$}
\ins{67pt}{55pt}{$P_{2}$}
\ins{116pt}{43pt}{$P_{3}$}
\ins{53pt}{25pt}{$X_{1}$}
\ins{93pt}{15pt}{$X_{2}$}
\ins{143pt}{5pt}{$X_{3}$}}
{figa2}
\vskip.3truecm
\line{\vtop{\line{\hskip1.3truecm\vbox{\advance\hsize by -2.0 truecm
\0{\css Fig. A2.1.}
{\ottorm Graphical representation of the sets $\scriptstyle X_{k}$,
$\scriptstyle k=1,2,3$. In the example $\scriptstyle X_{1}=\{1\}$, $\scriptstyle
X_{2}=\{1,2\}$ and $\scriptstyle X_{3}=\{1,2,3\}$. The $\scriptstyle \ell=(1\, 3)$
intersects the boundaries of 
$\scriptstyle X_1$ and of $\scriptstyle X_2$.}
} \hfill} }}
\*
\endinsert

From definition \equ(2A4.25) it follows:
$$ W_{X}(X_{1};t_{1}) = \sum_{j=2}^{s} t_{1} V_{1j} +
V_{11} + \sum_{1 < j' \le j  } V_{j'j} = 
(1-t_{1}) \left[ V(X_{1}) + V(X \setminus X_{1}) \right]
+ t_{1} V(X) \Eqa(2A4.28) $$
so that
$$ \eqalign{
e^{-V(X)} & = \int_{0}^{1} \der t_{1} \left[
{\partial\over\partial t_{1} }\, e^{-W_{X}(X_{1};t_{1}) }
\right] + e^{-W_{X}(X_{1};0) } \cr
& = - \sum_{ \ell_{1} \sim \partial X_{1} } V_{\ell_{1}} \,
\int_{0}^{1} \der t_{1} \, e^{-W_{X}(X_{1};t_{1}) } +
e^{-W_{X}(X_{1};0) } \; . \cr} \Eqa(2A4.29) $$

Again by definition we have:
$$ \eqalign{
& W_{X}( X_{1}, X_{2}; t_{1}, t_{2} )  =\cr 
& \qquad V_{11}+t_1V_{12}+t_1t_2\sum_{j=3}^sV_{1j}+V_{22}+t_2\sum_{j=3}^sV_{2j}
+\sum_{2<j'\le j}V_{j'j}=\cr
& \qquad =t_1t_2\sum_{j=2}^sV_{1j}+t_2V_{11}+t_2\sum_{1<j'\le j}V_{j'j}+
(1-t_2)\left[ V_{11}+t_1V_{12}+V_{22}+\sum_{2<j'\le j}V_{j'j}\right]=\cr
& \qquad =t_{2} W_{X} ( X_{1};t_{1} ) + 
(1-t_{2}) \left[ W_{X_{2}} ( X_{1};t_{1} ) +
V ( X \setminus X_{2} ) \right]\cr}\Eqa(2A4.30)$$
If we define $X_{2} \= X_{1} \cup \ell_{1}$, \ie
$X_{2} = \{ 1, \hbox{point connected by } \ell_{1} \hbox{ with 1}\}$, then:
$$ \eqalign{
e^{ - W_{X} ( X_{1};t_{1} ) } & =
\int_{0}^{1} \der t_{2} \, \left[ {\partial \over\partial t_{2}}
\, e^{-W_{X}(X_{1},X_{2};t_{1},t_{2}) } \right] +
e^{-W_{X}(X_{1},X_{2};t_{1},0)} \cr
& = - \sum_{\ell_{2} \sim \partial X_{2} } 
V_{\ell_{2}} \int_{0}^{1} \der t_2 \, t_{1} (\ell_{2})
\, e^{-W_{X}(X_{1},X_{2};t_{1},t_{2}) } +
e^{-W_{X}(X_{1},X_{2};t_{1},0)} \; . \cr} \Eqa(2A4.31) $$

Substituting \equ(2A4.31) in \equ(2A4.29) we get:
$$ \eqalign{ e^{-V(X)}
& = \sum_{ \ell_{1} \sim \partial X_{1} }
\sum_{ \ell_{2} \sim \partial X_{2} }
\int_{0}^{1} \der t_{1}
\int_{0}^{1} \der t_{2} \,
(-1)^{2} \, V_{\ell_{1}} \, V_{\ell_{2}} \,
t_{1}(\ell_{2}) \, e^{-W_{X}(X_{1},X_{2};t_{1},t_{2}) } 
\cr
& + \sum_{ \ell_{1} \sim \partial X_{1} } 
\int_{0}^{1} \der t_{1} \, (-1) \, V_{\ell_{1}} \,
e^{-W_{X}(X_{1},X_{2};t_{1},0) } + e^{-W_{X}(X_{1};0)}
\; . \cr} \Eqa(2A4.32) $$
A relation generalizing \equ(2A4.30) holds:
$$\eqalign{ & W_X(X_1,\ldots, X_{p+1}; t_1,\ldots, t_{p+1})= 
t_{p+1}W_X(X_1,\ldots, X_{p}; t_1,\ldots, t_{p})+\cr
& (1-t_{p+1})\left[
W_{X_{p+1}}(X_1,\ldots, X_{p}; t_1,\ldots, t_{p})+V(X\setminus X_{p+1})\right]
\cr}\Eqa(2A4.32a)$$
where $p<s$. In fact in the sum over $\ell$ in \equ(2A4.25)
we can distinguish two cases: either $\ell\sim X_{p+1}$ or $\ell\not\sim X_{p+1}$. 
In the former case $V_{\ell}$ is necessarily multiplied by $t_{p+1}$ 
and, if $\ell=(j'j)$, $j'\le p+1,\ j>p+1$; in the latter case 
$V_{\ell}$ is not multiplied by $t_{p+1}$ and either $j',j\le p+1$ or $j',j>p+1$.
Then, clearly:
$$\eqalign{ & W_X(X_1,\ldots, X_{p+1}; t_1,\ldots, t_{p+1})=\cr 
&  =t_{p+1}\left[ W_X(X_1,\ldots, X_{p}; t_1,\ldots, t_{p})-
W_{X_{p+1}}(X_1,\ldots, X_{p}; t_1,\ldots, t_{p})-W_{X\setminus X_{p+1}}
(X_1,\ldots, X_{p}; t_1,\ldots, t_{p})\right]+\cr
&  +W_{X_{p+1}}
(X_1,\ldots, X_{p}; t_1,\ldots, t_{p})+W_{X\setminus X_{p+1}}
(X_1,\ldots, X_{p}; t_1,\ldots, t_{p})\cr}\Eqa(A4.32b)$$
that is equivalent to \equ(2A4.32a).
We can iterate the procedure followed to get \equ(2A4.29) and 
\equ(2A4.32). In the general case we find: 
$$ \eqalign{
e^{-V(X)}
& = \sum_{r=0}^{s-1} \sum_{ \ell_{1} \sim \partial X_{1} }
\ldots
\sum_{ \ell_{r} \sim \partial X_{r} }
\int_{0}^{1} \der t_{1} \ldots
\int_{0}^{1} \der t_{r} \, 
(-1)^{r} \,
V_{\ell_{1}} \ldots V_{\ell_{r}} \cr
& \left(
\prod_{k=1}^{r-1} t_{1}(\ell_{k+1}) \ldots
t_{k}(\ell_{k+1}) \right)
e^{-W_{X}(X_{1}, \ldots , X_{r+1} ; 
t_{1}, \ldots, t_{r}, 0 ) } \; , \cr} \Eqa(2A4.33) $$
where the meaningless factors must be replaced by 1. Moreover,
from \equ(2A4.32a) we soon realize that
$$ \eqalign{ & W_{X}(X_{1}, \ldots , X_s ; t_{1}, \ldots, t_{s-1}, 0 ) =
W_{X}(X_{1}, \ldots , X_{s-1} ; t_{1}, \ldots, t_{s-1} )
\;  \cr
&  W_{X} ( X_{1}, \ldots , X_{r} ; t_{1}, \ldots, t_{r-1}, 0 ) =
W_{X_{r}} ( X_{1}, \ldots , X_{r-1} ; t_{1}, \ldots, t_{r-1} )
+ V ( X \setminus X_{r} ) \; \cr} \Eqa(2A4.35) $$
The last equation holds for $r>1$. If $r=1$:
$$W_{X} ( X_{1};0)=V ( X_1)+V ( X \setminus X_1 )\Eqa(2A4.35a)$$

Let $T$ be a tree graph connecting $X_1,\ldots, X_r$, such that:\\
\0(1) for all $k=1,\ldots,r$, $T$ is ``anchored''
to some point $(j,i)$, \ie $T$ contains a line incident with 
$(j,i)$, where $j\in X_{k}$ and $i\in\{1,\ldots,|P_{j}^{\pm}|\}$,\\
\0(2) each line $\ell\in T$ intersects at least one boundary $\dpr X_{k}$,\\
\0(3) the lines $\ell_{1},\ell_{2},\ldots$ are ordered in such a way that 
$\ell_{1}\sim \dpr X_{1},
\ell_{2}\sim \dpr X_{2},\ldots$,\\
\0(4) for each  $\ell\in T$ there exist two indexes $n(\ell)$ and
$n'(\ell)$ defined as follows:
$$ \eqalign{
n(\ell) & = \max \{ k : \ell \sim \partial X_{k} \} \; , \cr
n'(\ell) & = \min \{ k : \ell \sim \partial X_{k} \} \; . \cr}
\Eqa(2A4.37) $$
We shall say that $T$ is an {\it anchored tree}.

Using the definitions above, we can rewrite \equ(2A4.33) as:
$$ \eqalign{
e^{-V(X)} & = \sum_{r=1}^{s} 
\sum_{X_{r} \subset X } \; 
\sum_{ X_{2} \ldots X_{r-1} } \;
\sum_{T \; {\rm on} \; X_{r}}
(-1)^{r-1} \prod_{\ell \in T } V_{\ell} \cr
& \qquad \int_{0}^{1} \der t_{1} \ldots \int_{0}^{1} \der t_{r-1}
\left( \prod_{\ell \in T}
{ \prod_{k=1}^{r-1} t_{k}(\ell) \over
t_{n(\ell)} } \right) 
e^{-W_{X_{r}} (X_{1}, \ldots , X_{r-1} ; t_{1}, \ldots, t_{r-1} ) } 
\, e^{-V(X \setminus X_{r}) } \cr} \Eqa(2A4.38) $$
where ``$T \; {\rm on} \; X_{r}$'' means that $T$ is an anchored tree for the clusters
$P_j$ with $j\in X_r$.

Let us define
$$ \eqalign{
K(X_{r}) & = 
\sum_{ X_{2} \ldots X_{r-1} } \;
\sum_{T \; {\rm on} \; X_{r}}
\prod_{\ell \in T } V_{\ell} \cr
& \qquad \int_{0}^{1} \der t_{1} \ldots
\int_{0}^{1} \der t_{r-1} \left(
\prod_{\ell \in T}
{ \prod_{k=1}^{r-1} t_{k}(\ell) \over
t_{n(\ell)} } \right)
e^{-W_{X_{r}} (X_{1}, \ldots , X_{r-1} ; t_{1}, \ldots, t_{r-1} ) } 
\; , \cr} \Eqa(2A4.39) $$
so that \equ(2A4.38) can be rewritten as
$$ e^{-V(X)} = \sum_{Y \subset X \atop Y \ni \{ 1 \} }
(-1)^{|Y|-1} \, K(Y) \, e^{-V(X \setminus Y)} \; , \Eqa(2A4.40) $$
and, iterating,
$$ e^{-V(X)} = \sum_{ Q_{1},\ldots,Q_{m} } (-1)^{|X|} \,
(-1)^{m} \prod_{q=1}^{m} \, K(Q_q) \; . \Eqa(2A4.41) $$
The sets $Q_{1},\ldots,Q_{m}$ in \equ(2A4.41) are disjoint subsets of $X$, 
such that $\cup_{i=1}^mQ_i=X$.

Substituting \equ(2A4.41) in \equ(A2.13b), we find
$$\EE \left( \prod_{j=1}^{s} \tilde \phi(P_{j}) \right) =
\int {\cal D}\phi\sum_{(Q_1,\ldots,Q_m)}(-1)^{s+m}\prod_{q=1}^m K(Q_q)
\;,\Eqa(BB7)$$
where the sum is over the partitions $(Q_1,\ldots,Q_m)$ of $X$. It is easy to realize that
in the last equation $K(Q_q)$ (already defined in \equ(2A4.39)) can be rewritten as
$$\eqalign{&
K(Q)=\sum_{T\, {\rm on}\, Q}\sum_{X_2,\ldots, X_{|Q|-1}\atop{\rm fixed}\, T}
\prod_{\ell\in T}V_\ell\int_0^1 d t_1\cdots\int_0^1 d t_{|Q|-1}\cdot\cr
&\qquad\qquad\cdot
\prod_{\ell\in T}(t_{n'(\ell)}\cdots t_{n(\ell)-1})e^{-\sum_{\ell\in Q\times 
Q}t_{n'(\ell)}\cdots t_{n(\ell)}V_\ell}\cr}\Eqa(BB8)$$
Moreover, we can also rewrite \equ(BB7) as:
$$\EE \left( \prod_{j=1}^{s} \tilde \phi(P_{j}) \right) =
\sum_{(Q_1,\ldots,Q_m)}(-1)^{s+m}(-1)^\s 
\prod_{q=1}^m\int {\cal D}\phi_{Q_q} K(Q_q)\;,\Eqa(BB9)$$
where ${\cal D}\phi_{Q_q}=\prod_{j\in Q_q}\prod_{f\in P_j}d\phi^{\a(f)}_{
\xx(f),\o(f)}$ and $(-1)^\s$ is the sign of the permutation leading from the
ordering of the fields in $\DD\phi$ to the ones in $\prod_q\DD\phi_{Q_q}$.

Let us now consider the well known relation:
$$\EE \left( \prod_{j=1}^{s} 
\tilde\phi(P_j)\right) =
\sum_{(Q_{1},\ldots,Q_{m})} (-1)^{\s}
\EE^{T}\left(\tilde\phi(P_{j_{11}}),\ldots,\tilde\phi(P_{j_{1|Q_1|}})
\right)\ldots \EE^{T} \left( \tilde\phi(P_{j_{m1}}),\ldots,
\tilde\phi(P_{j_{m|Q_m|}})\right) \; , \Eqa(A4.48) $$
where the sum is over the partitions of $\{1,\ldots s\}$,
$Q_q=\{j_{q1},\ldots,j_{q|Q_q|}\}$ and $(-1)^\s$ is the parity of 
the permutation leading to the ordering on the r.h.s. from the one
on the l.h.s. (note that $\s$ is the same as in \equ(BB9)).
Comparing \equ(A4.48) with \equ(BB9) we get:
$$\EE^T(\tilde\phi(P_1),\ldots,\tilde\phi(P_s))=(-1)^{s+1}\sum_{T\,{\rm on}\,
X}\int\DD\phi\prod_{\ell\in T}V_\ell \int dP_T(\tt)e^{-V(\tt)}\;,\Eqa(BB10)$$
where we defined:
$$dP_T(\tt)=\sum_{ X_{2} \ldots X_{s-1} \atop {\rm fixed} \; T}
\prod_{\ell \in T}
\left( t_{n' ( \ell ) } \ldots t_{n( \ell )-1 } \right)
\prod_{q=1}^{s-1} \der t_{q}\Eqa(BB11)$$ 
and 
$$V(\tt) \equiv \sum_{\ell \in X\times X}
t_{n' ( \ell ) } \ldots t_{n( \ell )} \, V_{\ell}\;.\Eqa(BB12)$$ 
If in \equ(BB10) we integrate the Grassman fields appearing in the product
$$\prod_{\ell\in T}V_\ell=\prod_{(jj') \in T} \left( \overline V_{jj'}
+ \overline V_{jj'} \right)\;,\Eqa(BB13)$$ 
we obtain
$$ \EE^{T} \left( \tilde\phi(P_{1}), \ldots,
\tilde \phi(P_{s}) \right) = (-1)^{s+1}
\sum_{T\,{\rm on}\, P}\a_T
\prod_{\ell \in T } G_{f^1_\ell,f^2_\ell}
\int {\cal D}^*(\der\phi)
\int \der P_{T}(\tt) \, e^{-V^*(\tt)} \; , \Eqa(A2.47) $$
where $P=\cup_i P_i$, the sum $\sum_{T\,{\rm on}\,P}$ denotes the sum over the
graphs whose elements are lines connecting pairs of distinct 
points $\xx(f)$, $f\in P$ such that, if we identifie all the points in the 
clusters $P_j$, $j=1,\ldots,s$, then $T$ is a tree graph on $X$; moreover
$\a_T$ is a suitable sign and 
$${\cal D}^*(\der\phi)=\prod_{f\in P\atop f\not\in T}d \phi^{\a(f)}_{\xx(f),
\o(f)}\virg V^*(\tt)=\sum_{\ell \not\in T}
t_{n' ( \ell ) } \ldots t_{n( \ell )} \, V_{\ell}\;.\Eqa(BB14)$$
The term
$$ \int {\cal D}^*(\der\phi)
\int \der P_{T}(\tt) \, e^{-V^*(\tt)} \Eqa(A2.48) $$
in \equ(A2.47) is the Pfaffian of a suitable matrix
$G^{T}(\tt)$, with elements
$$ G^{T}_{f,f'}(\tt)=
t_{n'(\ell)} \ldots t_{n(\ell)}
G_{f,f'}\;,\Eqa(A2.49) $$
where $\ell=(j(f)j(f'))$, 
$j(f)\in X$ is s.t. $f\in P_{j(f)}$ and $G_{f,f'}$ was defined 
in \equ(A2.11). So \equ(4.12) is proven, with $t_{j,j'}=
t_{n'(jj')} \ldots t_{n(jj')}$. 

In order to complete the proof of the claims following \equ(4.12) we must prove 
that $dP_T(\tt)$ is a normalized, positive and $\s$--additive 
measure, so it can be interpreted as a probability measure in 
$\tt=(t_1,\ldots,t_{s-1})$; and that, moreover, we can find a family of versors 
${\bf u}_j\in\RRR^s$ such that 
$t_{j,j'}={\bf u}_j\cdot{\bf u}_{j'}$.

So, let us conclude this Appendix by proving the following Lemma.\\
\\
{\cs Lemma A2.1} 
{\it  $dP_T(\tt)$ is a normalized, positive and $\s$--additive 
measure on the natural $\s$--algebra of $[0,1]^{s-1}$. Moreover 
there exists a set of unit vectors ${\bf u}_j\in\RRR^s$, $j=1,\ldots,s$, 
such that $t_{j,j'}={\bf u}_j\cdot{\bf u}_{j'}$.}

\*

\proof Let us denote by $b_k$ the number of lines $\ell\in T$ exiting from
the points $x(j,i)$, $j\in X_k$, such that $\ell\sim X_k$. Let us consider
the integral 
$$ \sum_{ X_{2} \ldots X_{s-1} \atop  T\; {\rm fissato}}
\int_{0}^{1} \der t_{1} \ldots
\int_{0}^{1} \der t_{s-1} \,
\prod_{\ell \in T}
\left( t_{n' ( \ell ) } \ldots
t_{n( \ell )-1 } \right) = 1 \; , \Eqa(2A4.50) $$
and note that, by construction, the 
parameter $t_k$ inside the integral in the l.h.s. appears at the 
power $b_k-1$. In fact any line intersecting $\dpr X_k$ contributes 
by a factor $t_k$, except for the line connecting $X_k$ with the point 
in $X_{k+1}\setminus X_k$. See Fig. A2.2.

\midinsert
\*
\insertplotttt{285pt}{250pt}{
\ins{73pt}{115pt}{$X_{1}$}
\ins{66pt}{133pt}{$\ell_1$}
\ins{103pt}{107pt}{$X_{2}$}
\ins{103pt}{85pt}{$\ell_2$}
\ins{143pt}{80pt}{$X_{3}$}
\ins{152pt}{60pt}{$\ell_3$}
\ins{203pt}{45pt}{$X_{4}$}
\ins{170pt}{152pt}{$\ell_4$}
\ins{223pt}{25pt}{$X_{5}$}
\ins{237pt}{122pt}{$\ell_5$}
\ins{283pt}{5pt}{$X_{6}$}}
{figa3}
\vskip.3truecm
\line{\vtop{\line{\hskip1.3truecm\vbox{\advance\hsize by -2.0 truecm
\0{\css Fig. A2.1.}
{\ottorm The sets $\scriptstyle X_{1},\ldots,X_{6}$, the anchored tree
$\scriptstyle T$ and the lines $\scriptstyle \ell_1,\ldots,\ell_5$ belonging to 
$\scriptstyle T$. In the example, 
the coefficients $\scriptstyle b_1,\ldots, b_5$ are respectively 
equal to: $\scriptstyle 2,1,3,2,1$.}
} \hfill} }}
\*
\endinsert

Then
$$ \prod_{\ell \in T}
\left( t_{n'(\ell)} \ldots t_{n( \ell )-1 } \right) =
\prod_{k=1}^{s-1} t_{k}^{b_{k}-1} \; , \Eqa(2A4.51a) $$
and in \equ(2A4.50) the $s-1$ integrations are independent.
It holds: 
$$\int_{0}^{1} \der t_{1} \ldots
\int_{0}^{1} \der t_{s-1} \,
\prod_{\ell \in T}
\left( t_{n' ( \ell ) } \ldots
t_{n( \ell )-1 } \right) =\prod_{k=1}^{s-1} \left(
\int_{0}^{1} \der t_{k} \, t_{k}^{b_{k}-1} \right)  =
\prod_{k=1}^{s-1} {1 \over b_{k}} \; , \Eqa(2A4.51b) $$
that is well defined, since $b_{k}\ge 1$, $k=1, \ldots , m-2 $. 
Moreover we can write:
$$ \sum_{ X_{2} \ldots X_{s-1} \atop  T\;{\rm fixed}} =
\sum_{ X_{2} \atop  T,X_{1}\;{\rm fixed} }
\sum_{ X_{3} \atop  T,X_{1},X_{2}\;{\rm fixed} } \ldots
\sum_{ X_{s-1} \atop  T,X_{1},\ldots,X_{s-2}\;{\rm fixed} }
\; , \Eqa(2A4.51c) $$
where the number of possible choices in summing over $X_k$, once that
$T$ and the sets $X_1,\ldots, X_{k-1}$ are fixed, is exactly $b_{k-1}$.
In fact, if from $X_{k-1}$ there are $b_{k-1}$ exiting lines, then
$X_k$ is obtained by adding to
$X_{k-1}$ one of the $b_{k-1}$ points connected to $X_{k-1}$ through the 
tree lines. Then:
$$ \sum_{ X_{2} \ldots X_{s-1} \atop   T\;{\rm fissato}} 1
= b_{1} \ldots b_{s-2} \; , \Eqa(2A3.41d) $$
and, recalling that $b_{s-1}=1$,
$$ \sum_{ X_{2} \ldots X_{s-1} \atop  T\;{\rm fissato}}
\int_{0}^{1} \der t_{1} \ldots
\int_{0}^{1} \der t_{s-1} \,
\prod_{\ell \in T}
\left( t_{n' ( \ell ) } \ldots
t_{n( \ell )-1 } \right) =
\prod_{k=1}^{s-2} {b_{k} \over b_{k} } \; , \Eqa(2A4.51e) $$
yelding to $\int d P_T(\tt)=1$. The positivity and $\s$--addivity of 
$d P_T(\tt)$ is obvious by definition.

We are left with proving that we can find unit vectors ${\bf u}_j\in\RRR^s$ 
such that 
$t_{j,j'}={\bf u}_j\cdot{\bf u}_{j'}$.

For this aim, let us introduce a family of unit vectors in $\RRR^s$ defined as follows: 
$$ \cases{
\uu_{1} = \vv_{1} \; , & \cr
\uu_{j} = t_{j-1} \uu_{j-1} + \vv_{j}
\sqrt{1-t_{j-1}^2} \; , & $j=2,\ldots,s \; , $ \cr} \Eqa(2A4.60) $$
where $\{ {\bf v}_i\}_{i=1}^s$ is an orthonormal basis. 
Let us rename the sets  $P_i,\quad i=1,\ldots, s$ 
in such a way that $X_1=\{ 1\}$, $X_2\{ 1,2\}$, $\ldots$, 
$X_{s-1}=\{ 1,\ldots, s-1\}$. Then, for a given line $(jj')$, we have:
$$t_{j,j'}=t_{n'(jj')} \ldots t_{n(jj')}=t_j\ldots t_{j'-1}\Eqa(2A4.61)$$
From \equ(2A4.60) it follows
$$ \uu_{j} \cdot \uu_{j'} = t_{j} \ldots t_{j'-1}
\;  \Eqa(2A4.62) $$
as wanted.\qed

\*

\pagina
\setcap{appendix a3: Gram--Hadamard inequality.}
\appindex{A3}{Gram--Hadamard inequality.}
\vskip1.truecm
\appendix(A3, Gram--Hadamard inequality.)

In this Appendix we prove Gram--Hadamard inequality, that is the bound \equ(4.15).

Let ${\xx}_{1}, \ldots, {\xx}_{m}$ be $m$ vectors of a Hilbert space $\HH$
and let $E$ be their span.
We define the {\it Gram determinant} as
$$ \Gamma( {\xx}_{1}, \ldots, {\xx}_{m}) \= \det \G =
\det \left(
\matrix{
({\xx}_{1},{\xx}_{1}) & \ldots & ({\xx}_{1},{\xx}_{m}) \cr
\ldots & \ldots & \ldots \cr
({\xx}_{m},{\xx}_{1}) & \ldots & ({\xx}_{m},{\xx}_{m}) \cr}
\right) \; , \Eqa(A3.125) $$
where $( \cdot , \cdot )$ denotes the inner product in $\HH$.
The following results hold.

\*

\0{\cs Lemma A3.1.} {\it Given a Hilbert space $\HH$ and
$m$ vectors $\xx_1, \ldots,\xx_m$ in $\HH$, the Gram determinant
\equ(A3.125) satisfies
$$ \Gamma ( {\xx}_{1}, \ldots, {\xx}_{m} ) = 0 \; , \Eqa(A3.126) $$
if and only if the vectors $ {\xx}_{1}, \ldots, {\xx}_{m} $ 
are linearly independent. If the vectors $ {\xx}_{1}, \ldots, 
{\xx}_{m} $ are linearly independent then one has
$$ \Gamma ( {\xx}_{1}, \ldots, {\xx}_{m} ) > 0 \; . \Eqa(A3.127) $$
}

\*

\proof If the vectors ${\xx}_{1}, \ldots, {\xx}_{m}$ are
linearly dipendent then there exist $m$ coefficients
$c_1, \ldots, c_m$ not all vanishing such that the vector
$ \sum_{j=1}^{m} c_j {\xx}_j$ is vanishing. By
considering its inner product with the vectors
$ {\xx}_{1}, \ldots, {\xx}_{m}$, we obtain the system
$$
\matrix{
{c}_1 ({\xx}_{1},{\xx}_{1}) & + & \ldots & + &
{c}_m ({\xx}_{1},{\xx}_{m}) & = & 0 \cr
\ldots & & \ldots & & \ldots & & \ldots \cr 
{c}_1 ({\xx}_{m},{\xx}_{1}) & + & \ldots & + &
{c}_m ({\xx}_{m},{\xx}_{m}) & = & 0 \cr}
\Eqa(A3.128) $$
which is an homogeneous system admitting a nontrivial solution: therefore the determinant
of the matrix of the coefficients is zero, so implying \equ(A3.126).

{\it Vice versa} if \equ(A3.126) holds the system
\equ(A3.128) admits a nontrivial solution. If we multiply the $m$
equations defining the system by  $c_1, \ldots, c_m$, respectively,
then we sum them, we obtain
$$ \| c_1 {\xx}_1 + \ldots + c_m {\xx}_m \| = 0 \; , \Eqa(A3.129) $$
where $ \| \cdot \| $ is the norm induced by the inner product
$ ( \cdot, \cdot ) $. Therefore the vector $\sum_{j=1}^m c_j \xx_j$
has to be identically vanishing: as the coefficients
$c_1, \ldots, c_m$ are not all vanishing, then the vectors
$ {\xx}_{1}, \ldots, {\xx}_{m}$ have to be linearly dependent.

To prove \equ(A3.127) consider a non trivial
subset $S\subset E$, where $E$ is the span 
of $\xx_1,\ldots,\xx_m$,
and set, for any $\xx\in E$, ${\xx} = {\xx}_{S} + {\xx}_{N} $,
where ${\xx}_{S} \in S $ and $ {\xx}_{N}$ belonging
to the orthogonal complement to $S$. 
We can write $\xx_{N}$
as $\xx_N=c_1\xx_1+\cdots +c_p\xx_p$, where $p<m$ and $p=n-{\rm dim}(S)$
(now we are assuming that the vectors $\xx_1,\ldots, \xx_m$ are linearly independent). 
The vector
$$ \det \left( \matrix{
({\xx}_{1},{\xx}_{1}) & \ldots & ({\xx}_{1},{\xx}_{p}) & {\xx}_{1} \cr
\ldots & \ldots & \ldots & \ldots \cr
({\xx}_{p},{\xx}_{1}) & \ldots & ({\xx}_{p},{\xx}_{p}) & {\xx}_{p} \cr
({\xx},{\xx}_{1}) & \ldots & ({\xx},{\xx}_{p}) & {\xx}_{N} \cr} 
\right) \Eqa(A3.130) $$
is identically vanishing. In particular it follows that
$$ {\xx}_{N} = - {1 \over \det \G} \det \left(
\matrix{
& & & {\xx}_{1} \cr
& \Gamma & & \ldots \cr
& & & {\xx}_{p}\cr
({\xx},{\xx}_{1}) & \ldots & ({\xx},{\xx}_{p}) & {0} \cr} \right)
\; , \Eqa(A3.131) $$
and, analogoulsy,
$$ {\xx}_{S} \= {\xx} - {\xx}_{N} = 
{1 \over \det \G} \det \left(
\matrix{
& & & {\xx}_{1} \cr
& \Gamma & & \ldots \cr
& & & {\xx}_{p}\cr
({\xx},{\xx}_{1}) & \ldots & ({\xx},{\xx}_{p}) & {\xx} \cr}
\right) \quad , \Eqa(A3.132) $$
so that
$$ 0 \le h^{2} \= ( {\xx}_{S}, {\xx}) =
{1\over\det \Gamma}
\det \left(
\matrix{
& & & ( {\xx}_{1}, {\xx} ) \cr
& \Gamma & & \ldots \cr
& & & ( {\xx}_{p}, {\xx} ) \cr
({\xx},{\xx}_{1}) & \ldots & ({\xx},{\xx}_{p}) & ( {\xx}, {\xx} )
\cr} \right) =
{ \Gamma( {\xx}_{1}, \ldots, {\xx}_{p},{\xx} ) \over
\Gamma( {\xx}_{1}, \ldots, {\xx}_{p}) } \; . \Eqa(A3.133) $$
By setting $ {\xx} \= {\xx}_{p+1}$ and $ h^{2} = h_{p}^{2} $,
we can write \equ(A3.133) as
$$ { \Gamma( {\xx}_{1}, \ldots, {\xx}_{p}, {\xx}_{p+1} )
\over \Gamma( {\xx}_{1}, \ldots, {\xx}_{p}) } = h_{p}^{2} \ge 0
\; , \Eqa(A3.134) $$
where ${\xx}_{1}, \ldots, {\xx}_{p}$ are
$p$ linearly indepenedent vectors and 
${\xx}_{p+1} $ is arbitrary. The sign $=$ in \equ(A3.134)
can holds if and only if $\xx_{p+1}$ is a linear
combination of the vecors $ {\xx}_{1}, \ldots, {\xx}_{p} $
so that if ${\xx}_{1}, \ldots, {\xx}_{p}, {\xx}_{p+1}$
are linearly independent, then \equ(A3.134)
holds with the strict sign, \ie
$$ { \Gamma( {\xx}_{1}, \ldots, {\xx}_{p}, {\xx}_{p+1} )
\over \Gamma( {\xx}_{1}, \ldots, {\xx}_{p}) } = h_{p}^{2} > 0
\; . \Eqa(A3.135) $$
As $\Gamma({\xx}_{1}) = ( {\xx}_{1},
{\xx}_{1} ) = \| {\xx}_{1} \|^{2} > 0$
for $ {\xx}_{1} \neq {0} $, \equ(A3.135) implies \equ(A3.127). \qed

\*

\0{\cs Lemma A3.2 (Hadamard inequality).}
{\it The Gram determinant satisfies the inequality
$$ \Gamma( {\xx}_{1}, \ldots, {\xx}_{m}) \le
\Gamma( {\xx}_{1}) \ldots \Gamma ({\xx}_{m}) \; , \Eqa(A3.136) $$
where the sign $=$ holds if and only if the vectors
are orthogonal to each other.}

\*

\0{\it Proof.} By \equ(A3.135) and by using that
$ ( {\xx}_{S}, {\xx}_{S} ) \le ( {\xx}, {\xx} ) =
\Gamma({\xx}) $, we have
$$ \Gamma( {\xx}_{1}, \ldots, {\xx}_{m}, {\xx} ) \le
\Gamma( {\xx}_{1}, \ldots, {\xx}_{m}) \Gamma({\xx})
\; , \Eqa(A1.136a) $$
for any vectors ${\xx}_{1},\ldots,{\xx}_{m},{\xx} \in E$.
By iterating and recalling the arguments above
\equ(A3.136) follows. \qed

\*

Let $ {\xx}_{1}, \ldots, {\xx}_{m}$ be $m$
linearly independet vectors in $\HH$ and $E$ their span.
Let $ \{ {\ee}_{j} \}_{j=1}^{m}$ an orthonormal basis in $E$: set
$x_{jk} = ( {\ee}_{j}, {\xx}_{k} )$, so that
${\xx}_{k} = \sum_{j=1}^{m} x_{jk} {\ee}_{j}$, $k=1,\ldots,m$.
Then
$$ \eqalign{
\Gamma ( {\xx}_{1} , \ldots, {\xx}_{m} ) & =
\det \left( \matrix{
({\xx}_{1},{\xx}_{1}) & \ldots & ({\xx}_{1},{\xx}_{m}) \cr
\ldots & \ldots & \ldots \cr
({\xx}_{m},{\xx}_{1}) & \ldots & ({\xx}_{m},{\xx}_{m}) \cr}
\right) \cr 
& = \det \left(
\matrix{
\sum_{i j}\overline{x}_{i1} x_{j1} ({\ee}_{i},{\ee}_{j})
& \ldots & 
\sum_{i j}\overline{x}_{i1} x_{jm} ({\ee}_{i},{\ee}_{j}) \cr
\ldots & \ldots & \ldots \cr
\sum_{i j}\overline{x}_{im} x_{j1} ({\ee}_{i},{\ee}_{j})
& \ldots &
\sum_{i j}\overline{x}_{im} x_{jm} ({\ee}_{i},{\ee}_{j}) \cr}
\right) \cr
& = \det\left(
\matrix{
\sum_{i}\overline{x}_{i1} x_{i1}
& \ldots & \sum_{i}\overline{x}_{i1} x_{im} \cr
\ldots & \ldots & \ldots \cr
\sum_{i}\overline{x}_{im} x_{i1}
& \ldots & \sum_{i}\overline{x}_{im} x_{im}
\cr}  \right) \cr
& =
\det \left(
\matrix{
\overline{x}_{11} & \ldots & \overline{x}_{m1} \cr
\ldots & \ldots & \ldots \cr
\overline{x}_{1m} & \ldots & \overline{x}_{mm} \cr} \right) \;
\left(
\matrix{
{x}_{11} & \ldots & {x}_{1m} \cr
\ldots & \ldots & \ldots \cr
{x}_{m1} & \ldots & {x}_{mm} \cr} \right) \cr
& =
\det \overline{X}^{T} \,\det X  =| \det X |^{2}
\; , \cr} \Eqa(A3.137) $$
where the matrix $X$ is defined as
$$ X = \left(
\matrix{
x_{11} & x_{12} & \ldots x_{1m} \cr
x_{21} & x_{22} & \ldots x_{2m} \cr
\ldots & \ldots & \ldots \cr
x_{m1} & x_{m2} & \ldots x_{mm} \cr} \right)
\; . \Eqa(A3.138) $$
This yields that the Gram determinant \equ(A3.136) can be
written as
$$ \Gamma ( {\xx}_{1}, \ldots , {\xx}_{m} )
= | \det X|^{2} \; , \Eqa(A3.139) $$
so that from the lemma above the following result
follows immediately.

\*

\0{\cs Lemma A3.3.} {\it Given $m$ linearly independent vectors
of an Hilbert space $\HH$ and an orthonormal basis
$ \{ {\ee}_{j} \}_{j=1}^{m}$ on their span, and defining the matrix $X$ through
\equ(A3.137), one has
$$ \left| \det X \right|^2 \= \left| \det(\ee_i,\xx_j) \right|^2
\le \prod_{j=1}^m \|\xx_j\|^2 \; , \Eqa(A3.140) $$
where $(\ee_i,\xx_j)$ stands for the matrix with entries
$X_{ij}=(\ee_i,\xx_j)$.}

\*

The lemma above is simply a reformulation of the
preceeding Lemma: it implies the following inequality.

\*

\\{\cs Theorem A3.1 (Gram-Hadamard inequality).}
{\it Let $ \{ {\ff}_{j} \}_{j=1}^{m} $ and 
$ \{ {\ggg}_{j} \}_{j=1}^{m} $
two families of $m$ linearly independent vectors
in an Euclidean space $E$, and let
$( \cdot, \cdot ) $ an inner product in $E$ and $ \| \cdot \| $
the norm induced by that inner product. Then
$$ \left| \det ( {\ff}_{i}, {\ggg}_{j} ) \right|
\le \prod_{j=1}^{m} \| {\ff}_{j} \| \, \| {\ggg}_{j} \|
\; , \Eqa(A3.141) $$
where $(\ff_i,\ggg_j)$ stands for the $m\times m$ matrix
with entries $(\ff_i,\ggg_j)$.}

\*

\proof
If $ \{ {\ggg}_{j} \}_{j=1}^{m} $ is an orthogonal basis in
$E$ (so that $ \{ {\ee}_{j} \}_{j=1}^{m} $, with
$ {\ee}_{j} = \| {\ggg}_{j} \|^{-1}
{\ggg}_{j} $, is an orthonormal basis) then
\equ(A3.140) gives
$$ \left| \det({\ggg}_{i}, {\xx}_{j} ) \right| =
\left| \det({\ee}_{i}, {\xx}_{j} ) \right|
\; \prod_{j=1}^{m} \| {\ggg}_{j} \|
\le \prod_{j=1}^{m} \| {\ggg}_{j} \| \| {\xx}_{j} \|
\; , \Eqa(A3.142) $$
Now consider the case in which the only conditions
on the vectors $ \{ {\ggg}_{j} \}_{j=1}^{m} $ is that they are
linearly independent.
Set $\tilde{{\ggg}}_{j} =
\| {\ggg}_{j} \|^{-1} {\ggg}_{j}$, so that
$\| \tilde{{\ggg}}_{j} \|^{2} =1$,
and define inductively the family of vectors
$$ \eqalign{
\tilde{{\ee}}_{1} & \= \tilde{{\ggg}}_{1} \; , \cr
\tilde{{\ee}}_{2} & \= { \tilde{{\ggg}}_{2}
- ( \tilde{{\ggg}}_{2}, \tilde{{\ggg}}_{1} ) \tilde{{\ggg}}_{1} 
\over 1 - ( \tilde{{\ggg}}_{2}, \tilde{{\ggg}}_{1} )^{2} }
\; , \cr} \Eqa(A3.143) $$
and so on, in such a way that one has
$( \tilde{{\ee}}_{i}, \tilde{{\ee}}_{j} ) =
\delta_{i,j}$. The basis $\{\ee_{1},\ldots,\ee_{m}\}$,
with $\ee_j=\tilde{{\ee}}_{j}$ $\forall j=1,\ldots,m$
is by construction an orthonormal basis.

If $c_{2} = 1 - ( \tilde{{\ggg}}_{2}, \tilde{{\ggg}}_{1} )^{2}$,
with $0 \leq c_{2} \le 1$, one has
$$ \tilde{{\ggg}}_{2} = c_{2} \tilde{{\ee}}_{2} +
c_{2} ( \tilde{{\ggg}}_{2}, \tilde{{\ggg}}_{1} )
\tilde{{\ggg}}_{1} \; , 
\Eqa(A3.144) $$
\ie $\tilde{{\ggg}}_{2} \sim c_{2} \tilde{{\ee}}_{2}$,
if by $ \sim $ we mean that, by computing
$ \det ( \tilde{{\ggg}}_{i}, {\ff}_{j} ) $,
no difference is made by the fact that one has
the vector $ \tilde{{\ggg}}_{2}$ instead of
$ c_{2} \tilde{{\ee}}_{2} $: in fact the contributions
arising from the remaining part in \equ(A3.142)
sum up to zero.

We can reason analogously for the terms with
$ j=3, \ldots, m $, and we find $\tilde{{\ggg}}_{j}
\sim c_{j} \tilde{{\ee}}_{j}$, where $\sim$ is meant
as above and the coefficients $c_{j}$ are such that
$0 \leq c_{j} \le 1$ $\forall j=1,\ldots,m$. In conclusion:
$$ \eqalign{
\left| \det({\ggg}_{i}, {\ff}_{j} ) \right| & = 
\left| \det(\tilde{{\ggg}}_{i}, {\ff}_{j} ) \right| \;
\prod_{j=1}^{m} \| {\ggg}_{j} \| =
\left| \det({\ee}_{i}, {\ff}_{j} ) \right| \;
\prod_{j=1}^{m} c_{j} \, \| {\ggg}_{j} \| \cr
& = \prod_{j=1}^{m} c_{j} \, \| {\ggg}_{j} \| \,
\| {\ff}_{j} \| \le
\prod_{j=1}^{m} \| {\ggg}_{j} \| \,
\| {\ff}_{j} \| \quad , \cr} \Eqa(A3.145) $$
so that \equ(A3.141) follows. \qed

\pagina
\setcap{appendix a4: Proof of Lemma 5.3.}
\appindex{A4}{Proof of Lemma 5.3.}
\vskip1.truecm
\appendix(A4, Proof of Lemma 5.3.)

The propagators $g_{\underline a,\underline a'}^{(h)}(\xx)$
can be written in terms of the propagators
$g_{\o,\o'}^{(j,h)}(\xx)$, $j=1,2$,
see \equ(5.210b) and following lines; $g_{\o,\o'}^{(j,h)}(\xx)$
are given by
$$\eqalign{&g^{(j,h)}_{\o,\o}(\xx-\yy)=\cr
&\qquad\qquad={2\over M^2}
\sum_\kk e^{-i\kk(\xx-\yy)}\widetilde f_h(\kk) 
{-i\sin k+\o\sin k_0+a^{-(j)}_{h-1}(\kk)\over 
\sin^2 k+\sin^2 k_0+\big(\lis m^{(j)}_{h-1}(\kk)\big)^2+
\d B^{(j)}_{h-1}(\kk)}\cr 
&g^{(j,h)}_{\o,-\o}(\xx-\yy)=\cr
&\qquad\qquad={2\over M^2}
\sum_\kk e^{-i\kk(\xx-\yy)} 
\widetilde f_h(\kk)
{-i\o \lis m^{(j)}_{h-1}(\kk)
\over \sin^2 k+\sin^2 k_0+\big(\lis m^{(j)}_{h-1}(\kk)
\big)^2+\d B^{(j)}_{h-1}(\kk)}
\;,\cr}\Eqa(A4.24)$$
where
$$\eqalign{&a^{\o(j)}_{h-1}(\kk)\defin -a^\o_{h-1}(\kk)+(-1)^j
b^\o_{h-1}(\kk)\virg
c^{(j)}_{h-1}(\kk)\defin c_{h-1}(\kk)+(-1)^jd_{h-1}(\kk)\cr
&m_{h-1}^{(j)}(\kk)\defin \s_{h-1}(\kk)+(-1)^j\m_{h-1}(\kk)\virg
\lis m_{h-1}^{(j)}(\kk)\defin m_{h-1}^{
(j)}(\kk)+c^{(j)}(\kk)\cr
&\d B^{(j)}_{h-1}(\kk)\defin\sum_\o\big[a_{h-1}^{\o(j)}(\kk)
(i\sin k-\o\sin k_0)+a^{\o(j)}_{h-1}(\kk)
a^{-\o(j)}_{h-1}(\kk)/2\big]\;.\cr}\Eqa(A4.24y)$$ 
In order to bound the propagators defined above, we need estimates
on $\s_h(\kk),\m_h(\kk)$ and on the ``corrections''
$a^\o_{h-1}(\kk)$, $b^\o_{h-1}(\kk)$, $c_{h-1}(\kk)$, $d_{h-1}(\kk)$.
As regarding $\s_h(\kk)$ and $\m_h(\kk)$, it is easy to realize that, 
on the support of $f_h(\kk)$, for some $c$,
$c^{-1}|\s_h|\le|\s_{h-1}(\kk)|\le c|\s_h|$ and 
$c^{-1}|\m_h|\le|\m_{h-1}(\kk)|\le c|\m_h|$, see Proof
of Lemma 2.6 in [BM].
Note also that, if $h\ge \bar h$, using the first two of \equ(5.40z), we have
${|\s_h|+|\m_h|\over \g^h}\le 2C_1$.
As regarding the corrections, 
using their iterative definition \equ(5.17a), the asymptotic estimates
near $\kk={\bf 0}$ of the corrections on scale $h=1$ (see item (2)
in Theorem 4.1) and the hypothesis \equ(5.40z), we easily find that,
on the support of $f_h(\kk)$:
$$\eqalign{&
a_{h-1}^\o(\kk)= O(\s_h\g^{(1-2c|\l|)h})+O(\g^{(3-c|\l|^2)h})\virg
b_h^\o(\kk)=O(\m_h\g^{(1-2c|\l|)h})+O(\g^{(3-c|\l|^2)h})\;,\cr
& c_h(\kk)=O(\g^{(2-c|\l|^2)h})\virg d_h(\kk)=O(\m_h\g^{(2-2c|\l|)h})\;.\cr}
\Eqa(A4.c.5)$$
The bounds on the propagators follow from the remark that, as a consequence
of the estimates discussed above,
the denominators in \equ(A4.24) are $O(\g^{2h})$
on the support of $f_h$.\\

\pagina
\setcap{appendix a5: Proof of {\equ(5.53)}.}
\appindex{A5}{Proof of {\equ(5.53)}.}
\vskip1.truecm
\appendix(A5, Proof of {\equ(5.53)}.)

We have, by definition 
$\Pf G=(2^k k!)^{-1}\sum_{\pp}(-1)^\pp G_{p(1)p(2)}\cdots
G_{p(2k-1)p(2k)}$,
where $\pp=(p(1),\ldots$ $\ldots,p(|J|))$ 
is a permutation of the indeces $f\in J$ (we suppose $|J|=2k$)
and $(-1)^\pp$ its sign. 

If we apply $\SS_1=1-\PP_0$ to $\Pf G$ and we call 
$G^0_{f,f'}\defin \PP_0G_{f,f'}$, we find that $\SS_1\Pf G$
is equal to
$$\eqalign{&{1\over 2^k k!}\sum_{\pp}(-1)^\pp \Big[G_{p(1)p(2)}
\cdots
G_{p(2k-1)p(2k)}-G^0_{p(1)p(2)}\cdots
G^0_{p(2k-1)p(2k)}\Big]={1\over 2^k k!}\sum_{\pp}(-1)^\pp\sum_{j=1}^k\cdot\cr
&\cdot\Big(G^0_{p(1)p(2)}\cdots G^0_{
p(2j-3)p(2j-2)}\Big)\SS_1 G_{p(2j-1)p(2j)} \Big(G_{p(2j+1)p(2j+2)}\cdots
G_{p(2k-1)p(2k)}\Big)\;,\cr}\Eqa(a4.2)$$
where in the last sum the meaningless factors must be put equal to 1.
We rewrite the two sums over $\pp$ and $j$ in the following way:
$$\sum_\pp\sum_{j=1}^k=\sum_{j=1}^k\sum_{f_1,f_2\in J\atop f_1\not=
f_2}
\sum_{J_1,J_2}^*\sum_{\pp}^{**}\;,\Eqa(a4.3)$$
where: the $*$ on the second sum means that the sets $J_1$ and $J_2$ are s.t.
$(f_1,f_2,J_1,J_2)$ is a partition of $J$; the $**$ on the second
sum means that $p(1),\ldots,p(2j-2)$ belong to $J_1$, $(p(2j-1),p(2j))=
(f_1,f_2)$ and $p(2j+1),\ldots,p(2k)$ belong to $J_2$. 
Using \equ(a4.3) we can rewrite \equ(a4.2) as 
$$\eqalign{&
\SS_1\Pf G={1\over 2^k k!}\sum_{j=1}^k\sum_{f_1,f_2\in J\atop f_1\not=
f_2}(-1)^\p\SS_1 G_{f_1,f_2}\sum_{J_1,J_2}^{*}\cdot\cr
&\qquad\cdot\sum_{\pp_1,\pp_2}(-1)^{\pp_1+\pp_2}
\Big(G^0_{p_1(1)p_1(2)}\cdots G^0_{
p_1(2k_1-1)p(2k_1)}\Big)\Big(G_{p_2(1)p_2(2)}\cdots
G_{p_2(2k_2-1)p(2k_2)}\Big)\;,\cr}\Eqa(a4.4)$$
where: $(-1)^\p$ is the sign of the permutation leading
from the ordering $J$ to the ordering $(f_1,f_2,$ $J_1,J_2)$;
$\pp_i$, $i=1,2$ is a permutation of the labels in $J_i$ (we suppose $|J_i|=
2k_i$) and $(-1)^{\pp_i}$
is its sign. It is clear that \equ(a4.4) is equivalent to \equ(5.55).

\pagina
\setcap{appendix a6: Vanishing of the Beta function.}
\appindex{A6}{Vanishing of the Beta function.}
\vskip1.truecm
\appendix(A6, Vanishing of the Beta function.)

In this Appendix we want to prove the first bound in \equ(6.5.1bbb),
also called the {\it vanishing of the Luttinger model Beta function};
we reproduce the proof proposed in [BM1].

We will consider the reference model for our system, that is a model
with propagator given by \equ(6.gL), a local quartic interaction and
with both ultraviolet and infrared cutoffs. The model is similar (but
not the same) to the Luttinger model, but it is not exactly solvable.
However its Beta function, coinciding with the first term in the 
r.h.s. of \equ(6.5.1aaa), also coincide with the infrared part of the Luttinger
model Beta function. 

The reference model formally satisfies chiral gauge invariance, in the sense 
that, {\it neglecting the UV and IR cutoffs}, it is invariant under
the transformations 
$$\psi^{\pm}_{\xx,\o}\to e^{\pm i\a_{\xx,\o}}
\psi^{\pm}_{\xx,\o},\qquad (\dpr_0+i\o\dpr_1)\to
(\dpr_0+i\o\dpr_1)+i\big[(\dpr_0+i\o\dpr_1)\a_{\xx,\o}\big]\;.$$ 
Using the invariance of the 
Schwinger functions generating functional under these transformations, one
gets a hierarchy of {\it Ward identities}, which differ from the 
formal ones by terms which formally vanish when the cutoffs are removed.
However these terms could give no trivial contributions
to the correlation functions, because they must be included in the multiscale
integration and the cutoffs must be removed after the integration procedure
is finished. This is in fact the case, and the result can be expressed in terms
of some correction identities, relating the corrections to the formal Ward 
identities to the 2 or 4 legs Schwinger functions. 
The exact Ward identities differ from the formal ones, even when the UV and
IR cutoffs are removed. This is called {\it breaking of chiral symmetry}. 

The Ward identities together with the so called Dyson equation, allow to 
express $\l_h$ in terms of $\l$ and of Schwinger functions
satisfying the ``right'' dimensional bounds. The conclusion will be that,
keeping the UV cutoff fixed at scale $0$, $\l_h=\l+O(\l^2)$, uniformely
in $h<0$. This implies that the IR cutoff can be removed and 
that the infrared part of the Beta function satisfies the first of 
\equ(6.5.1bbb). The bound \equ(6.5.1bbb) is an easy consequence of the bound
$\l_h=\l+O(\l^2)$, and the proof of this will be presented below for 
completeness, see \equ(van1)--\equ(A6.5.3dhk).
\\
\\
\asub(6.1.1){\bf The reference model}
\\
The reference model is defined by the interaction
$$V(\psi)=\l \int d\xx\; \psi^{+}_{\xx,+} \psi^{-}_{\xx,+}
\psi^{+}_{\xx,-} \psi^{-}_{\xx,-}\Eqa(A6.1.2)$$
where $\int d\xx$ is a
shorthand for ``$\sum_{\xx\in\L_M}$'',
and by the free ``measure''
$$P(d\psi) = \NN^{-1} \DD\psi \cdot\;\exp \left\{-{1\over M^2}
\sum_{\o=\pm 1} \sum_{\kk} C_{h,0}(\kk)(-i k_0+\o k)
\hat\psi^{+}_{\kk,\o} \hat\psi^{-}_{\kk,\o}\right\}\;,\Eqa(A6.1.3)$$
where the summation over $\kk$ is over the momenta allowed by the antiperiodic
boundary conditions, $\NN=\prod_{\kk\in
\DD}[(L\b)^{-2}(-k_0^2-k^2)C_{h,0}(\kk)^2]$ and $[C_{h,0}(\kk)]^{-1}\defin
\sum_{k=h}^0 f_h(\kk)\=\c_{h,0}(\kk)$. 
We read the presence of $C_{h,0}(\kk)$ by saying that
an ultraviolet cutoff on scale $0$ and an infrared cutoff on scale $h$
are imposed.
We introduce the {\it generating functional}
$$\WW(\phi,J)= \log \int P(d\psi) e^{-V(\psi)+ \sum_\o \int d\xx
\left[J_{\xx,\o}\psi^{+}_{\xx,\o}\psi^{-}_{\xx,\o}+
\phi^+_{\xx,\o}\psi^{-}_{\xx,\o}+ \psi^{+}_{\xx,\o}\phi^-_{\xx,\o}\right]}
\;.\Eqa(A6.1.7)$$
The interaction $V$ and the density operators appearing at the exponent
of \equ(A6.1.7) can be represented as in Fig \graf(1aa).

\midinsert
\*
\insertplotbm{300pt}{80pt}%
{\ins{30pt}{60pt}{$+$}
\ins{60pt}{60pt}{$+$}
\ins{20pt}{40pt}{$-$}
\ins{70pt}{40pt}{$-$}

\ins{120pt}{60pt}{$\o$}
\ins{170pt}{60pt}{$\o$}

}%
{vertici}{}
\vskip.1truecm
\line{\vtop{\line{\hskip1.3truecm\vbox{\advance\hsize by -2.0 truecm
\0{\css \eqg(1aa).}
{\ottorm  Graphical representation of the interaction ${\scriptstyle V(\psi)}$
and the density ${\scriptstyle \psi^{+}_{\xx,\o}\psi^{-}_{\xx,\o}}$} 
\hfill} }}}
\*
\endinsert
%

The Schwinger functions can be obtained by functional
derivatives of \equ(A6.1.7); for instance
$$G^{2,1}_\o(\xx;\yy,\zz)={\dpr\over\dpr J_{\xx,\o}}
{\dpr^2\over\dpr\phi^+_{\yy,+}\dpr\phi^-_{\zz,+}}
\WW(\phi,J)|_{\phi=J=0}\;,\Eqa(A6.1.8)$$
$$G_{\o}^{4,1}(\xx;\xx_1,\xx_2,\xx_3,\xx_4)={\dpr\over\dpr J_{\xx,\o}}
{\dpr^2\over\dpr\phi^+_{\xx_1,\o}\dpr\phi^-_{\xx_2,\o}}
{\dpr^2\over\dpr\phi^+_{\xx_3,-\o}\dpr\phi^-_{\xx_4,-\o}}
\WW(\phi,J)|_{\phi=J=0}\;,\Eqa(A6.1.9)$$
$$G_{\o}^{4}(\xx_1,\xx_2,\xx_3,\xx_4)=
{\dpr^2\over\dpr\phi^+_{\xx_1,\o}\dpr\phi^-_{\xx_2,\o}}
{\dpr^2\over\dpr\phi^+_{\xx_3,-\o}\dpr\phi^-_{\xx_4,-\o}}
\WW(\phi,J)|_{\phi=J=0}\;,\Eqa(A6.1.10)$$

$$G_{\o}^{2}(\yy,\zz)= {\dpr^2\over\dpr\phi^+_{\yy,\o}\dpr\phi^-_{\zz,\o}}
\WW(\phi,J)|_{\phi=J=0}\;.\Eqa(A6.1.11a)$$

\midinsert
\*
\insertplotbm{300pt}{150pt}%
{\ins{25pt}{82pt}{$G^{2,1}_\o$}
\ins{28pt}{32pt}{$\xx$}
\ins{-5pt}{110pt}{$\yy$}
\ins{58pt}{110pt}{$\zz$}
\ins{10pt}{97pt}{$\o$}
\ins{45pt}{97pt}{$\o$}
\ins{20pt}{50pt}{$\o$}
\ins{35pt}{50pt}{$\o$}
\ins{105pt}{82pt}{$G^{4,1}_\o$}
\ins{88pt}{32pt}{$\xx$}
\ins{128pt}{32pt}{$\xx_4$}
\ins{80pt}{120pt}{$\xx_1$}
\ins{108pt}{122pt}{$\xx_2$}
\ins{130pt}{120pt}{$\xx_3$}
\ins{80pt}{105pt}{$\o$}
\ins{102pt}{107pt}{$\o$}
\ins{118pt}{110pt}{$-\o$}
\ins{112pt}{45pt}{$-\o$}
\ins{83pt}{50pt}{$\o$}
\ins{98pt}{50pt}{$\o$}
\ins{195pt}{82pt}{$G^4_\o$}
\ins{165pt}{120pt}{$\xx_1$}
\ins{228pt}{120pt}{$\xx_2$}
\ins{168pt}{38pt}{$\xx_3$}
\ins{228pt}{38pt}{$\xx_4$}
\ins{165pt}{105pt}{$\o$}
\ins{218pt}{108pt}{$\o$}
\ins{162pt}{55pt}{$-\o$}
\ins{210pt}{45pt}{$-\o$}
\ins{275pt}{82pt}{$G^2_\o$}
\ins{279pt}{114pt}{$\zz$}
\ins{279pt}{38pt}{$\yy$}
\ins{272pt}{105pt}{$\o$}
\ins{272pt}{45pt}{$\o$}
}%
{schw}{}
\vskip.1truecm
\line{\vtop{\line{\hskip1.3truecm\vbox{\advance\hsize by -2.0 truecm
\0{\css \eqg(1a).}
{\ottorm Graphical representation of the Schwinger functions
${\scriptstyle G^{2,1}_\o, G^{4,1}_\o, G^{4}_\o, G^2_\o}$.}
\hfill} }}}
\*
\endinsert

The generating functional and the Schwinger functions introduced above can be 
studied by a multiscale analysis similar to that described in Chapter 5,
with some complications, due to the presence of the density fields $J$,
and, from the other side, with some simplifications, due
to the absence of mass terms in our action (it can be easily seen by 
symmetry that mass terms analogue of $F_\s$ or $F_\m$ cannot even be
generated by the multiscale expansion). We will sketch the expansion below.\\
\\
\asub(1.3){\bf The Dyson equation.}
\\
Let us consider the four legs Schwinger function $G_{\o}^{4}$ in \equ(A6.1.10),
computed at a momentum scale $=h$, which is proportional to $\l_h$, 
as it is easy to realize. Since we want to connect 
$\l_h$ with the ``bare coupling'' $\l$, 
it is natural to write a {\it Dyson}
equation for $\hat G^4$:
$$\eqalign{
-\hat G^{4}_+(\kk_1,\kk_2,\kk_3,\kk_4) &= \l \hat g_-(\kk_4) \Big[
\hat G^2_-(\kk_3) \hat G^{2,1}_+(\kk_1-\kk_2,\kk_1,\kk_2)+\cr
&+ {1\over M^2} \sum_\pp
G^{4,1}_+(\pp;\kk_1,\kk_2,\kk_3,\kk_4-\pp)\Big]\;,}\Eqa(A6.2.11a)$$
relating the correlations in
\equ(A6.1.8),\equ(A6.1.9),\equ(A6.1.10),\equ(A6.1.11a); see Fig. \graf(3).

\midinsert
\*
\insertplotbm{300pt}{150pt}%
{\ins{45pt}{82pt}{$\hat G^{4}_+$}
\ins{30pt}{110pt}{$\kk_1$}
\ins{15pt}{110pt}{$+$}
\ins{65pt}{115pt}{$\kk_2$}
\ins{80pt}{105pt}{$+$}
\ins{65pt}{45pt}{$\kk_3$}
\ins{20pt}{55pt}{$-$}
\ins{30pt}{45pt}{$\kk_4$}
\ins{75pt}{55pt}{$-$}
\ins{95pt}{75pt}{$=$}
\ins{143pt}{123pt}{$\hat G^{2,1}_+$}
\ins{147pt}{50pt}{$\hat G^{2}_-$}
\ins{130pt}{140pt}{$\kk_1$}
\ins{120pt}{132pt}{$+$}
\ins{165pt}{145pt}{$\kk_2$}
\ins{180pt}{135pt}{$+$}
\ins{130pt}{85pt}{$\kk_4$}
\ins{120pt}{75pt}{$-$}
\ins{155pt}{90pt}{$\kk_1-\kk_2$}
\ins{140pt}{95pt}{$+$}
\ins{155pt}{97pt}{$+$}
\ins{155pt}{70pt}{$\kk_3$}
\ins{143pt}{68pt}{$-$}
\ins{155pt}{20pt}{$\kk_3$}
\ins{143pt}{28pt}{$-$}
\ins{200pt}{75pt}{$+$}
\ins{245pt}{108pt}{$\hat G^{4,1}_+$}
\ins{220pt}{130pt}{$\kk_1$}
\ins{220pt}{142pt}{$+$}
\ins{240pt}{135pt}{$\kk_2$}
\ins{252pt}{138pt}{$+$}
\ins{275pt}{125pt}{$\kk_3$}
\ins{275pt}{142pt}{$-$}
\ins{255pt}{50pt}{$\kk_4$}
\ins{240pt}{40pt}{$-$}
\ins{260pt}{70pt}{$\kk_4-\pp$}
\ins{260pt}{80pt}{$-$}
\ins{235pt}{70pt}{$\pp$}
\ins{235pt}{80pt}{$+$}
}%
{dyson}{}
\vskip.1truecm
\line{\vtop{\line{\hskip1.3truecm\vbox{\advance\hsize by -2.0 truecm
\0{\css \eqg(3).}
{\ottorm Graphical representation of the Dyson equation
\equ(A6.2.11a); the dotted line represents the ``bare'' propagator
$g(\kk_4)$}
\hfill} }}}
\*
\endinsert

The Dyson equation can be derived as follows. 

We define
$$G^{4,1}_{\o}(\zz,\xx_1,\xx_2,\xx_3,\xx_4)=
<\r_{\zz,\o};\psi^-_{\xx_1,+};\psi^+_{\xx_2,+};
\psi^-_{\xx_3,-};\psi^+_{\xx_4,-}>^T\;,\Eqa(A6.2.5)$$
$$G^{4}_+(\xx_1,\xx_2,\xx_3,\xx_4)=
<\psi^-_{\xx_1,+}\psi^+_{\xx_2,+};
\psi^-_{\xx_3,-}\psi^+_{\xx_4,-}>^T\;,\Eqa(A6.2.6)$$
where
$$\r_{\xx,\o}=\psi^+_{\xx,\o}\psi^-_{\xx,\o}\;.\Eqa(A6.2.10)$$
Moreover, we shall denote by $\hat G^{4,1}_{+\o}(\pp; \kk_1, \kk_2,
\kk_3, \kk_4)$ and $\hat G^{4}_+(\kk_1,\kk_2,\kk_3,\kk_4)$ the
corresponding Fou\-rier transforms, deprived of the momentum
conservation delta. Note that, if
the $\psi^+$ momenta are interpreted as ``ingoing momenta'' in the
usual graph pictures, then the $\psi^-$ momenta are ``outgoing
momenta''; our definition of Fourier transform is such that even
$\pp$, the momentum associated with the $\rho$ field, is an
ingoing momentum. Hence, the momentum conservation implies that
$\kk_1 +\kk_3 =\kk_2 +\kk_4+\pp$, in the case of $\hat
G^{4,1}_\o(\pp; \kk_1, \kk_2, \kk_3, \kk_4)$ and $\kk_1 +\kk_3 =
\kk_2 +\kk_4$ in the case of $\hat G^{4}_+(\kk_1, \kk_2, \kk_3,
\kk_4)$.

If $Z=\int P(d\psi) \exp \{-V(\psi) \}$ and $<\cdot>$ denotes the
expectation with respect to $Z^{-1} \int P(d\psi)$
$\exp\{-V(\psi)\}$, by the definition of truncated expectation it
follows:
$$G^4_+(\xx_1,\xx_2,\xx_3,\xx_4) = <\psi^-_{\xx_1,+} \psi^+_{\xx_2,+}
\psi^-_{\xx_3,-}\psi^+_{\xx_4,-}> -
G^2_+(\xx_1,\xx_2) G^2_-(\xx_3,\xx_4)\;,\Eqa(A6.2.7)$$
where we used the fact that $<\psi^-_{\xx,\o}\psi^+_{\yy,-\o}>=0$.

Let $g_\o(\xx)$ be the free propagator, whose Fourier transform is
$g_\o(\kk)=\chi_{h,0}(\kk)/(-ik_0+\o k)$. Then, we
can write the last equation as
$$\eqalign{
&G^4_+(\xx_1,\xx_2,\xx_3,\xx_4)=
-\l \int d\zz\; g_{-}(\zz-\xx_4) <\psi^-_{\xx_1,+} \psi^+_{\xx_2,+}
\psi^-_{\xx_3,-} \psi^+_{\zz,-} \psi^+_{\zz,+}\psi^-_{\zz,+}> +\cr
& +\l\; G^2_+(\xx_1,\xx_2) \int d\zz\; g_{-}(\zz-\xx_4)
<\psi^-_{\xx_3,-} \psi^+_{\zz,-} \psi^+_{\zz,+}\psi^-_{\zz,+}>=\cr
&= -\l \int d\zz g_{-1}(\zz-\xx_4) <[\psi^-_{\xx_1,+}
\psi^+_{\xx_2,+}]\,;[\psi^-_{\xx_3,-}
\psi^+_{\zz,-}\psi^+_{\zz,+}\psi^-_{\zz,+}]>^T\;.\cr}\Eqa(A6.2.8)$$
Again, by definition of truncated expectations, we have:
$$<\psi^-_{\xx_1,+}; \psi^+_{\xx_2,+}; \r_{\zz,+}>^T=
<\psi^-_{\xx_1,+} \psi^+_{\xx_2,+} \r_{\zz,+}>-
<\psi^-_{\xx_1,+}\psi^+_{\xx_2,+}> <\r_{\zz,+}>\;,\Eqa(trunc1)$$
and
$$\eqalign{&
<\r_{\zz,+}; \psi^-_{\xx_1,+}; \psi^+_{\xx_2,+};
\psi^-_{\xx_3,-}; \psi^+_{\zz,-}>^T=
<\r_{\zz,+} \psi^-_{\xx_1,+} \psi^+_{\xx_2,+}
\psi^-_{\xx_3,-} \psi^+_{\zz,-}>-\cr
&-<\r_{\zz,+} \psi^-_{\xx_1,+} \psi^+_{\xx_2,+}>
<\psi^-_{\xx_3,-} \psi^+_{\zz,-}>-
<\r_{\zz,+}> <\psi^-_{\xx_1,+} \psi^+_{\xx_2,+}
\psi^-_{\xx_3,-} \psi^+_{\zz,-}>-\cr
&-<\psi^-_{\xx_1,+} \psi^+_{\xx_2,+}>
<\r_{\zz,+}\psi^-_{\xx_3,-} \psi^+_{\zz,-}>
+2<\r_{\zz,+}> < \psi^-_{\xx_1,+} \psi^+_{\xx_2,+}>
<\psi^-_{\xx_3,-} \psi^+_{\zz,-}>\;.\cr}\Eqa(trunc)$$
Using the last two equations, together with 
\equ(A6.2.7), we can rewrite \equ(A6.2.8) as:
$$\eqalign{
&-G^4_+(\xx_1,\xx_2,\xx_3,\xx_4)= \l\int d\zz
g_{-}(\zz-\xx_4)<\psi^-_{\xx_1,+}; \psi^+_{\xx_2,+}; \r_{\zz,+}>^T
<\psi^-_{\xx_3,-} \psi^+_{\zz,-}>+\cr &\quad +\l\int d\zz
g_{-}(\zz-\xx_4) <\r_{\zz,+}; \psi^-_{\xx_1,+}; \psi^+_{\xx_2,+};
\psi^-_{\xx_3,-}; \psi^+_{\zz,-}>^T+ \cr &\quad +\l\int
d\zz g_{-}(\zz-\xx_4)<\psi^-_{\xx_1,+}; \psi^+_{\xx_2,+};
\psi^-_{\xx_3,-}; \psi^+_{\zz,-}>^T<\r_{\zz,+}>\;.\cr}\Eqa(dys)$$
The last addend is vanishing, since $<\r_{\zz,\o}>=0$ by the
propagator parity properties. In terms of Fourier transforms, we
get the {\it Dyson equation} \equ(A6.2.11a).

The l.h.s. of the Dyson equation computed at the cutoff scale
is indeed proportional to the
effective interaction $\l_h$ (see \equ(A6.1.47)
below), while the r.h.s. is proportional to
$\l$. {\it If one does not take into account cancellations in
\equ(A6.2.11a)}, this equation only allows us to prove that
$|\l_h|\le C_h |\l|$, with $C_h$ diverging as $h\to-\io$. However,
inspired by the analysis in the physical literature, see [DL][So][DM],
we can try to express $\hat G^{2,1}_\o$ and $\hat G^{4,1}_\o$, in the
r.h.s. of \equ(A6.2.11a), in terms of $\hat G^{2}_\o$ and $\hat
G^{4}_\o$ by suitable {\it Ward identities} and {\it correction identities}.\\
\\
\asub(1.4){\bf Ward identities and the first addend of \equ(A6.2.11a)}
\\
To begin with, we consider the first addend in the r.h.s. of the Dyson
equation \equ(A6.2.11a). A remarkable identity relating $\hat G^{2,1}_+$ to
$\hat G^2_+$ can be obtained by the chiral Gauge transformation
$\psi^\pm_{\xx,+}\to e^{\pm i\a_{\xx}} \psi^\pm_{\xx,+}$,
$\psi^\pm_{\xx,-}\to \psi^\pm_{\xx,-}$ in the generating functional
\equ(A6.1.7); one obtains the following identity, represented pictorially
in Fig. \graf(4), with $D_\o(\pp)=-i p_0+\o p$:
$$D_+(\pp)\hat G^{2,1}_{+}(\pp,\kk) = G^{2}_{+}(\qq)-
G^{2}_{+}(\kk)+\hat\D^{2,1}_{+}(\pp,\kk)\;,\Eqa(A6.2.12)$$
with $\hat\D^{2,1}_{+}$ the Fourier transform of 
$\D^{2,1}_{+}(\xx;\yy,\zz)$:
$$\D^{2,1}_{+}(\xx;\yy,\zz)={1\over M^4}\sum_{\kk,\pp} e^{i\pp\xx-i\kk\yy
+i(\kk-\pp)\zz}\hat\D^{2,1}_{+}(\pp,\kk)=<\psi^-_{\yy,+};\psi^+_{\zz,+};
\d T_{\xx,+}>^T\Eqa(A6.2.13)$$
and
$$\eqalign{&\d T_{\xx,\o}={1\over M^2}\sum_{\kk_+\not =\kk_-}
e^{i(\kk_+-\kk_-)\xx}
C_\o(\kk_+,\kk_-)\hat\psi^+_{\kk_+,\o}\hat\psi^-_{\kk_-,\o}\;,\cr
&C_\o(\kk^+,\kk^-)=[C_{h,0}(\kk^-)-1]D_\o(\kk^-)
-[C_{h,0}(\kk^+)-1]D_\o(\kk^+)\;.\cr}\Eqa(A6.2.14)$$

\midinsert
\*
\insertplotbm{300pt}{150pt}%
{\ins{53pt}{83pt}{$\hat G^{2,1}_+$}
\ins{5pt}{83pt}{$D_+(\pp)$}
\ins{45pt}{110pt}{$\kk$}
\ins{70pt}{110pt}{$\qq$}
\ins{42pt}{25pt}{$\pp=\kk-\qq$}
\ins{97pt}{80pt}{$=$}
\ins{125pt}{81pt}{$\hat G^2_+$}
\ins{135pt}{50pt}{$\qq$}
\ins{135pt}{105pt}{$\qq$}
\ins{155pt}{80pt}{$-$}
\ins{185pt}{81pt}{$\hat G^2_+$}
\ins{195pt}{50pt}{$\kk$}
\ins{195pt}{105pt}{$\kk$}
\ins{215pt}{80pt}{$+$}
\ins{248pt}{83pt}{$\hat \D^{2,1}_+$}
\ins{240pt}{110pt}{$\kk$}
\ins{265pt}{110pt}{$\qq$}
\ins{253pt}{25pt}{$\pp$}
}%
{ward1}{}
\vskip.1truecm
\line{\vtop{\line{\hskip1.3truecm\vbox{\advance\hsize by -2.0 truecm
\0{\css \eqg(4).}
{\ottorm Graphical representation of the Ward identity
\equ(A6.2.12); the small circle in $\hat \D^{2,1}_+$ represents the
function $C_+$ of \equ(A6.2.14).}
\hfill} }}}
\*
\endinsert

The above Ward identity can be derived as follows. Consider the chiral
gauge transformation
$$\psi^\pm_{\xx,+}\to e^{i\pm\a_{\xx}}\psi^\pm_{\xx,+}\;,\qquad
\psi^\pm_{\xx,-}\to \psi^\pm_{\xx,-}\;,\Eqa(chir)$$
and notice that $\WW(\phi,J)$, as defined by \equ(A6.1.7), is
invariant under this change of variables. Then we can rewrite
$$\eqalign{&\WW(\phi,J)= \log \int P(d\psi) \exp\Big\{-\int d\xx
\psi^+_{\xx,+}\Big(e^{i\a_\xx}D_+^{[h,0]}e^{-i\a_\xx}-D_+^{[h,0]}
\Big)\psi^-_{\xx,+}
\Big\}\cdot\cr
&\cdot\exp\Big\{-V(\psi)+ \int d\xx
\left[\sum_\o J_{\xx,\o}\psi^{+}_{\xx,\o}\psi^{-}_{\xx,\o}+
e^{-i\a_\xx}\phi^+_{\xx,+}\psi^{-}_{\xx,+}+ e^{i\a_\xx}\psi^{+}_{\xx,+}
\phi^-_{\xx,+}+\phi^+_{\xx,-}\psi^{-}_{\xx,-}+ \psi^{+}_{\xx,-}
\phi^-_{\xx,-}\right]\Big\}\;.\cr}\Eqa(chir2)$$
where $D_\o^{[h,0]}$, $\o=\pm$, is the pseudo differential operator defined by
$$D_\o^{[h,0]}\psi^\a_{\xx,\o}={1\over M^2}\sum_\kk e^{i\a\kk\xx}
(i\a k_0-\o\a k)\hat \psi^\a_{\kk,\o}\;,\qquad D_\o^{[h,0]}\psi^\a_{\xx,-\o}=0
\;.\Eqa(chir3)$$
As we have just remarked, the l.h.s. of \equ(chir2) is independent
of $\a_\xx$. Then, by differentiating both sides w.r.t. $\a_\xx$
and posing $\a_\xx\=0$, we find:
$$0=<-D_+(\psi^+_{\xx,+}\psi^-_{\xx,+})-\d T_{\xx,+}
-\phi^+_{\xx,+}\psi^{-}_{\xx,+}+ \psi^{+}_{\xx,+}
\phi^-_{\xx,+}>_{\phi,J}\;,\Eqa(chir4)$$
where 
$$<\cdot>_{\phi,J}\defin e^{-\WW(\phi,J)}
\int P(d\psi) e^{-V(\psi)+ \sum_\o \int d\xx
\left[J_{\xx,\o}\psi^{+}_{\xx,\o}\psi^{-}_{\xx,\o}+
\phi^+_{\xx,\o}\psi^{-}_{\xx,\o}+ \psi^{+}_{\xx,\o}\phi^-_{\xx,\o}\right]}
\cdot\;,\Eqa(chir4a)$$
$D_\o$ is defined as the Fourier transform of $D_\o(\pp)$:
$$D_\o(\psi^+_{\xx,\o}\psi^+_{\xx,\o})={1\over M^4}\sum_{\pp,\kk}
D_\o(\pp)e^{-i\pp\xx}\hat\psi^+_{\kk,\o}\hat\psi^-_{\kk-\pp,\o}\;,
\Eqa(chir4b)$$
and the corrections $\d T_{\xx,+}$ and $C_\o(\kk_+,\kk_-)$ were 
defined by \equ(A6.2.14).

Now, differentiating \equ(chir4) w.r.t. $\phi^+_{\yy,+}$ and $\phi^-_{\zz,+}$
and setting $\phi=J=0$, we find:
$$-D_+ G^{2,1}_+(\xx;\yy,\zz)=\d(\xx-\yy)G^{2}_+(\xx,\zz)-
\d(\xx-\zz)G^{2}_+(\yy,xx)+\D^{2,1}_+(\xx;\yy,\zz)\;,\Eqa(chir5)$$
whose Fourier transform gives \equ(A6.2.12).

The use of Ward identities is to provide relations between Schwinger
functions, but the correction terms (due to the cutoffs) substantially affect
the Ward identities and apparently spoil them of their utility. However there
are other remarkable relations connecting the correction terms to the
Schwinger functions; such {\it correction identities} can be proved by
performing a careful analysis of the renormalized expansion for the
correction terms, and come out of the peculiar properties of the function
$C_+(\kk,\kk-\pp)$, see next section. The {\it
correction identity} for $\hat \D_+^{2,1}$ is the following, see Fig.\graf(5).
$$\hat \D_+^{2,1}(\pp,\kk)=D_+(\pp)\Big[ \n_+  
\hat G_+^{2,1}(\pp,\kk)+\n_- {D_-(\pp)\over D_+(\pp)}
\hat G_-^{2,1}(\pp,\kk)+
\hat H_+^{2,1}(\pp,\kk)\Big]\Eqa(A6.1.3b)$$
where $\n_+,\n_-$ are $O(\l)$ and weakly dependent on $h$, once we prove that
$\l_j$ is small enough for $j\ge h$, and $\hat H_+^{2,1}(\pp,\kk,\qq)$ can be
obtained through the analogue of \equ(A6.1.8), with $\WW(\phi,J)$ replaced by 
$$
\WW_\D(\phi,J)=\log\int P(d\psi) e^{-V(\psi)+\int d\xx [J_{\xx,+} T_\xx
+\sum_\o(-\n_\o J_{\xx,+} T_{\xx,\o}^\n+\phi^+_{\xx,\o}\psi^-_{\xx,\o}+
\psi^+_{\xx,\o}\phi^-_{\xx,\o})]}\;,\Eqa(wdelta)$$
where
$$\eqalign{
&T_\xx={1\over M^4}\sum_{\kk_+\not =\kk_-}e^{i(\kk_+-\kk_-)\xx}{C_+(\kk_+,
\kk_-)\over D_+(\kk_+-\kk_-)}\hat\psi^+_{\xx,+}\hat\psi^-_{\xx,+}\;,\cr
&T_{\xx,\o}^\n={1\over M^4}\sum_{\kk_+\not =\kk_-}
e^{i(\kk_+-\kk_-)\xx}{D_\o(\kk_+-
\kk_-)\over D_+(\kk_+-\kk_-)}\hat\psi^+_{\xx,\o}\hat\psi^-_{\xx,\o}\;,\cr}
\Eqa(wdelta1)$$
\midinsert
\*
\insertplotbm{300pt}{140pt}%
{\ins{33pt}{130pt}{$\hat \D^{2,1}_+$}
\ins{67pt}{80pt}{$=$}
\ins{105pt}{130pt}{$\n_+ D_+\hat G^{2,1}_+$}
\ins{155pt}{80pt}{$+$}
\ins{185pt}{130pt}{$\n_- D_-\hat G^{2,1}_-$}
\ins{235pt}{80pt}{$+$}
\ins{258pt}{130pt}{$\hat H^{2,1}_+$}
}%
{ward3}{}
\vskip.1truecm
\line{\vtop{\line{\hskip1.3truecm\vbox{\advance\hsize by -2.0 truecm
\0{\css \eqg(5).}
{\ottorm Graphical representation of the correction identity
\equ(A6.1.3b); the filled point in the last term represents
${\scriptstyle J_{\xx,+}(T_\xx-\sum_\o\n_\o T^\n_{\xx,\o})}$.}
\hfill} }}}
\*
\endinsert

The crucial point is that if $\n_\pm$ are suitably chosen, 
$\hat H_+^{2,1}$, when computed for momenta at the
cut-off scale, {\it is $O(\g^{\th h})$ smaller, with $0<\th<1$ a positive
constant, with respect to the first 
two addends of the r.h.s. of \equ(A6.1.3b)}.
In other words the correction identity \equ(A6.1.3b) says that the correction
term $\hat \D_+^{2,1}$, which is usually neglected in the physical
literature, can be written in terms
of the Schwinger functions $\hat G_+^{2,1}$ and $\hat G_-^{2,1}$ up to the
exponentially smaller term $\hat H_+^{2,1}$.

Inserting the correction identity \equ(A6.1.3b) 
in the Ward identity \equ(A6.2.12), we obtain the new identity
$$(1-\n_+)D_+(\pp)\hat G^{2,1}_{+}(\pp,\kk,\qq)-
\n_- D_-(\pp)\hat G^{2,1}_{-}(\pp,\kk,\qq)=\hat G^{2}_{+}(\qq)- \hat
G^{2}_{+}(\kk)+\hat H^{2,1}_{+}(\pp,\kk,\qq)\;.\Eqa(A6.2.12b)$$
In the same way one can show that the formal Ward identity $D_-(\pp)\hat
G^{2,1}_{-}(\pp,\kk,\qq)=\hat G^2_-(\qq)-\hat G^2_-(\kk)$ becomes, 
if the cutoffs are taken into account:
$$(1-\n'_-)D_-(\pp)\hat G^{2,1}_-(\pp,\kk,\qq)-
\n'_+ D_+(\pp)\hat G^{2,1}_+(\pp,\kk,\qq)=\hat
H^{2,1}_-(\pp,\kk,\qq)\;,\Eqa(A6.2.12c)$$
where, again $\n'_\pm=O(\l)$ and $H^{2,1}_-$ satisfies a
bound similar to that of $H^{2,1}_+$, when computed for momenta at the cutoff
scale.

The identities \equ(A6.2.12b) and \equ(A6.2.12c) allow us to write $\hat
G^{2,1}_{+}$ in terms of $\hat G^2_\pm$ and $\hat H_\pm^{2,1}$; one finds:
$$D_+(\pp)\hat G^{2,1}(\pp,\kk)=\Big[1-\n_+-{\n_-\n_+'\over 1-\n_-}\Big]^{-1}
\Big\{
\hat G^2_+(\qq)-\hat G^2_+(\kk)+{\n_-\over 1-\n_-'}\hat H^{2,1}_-(
\pp,\kk)+\hat H^{2,1}_+(
\pp,\kk)\Big\}\;,\Eqa(chir6)$$
In order to bound $\hat G^{2,1}$, we can use the dimensional bounds for 
the two and four legs Schwinger functions, 
easily proved by repeating for $\WW(\phi,J)$
an iterative construction similar to that exposed 
in Chapter 5 and performing the bounds as explained in \sec(5.5).
The expansion involves the definition of a more involved localization operator,
also acting on the kernels of the monomials involving the external fields
and can be found in many review papers [BGPS][GM][BM]; it is very similar
(and even simpler) to the expansion for $\WW_\D(\phi,J)$, that 
will be described in next section.

The result we need is the following.\\
\\
{\cs Theorem A6.1} {\it There exists $\e_0$ such that, if $\bar\l_h
\defin\max_{k> h}|\l_k|\le
\e_0$ and $|\bar \kk|=\g^h$, then
$$\hat G^{2,1}_\o(2\bar\kk,\bar\kk) =-{\g^{\h_{2,1}h}\over Z_h 
D_\o(\bar\kk)^2} [1 +O(\bar\l_h^2)]\;,\Eqa(A6.1.45)$$
where $\h_{2,1}(\l)$ is an exponent $O(\l)$ and
$$\hat G^{2}_\o(\bar\kk)={1\over Z_h D_\o(\bar\kk)}
[1 +O(\bar\l_h^2)]\;,\qquad\hat G^{4}_+(\bar\kk,-\bar\kk,-\bar\kk)=
Z_h^{-2} |\bar\kk|^{-4} [-\l_h +O(\bar\l_h^2)]\;.\Eqa(A6.1.47)$$ } \*
\\
{\it Remarks}
\\
\01 - The proof of Theorem A6.1 follows by a repetition of the estimates 
of Chapter 5. For some references: \equ(A6.1.45) follows
by the analysis in [BM2]; the first of \equ(A6.1.47) can be proven as
explained in [BGPS][GM]; the second \equ(A6.1.47) follows as 
a combination of the first of \equ(A6.1.47) and of the results
in Chapter 5.\\
\02 - 
A posteriori, it will result that $\h_{2,1}(\l)=0$. For the moment 
we just need \equ(A6.1.45) to bound $|\hat H^{2,1}_+|$
with a constant times $\g^{\th h}$ times the r.h.s. of \equ(A6.1.45),
as explained above, that is 
$$|\hat H^{2,1}_+(2\bar\kk,\bar\kk)|\le C\g^{\th h}{\g^{\h_{2,1}h}\over Z_h 
D_\o(\bar\kk)^2}\le C \g^{(\th/2) h}\g^{-2h}\Eqa(stimacca)$$
\\

Substituting the preceding bounds into \equ(chir6), 
we soon find that $\hat G^{2,1}(\bar\pp,\bar\kk)$,
with $|\bar\pp|=|\bar\kk|=\g^h$, can be bounded as
$$|\hat G^{2,1}_\o(\bar\pp,\bar\kk)|\le C {\g^{-2 h}\over Z_h}
\;,\Eqa(A6.1.45a)$$
where, as in Theorem A6.1, we assumed $\bar\l_h\le\e_0$.

Substituing the last bound into
the first addend of \equ(A6.2.11a), with the arguments set on scale $h$,
we soon find
$$\Big|\l \hat g_-(\kk_4) 
\hat G^2_-(\kk_3) \hat G^{2,1}_+(\kk_1-\kk_2,\kk_1,\kk_2)\Big|\le C |\l|\g^{-h}
{\g^{-h}\over Z_h}{\g^{-2h}\over Z_h}\;,\Eqa(chir7)$$
that is the ``right'' dimensional bound. In fact the l.h.s. of \equ(A6.2.11a)
can be 
bounded as in \equ(A6.1.47) so that, if we could neglet the second term
in the r.h.s. \equ(A6.2.11a), we would soon find $|\l_h|\le C |\l|$.

Aim of the next sections will be first to prove the correction 
identity \equ(A6.1.3b); then to describe a strategy which will allow
us to find a ``right'' dimensional bound also for the second term in
the r.h.s. \equ(A6.2.11a).\\
\\
\asub(A6.4){\bf The first correction identity}
\\
We start from the generating function \equ(wdelta) and we perform 
iteratively the integration of the $\psi$ variables, to be defined iteratively
in the following way. After the fields $\psi^{(0)},\cdots,\psi^{(j)}$ have been
integrated, we can write
$$e^{\WW_\Delta(\phi,J)}=e^{-M^2 E_j}
\int P_{\tilde Z_j,C_{h,j}} (d\psi^{[h,j]})
e^{-\VV^{(j)}(\sqrt{Z_j}\psi^{[h,j]})+K^{(j)}
(\sqrt{Z_j}\psi^{[h,j]},\phi,J)}\,,\Eqa(A6.1.11)$$
with $\VV^{(j)}(0)=0$, $Z_j=\max_\kk \tilde Z_j(\kk)$,

\01) $P_{\tilde Z_j,C_{h,j}}(d\psi^{[h,j]})$ is the {\it effective
Grassmannian measure at scale $j$}, equal to
$$\eqalign{
P_{\tilde Z_j,C_{h,j}}(d\psi^{[h,j]}) &= \prod_{\kk:C_{h,j}(\kk)>0}
\prod_{\o=\pm1}
{d\hat\psi^{[h,j])+}_{\kk,\o}d\hat\psi^{[h,j]-}_{\kk,\o}\over \NN_j(\kk)}
\cdot\cr &\cdot\; \exp \left\{-{1\over L\b} \sum_{\kk} \, C_{h,j}(\kk)
\tilde Z_j(\kk)\sum_{\o\pm1} \hat\psi^{[h,j]+}_{\o} D_\o(\kk)
\hat\psi^{[h,j]-}_{\kk,\o}\right\}\;,\cr} \Eqa(A6.1.12)$$
$$\NN_j(\kk)=(L\b)^{-1} C_{h,j}(\kk) \tilde Z_j(\kk)
[-k_0^2-k^2]^{1/2}\;,\Eqa(A6.1.13)$$
$$C_{h,j}(\kk)^{-1}=\sum_{r=h}^j f_r(\kk) \=\c_{h,j}(\kk) \virg
D_\o(\kk)=-ik_0+\o k\;;\Eqa(A6.1.14)$$

\02) the {\it effective potential on scale $j$}, $\VV^{(j)}(\psi)$,
is a sum of monomial of Grassmannian variables multiplied by
suitable kernels, as in \equ(5.6). The localization operator
acts on the kernels of $\VV^{(j)}(\psi)$ as described in \sec(5.2).
Note however that in the present case (\ie for the reference model)
the terms proportional to $F_\s$ and $F_\s$ are automatically venishing, 
by symmetry: only the terms proportional to $F_\l$ and $F_\z$ survive.

\03) the {\it effective source term at scale $j$},
$K^{(j)}(\sqrt{Z_j}\psi, \phi,J)$, is a sum of monomials of
Grassmannian variables and $\phi^\pm,J$ field, with at least one
$\phi^\pm$ or one $J$ field; we shall write it in the form
$$K^{(j)}(\sqrt{Z_j}\psi, \phi,J) = \BB_\phi^{(j)}(\sqrt{Z_j}\psi)+
K_J^{(j)}(\sqrt{Z_j}\psi) + W_R^{(j)}(\sqrt{Z_j}\psi,\phi,J)\;,\Eqa(A6.1.16)$$
where $\BB_\phi^{(j)}(\psi)$ and $K_J^{(j)}(\psi)$ denote the sums over
the terms containing only one $\phi$ or $J$ field, respectively.

Of course \equ(A6.1.11) is true for $j=0$, with
$$\eqalign{
&\tilde Z_0(\kk)=1,\qquad E_0=0,\qquad \VV^{(0)}(\psi)=V(\psi),\qquad
W_R^{(0)}=0,\cr
&\BB_\phi^{(0)}(\psi)=\sum_\o \int d\xx [
\phi^+_{\xx,\o}\psi^{-}_{\xx,\o}+ \psi^{+}_{\xx,\o}\phi^-_{\xx,\o}],\qquad
K_J^{(0)}(\psi)=\int d\xx
J_{\xx,+}\Big(T_\xx-\sum_\o\n_\o T^\n_{\xx,\o}\Big)\;.\cr}\Eqa(A6.1.17)$$
Let us now assume that \equ(A6.1.11) is satisfied for a certain $j\le 0$ and
let us show that it holds also with $j-1$ in place of $j$.

In order to perform the integration corresponding to $\psi^{(j)}$,
we write the effective potential and the effective source
as sum of two terms, according to the following rules.

We split the effective potential
$\VV^{(j)}$ as $\LL \VV^{(j)}+\RR \VV^{(j)}$, with 
$\LL$ acting on $\VV^{(j)}$ as explained in \sec(5.2).

Analogously we write $K^{(j)}=\LL K^{(j)}+\RR K^{(j)}$,
$\RR=1-\LL$, according to the following definition. First of all,
we put $\LL W_R^{(j)}=W_R^{(j)}$. 

Let us consider now 
$\BB_\phi^{(j)}(\sqrt{Z_j}\psi)$; we want to show that,
by a suitable choice of the localization procedure,
if $j\le -1$, it
can be written in the form
$$\eqalign{
&\qquad\qquad \BB_\phi^{(j)}(\sqrt{Z_j}\psi) =
\sum_\o \sum_{i=j+1}^0 \int d\xx d\yy\;\cdot\cr &\cdot\;\left[
\phi^+_{\xx,\o} g^{Q,(i)}_{\o}(\xx-\yy){\dpr\over \dpr\psi^+_{\yy\o}}
\VV^{(j)}(\sqrt{Z_j}\psi) + {\dpr\over \dpr\psi^-_{\yy,\o}}
\VV^{(j)}(\sqrt{Z_{j}}\psi) g^{Q,(i)}_{\o}(\yy-\xx)\phi^-_{\xx,\o} \right]+\cr
&+ \sum_\o\int {d\kk\over (2\p)^2} \left[ \hat\psi^{[h,j]+}_{\kk,\o}
\hat Q^{(j+1)}_{\o}(\kk)
\hat\phi_{\kk,\o}^- +\hat\phi^+_{\kk,\o} \hat Q^{(j+1)}_{\o}(\kk)
\hat\psi^{[h,j]-}_{\kk,\o} \right]\;,\cr}\Eqa(A6.1.28)$$
where $\hat g^{Q,(i)}_{\o}(\kk)= \hat g^{(i)}_{\o}(\kk) \hat
Q^{(i)}_{\o}(\kk)$, with
$$\hat g_\o^{(j)}(\kk)={1\over Z_{j-1}}{\tilde f_j(\kk) \over D_\o(\kk)}\;,
\Eqa(A6.1.35)$$
$\tilde f_j(\kk)=f_j(\kk) Z_{j-1} [\tilde Z_{j-1}(\kk)]^{-1}$ and
$Q^{(j)}_{\o}(\kk)$ defined inductively by the relations
$$\hat Q^{(j)}_{\o}(\kk)=\hat Q^{(j+1)}_{\o}(\kk) - z_j Z_j
D_\o(\kk) \sum_{i=j+1}^0 \hat g^{Q,(i)}_{\o}(\kk)\;,
\quad \hat Q^{(0)}_{\o}(\kk)=1\;.\Eqa(A6.1.29)$$
Note that $\hat g_\o^{(j)}(\kk)$ does not depend on the infrared
cutoff for $j>h$ and that (even for $j=h$) $\hat g^{(j)}(\kk)$ is
of size $Z_{j-1}^{-1}\g^{-j}$, see discussion in \S3 of [BM3],
after eq. (60). Moreover the propagator $\hat g^{Q,(i)}_{\o}(\kk)$
is equivalent to $\hat g^{(i)}_{\o}(\kk)$, as concerns the
dimensional bounds.

The $\LL$ operation for $\BB^{(j)}_\phi$
is defined by decomposing $\VV^{(j)}$ in the r.h.s. of \equ(A6.1.28)
as $\LL \VV^{(j)}+\RR \VV^{(j)}$.

Finally we have to define $\LL$ for $K_J^{(j)}(\sqrt{Z_j}\psi)$. 
It is easy to see that the field
$J$ is equivalent, from the point of view of dimensional
considerations, to two $\psi$ fields. Hence, the only terms which
need to be renormalized are those of second order in $\psi$, which
are indeed marginal; let us denote their sum with $K_J^{(j,2)}$. 
Let us start with defining the $\LL$ operation 
on $K_J^{(0)}$ as the identity. Let us now analyze the structure of 
$K_J^{(-1,2)}(\sqrt{Z_{-1}}\psi^{[h,-1]})$, as it appears after integrating 
the $\psi^{(0)}$ field and rescaling $\psi^{[h,-1]}$. We have 
$$\eqalign{&K_J^{(-1,2)}(\psi)={1\over Z_{-1}}\int d\xx J_{\xx,+}
\Big\{ T_\xx
+\sum_\o\int d\yy d\zz \big[F_{2,+,\o}^{(-1)}(\xx,\yy,\zz)
+F_{1,+}^{(-1)}(\xx,\yy,\zz)\d_{+,\o}\big]
\psi^+_{\yy,\o}\psi^-_{\zz,\o}\Big\}\cr}\Eqa(A6.132)$$
$F_{2,+,\o}^{(-1)}$ denotes the sum of all Feynman diagrams
containing a $T^\n_{\xx,\o}$ vertex or those obtained by contracting
both $\psi$ fields of a $T_\xx$ vertex (the index $\o$ refers to the 
$\o$ index of the two left--over external $\psi$ fields). 
$F_{1,+}^{(-1)}$ represents the sum over the diagrams
built by leaving external one of these fields of $T_\xx$. 

Now, if $\bS^{(0)}_\o$ is defined as in Appendix A7, it is easy
to see that the Fourier transform of $F^{(-1)}_{2,+,\o}$ can be 
written as 
$$\hat F^{(-1)}_{2,+,\o}(\kk_+,\kk_-)={\pp\over D_+(\pp)}\int
d\tilde\kk_+ \bS^{(0)}_+(\tilde\kk_+,\tilde\kk_+-\pp)G_{+,\o}^{(-1)}(
\tilde\kk_+,\kk_+,\kk_-)\;,\Eq(A6.133)$$
where $\pp=\kk_+-\kk_-$ and $G_{+,\o}^{(-1)}(
\tilde\kk_+,\kk_+,\kk_-)$ is of the form 
$$G_{+,\o}^{(-1)}(
\tilde\kk_+,\kk_+,\kk_-)=G_{0}(\tilde\kk_+,\kk_+,\kk_-)+G_1(\kk_+)
G_2(\kk_-)\d(\tilde\kk_+-\kk_+)\;,\Eqa(A6.133a)$$
where $G_0$ represents a suitable 
sum over connected graphs with four external lines, while $G_1$ and $G_2$
represent suitable 
sums over connected graphs with two external lines.

Using the symmetry of the propagator $D_\o(\kk)=i\o D_\o(\kk^*)$,
where, if $\kk=(k_0,k)$, $\kk^*=(k,-k_0)$, one easily gets the following
symmetry
properties for the functions appearing in \equ(A6.133):
$$G_{+,\o}^{(-1)}(
\tilde\kk_+,\kk_+,\kk_-)=-\o G_{+,\o}^{(-1)}(
\tilde\kk_+^*,\kk_+^*,\kk_-^*)\;,\qquad\pp\cdot\bS^{(0)}_+(\kk_+,\kk_-)=-i
\pp^*\cdot\bS^{(0)}_+(\kk_+^*,\kk_-^*)\;.\Eqa(A6.136)$$
The last equation implies that 
$$\hat F^{(-1)}_{2,+,\o}(\kk_+,\kk_-)={1\over D_+(\pp)}
[p_0 \hat A_{+,\o,0}^{(-1)}(\kk_+,\kk_-)+
p \hat A_{+,\o,1}^{(-1)}(\kk_+,\kk_-)]\;,\Eqa(A6.137)$$
where $\hat A_{+,\o,i}^{(-1)}$ are smooth functions satisfying
$$\hat A_{+,\o,1}^{(-1)}(\kk_+,\kk_-)=i\o \hat A_{+,\o,1}^{(-1)}
(\kk_+^*,\kk_-^*)\;.\Eqa(A6.139)$$
It follows that, if we define 
$$\LL\hat F^{(-1)}_{2,+,\o}(\kk_+,\kk_-)={1\over D_+(\pp)}
[p_0 \hat A_{+,\o,0}^{(-1)}(\V0,\V0)+
p \hat A_{+,\o,1}^{(-1)}(\V0,\V0)]\;,\Eqa(A6.140)$$
then
$$\LL\hat F^{(-1)}_{2,+,+}(\kk_+,\kk_-)=\n_{-1}^+\;,\qquad
\LL\hat F^{(-1)}_{2,+,-}(\kk_+,\kk_-)=\n_{-1}^-{D_-(\pp)\over D_+(\pp)}\;,
\Eqa(A6.141)$$
where $\n_{-1}^\pm$ are real constants, as it can be verified by symmetry.

We now consider the contribution $F^{(-1)}_{1,+}$ in \equ(A6.132). Its 
Fourier transform has two contributions, the first of the form
$$\hat F^{(-1)}_{1,+}(\kk_+,\kk_-)={[C_{h,0}(\kk_-)-1]D_+(\kk_-)
\hat g_+^{(0)}(\kk_+)-u_0(\kk_+)\over D_+(\pp)}G^{(2)}_+(\kk_+)\;,
\Eqa(A6.145)$$
where $u_0$ is defined in Appendix A7, see \equ(A7.4.8),
and the second possible contribution has the same form of \equ(A6.145),
with $\kk_+$ and $\kk_-$ interchanged. The natural way
to regularize \equ(A6.145) is to define
$$\LL\hat F^{(-1)}_{1,+}(\kk_+,\kk_-)={[C_{h,0}(\kk_-)-1]D_+(\kk_-)
\hat g_+^{(0)}(\kk_+)-u_0(\kk_+)\over D_+(\pp)}G^{(2)}_+(\V0)\;.\Eqa(A6.146)$$
Note however that $G_\o^{(2)}(\V0)=0$, by parity, so the local part
in \equ(A6.146) is vanishing. In other words the dimensional
gain here is obtained without the introduction of a renormalization constant.
The same procedure can be defined for the term obtained by interchanging 
$\kk_+$ and $\kk_-$ in \equ(A6.145).

We can summarize the previous discussion by defining 
$$\eqalign{&\LL K_J^{(-1,2)}(\psi)={1\over Z_{-1}}\int d\xx \Big\{J_{\xx,+}
\Big[ T_\xx + \n_{-1}^+\psi^+_{\xx,+}\psi^-_{\xx,+}\Big]
+\n_{-1}^- J_{\xx,+}^{(-)}\psi^+_{\xx,-}\psi^-_{\xx,-}\Big\}\cr}\Eqa(A6.132)$$
where, if $\hat J_{\pp,+}$ is the Fourier transform of $J_{\xx,+}$,
$J_{\xx,+}^{(-)}$ is the Fourier transform of $\hat J_{\pp,+}
D_-(\pp)/D_+(\pp)$.

We are now ready to describe the general step, by defining the action 
of $\LL$ over $K_J^{j,2}$, which can be written, if $j<1$, after rescaling 
$\psi^{[h,j]}$, as
$$\eqalign{&K_J^{(j,2)}(\psi)={1\over Z_{j}}\int d\xx \Big\{J_{\xx,+}
T_\xx+\sum_\o\int d\yy d\zz\Big[
J_{\xx,+}F^{(j)}_{\n^+,+,\o}(\xx,\yy,\zz)+
J_{\xx,+}^{(-)}F^{(j)}_{\n^-,+,\o}(\xx,\yy,\zz)+\cr
&+J_{\xx,+}F^{(j)}_{2,+,\o}(\xx,\yy,\zz)+\d_{+,\o}
J_{\xx,+}F^{(j)}_{1,\o}(\xx,\yy,\zz)\Big]\psi^+_{\yy,\o}\psi^-_{\zz,\o}
\Big\}\cr}\Eqa(A6.149)$$
where $F^{(j)}_{\n^\pm,+,\o}(\xx,\yy,\zz)$ represent the sum 
over all graphs with one vertex of type $\n^\pm$ and two $\psi$ 
external fields of type $\o$, $F^{(j)}_{2,+,\o}$ is the sum
over the same kind of graphs with one vertex $T_\xx$, whose $\psi$ 
fields are both contracted and $F^{(j)}_{1,\o}$ is the sum over 
the graphs with one vertex $T_\xx$, such that one of its fields is 
external. 

It is important to stress that, thanks to the support properties
of $C_\o(\kk_+,\kk_-)$, given a graph contributing to $F^{(j)}_{2,+,\o}$,
at least one of the $\psi$ fields belonging to $T_\xx$ is contracted 
on scale $0$. This property will give crucial dimensional gains (through
the short memory property) for the contributions to the Beta function 
for $\n^\pm$ coming from $F^{(j)}_{2,+,\o}$. It is clear that, because
of this property, $F^{(j)}_{2,+,\o}$ can be rewritten as
$$\hat F^{(j)}_{2,+,\o}(\kk_+,\kk_-)={\pp\over D_+(\pp)}\sum_{i=j}^0\int
d\tilde\kk_+ \tilde\bS^{(j)}_+(\tilde\kk_+,\tilde\kk_+-\pp)G_{+,\o}^{(j)}(
\tilde\kk_+,\kk_+,\kk_-)\;,\Eqa(A6.150)$$
for suitable functions $\tilde\bS^{(j)}_+$ and $G_{+,\o}^{(j)}$
satisfying the same symmetry properties of \equ(A6.136). Then,
the action of $\LL$ over $\hat F^{(j)}_{2,+,\o}(\kk_+,\kk_-)$
is defined exactly as for $j=-1$. Moreover we define
$$\LL\hat F^{(j)}_{\n^\pm,+,\o}(\kk_+,\kk_-)=
\hat F^{(j)}_{\n^\pm,+,\o}(\V0,\V0)\;,\Eqa(A6.151)$$
and, finally, we note that, for the same reasons as in the $j=-1$ case
$$\hat F^{(j)}_{\n^+,+,-}(\V0,\V0)=\hat F^{(j)}_{\n^-,+,+}(\V0,\V0)=
\hat F^{(j)}_{1,+}(\V0,\V0)=0\;,\Eqa(A6.152)$$
so that the corresponding kernels are automatically regularized,
without need of defining any non trivial action of $\LL$.

It follows that we can write
$$\eqalign{&\LL K_J^{(j,2)}(\psi)={1\over Z_{j}}\int d\xx \Big\{J_{\xx,+}
\Big[ T_\xx + \n_{j}^+\psi^+_{\xx,+}\psi^-_{\xx,+}\Big]
+\n_{j}^- J_{\xx,+}^{(-)}\psi^+_{\xx,-}\psi^-_{\xx,-}\Big\}\cr}\Eqa(A6.153)$$
which defines the renormalization constants $\n_j^\pm$. Note that 
$\n_j^\pm$ is built by contribution that either contain another 
constant $\n_k^\pm$, $k>j$, or contain a $T$ vertex, which is on scale 
$0$, in the sense explained before \equ(A6.150).\\
\\
After writing $\VV^{(j)}=\LL \VV^{(j)}+\RR\VV^{(j)}$
and $K^{(j)}=\LL K^{(j)}+\RR K^{(j)}$, the next step is
to {\sl renormalize} the free measure $P_{\tilde Z_j,C_{h,j}}
(d\psi^{[h,j]})$, by adding to it part of $\LL\VV^{(j)}$. We get
$$\eqalign{
& \int P_{\tilde Z_j,C_{h,j}}(d\psi^{[h,j]}) \, e^{-\VV^{(j)}
(\sqrt{Z_j}\psi^{[h,j]})+K^{(j)}
(\sqrt{Z_j}\psi^{[h,j]})}=\cr
&\quad =e^{-M^2 t_j}\int P_{\tilde Z_{j-1},C_{h,j}}(d\psi^{[h,j]}) \,
e^{-\tilde\VV^{(j)}(\sqrt{Z_j}\psi^{[h,j]})+\tilde K^{(j)}
(\sqrt{Z_j}\psi^{[h,j]})}\;,\cr}\Eqa(A6.1.30)$$
where
$$\eqalign{&\tilde Z_{j-1}(\kk)=Z_j[1+\c_{h,j}(\kk) z_j]\;,\cr
&\tilde\VV^{(j)}(\sqrt{Z_j}\psi^{[h,j]}) = \VV^{(j)}(\sqrt{Z_j}\psi^{[h,j]})
-z_j Z_j F_\z^{[h,j]}\,,\cr}\Eqa(A6.1.32)$$
where $F_\z^{[h,j]}$ is defined as in \equ(5.11) 
and the factor $\exp(-M^2 t_j)$ in \equ(A6.1.30) takes into account 
the different normalization of the two measures. Moreover
$$\tilde K^{(j)}(\sqrt{Z_j}\psi^{[h,j]})=
\tilde\BB^{(j)}_\phi(\sqrt{Z_j}\psi^{[h,j]})
+K^{(j)}_J(\sqrt{Z_j}\psi^{[h,j]})+W^{(j)}_R\;,\Eqa(A6.1.33)$$
where $\tilde\BB^{(j)}_\phi$ is obtained from $\BB^{(j)}_\phi$ by inserting
\equ(A6.1.32) in the second line of \equ(A6.1.28) and by absorbing the terms
proportional to $z_j$ in the terms in the third line of \equ(A6.1.28).

If $j>h$, the r.h.s of \equ(A6.1.30) can be written as
$$e^{-M^2 t_j} \int P_{\tilde Z_{j-1},C_{h,j-1}} (d\psi^{[h,j-1]})
\int P_{Z_{j-1},\tilde f_j^{-1}}(d\psi^{(j)}) 
e^{-\tilde \VV^{(j)}\big(\sqrt{Z_j}[\psi^{[h,j-1]} + \psi^{(j)}]\big)+
\tilde K^{(j)} \big(\sqrt{Z_j}[\psi^{[h,j-1]} + \psi^{(j)}]
\big)}\;,\Eqa(A6.1.34)$$
where $P_{Z_{j-1},\tilde f_j^{-1}}(d\psi^{(j)})$ is the
integration with propagator $\hat g_\o^{(j)}(\kk)$.

We now {\it rescale} the field so that
$$\tilde\VV^{(j)}(\sqrt{Z_j}\psi^{[h,j]})= \hat\VV^{(j)}(\sqrt{Z_{j-1}}
\psi^{[h,j]})\quad,\quad \tilde K^{(j)}(\sqrt{Z_j}\psi^{[h,j]})=
\hat K^{(j)}(\sqrt{Z_{j-1}}
\psi^{[h,j]})\;;\Eqa(A6.1.36)$$
it follows that
$$\LL\hat\VV^{(j)}(\psi^{[h,j]})=\l_j F_\l^{[h,j]}\;,
\Eqa(A6.1.37)$$
where $\l_j=(Z_j Z_{j-1}^{-1})^2 l_j$. If we now define
$$\eqalign{
&e^{-\VV^{(j-1)}\big(\sqrt{Z_{j-1}} \psi^{[h,j-1]} \big)+
K^{(j-1)}\big(\sqrt{Z_{j-1}}\psi^{[h,j-1]}\big)-M^2 E_j}=\cr
&=\int P_{Z_{j-1},\tilde f_j^{-1}}(d\psi^{(j)}) \, e^{-\hat
\VV^{(j)} \big(\sqrt{Z_{j-1}}[\psi^{[h,j-1]} + \psi^{(j)}]\big)+
\hat K^{(j)} \big(\sqrt{Z_{j-1}}[\psi^{[h,j-1]} + \psi^{(j)}]
\big)}\;,\cr}\Eqa(A6.1.38)$$
it is easy to see that $\VV^{(j-1)}$ and $K^{(j-1)}$ are of the
same form of $\VV^{(j)}$ and $K^{(j)}$ and that the procedure can
be iterated. Note that the above procedure
allows, in particular, to write the running coupling constant $\l_{j-1}$,
$0> j-1\ge h$, in terms of $\l_{j'}$, $0\ge j'\ge j$:
$$\l_{j-1}=\l_j+ \b_\l^j( \l_{j},\ldots,\l_0) \virg \l_0=\l,
\Eqa(A6.1.39)$$
and the renormalization constants $\n_{j-1}^\pm$ in terms of $\l_k,\n^\pm_k$,
$k\ge j$:
$$\n_{j-1}^\a=\n_j^\a +\b_j^{\n,\a}(\l_j,\n_j^\pm;\ldots;
\l_0,\n_0^\pm)\virg \n_0^\a=\n_\a \virg \a=\pm\;.\Eqa(A6.1.39a)$$
The functions $\b_j^\l(\l_{j+1},\ldots,\l_0)$ 
and $\b_j^{\n,\a}(\l_j,\n_j^\pm;\ldots;
\l_0,\n_0^\pm)$ are called
the $\l$ and the $\n^\a$ component of 
the {\it Beta function}, respectively. 
Both functions can be represented by a tree expansion 
similar to that exposed in Chapter 5, and we do not repeat here the details.

We now want to 
show that, if $\bar\l_h\le \e_0$ is small enough, it is possible to choose
$\n^\pm$ as suitable functions of $\l$, in such a way
that $|\n_j^\pm|\le C\e_0 \g^{\th j}$, for some $\th>0$. If we manage
to prove this, it will soon follow that $\hat H^{2,1}_+$, when computed 
on the IR cutoff scale, can be bounded by \equ(stimacca), that 
is by the dimensional bound for $\hat G^{2,1}$ times an exponentially
small factor $\g^{\th h}$. In fact the renormalized expansion 
for $\hat H^{2,1}_+$ contain contributes that either contain a $T_\xx$
vertex on scale $0$, or a $\n^\pm_j$, with $h<j\le 0$. If 
$|\n_j^\pm|\le C\e_0 \g^{\th j}$, using the short memory property, it is 
immediate to verify that both contributes are exponentially
small w.r.t. the dimensional bound for $\hat G^{2,1}$.

So, let us prove the bound on the renormalization constants $\n_j^\pm$.
We rewrite $\b_j^{\n,\a}$ by distinguishing the contributions
independent of $n_k^\pm$ (which necessarely contain a $T$ vertex on 
scale 0) and the contribution linear in $\n_k^\pm$, $k\ge j$:
$$\b_j^{\n,\a}(\l_j, \n_{j}^\pm;
\ldots; \l_0,\n_{0}^\pm) = \b_{j,1}^{\n,\a}(\l_j, \ldots, \l_0) 
+ \sum_{k=j}^0\sum_{\o=\pm } \n_{k}^\o \b_{j,k}^{\n,\a,\o}
(\l_j, \ldots, \l_0) \;.\Eqa(A6.3.32)$$
Moreover, by the short memory property, there exists
$0<\th<1/4$ and positive constants $c_1$ and $c_2$ such that
$$|\b_{j,1}^{\n,\a}(\l_j, \ldots, \l_0)| \le c_1
\bar\l_h \g^{2\th j} \virg |\b_{j,k}^{\n,\a,\o}
(\l_j, \ldots, \l_0)| \le c_2 \bar\l_h^2 \g^{2\th(j-j')}\;.\Eqa(A6.3.33)$$
By iterating \equ(A6.1.39a), we find:
$$\n_{j-1}^\a=\n_0^\a+\sum_{k=j}^{0} \b_k^{\n,\a}(\l_{k}, \n_{k}^\pm;
\ldots; \l_0,\n_{0}^\pm)\;,\Eqa(A6.3.33a)$$
so that, imposing the condition $\n_h^\pm\=0$, we get:
$$\n_{0}^\a=-\sum_{k=h+1}^{0} \b_k^{\n,\a}(\l_{k}, \n_{k}^\pm;
\ldots; \l_0,\n_{0}^\pm)\;.\Eqa(A6.3.33b)$$
Inserting the last equation into \equ(A6.3.33a) we get:
$$\n_j^\a = -\sum_{k=h+1}^j \b_k^{\n,\a}(\l_{k},
\n_{k}^\pm;\ldots;\l_0,\n_0^\pm)\;.\Eqa(A6.3.34)$$
In other words, the condition $\n_h^\pm\=0$ can be satisfied iff 
it can be found a sequence $\un=\{\n_j^\o\}_{h\le j\le 0}^{\o=\pm}$
satisfying \equ(A6.3.34). In order to prove that this is possible,
we introduce the space $\MMM_\th$ of the
sequences $\un$ such that $\max_\o|\n_j^\o|\le c
\bar\l_h\g^{\th j}$, for some $c$; we shall think $\MMM_\th$ as a
Banach space with norm $||\un||_\th = \sup_{h+1\le j\le
0}\max_\o|\n_j^\o|\g^{-\th j} \bar\l_h^{-1}$. We then look for a fixed point
of the operator $\bT:\MMM_{\th}\to\MMM_{\th}$ defined as:
$$(\bT\un)_j^\a = -\sum_{k=h+1}^j \b_k^{\n,\a}(\l_{k},
\n_{k}^\pm;\ldots;\l_0,\n_0^\pm)\;.\Eqa(A6.3.35)$$
Note that, if $\bar\l_h$ is sufficiently small, then $\bT$ leaves
invariant the ball $\BBB_\th$ of radius $c_0=2c_1\sum_{n=0}^\io
\g^{-\th n}$ of $\MMM_\th$, $c_1$ being the constant in \equ(A6.3.33). In
fact, by
\equ(A6.3.32) and \equ(A6.3.33), if $||\un||_\th \le c_0$, then
$$|(\bT\un)_j^\a| \le \sum_{k=h+1}^j
c_1 \bar\l_h \g^{2\th k} + 2\sum_{k=h+1}^j \sum_{i=k}^0 c_0
\bar\l_h \g^{\th i} c_2 \bar\l_h^2 \g^{2\th (k-i)} \le c_0
\bar\l_h \g^{\th j}\;,\Eqa(A6.3.35a)$$
if $4c_2 \bar\l_h^2 (\sum_{n=0}^\io \g^{-\th n})^3 \le 1$.

$\bT$ is a also a contraction on $\BBB_\th$, if $\bar\l_h$ is
sufficiently small; in fact, if $\un,\un' \in \MMM_\th$,
$$\eqalign{
&|(\bT \n)_j^\a-(\bT\un')_j^\a|\le \sum_{k=h+1}^j |
\b_k^{\n,\a}(\l_{k}, \n_{k}^\pm;\ldots;\l_0,\n_0^\pm)
-\b_k^{\n,\a}(\l_{k}, {\n'}_{k}^\pm;\ldots;\l_0,{\n'}_0^\pm)
\le\cr
&\le 2\sum_{k=h+1}^j \sum_{i=k}^0  ||\un-\un'||_\th \bar\l_h
\g^{\th i} c_2 \bar\l_h^2 \g^{2\th (k-i)} \le {1\over 2}
||\un-\un'||_\th \bar\l_h \g^{\th j}\;, \cr }\Eqa(A6.3.36)$$
if $4c_2 \bar\l_h^2 (\sum_{n=0}^\io \g^{-\th n})^3 \le 1$, as above.
Hence, by the contraction principle, there is a unique fixed point $\un^*$
of $\bT$ on $\BBB_\th$. This concludes the proof of the exponential 
decay of $\n_j^\pm$ and, as discussed above, of the first correction 
identity.\Halmos
\\
\asub(2.3j){\bf Ward identities and the second addend in {\equ(A6.2.11a)}}
\\
Starting from the present section, we begin to deal with the 
second term in the r.h.s. of \equ(A6.2.11a), with the aim of showing
that it admits a good dimensional bound, as discussed for the first one.
The vanishing of the $\l$ component of the Beta function will
be an easy consequence of such a good dimensional bound. 

The strategy will be the following: in the present section we will
first describe two more Ward identities 
connecting $\hat G^{4,1}$ with $\hat G^4$.
As the first Ward identity considered above, they will have a correction
due to the cutoff function, and these corrections will satisfy new 
correction identities, presented below in this section. 
The proof of the new correction identities, which is the main difficulty of 
all the proof of the present Appendix, will be presented in next section.
The present section will be concluded
with the proof of the vanishing of the Beta function, obtained by a
careful use of the new Ward 
identities together with the new corretion identities.\\

The new pair of Ward identities we need here is the following, see Fig. 
\graf(6).
$$\eqalign{
&D_+(\pp)\hat G^{4,1}_{+}(\pp,\kk_1,\kk_2,\kk_3,\kk_4-\pp)
=\hat G^{4}_{+}(\kk_1-\pp,\kk_2,\kk_3,\kk_4-\pp)-
\hat G^{4}_{+}(\kk_1,\kk_2+\pp,\kk_3,\kk_4-\pp)+\D^{4,1}_{+}\;,\cr
&D_-(\pp)\hat G^{4,1}_{-}(\pp,\kk_1,\kk_2,\kk_3,\kk_4-\pp)
=\hat G^{4}_{+}(\kk_1,\kk_2,\kk_3-\pp,\kk_4-\pp)-
\hat G^{4}_{+}(\kk_1,\kk_2,\kk_3,\kk_4)+\hat \D^{4,1}_{-}\;,\cr}
\Eqa(A6.2.14a)$$
where $\hat \D^{4,1}_\pm$ are the ``correction terms''
$$\hat \D^{4,1}_{\pm }(\pp,\kk_1,\kk_2,\kk_3)={1\over M^2}\sum_\kk
C_\pm(\kk,\kk-\pp) <\hat\psi^+_{\kk,\pm}\hat\psi^-_{\kk-\pp,\pm};
\hat\psi^-_{\kk_1,+};\hat\psi^+_{\kk_2,+}; \hat\psi^-_{\kk_3,-};
\hat\psi^+_{\kk_4-\pp,-}>^T\;.\Eqa(A6.2.15)$$
\midinsert
\*
\insertplotbm{300pt}{150pt}%
{\ins{53pt}{83pt}{$\hat G^{4,1}_+$}
\ins{5pt}{83pt}{$D_+(\pp)$}
\ins{15pt}{110pt}{$\kk_1 +$}
\ins{40pt}{120pt}{$\kk_2 +$}
\ins{66pt}{120pt}{$\kk_3 -$}
\ins{75pt}{110pt}{$\kk_4-\pp$}
\ins{90pt}{100pt}{$-$}
\ins{68pt}{40pt}{$\pp$}
\ins{97pt}{80pt}{$=$}
\ins{125pt}{81pt}{$\hat G^{4}_+$}
\ins{110pt}{115pt}{$\kk_1-\pp$}
\ins{120pt}{102pt}{$+$}
\ins{150pt}{115pt}{$\kk_2$}
\ins{145pt}{95pt}{$+$}
\ins{140pt}{50pt}{$\kk_3$}
\ins{145pt}{60pt}{$-$}
\ins{99pt}{45pt}{$\kk_4-\pp$}
\ins{112pt}{63pt}{$-$}
\ins{165pt}{80pt}{$-$}
\ins{195pt}{81pt}{$\hat G^{4}_+$}
\ins{175pt}{115pt}{$\kk_1$}
\ins{190pt}{102pt}{$+$}
\ins{200pt}{115pt}{$\kk_2+\pp$}
\ins{215pt}{95pt}{$+$}
\ins{210pt}{50pt}{$\kk_3$}
\ins{215pt}{60pt}{$-$}
\ins{175pt}{45pt}{$\kk_4-\pp$}
\ins{180pt}{60pt}{$-$}
\ins{235pt}{80pt}{$+$}
\ins{268pt}{83pt}{$\hat \D^{4,1}_+$}
\ins{235pt}{110pt}{$\kk_1 +$}
\ins{255pt}{120pt}{$\kk_2 +$}
\ins{281pt}{120pt}{$\kk_3 -$}
\ins{292pt}{110pt}{$\kk_4-\pp$}
\ins{303pt}{100pt}{$-$}
\ins{283pt}{40pt}{$\pp$}
\ins{100pt}{20pt}{$0=\kk_1+\kk_3-\kk_2-\kk_4$}
}%
{ward2}{}
\vskip.1truecm
\line{\vtop{\line{\hskip1.3truecm\vbox{\advance\hsize by -2.0 truecm
\0{\css \eqg(6).}
{\ottorm Graphical representation of the Ward identity
\equ(A6.2.14)}
\hfill} }}}
\*
\endinsert
The Ward identities \equ(A6.2.14a) can be derived from \equ(chir4)
by deriving four times w.r.t. the external $\phi$ fields. 
By adding and subtracting suitable
counterterms \footnote{${}^1$}{\nota 
with an abuse of notation, here we call the counterterms
with the same simbols as those used for the first correction identity;
note that here the new counterterms are different from those of previous 
section.}
$\n_\pm$, the first of \equ(A6.2.14a) can be rewritten as
$$\eqalign{
&(1-\n_+) D_{+}(\pp) \hat G_{+}^{4,1}
(\pp,\kk_1,\kk_2,\kk_3,\kk_4-\pp)-\n_{-} D_{-}(\pp) \hat
G_{-}^{4,1}(\pp,\kk_1,\kk_2,\kk_3,\kk_4-\pp)\cr &= \hat
G_{+}^{4}(\kk_1-\pp,\kk_2,\kk_3,\kk_4-\pp)- \hat
G_{+}^{4}(\kk_1,\kk_2+\pp,\kk_3,\kk_4-\pp) +\hat 
H^{4,1}_{+}(\pp,\kk_1,\kk_2,\kk_3,\kk_4-\pp)\;,\cr}\Eqa(A6.2.16)$$
where by definition
$$\eqalign{&\hat H^{4,1}_{+}(\pp,\kk_1,\kk_2,\kk_3,\kk_4-\pp)=
{1\over M^2}\sum_\kk C_+(\kk,\kk-\pp)
<\hat\psi^+_{\kk,+}\hat\psi^-_{\kk-\pp,+};
\hat\psi^-_{\kk_1,+};\hat\psi^+_{\kk_2,+}; \hat\psi^-_{\kk_3,-};
\hat\psi^+_{\kk_4-\pp,-}>^T -\cr
&- {1\over M^2}\sum_\kk\sum_\o\nu_{\o} D_\o(\pp)
<\hat\psi^+_{\kk,\o}\hat\psi^-_{\kk-\pp,\o};
\hat\psi^-_{\kk_1,+};\hat\psi^+_{\kk_2,+}; \hat\psi^-_{\kk_3,-};
\hat\psi^+_{\kk_4-\pp,-}>^T\;.\cr}\Eqa(A6.2.16a)$$
In the same way, in terms of new counterterms $\n'_\pm$,
the second of \equ(A6.2.14a) can be written as
$$\eqalign{
& (1-\n'_-)D_{-}(\pp) \hat G_{-}^{4,1}
(\pp,\kk_1,\kk_2,\kk_3,\kk_4-\pp)-\n'_{+} D_{+}(\pp) \hat
G_{+}^{4,1}(\pp,\kk_1,\kk_2,\kk_3,\kk_4-\pp) =\cr
&= \hat
G_{+}^{4}(\kk_1,\kk_2,\kk_3-\pp,\kk_4-\pp)- \hat
G_{+}^{4}(\kk_1,\kk_2,\kk_3,\kk_4) + \hat
H^{4,1}_{-}(\pp,\kk_1,\kk_2,\kk_3,\kk_4-\pp)\;,\cr}\Eqa(A6.2.17)$$
where
$$\eqalign{&
\hat H^{4,1}_{-}(\pp,\kk_1,\kk_2,\kk_3,\kk_4-\pp)={1\over M^2}\sum_\kk
C_-(\kk,\kk-\pp) <\hat\psi^+_{\kk,-}\hat\psi^-_{\kk-\pp,-};
\hat\psi^-_{\kk_1,+};\hat\psi^+_{\kk_2,+}; \hat\psi^-_{\kk_3,-};
\hat\psi^+_{\kk_4-\pp,-}>^T-\cr
&-{1\over M^2}\sum_\kk\sum_\o\nu'_{\o} D_\o(\pp)
<\hat\psi^+_{\kk,\o}\hat\psi^-_{\kk-\pp,\o};
\hat\psi^-_{\kk_1,+};\hat\psi^+_{\kk_2,+}; \hat\psi^-_{\kk_3,-};
\hat\psi^+_{\kk_4-\pp,-}>^T\;.\cr}\Eqa(A6.2.17a)$$
If we insert in the r.h.s. of \equ(A6.2.16) the value of $\hat
G_-^{4,1}$ taken from \equ(A6.2.17), we get
$$\eqalign{
&(1+A) D_{+}(\pp) \hat G_{+}^{4,1}
(\pp,\kk_1,\kk_2,\kk_3,\kk_4-\pp)= \hat
G_{+}^{4}(\kk_1-\pp,\kk_2,\kk_3,\kk_4-\pp)- \cr 
&-\hat G_{+}^{4}(\kk_1,\kk_2+\pp,\kk_3,\kk_4-\pp) + B\Big[\hat
G_{+}^{4}(\kk_1,\kk_3-\pp,\kk_4-\pp)- \hat
G_{+}^{4}(\kk_1,\kk_2,\kk_3,\kk_4)\Big] +\hat H^{4,1}_{+}+ B \hat
H^{4,1}_{-}\;,\cr}\Eqa(A6.2.18)$$
where
$$A=-\n_+-{\n_{-}\n'_{+}\over 1-\n'_{-}} \virg B={\n_{-}\over
1-\n'_-}\;.\Eqa(A6.2.19)$$

Let us now consider the second term in the r.h.s. of \equ(A6.2.11a)
and let us rewrite it as:
$$\l \hat g_-(\kk_4)\Big[{1\over M^2}\sum_\pp \c_M(\pp)\hat G^{4,1}_+(\pp;
\kk_1,\kk_2,\kk_3,\kk_4-\pp)+
{1\over M^2}\sum_\pp \tilde\c_M(\pp)\hat G^{4,1}_+(\pp;
\kk_1,\kk_2,\kk_3,\kk_4-\pp)\Big]\;,\Eqa(A6.2.17aa)$$
where: $\c_M(\pp)$ is a cutoff function vanishing for scales bigger than 
$h+\log_\g 2$ (\ie the presence of $\c_M(\pp)$ constraints
the transferred momentum to be $\le O(\g^h)$); 
$\tilde\c_M(\pp)$ is a cutoff function vanishing for scales 
smaller than $h+\log_\g 2$ and bigger than $\log_\g 2$ (\ie 
the presence of $\tilde\c_M(\pp)$ constraints
the transferred momentum to be $O(\g^h)\le |\pp|\le O(1)$). 
The two functions
are chosen so that they sum up to 1 in the scales range between $h$ and $0$. 

If we insert in the last term of \equ(A6.2.17a) the
value of $\hat G_{+}^{4,1}$ taken from \equ(A6.2.18), we get
$$\eqalign{
&\l \hat g_-(\kk_4){1\over M^2}\sum_\pp\chi_M(\pp)
G^{4,1}_+(\pp;\kk_1,\kk_2,\kk_3,\kk_4-\pp) +\cr
&+ {\l \hat g_-(\kk_4)\over (1+A)} {1\over M^2} \sum_\pp
\tilde\chi_M(\pp) {\hat G_+^{4}(\kk_1-\pp,\kk_2,\kk_3,\kk_4-\pp)-
\hat G_+^{4}(\kk_1,\kk_2+\pp,\kk_3,\kk_4-\pp)\over D_+(\pp)}\;+\cr
&+{\l \hat g_-(\kk_4)\over (1+A)} {1\over M^2} \sum_\pp
\tilde\chi_M(\pp) {\hat G_+^{4}(\kk_1,\kk_2,\kk_3-\pp,\kk_4-\pp)-
\hat G_+^{4}(\kk_1,\kk_2,\kk_3,\kk_4)\over D_+(\pp)}\;+ \cr
&+{\l \hat g_-(\kk_4)\over (1+A)} {1\over M^2} \sum_\pp
\tilde\chi_M(\pp) {\hat
H^{4,1}_{+}(\pp;\kk_1,\kk_2,\kk_3,\kk_4-\pp)+ B \hat
H^{4,1}_{-}(\pp;\kk_1,\kk_2,\kk_3,\kk_4-\pp)\over D_+(\pp)}
\;.\cr}\Eqa(A6.2.20)$$
Note that
$${1\over M^2} \sum_\pp \tilde\chi_M(\pp) {\hat
G_+^{4}(\kk_1,\kk_2,\kk_3,\kk_4)\over D_+(\pp)}=0\;,\Eqa(A6.2.21)$$
since $D_+(\pp)$ is odd. Then, \equ(A6.2.20)
computed with the momenta equal to
$$\kk_i=\bar\kk_i \virg \bar\kk_1= \bar\kk_4=
-\bar\kk_2= -\bar\kk_3= \bar\kk \virg
|\bar\kk|=\g^h\;,\Eqa(A6.2.21a)$$
is equivalent to
$$\eqalign{
&\l\hat g_-(\bar\kk_4){1\over M^2}\sum_\pp\chi_M(\pp)
G^{4,1}_+(\pp;\bar\kk_1,\bar\kk_2,\bar\kk_3,\bar\kk_4-\pp)+ \cr
&+ {\l \hat g_-(\bar\kk_4)\over (1+A)} {1\over M^2} \sum_\pp
\tilde\chi_M(\pp) {\hat
G_+^{4}(\bar\kk_1-\pp,\bar\kk_2,\bar\kk_3,\bar\kk_4-\pp)- \hat
G_+^{4}(\bar\kk_1,\bar\kk_2+\pp,\bar\kk_3,\bar\kk_4-\pp)\over
D_+(\pp)}\;+\cr
&+{\l \hat g_-(\bar\kk_4)\over (1+A)} {1\over M^2} \sum_\pp
\tilde\chi_M(\pp) {\hat
G_+^{4}(\bar\kk_1,\bar\kk_2,\bar\kk_3-\pp,\bar\kk_4-\pp)\over
D_+(\pp)}\;+ \cr
&+{\l \hat g_-(\bar\kk_4)\over (1+A)} {1\over M^2} \sum_\pp
\tilde\chi_M(\pp) {\hat
H^{4,1}_{+}(\pp;\bar\kk_1,\bar\kk_2,\bar\kk_3,\bar\kk_4-\pp)+ B
\hat
H^{4,1}_{-}(\pp;\bar\kk_1,\bar\kk_2,\bar\kk_3,\bar\kk_4-\pp)\over
D_+(\pp)} \;.\cr}\Eqa(A6.2.20a)$$

All the terms appearing in the above equation can be expressed in
terms of convergent tree expansions, via a recursive expansion similar
to that described in the last section. Dimensional
bounds for the terms in the first three lines can be easily derived,
in analogy with the results of Theorem A6.1. In Appendix A1 of [BM2]
the following bound is proven:
$$\left| \l\hat g_-(\bar\kk_4){1\over M^2}\sum_\pp\chi_M(\pp)
G^{4,1}_+(\pp;\bar\kk_1,\bar\kk_2,\bar\kk_3,\bar\kk_4-\pp) \right|
\le C\bar \l_h^3\g^{\h_{2,1} h} {\g^{-4 h}\over Z_h^2} \Eqa(A6.2.25)$$
where the exponent $\h_{2,1}$ is the same of \equ(A6.1.45). However,
from the result of previous section (\ie from the validity of the first
correction identity) it follows that $\h_{2,1}=0$, so that the 
r.h.s. of \equ(A6.2.25) is the right dimensional bound we need.
As regarding the terms in the second and the third line of \equ(A6.2.20), 
following a procedure similar to that leading to the second of \equ(A6.1.47)
(see again Appendix A1 of [BM2]), we find
$$\eqalign{
&\left| {1\over M^2} \sum_\pp \tilde\chi_M(\pp) {\hat
G_+^{4}(\bar\kk_1-\pp,\bar\kk_2,\bar\kk_3,\bar\kk_4-\pp)- \hat
G_+^{4}(\bar\kk_1,\bar\kk_2+\pp,\bar\kk_3,\bar\kk_4-\pp)\over
D_+(\pp)} \right.\;+\cr
&+ \left. {1\over M^2} \sum_\pp \tilde\chi_M(\pp) {\hat
G_+^{4}(\bar\kk_1,\bar\kk_2,\bar\kk_3-\pp,\bar\kk_4-\pp)\over
D_+(\pp)} \right| \le C \bar\l_{h}  {\g^{-3 h}\over
Z_h^2}\;,}\Eqa(A6.2.24)$$
that, again, is the right dimensional bound.

The bound on the term in the last line of \equ(A6.2.20) is more involved,
and require an analysis similar (but more complicated) to that of previous 
section. We will prove in next section that\\
\\
{\it there exists $\e_1\le \e_0$ and four $\l$-functions
$\n_+,\n_-,\n'_+,\n'_-$ of order $\l$ (uniformly in $h$), such
that, if $\bar\l_{h}\le \e_1$,
$$\left| \l\hat  g_-(\bar\kk_4) {1\over M^2} \sum_\pp
\tilde\chi_M(\pp) {\hat
H^{4,1}_{\pm}(\pp;\bar\kk_1,\bar\kk_2,\bar\kk_3,\bar\kk_4-\pp) \over
D_+(\pp)} \right| \le C\bar\l_{h}^2{\g^{-4 h}\over Z_h^2}
\Eqa(A6.2.26)$$
}

Substituting the bounds \equ(A6.2.26), \equ(A6.2.25), \equ(A6.2.24)
together with \equ(chir7) and the second of \equ(A6.1.47) in 
the Dyson equation \equ(A6.2.11a), we finally get
$$|\l_h|\le c|\l|(1+O(\bar\l_h))\;,\Eqa(A6.2.26a)$$
which implies\\
\\
{\cs Theorem A6.2} {\it The model \equ(A6.1.7) is well defined in the
limit $h\to-\io$. In fact there are constants $\e_1$ and $c_2$
such that $|\l|\le \e_1$ implies $\bar\l_j\le c_2\e_1$, for any
$j< 0$.}\\
\\
Finally, a standard argument shows that, as a consequence of 
Theorem A6.2, the first bound in \equ(6.5.1bbb) holds, that is
$$|\b_\l^{h}(\l_h,\ldots,\l_h)|
\le C |\bar\l_h|^2 \g^{\th h} \;,\Eqa(van1)$$
The proof is by contradiction. Consider the Taylor expansion
of $\b_\l^{h}(\l_h,\ldots,l_h)$ in $\l_h$ (which is convergent for $\l_h$
small enough) and let us call $b_r^{(h)}$
the coefficient of $(\l_h)^r$. Let us also call $b_r\defin\lim_{h\to -\io}
b_r^{(h)}$. By performing the bounds on the 
trees representing $b_r^{(h)}$, in the same way explained in Chapter 5, 
we find that necessarely $b_r^{(h)}=b_r+O(\g^{\th h})$, for some
$\th >0$. Now, let us assume by contradiction that, for some $r\ge 2$,
$$\b_\l^{h}(\l_h,\ldots,\l_h)=b_r (\l_h)^r
+O(|\l_h|^{r+1})+O(\l_h^2\g^{\th h})\;,\Eqa(A6.5.60z)$$
with $b_r$ a non vanishing constant. By the discussion above and 
Theorem 5.1, 
the running coupling constants $\l_h$ are analytic functions
of $\l$:
$$\l_h=\l+\sum_{n=2}^r c^{(h)}_n \l^n+O(\l^{r+1})\Eqa(A6.5.78c)$$
and for any fixed $h$ the sequence $c_n^{(h)}$ is a bounded sequence.

Consider the flow equation \equ(A6.1.39) and rewrite it as 
$$\l_{h-1}=\l_h+\b_\l^{h}(\l_h,\ldots,\l_h)+\sum_{k\ge h}D_\l^{h,k}\;,
\Eqa(A6.5.78cc)$$
where 
$$D^{h,k}_\l=\b^h_{\l}
(\l_h,\ldots,\l_h,\l_k,\l_{k+1},
\ldots,\l_1)-
\b^h_{\l}(\l_h,\ldots,\l_h,\l_h,\l_{k+1},
\ldots,\l_1)\;.\Eqa(A6.5.3dhk)$$
and, by using the short memory property, it is easy to show that 
$|D^{h,k}_\l|\le c\bar\l_h\g^{\th (h-k)}|\l_h-\l_k|$, for some $\th>0$

Inserting \equ(A6.5.60z) and \equ(A6.5.78c) into
\equ(A6.5.78cc) and keeping at both sides the terms
of order $r$, we find:
$$c_r^{(h-1)}=c_r^{(h)}+b_r+O(\l^2\g^{\th h})\;.\Eqa(A6.5.3dhkj)$$
This would mean that $c_r^{(h)}$ is a sequence diverging for $h\to -\io$,
which is in contradiction with the fact, following from Theorem A6.2
and the discussin above, that $\l_h$ is an analytic function of $\l$, 
uniformely in $h$.\Halmos
\\
\asub(A6.33){\bf Proof of {\equ(A6.2.26)}}
\\
In this final section we prove the bound \equ(A6.2.26), that is we 
conclude the proof of the second pair of correction identities and, with
this, we conclude the proof of Theorem A6.2, that is the main result of 
this Appendix.
We shall prove first the bound \equ(A6.2.26) for $\hat H^{4,1}_+$; the
bound for $\hat H_-^{4,1}$ is done essentially in the same way and will
be briefly discussed later. By using \equ(A6.2.16a), we get
$$\eqalign{& \hat g_-(\kk_4) {1\over M^2}\sum_{\pp} \tilde\chi_M(\pp)
D_+^{-1}(\pp) \hat H^{4,1}_{+}(\pp;\kk_1,\kk_2,\kk_3,\kk_4-\pp) =\cr
&=\hat g_-(\kk_4) {1\over M^2}\sum_{\pp} \tilde\chi_M(\pp)
{1\over M^2}\sum_{\kk} {C_+(\kk,\kk-\pp)\over D_+(\pp)}
<\hat\psi^+_{\kk,+}\hat\psi^-_{\kk-\pp,+};
\hat\psi^-_{\kk_1,+};\hat\psi^+_{\kk_2,+}; \hat\psi^-_{\kk_3,-};
\hat\psi^+_{\kk_4-\pp,-}>^T-\cr
&-\nu_-\hat g_-(\kk_4) {1\over M^2}\sum_{\pp} \tilde\chi_M(\pp)
{1\over M^2}\sum_{\kk} {D_-(\pp)\over D_+(\pp)}
<\hat\psi^+_{\kk,-}\hat\psi^-_{\kk-\pp,-};
\hat\psi^-_{\kk_1,+};\hat\psi^+_{\kk_2,+}; \hat\psi^-_{\kk_3,+};
\hat\psi^+_{\kk_4-\pp,-}>^T-\cr
&-\nu_+\hat g_-(\kk_4) {1\over M^2} \sum_{\pp} \tilde\chi_M(\pp)
 {1\over M^2}\sum_{\kk} <\hat\psi^+_{\kk,+}\hat\psi^-_{\kk-\pp,+};
\hat\psi^-_{\kk_1,+};\hat\psi^+_{\kk_2,+}; \hat\psi^-_{\kk_3,-};
\hat\psi^+_{\kk_4-\pp,-}>^T\;.\cr} \Eqa(A6.3.1)$$
Let us define
$$\widetilde G^4_+(\kk_1, \kk_2,\kk_3, \kk_4) = \left.
{\partial^4\over \partial \phi^+_{\kk_1,+} \partial
\phi^-_{\kk_2,+} \partial \phi^+_{\kk_3,-}
\partial J_{\kk_4}} \widetilde\WW \right|_{\phi=0}\;,\Eqa(A6.3.2)$$
where
$$\widetilde\WW = \log \int P(d\hat\psi)e^{-T_1(\psi) + \n_+T_+(\psi)
+ \n_-T_-(\psi)} e^{-V(\hat\psi)+ \sum_\o\int d\xx
[\phi^+_{\xx,\o}\hat\psi^{-}_{\xx,\o}+
\hat\psi^{+}_{\xx,\o}\phi^-_{\xx,\o}]} \;,\Eqa(A6.3.3)$$
and
$$\eqalign{&T_1(\psi) = {1\over M^2} \sum_{\pp} \tilde\chi_M(\pp)
{1\over M^2} \sum_{\kk} {C_+(\kk, \kk-\pp) \over D_+(\pp)}
(\hat\psi_{\kk,+}^+ \hat\psi_{\kk-\pp,+}^-)
\hat\psi^+_{\kk_4-\pp,-} \hat J_{\kk_4} \hat
g_-(\kk_4)\;,\cr
&T_+(\psi)= {1\over M^2} \sum_{\pp} \tilde\chi_M(\pp)
{1\over M^2} \sum_{\kk} (\hat\psi_{\kk,+}^+
\hat\psi_{\kk-\pp,+}^-) \hat\psi^+_{\kk_4-\pp,-} \hat J_{\kk_4}
\hat g_-(\kk_4)\;,\cr
&T_-(\psi) = {1\over M^2} \sum_{\pp} \tilde\chi_M(\pp)
{1\over M^2} \sum_{\kk} {D_-(\pp) \over D_+(\pp)}
(\hat\psi_{\kk,-}^+ \hat\psi_{\kk-\pp,-}^-)
\hat\psi^+_{\kk_4-\pp,-} \hat J_{\kk_4} \hat
g_-(\kk_4)\;.\cr}\Eqa(A6.3.6)$$
The vertices $T_1$, $T_+$ and $T_-$ can be graphically represented as in 
Fig. \graf(1v). The wavy lines represent the cutoff
function $\tilde\c_M(\pp)$, constraining the transferred momentum
to be $|\pp|\ge O(\g^h)$.

\midinsert
\*
\insertplotbm{300pt}{100pt}%
{\ins{45pt}{0pt}{$T_1$}
\ins{145pt}{0pt}{$T_+$}
\ins{245pt}{0pt}{$T_-$}
}%
{verticiT}{}
\vskip.5truecm
\line{\vtop{\line{\hskip1.3truecm\vbox{\advance\hsize by -2.0 truecm
\0{\css \eqg(1v).}
{\ottorm Graphical representation of ${\scriptstyle T_1,T_+,T_-}$; the dotted
line carries momentum ${\scriptstyle \bar\kk_4}$, 
the empty circle represents ${\scriptstyle C_+}$, 
the filled one ${\scriptstyle D_-(\pp)/D_+(\pp)}$}
\hfill} }}}
\*
\endinsert
It is easy to see that $\widetilde G^4_+$ is related to \equ(A6.3.1)
by an identity similar to \equ(A6.2.8). In fact we can write
$$\eqalign{& -\widetilde G_+^4(\kk_1,\kk_2,\kk_3,\kk_4)= \cr
&= \hat g_-(\kk_4) {1\over M^2} \sum_{\pp} \tilde\chi_M(\pp)
{1\over M^2}\sum_{\kk} {C_+(\kk,\kk-\pp) \over D_+(\pp)}
<(\hat\psi^+_{\kk,+}\hat\psi^-_{\kk-\pp,+})\hat\psi^+_{\kk_4-\pp,-};
\hat\psi^-_{\kk_1,+};\hat\psi^+_{\kk_2,+}; \hat\psi^-_{\kk_3,-}>^T-\cr
&-\nu_-\hat g_-(\kk_4) {1\over M^2} \sum_{\pp} \tilde\chi_M(\pp)
{1\over M^2} \sum_{\kk} {D_-(\pp)\over D_+(\pp)}
<(\hat\psi^+_{\kk,+}\hat\psi^-_{\kk-\pp,+})\hat\psi^+_{\kk_4-\pp,-};
\hat\psi^-_{\kk_1,+};\hat\psi^+_{\kk_2,+}; \hat\psi^-_{\kk_3,-}>^T
-\cr
&-\nu_+ \hat g_-(\kk_4) {1\over M^2} \sum_{\pp} \tilde\chi_M(\pp)
{1\over M^2} \sum_{\kk}
<(\hat\psi^+_{\kk,+}\hat\psi^-_{\kk-\pp,+})\hat\psi^+_{\kk_4-\pp,-};
\hat\psi^-_{\kk_1,+};\hat\psi^+_{\kk_2,+}; \hat\psi^-_{\kk_3,-}>^T
\;.\cr}\Eqa(A6.3.7)$$
If we introduce the definitions
$$\d\r_{\pp,+}={1\over M^2}\sum_\kk{ C_+(\pp,\kk)\over D_+(\pp)}
(\hat\psi_{\kk,+}^+\hat\psi_{\kk-\pp,+}^-)\;,\qquad
\r_{\pp,+}={1\over M^2}\sum_\kk
(\hat\psi_{\kk,+}^+\hat\psi_{\kk-\pp,+}^-)\;,\Eqa(A6.3.8)$$
we can rewrite 
$$\eqalign{&{1\over M^2}\sum_{\kk} {C_+(\kk,\kk-\pp) \over D_+(\pp)}
<(\hat\psi^+_{\kk,+}\hat\psi^-_{\kk-\pp,+})\hat\psi^+_{\kk_4-\pp,-};
\hat\psi^-_{\kk_1,+};\hat\psi^+_{\kk_2,+}; \hat\psi^-_{\kk_3,-}>^T=\cr
&=-<\d\r_{\pp,+};
\hat\psi^-_{\kk_1,+};\hat\psi^+_{\kk_2,+}; \hat\psi^-_{\kk_3,-};
\hat\psi^+_{\kk_4-\pp,-}>^T-<\d\r_{\pp,+};
\hat\psi^-_{\kk_1,+};\hat\psi^+_{\kk_2,+}>^T <\hat\psi^-_{\kk_3,-};
\hat\psi^+_{\kk_4-\pp,-}>\cr}\Eqa(A6.3.8a)$$
and 
$$\eqalign{&{1\over M^2}\sum_{\kk}
<(\hat\psi^+_{\kk,+}\hat\psi^-_{\kk-\pp,+})\hat\psi^+_{\kk_4-\pp,-};
\hat\psi^-_{\kk_1,+};\hat\psi^+_{\kk_2,+}; \hat\psi^-_{\kk_3,-}>^T=\cr
&=-<\r_{\pp,+};
\hat\psi^-_{\kk_1,+};\hat\psi^+_{\kk_2,+}; \hat\psi^-_{\kk_3,-};
\hat\psi^+_{\kk_4-\pp,-}>^T-<\r_{\pp,+};
\hat\psi^-_{\kk_1,+};\hat\psi^+_{\kk_2,+}>^T <\hat\psi^-_{\kk_3,-};
\hat\psi^+_{\kk_4-\pp,-}>\;,\cr}\Eqa(A6.3.8b)$$
where we used the fact that $\pp\not=0$ in the support of
$\tilde\chi_M(\pp)$ and $<\d \r_{\pp,+}> =<\r_{\pp,+}>=0$ for $\pp\not=0$. 

Substituting \equ(A6.3.8a) and \equ(A6.3.8b)
into \equ(A6.3.7), we get
$$\eqalign{
&\widetilde G^4_+(\kk_1,\kk_2,\kk_3,\kk_4)= g_-(\kk_4){1\over M^2}
\sum_{\pp} \tilde\chi_M(\pp)
{H_+^{4,1}(\pp;\kk_1,\kk_2,\kk_3,\kk_4-\pp)\over D_+(\pp)} +\cr
&+ \tilde\chi_M(\kk_1-\kk_2) g_-(\kk_4) G^2_-(\kk_3) \left[
<\hat\psi^-_{\kk_1,+}; \hat\psi^+_{\kk_2,+}; \d
\r_{\kk_1-\kk_2,+}>^T -\right. \cr
&\left. -\n_+ <\hat\psi^-_{\kk_1,+}; \hat\psi^+_{\kk_2,+};
\r_{\kk_1-\kk_2,+}>^T -\n_- {D_-(\kk_1-\kk_2)\over
D_+(\kk_1-\kk_2)} <\hat\psi^-_{\kk_1,+}; \hat\psi^+_{\kk_2,+};
\r_{\kk_1-\kk_2,-}>^T \right]\cr}\Eqa(A6.3.7b)$$

We now put $\kk_i=\bar\kk_i$, see \equ(A6.2.21a). Since
$|\bar\kk_1-\bar\kk_2|=2\g^h$,
$\tilde\chi_M(\bar\kk_1- \bar\kk_2)=0$, hence we get
$$\widetilde G^4_+(\bar\kk_1,\bar\kk_2,\bar\kk_3,\bar\kk_4)= 
g_-(\bar\kk_4){1\over M^2}
\sum_{\pp} \tilde\chi_M(\pp)
{H_+^{4,1}(\pp;\bar\kk_1,\bar\kk_2,\bar\kk_3,\bar\kk_4-\pp)\over
D_+(\pp)}\;.\Eqa(A6.3.7c)$$
\\ 
\\{\it Remark.} \equ(A6.3.7c) says that the last line of
equation \equ(A6.2.20) can be written as a functional integral very
similar to the one for $G^4_+$ except that the interaction $V$
\equ(A6.1.3) is replaced by $\VV+T_1-\n_+ T_+ -\n_- T_-$; we will
evaluate it via a multiscale integration procedure similar to the
one for $G^4_+$, and in the expansion additional running coupling
constants will appear; the expansion is convergent again if such
new running couplings will remain small uniformly in the infrared
cutoff.\\
\\
\asub(A6.3.3) 
The calculation of $\widetilde G_+^4
(\bar\kk_1,\bar\kk_2,\bar\kk_3,\bar\kk_4)$ is done via a multiscale
expansion; we shall concentrate on the differences with respect to
that described in \sec(A6.4), due to the presence in the potential of
the terms $T_1(\psi)$ and $T_\pm(\psi)$. 

At each step we will write the effective potential $\widetilde\WW$ as:
$$e^{\widetilde\WW(\phi,J)}=e^{-M^2 E_j}
\int P_{\tilde Z_j,C_{h,j}} (d\psi^{[h,j]})
e^{-\VV^{(j)}(\sqrt{Z_j}\psi^{[h,j]})+\BB_\phi^{(j)}(\sqrt{Z_j}\psi^{[h,j]})
+K^{(j)}(\psi^{[h,j]},\phi,J)}\,,\Eqa(A6.1.11yy)$$
where $\VV^{(j)}$ and $\BB_\phi^{(j)}$ are defined as in \sec(A6.4), while
$$K^{(j)}(\psi, \phi,J) = 
\bar\VV_J^{(j)}(\psi) + W_R^{(j)}(\psi,\phi,J)\;,\Eqa(A6.1.16yy)$$
with $\bar\VV^{(j)}_J(\psi^{[h,-1]})$ the sum over the terms
containing exactly one $J$ field and no $\phi$ fields and 
$W_R^{(j)}$ the rest (not involved in the construction of 
$\widetilde G_4(\bar\kk_1,\bar\kk_2,\bar\kk_3,\bar\kk_4)$). 

The iterative construction of \equ(A6.1.11yy) is defined through
the analogue of \equ(A6.1.38):
$$\eqalign{
&e^{-\VV^{(j-1)}\big(\sqrt{Z_{j-1}} \psi^{[h,j-1]} \big)+
\BB_\phi\big(\sqrt{Z_{j-1}} \psi^{[h,j-1]}\big)+
K^{(j-1)}\big(\psi^{[h,j-1]},\phi,J\big)-L\b E_j}=\cr
&=\int P_{Z_{j-1},\tilde f_j^{-1}}(d\psi^{(j)}) \, e^{-\hat
\VV^{(j)} \big(\sqrt{Z_{j-1}}[\psi^{[h,j-1]} + \psi^{(j)}]\big)+
\hat \BB_\phi \big(\sqrt{Z_{j-1}}[\psi^{[h,j-1]} + \psi^{(j)}]\big)
+K^{(j)} \big([\psi^{[h,j-1]} + \psi^{(j)}]
\big)}\;,\cr}\Eqa(A6.1.38)$$
Note that
in \equ(A6.1.11yy) we chose {\it not to rescale} the fields
in $K^{(j)}(\psi, \phi,J)$. 

In order to define the action of $\LL$ over 
$\bar\VV^{(j)}_J(\psi^{[h,-1]})$, let us first consider in detail
the first step of the iterative integration
procedure, the integration of the field $\psi^{(0)}$. We write
$$\bar\VV_J^{(-1)}(\psi^{[h,-1]}) = \bar\VV^{(-1)}_{J,a,1}(\psi^{[h,-1]})
+ \bar\VV^{(-1)}_{J,a,2}(\psi^{[h,-1]}) +
\bar\VV^{(-1)}_{J,b,1}(\psi^{[h,-1]}) +
\bar\VV^{(-1)}_{J,b,2}(\psi^{[h,-1]}) \;,\Eqa(A6.3.19)$$
where $\bar\VV^{(-1)}_{J,a,1} + \bar\VV^{(-1)}_{J,a,2}$ is the sum of
the terms in which the field $\hat\psi^+_{\bar\kk_4-\pp,-}$
appearing in the definition of $T_1(\psi)$ or $T_\pm(\psi)$ is
contracted, $\bar\VV^{(-1)}_{J,a,1}$ and $\bar\VV^{(-1)}_{J,a,2}$
denoting the sum over the terms of this type containing a $T_1$ or
a $T_\pm$ vertex, respectively; $\bar\VV^{(-1)}_{J,b,1} +
\bar\VV^{(-1)}_{J,b,2}$ is the sum of the other terms, that is those
where the field $\hat\psi^+_{\bar\kk_4-\pp,-}$ is an external
field, the index $i=1,2$ having the same meaning as before.

Note that the condition \equ(A6.2.21a) on the external momenta
$\kk_i$ forbids the presence of vertices of type $\phi$, if $h<0$,
as we shall suppose. Hence, all graphs contributing to
$\bar\VV_J^{(-1)}$ have, besides the external field of type $J$, an
odd number of external fields of type $\psi$.

\*

Let us consider first $\bar\VV^{(-1)}_{J,a,1}$; we shall still
distinguish different group of terms, those where both fields
$\hat\psi_{\kk,+}^+$ and $\hat\psi_{\kk-\pp,+}^-$ are contracted,
those where only one among them is contracted and those where no
one is contracted.

If no one of the fields $\hat\psi_{\kk,+}^+$ and
$\hat\psi_{\kk-\pp,+}^-$ is contracted, we can only have terms
with at least four external lines; for the properties of
$\D^{(i,j)}$, see Appendix A7, 
at least one of the fields $\hat\psi_{\kk,+}^+$ and
$\hat\psi_{\kk+\pp,+}^-$ must be contracted at scale $h$. If one
of these terms has four external lines, hence it is marginal, it
has the following form
$$\int d\pp \tilde\chi_M(\pp) \hat\psi^+_{\bar\kk_4-\pp,-}
G^{(0)}_2(\bar\kk_4-\pp) \hat g^{(0)}_-(\bar\kk_4-\pp) \hat
g_-(\bar\kk_4) \hat J_{\bar\kk_4} \int d\kk { C(\kk,\kk-\pp) \over
D_+(\pp)} \hat\psi^+_{\kk,+}\hat\psi^-_{\kk-\pp,+}\;, \Eqa(A6.3.20)$$
where $G^{(0)}_2(\kk)$ is a suitable function which can be
expressed as a sum of graphs with an odd number of propagators,
hence it vanishes at $\kk=0$. This implies that $G^{(0)}_2(0)=0$,
so that we can regularize it without introducing any running
coupling.
\midinsert
\*
\insertplotbm{300pt}{100pt}%
{\ins{90pt}{80pt}{$G^{(0)}_2$}
}%
{G20}{}
\vskip.5truecm
\line{\vtop{\line{\hskip3.3truecm\vbox{\advance\hsize by -4.0 truecm
\0{\css \eqg(2v).}
{\ottorm Graphical representation of \equ(A6.3.20)}
\hfill} }}}
\*
\endinsert
If both $\hat\psi_{\kk,+}^+$ and $\hat\psi_{\kk-\pp,+}^-$ in
$T_1(\psi)$ are contracted, we get terms of the form
$$\widetilde W_{n+1}^{(-1)} (\bar\kk_4,\kk_1,\ldots,\kk_n) \hat
g_-(\bar\kk_4) \hat J_{\bar\kk_4} \prod_{i=1}^n
\hat\psi^{\e_i}_{\kk_i}\;, \Eqa(A6.3.21)$$
where $n$ is an odd integer. We want to define an $\RR$ operation
for such terms. There is apparently a problem, as the $\RR$
operation involves derivatives and in $\widetilde W^{(-1)}$ appears
the function $\D^{(0,0)}$ of the form \equ(A7.4.9) and the cutoff
function $\tilde\chi_M(\pp)$, with support on momenta of size
$\g^h$. Hence one can worry about the derivatives of the factor
$\tilde\chi_M(\pp) \pp D_+(\pp)^{-1}$. However, as the line of
momentum $\bar\kk_4-\pp$ is necessarily at scale $0$ (we are
considering terms in which it is contracted), then $|\pp|\ge
\g^{-1} - \g^h\ge \g^{-1}/2$ (for $|h|$ large enough), so that 
no bad factors can be produced by the derivatives acting on 
$\tilde\chi_M(\pp) \pp D_+(\pp)^{-1}$.
We can define the $\LL$ operation in the usual way:
$$\eqalign{&\LL\widetilde W_4^{(-1)} (\bar\kk_4,\kk_1,\kk_2,\kk_3) =
\widetilde W_4^{(-1)}(\V0,\ldots,\V0)\;,\cr
&\LL\widetilde W_2^{(-1)} (\bar\kk_4)=\widetilde W_2^{(-1)}(\V0)+
\bar\kk_4\partial_\kk \widetilde W_2^{(-1)}(\V0)\;.\cr} \Eqa(A6.3.23)$$
Note that by parity the first term in the second equation of
\equ(A6.3.23) is vanishing;
this means that there are only marginal terms. Note also that the
local term proportional to $\hat J_{\bar\kk_4}
\hat\psi^+_{\bar\kk_4,-}$ is such that the field
$\hat\psi^+_{\bar\kk_4,-}$ can be contracted only at the last
scale $h$; hence it does not have any influence on the integrations of all
the scales $>h$.
\midinsert
\*
\insertplotbm{300pt}{100pt}%
{\ins{45pt}{15pt}{$\widetilde W_4^{(-1)}$}
\ins{215pt}{15pt}{$\widetilde W_2^{(-1)}$}
}%
{W4W2}{}
\vskip.5truecm
\line{\vtop{\line{\hskip3.3truecm\vbox{\advance\hsize by -4.0 truecm
\0{\css \eqg(3v).}
{\ottorm Graphical representation of ${\scriptstyle \widetilde W_4^{(-1)}}$ and
${\scriptstyle \widetilde W_2^{(-1)}}$.}
\hfill} }}}
\*
\endinsert
If only one among the fields $\hat\psi_{\kk,+}^+$ and
$\hat\psi_{\kk-\pp,+}^-$ in $T_1(\psi)$ is contracted, we note
first that we cannot have terms with two external lines (including
$\hat J_{\kk_4})$; in fact in such a case there is an external
line with momentum $\bar\kk_4$ with $\o=-$ and the other has
$\o=+$; this is however forbidden by global gauge invariance.
Moreover, for the same reasons as before, we do not have to worry
about the derivatives of the factor $\tilde\chi_M(\pp) \pp
D_+(\pp)^{-1}$, related with the regularization procedure of the
terms with four external lines, which have the form
$$\eqalign{
& \int d\kk^+ \hat\psi^+_{\kk_1,+} \hat\psi^-_{\kk^-,+}
\hat\psi^+_{\kk^- +\bar\kk_4- \kk_1,-} \hat g_-(\bar\kk_4) \hat
J_{\bar\kk_4} \tilde\chi_M(\kk^+ -\kk^-) \hat g^{(0)}_-(\bar\kk_4-
\kk^+ +\kk^-)\;\cdot\cr
& \cdot\; G_4^{(0)}(\kk^+,\kk_1,\bar\kk_4-\kk^++\kk^-) \left\{
{[C_{h,0}(\kk^-)-1] D_{+}(\kk^-) \hat g^{(0)}_+(\kk^+)\over
D_+(\kk^+-\kk^-)} - {u_0(\kk^+) \over D_+(\kk^+-\kk^-)}
\right\}\;, \cr}\Eqa(A6.3.24)$$
or the similar one with the roles of $\kk^+$ and $\kk^-$
exchanged.
\midinsert
\*
\insertplotbm{300pt}{100pt}%
{\ins{145pt}{60pt}{$G^{(0)}_4$}
}%
{G40}{}
\vskip.5truecm
\line{\vtop{\line{\hskip3.3truecm\vbox{\advance\hsize by -4.0 truecm
\0{\css \eqg(4v).}
{\ottorm Graphical representation of a single addend in
\equ(A6.3.24).}
\hfill} }}}
\*
\endinsert
The two terms in \equ(A6.3.24) must be treated differently, as concerns the
regularization procedure. The first term is such that one of the external
lines is associated with the operator $[C_{h,0}(\kk^-)-1] D_+(\kk^-)
D_+(\pp)^{-1}$. We define $\RR=1$ for such terms; in fact, when 
the $\psi^-_{\kk^-,+}$ external line 
is contracted (and this can happen only at scale $h$), the factor
$D_+(\kk^-) D_+(\pp)^{-1}$ produces an extra factor $\g^{h}$ in the bound,
with respect to the dimensional one. This claim simply follows by the
observation that $|D_+(\pp)|\ge 1-\g^{-1}$ as $\pp=\kk^+-\kk^-$ and $\kk^+$
is at scale $0$, while $\kk^-$, as we said, is at scale $h$. This factor has
the effect that all the marginal terms in the tree path connecting $v_0$ with
the end-point to which is associated the $T_1$ vertex acquires negative
dimension.

The second term in \equ(A6.3.24) can be regularized as above, by
subtracting the value of the kernel computed at zero external
momenta, \ie for $\kk^-=\bar\kk_4=\kk_1=0$. Note that such local
part is given by
$$\int d\kk^+ \tilde\chi_M(\kk^+)
\hat g^{(0)}_-(\kk^+) G_4^{(0)}(\kk^+,\V0,-\kk^+) {u_0(\kk^+) \over
D_+(\kk^+)} \;,\Eqa(A6.3.25)$$
and there is no singularity associated with the factor
$D_+(\kk^+)^{-1}$, thanks to the support on scale $0$ of the
propagator $\hat g^{(0)}_-(\kk^+)$.

A similar (but simpler) analysis holds for the terms contributing
to $\bar\VV^{(-1)}_{J,a,2}$, which contain a vertex of type $T_+$ or
$T_-$ and are of order $\l\n_\pm$. Now, the only thing to analyze
carefully is the possible singularities associated with the
factors $\tilde\chi_M(\pp)$ and  $\pp D_+(\pp)^{-1}$. However,
since in these terms the field $\hat\psi^+_{\bar\kk_4-\pp,-}$ is
contracted, $|\pp|\ge \g^{-1}/2$, for $|h|$ large enough, a
property already used before; hence the regularization procedure
can not produce bad dimensional bounds.

\*

We will define $\tilde z_{-1}$ and $\tilde\l_{-1}$, so that
$$\LL[\bar\VV^{(-1)}_{J,a,1} +
\bar\VV^{(-1)}_{J,a,2}](\psi^{[h,-1]}) = \left[ \tilde\l_{-1}Z_{-2}^2 \bar
F_\l^{[h,-1]}(\psi^{[h,-1]}) + \tilde z_{-1}
\hat\psi^{[h,-1]+}_{\bar\kk_4,-} D_-(\bar\kk_4) \right] \hat
g_-(\bar\kk_4) \hat J_{\bar\kk_4}\;,\Eqa(A6.3.25a)$$
where we used the definition
$$\bar F_\l^{[h,j]}(\psi^{[h,j]}) = {1\over
(M^2)^4}\sum_{\kk_1,\kk_2,\kk_3:C^{-1}_{h,j}(\kk_i)>0}
\hat\psi^{[h,j]+}_{\kk_1,+} \hat\psi^{[h,j]-}_{\kk_2,+}
\hat\psi^{[h,j]+}_{\kk_3,-}
\d(\kk_1-\kk_2+\kk_3-\bar\kk_4)\;.\Eqa(A6.1.24a)$$
%
Note that the fields in the 
monomial $\bar F_\l^{[h,j]}(\psi^{[h,-1]})
g_-(\bar\kk_4) \hat J_{\bar\kk_4}$ associated to the coupling $\tilde\l_{-1}$
have no constraint on the transferred momentum, in particular transferred
momenta $|\pp|\le O(\g^h)$ are allowed: this deeply distinguish 
the term associated to $\tilde\l_{-1}$ 
from a term like $T_+(\psi)$ in \equ(A6.3.6): the 
transferred momentum associated to the fields
in $T_+(\psi)$ has instead a lower cutoff $\sim \g^h$.
\\

Let us consider now the terms contributing to
$\bar\VV^{(-1)}_{J,b,1}$, that is those where
$\hat\psi^+_{\bar\kk_4-\pp}$ is not contracted and there is a
vertex of type $T_1$.

Besides the term of order $0$ in $\l$ and $\n_\pm$, equal to
$T_1(\psi^{[h,-1]})$, there are the terms containing at least one
vertex $\l$; among these terms, the only marginal ones (those
requiring a regularization) have four external lines (including
$\hat J_{\kk_4}$), since the oddness of the propagator does not
allow tadpoles. These terms are of the form
$$\eqalign{
&\sum_{\tilde \o} \int d\pp \tilde\chi_M(\pp)
\hat\psi^+_{\kk^+,\tilde\o} \int d\kk^+
\hat\psi^-_{\kk^+-\pp,\tilde\o} \hat\psi^+_{\bar\kk_4-\pp,-} \hat
g_-(\bar\kk_4) \hat J_{\bar\kk_4} 
\left[\hat F^{(-1)}_{2,+,\tilde\o}(\kk^+, \kk^+-\pp) +
\hat F^{(-1)}_{1,+}(\kk^+, \kk^+-\pp) \d_{+,\tilde\o} \right]\;,\cr}
\Eqa(A6.3.28)$$
where $\hat F^{(-1)}_{2,+,\tilde\o}$ and $\hat F^{(-1)}_{1,+}$ are defined
as in \equ(A6.132); they represent the terms in which both
or only one of the fields in $\d \r_{\pp,+}$, respectively, are
contracted. Both contributions to the r.h.s. of \equ(A6.3.28) are
dimensionally marginal; however, the regularization of
$F^{(-1)}_{1,+}$ is trivial, as the latter is of the form \equ(A6.145)
or the similar one, obtained exchanging $\kk^+$ with $\kk^-$.
\midinsert
\*
\insertplotbm{300pt}{80pt}%
{\ins{150pt}{30pt}{$+$}
\ins{80pt}{5pt}{$\hat F^{(-1)}_{2,+,\tilde\o}$}
\ins{230pt}{5pt}{$\hat F^{(-1)}_{1,+}$}
}%
{F2F1}{}
\vskip.5truecm
\line{\vtop{\line{\hskip3.3truecm\vbox{\advance\hsize by -4.0 truecm
\0{\css \eqg(5v).}
{\ottorm Graphical representation of \equ(A6.3.28)}
\hfill} }}}
\*
\endinsert
As already discussed above, by the oddness of the propagator in the momentum,
$G_+^{(2)}(\V0)=0$, hence we can regularize such term without
introducing any local term; the action of $\RR$ on it is defined to be the
identity.

Moreover, $\hat F^{(-1)}_{2,+,\tilde\o}$ satisfies the symmetry properties
\equ(A6.137)--\equ(A6.139), so that, defining the action of $\LL$ 
on $\hat F^{(-1)}_{2,+,\tilde\o}$ as in \equ(A6.140), we get
$$\LL F^{-1}_{2,+,+}=Z_{-1}^{3,+} \virg \LL F^{-1}_{2,+,-}=
{D_{-}(\pp)\over D_+(\pp)} Z_{-1}^{3,-}\;,\Eqa(A6.3.30)$$
where  $Z_{-1}^{3,+}$ and $Z_{-1}^{3,-}$ are suitable real
constants. Hence the local part of the marginal term \equ(A6.3.28)
is, by definition, equal to
$$Z_{-1}^{3,+} T_+(\psi^{[h,-1]}) + Z_{-1}^{3,-}
T_-(\psi^{[h,-1]})\;.\Eqa(A6.3.30a)$$

\*

Let us finally consider the terms contributing to
$\bar\VV^{(-1)}_{J,b,2}$, that is those where
$\hat\psi^+_{\bar\kk_4-\pp}$ is not contracted and there is a
vertex of type $T_+$ or $T_-$. If even this vertex is not
contracted, we get a contribution similar to \equ(A6.3.30a), with
$\n_\pm$ in place of $Z_{-1}^{3,\pm}$. Among the terms with at
least one vertex $\l$, there is, as before, no term with two
external lines; hence the only marginal terms have four external
lines and can be written in the form
$$\eqalign{
&\int d\pp \tilde\chi_M(\pp) \hat J_{\kk_4} \hat g_-(\kk_4) \int
d\kk^+ \sum_{\tilde\o} \hat\psi_{\kk^+,\tilde\o}^+ \hat\psi_{\kk^+
-\pp,\tilde\o}^-
\left[ \n_+ G^{(0)}_{+,\tilde\o}(\kk^+, \kk^+-\pp) + \n_-
{D_-(\pp) \over D_+(\pp)} G^{(0)}_{-,\tilde\o}(\kk^+, \kk^+-\pp)
\right] \;.\cr}\Eqa(A6.3.30b)$$
By using the symmetry property $D_\o(\kk)=i\o D_\o(\kk^*)$
discussed in the lines above \equ(A6.136), it
is easy to show that $G^{(0)}_{\o,-\o}(\V0,\V0)=0$. Hence, if we
regularize \equ(A6.3.30b) by subtracting $G^{(0)}_{\o,\tilde\o}(\V0,\V0)$
to $G^{(0)}_{\o,\tilde\o}(\kk^+, \kk^+-\pp)$, we still get a local
term of the form \equ(A6.3.30a).

By collecting all the local term, we can write
$$\LL[\bar\VV^{(-1)}_{J,b,1} + \bar\VV^{(-1)}_{J,b,2}](\psi^{[h,-1]}) =
\n_{-1,+} T_+(\psi^{[h,-1]}) + \n_{-1,-}
T_-(\psi^{[h,-1]})\;,\Eqa(A6.3.31)$$
where $\n_{-1,\o} = \n_\o + Z_{-1}^{3,\o} + G^{(0)}_{\o,\o}(\V0,\V0)$.
Hence
$$\eqalign{
&\bar\VV_J^{(-1)}(\psi^{[h,-1]}) = T_1(\psi^{[h,-1]}) + \n_{-1,+}
T_+(\psi^{[h,-1]}) + \n_{-1,-} T_-(\psi^{[h,-1]}) +\cr
&+ \left[ \tilde\l_{-1} Z_{-2}^2\bar F_\l^{[h,-1]}(\psi^{[h,-1]}) + \tilde
z_{-1} \hat\psi^{[h,-1]+}_{\bar\kk_4,-} D_-(\bar\kk_4) \right]
\hat g_-(\bar\kk_4) \hat J_{\bar\kk_4} +
\bar\VV^{(-1)}_{J,R}(\psi^{[h,-1]})\;,\cr}\Eqa(A6.3.31a)$$
where $\bar\VV^{(-1)}_{J,R}(\psi^{[h,-1]})$ is the sum of all
irrelevant terms linear in the external field $J$.\\
\\
{\it Remark.} Note that, as already commented after \equ(A6.1.24a),
the structure of the field monomials associated to
$\tilde\l_{-1}$ and to $\n_{-1,+}$ respectively
are deeply different, because of the 
presence of the cutoff function $\tilde\c_M$ in the definition of 
$T_+(\psi^{[h,-1]})$. This implies that the coupling constant 
$\tilde\l_{-1}$ {\it cannot be included} in the definition of $\n_{-1,+}$
and is really a different marginal coupling.
\\
\\
\asub(A6.3.6) 
We now consider the integration of the higher scales.
The integration of the field $\psi^{(-1)}$ is done in a similar
way; we shall call $\bar\VV_J^{(-2)}(\psi^{[h,-2]})$ the sum over
all terms linear in $J$. As before, the condition
\equ(A6.2.21a) on the external momenta $\kk_i$ forbids the presence
of vertices of type $\phi$, if $h<-1$, as we shall suppose.

The main difference is that there is no contribution obtained by
contracting both field variables belonging to $\d\r$ in
$T_1(\psi)$ at scale $-1$, because of \equ(A7.4.6). It is instead
possible to get marginal terms with four external lines (two is
impossible), such that one of these fields is contracted at scale
$-1$. However, in this case, the second field variable will be
necessarily contracted at scale $h$, so that we can put $\RR=1$
for such terms. In fact, after the integration of the last scale
field, an extra factor $\g^{-(-1-h)}$ comes out 
from a bound similar to that described after \equ(A6.3.24).
Such factor has the effect of automatically regularize these terms, and even
the terms containing them as subgraphs.

The terms with a $T_1$ vertex, such that the field variables
belonging to $\d\r$ are not contracted, can be treated as in
\sec(A6.3.3), hence do not need a regularization.

It follows that, if the irrelevant part $\bar\VV^{(-1)}_{J,R}$ were
absent in the r.h.s. of \equ(A6.3.31a), then the regularization
procedure would not produce any local term proportional to $\bar
F_\l^{[h,-1]}(\psi^{[h,-2]})$, starting from a graph containing a
$T_1$ vertex.

It is easy to see that all other terms containing a vertex of type
$T_1$ or $T_\pm$ can be treated as in \sec(A6.3.3). Moreover, the
support properties of $\hat g_-(\bar\kk_4)$ immediately implies
that it is not possible to produce a graph contributing to
$\bar\VV_J^{(-2)}$, containing the $\tilde z_{-1}$ vertex. Hence, in
order to complete the analysis of $\bar\VV_J^{(-2)}$, we still have
to consider the marginal terms containing the $\tilde\l_{-1}$
vertex, for which we simply apply the localization procedure
defined in \equ(A6.3.23). We shall define two new
constants $\tilde\l_{-2}$ and $\tilde z_{-2}$, so that
$\tilde\l_{-2} (Z_{-3})^2$ is the coefficient of the local term
proportional to $\bar F_\l^{[h,-1]}(\psi^{[h,-2]})$, while $\tilde
z_{-2} Z_{-2} \hat\psi^{[h,-2]+}_{\bar\kk_4,-} D_-(\bar\kk_4) \hat
g_-(\bar\kk_4) \hat J_{\bar\kk_4}$ denotes the sum of all local
terms with two external lines produced in the second integration
step.

The above procedure can be iterated up to scale $h+1$, without any
important difference. In particular, for all marginal terms
(necessarily with four external lines) such that one of the field
variables belonging to $\d\r$ in $T_1(\psi)$ is contracted at
scale $i\ge j$, we put $\RR=1$. We can do that, because, in this
case, the second field variable belonging to $\d\r$ has to be
contracted at scale $h$, so that an extra factor $\g^{-(i-h)}$ 
(coming out from a discussion similar to that following \equ(A6.3.24))
has the effect of automatically regularize their
contribution to the tree expansion of $\widetilde G^4_+(\bar\kk_1,
\bar\kk_2, \bar\kk_3, \bar\kk_4)$ (that it similar to that descibed in Chapter
5, with the obvious modifications induced by the presence of new kind
of vertices and of a different definition of the $\LL$ operator).

Note that, as in the case $j=-1$, there is no problem connected
with the presence of the factors $\tilde\chi(\pp)$ and $D_-(\pp)
D_+(\pp)^{-1}$. In fact, if the field
$\hat\psi^+_{\bar\kk_4-\pp,-}$ appearing in the definition of
$T_1(\psi)$ or $T_\pm(\psi)$ is contracted on scale $j$, each
momentum derivative related with the regularization procedure
produces the right $\g^{-j}$ dimensional factor, since $\pp$ is of
order $\g^j$ and the derivatives of $\tilde\chi(\pp)$ are
different from $0$ only for momenta of order $\g^h$. If, on the
contrary, the field $\hat\psi^+_{\bar\kk_4-\pp,-}$ is not
contracted, then the renormalization procedure is tuned so that
$\tilde\chi(\pp)$ and $D_-(\pp) D_+(\pp)^{-1}$ are not affected by
the regularization procedure.

At step $-j$, we get an expression of the form
$$\eqalign{
&\bar\VV_J^{(j)}(\psi^{[h,j]}) = T_1(\psi^{[h,j]}) + \n_{j,+}
T_+(\psi^{[h,j]}) + \n_{j,-} T_-(\psi^{[h,j]}) +\cr
&+ \left[ \tilde\l_j Z_{j-1}^2 \bar F_\l^{[h,j]}(\psi^{[h,j]}) +
\sum_{i=j}^{-1} \tilde z_i Z_i \hat\psi^{[h,j]+}_{\bar\kk_4,-}
D_-(\bar\kk_4) \right] \hat g_-(\bar\kk_4) \hat J_{\bar\kk_4} +
\bar\VV^{j}_{J,R}(\psi^{[h,-1]})\;,\cr}\Eqa(A6.3.31b)$$
where $\bar\VV^{j}_{J,R}(\psi^{[h,-1]})$ is thought as a convergent
tree expansion (under the hypothesis that $\bar\l_h$ is small
enough). Since $Z_{-1}=1$, this
expression is in agreement with \equ(A6.3.31a).

The expansion of $\widetilde G^4_+(\bar\kk_1, \bar\kk_2, \bar\kk_3,
\bar\kk_4)$ is obtained by building all possible graphs with four
external lines, which contain one term taken from the expansion of
$\bar\VV_J^{(h)}(\psi^{(h)})$, two terms from 
$\BB_\phi^{(h)}$ and an arbitrary number of terms taken
from the effective potential $\VV^{(h)}(\psi^{(h)})$. One of the
external lines is associated with the free propagator
$g_-(\bar\kk_4)$, the other three are associated with propagators
of scale $h$ and momenta $\bar\kk_i$, $i=1,2,3$.\\
\\
{\it Remark.} 
With respect to the expansion for $G^{4}_+$,
there are three additional quartic running coupling constants,
$\n_{j,+},\n_{j,-}$ and $\tilde\l_j$. Note that they are all
$O(\l)$, despite of the fact that the interaction $T_1$ has a
coupling $O(1)$; this is a crucial property, which follows from
the discussion above, implying that either $T_1$ is
contracted at scale $0$, or it gives no contribution to the
running coupling constants. At a first sight, it seems that now we
have a problem more difficult than the initial one; we started
from the expansion for $G_+^4$, which is convergent if the running
coupling $\l_j$ is small, and we have reduced the problem to that
of controlling the flow of four running coupling constants,
$\n_{+,j}$, $\n_{j,-}$, $\l_j$, $\tilde\l_j$. However, we will see
that, under the hypothesis $\bar\l_h\le \e_1$, also the flow of
$\n_{j,+}$, $\n_{j,-}$, $\tilde\l_j$ is bounded. In fact one can use the
counterterms $\n_+,\n_-$ (this is the reason why we introduced
them) to impose that $\n_{+,j},\n_{j,-}$ are
decreasing and vanishing at $j=h$; moreover it can be verified that the beta
functions for $\tilde\l_j$ and $\l_j$ are identical up to
exponentially decaying $O(\g^{\t j})$ terms.\\
\\
\asub(A6.3.8) Now we will describe the flow of the 
new effective constants $\n_{j,\o}$, $\tilde\l_j$ and $\tilde z_j$.

First, let us consider $\n_{j,\pm}$. Note that 
the definitions of the previous sections imply that there is no
contribution to $\n_{j,\pm}$, coming from trees with a special
endpoint of type $\tilde\l$ or $\tilde z$. Then the contributions
to $\n_{j,\pm}$ either contain a constant $\n_{k,\pm}$, $k>j$,
or an endpoint
of type $T_1$, that mustbe on scale $0$. Then, by inductively
suppose that the size of $\n_{k,\pm}$, $k>j$ is exponentially small, 
and using the short memory property, one can show that 
$\n_{j,\pm}$ is exponentially small, that is 
$|\n_{j,\pm}|\le c\bar\l_h\g^{\th j}$, for some constants $c,\th>0$. 
The formal proof can be done using
a fixed point argument, following step by step the analogous
analysis used to prove that \equ(A6.1.39) admits as a solution an exponentially
decreasing sequence.

Let us now focus on $\tilde\l_j$.
We start noting that the beta function equation for $\l_j$
can be written as
$$\l_{j-1}= \left( {Z_{j-1}\over Z_{j-2}} \right)^2 \l_j +
\b_j + \b^{(0)}_{j}\;,\Eqa(A6.3.37)$$
where $\b_j$ is the sum over the local parts of the trees with at
least two endpoints and no endpoint of scale index $0$, while
$\b^{(0)}_{j}$ is the similar sum over the trees with at least one
endpoint of scale index $0$.

On the other hand we can write
$$\tilde\l_{j-1}= \left( {Z_{j-1}\over Z_{j-2}} \right)^2
\tilde\l_j + \tilde\b_j + \tilde\b^{(0)}_{j} + \tilde\b^{(T)}_{j}
+ \tilde\b^{(\n)}_{j}\;,\Eqa(A6.3.38)$$
where:\\
\0 1) $\tilde\b_j$ is the sum over the local parts of the trees
with at least two endpoints, no endpoint of scale index $0$ and
one special endpoint of type $\tilde\l$.\\
\0 2) $\tilde\b^{(0)}_{j} + \tilde\b^{(T)}_{j}$ is the sum over
the trees with at least one endpoint of scale index $0$; 
$\tilde\b^{(0)}_{j}$ and $\tilde\b^{(T)}_{j}$ are, respectively,
the sum over the trees with the special endpoint of type $\tilde\l$ or
$T_1$.\\
\0 3) $\tilde\b^{(\n)}_{j}$ is the sum over the trees with at
least two endpoints, whose special endpoint is of type $T_\pm$.\\
\\
A crucial role in the proof has the following Lemma.\\
\\
{\cs Lemma A6.1} 
{\it Let $\a=\tilde\l_h/\l_h$; then if $\bar\l_h$ is
small enough, there exists a constant $c$, independent of $\l$,
such that $|\a| \le c$ and
$$|\tilde\l_j - \a\l_j| \le c \bar\l_h \g^{\th j} \virg h+1\le j
\le -1\;.\Eqa(A6.3.38a)$$ }

\*

\proof The main point is the remark that there is a one to one
correspondence between the trees contributing to $\b_j$ and the
trees contributing to $\tilde\b_j$. In fact the trees contributing
to $\tilde\b_j$ have only endpoints of type $\l$, besides the
special endpoint $v^*$ of type $\tilde\l$, 
and the external field with $\o=-$ and
$\s=-$ has to belong to $P_{v^*}$. It follows that we can
associate uniquely with any tree contributing to $\tilde\b_j$ a
tree contributing to $\b_j$, by simply substituting the special
endpoint with a normal endpoint, without changing any label. This
correspondence is surjective, since we have imposed the condition
that the trees contributing to $\tilde\b_j$ and $\b_j$ do not have
endpoints of scale index $0$. Hence, we can write
$$\left[\left( {Z_{j-1}\over Z_{j-2}} \right)^2 -1\right]
(\tilde\l_j - \a\l_j) + \tilde\b_j - \a \b_j = \sum_{i=j}^{-1}
\b_{j,i} (\tilde\l_i - \a\l_i)\;, \Eqa(A6.3.43)$$
where, thanks to the ``short memory property'' and the fact that
$Z_{j-1}/Z_{j-2}=1 +O(\bar\l_j^2)$, the constants $\b_{j,i}$ satisfy
the bound $|\b_{j,i}|\le C \bar\l_j\g^{2\th(j-i)}$, with $\th>0$.

Among the four last terms in the r.h.s. of \equ(A6.3.38), the only
one depending on the $\tilde\l_j$ is $\tilde\b^{(0)}_{j}$, which
can be written in the form
$$\tilde\b^{(0)}_{j} = \sum_{i=j}^{-1} \b'_{j,i}
\tilde\l_i\;,\Eqa(A6.3.43a)$$
the $\b'_{j,i}$ being constants which satisfy the bound
$|\b'_{j,i}| \le C \bar\l_j \g^{2\th j}$, since they are related
to trees with an endpoint of scale index $0$. For the same
reasons, we have the bounds $|\tilde\b^{(T)}_{j}| \le C \bar\l_j
\g^{2\th j}$, $|\b^{(0)}_{j}| \le C \bar\l_j^2 \g^{2\th j}$.
Finally, by using the exponential decay of the $\n_{j,\o}$, we see that
$|\tilde\b^{(\n)}_{j}| \le C \bar\l_j \bar\l_h \g^{2\th j}$.

We now choose $\a$ so that
$$\tilde\l_h - \a\l_h=0\;,\Eqa(A6.3.43b)$$
and we put
$$x_j = \tilde\l_j - \a\l_j \virg h+1 \le j \le -1\;.\Eqa(A6.3.43c)$$
We can write
$$x_{j-1} = x_{-1} + \sum_{j'=j}^{-1} \left[ \sum_{i=j'}^{-1}
\b_{j',i} x_i + \sum_{i=j'}^{-1} \b'_{j',i} (x_i + \a\l_i) +
\tilde\b^{(T)}_{j'} + \tilde\b^{(\n)}_{j'} - \a \b^{(0)}_{j}
\right]\;.\Eqa(A6.3.39)$$
On the other hand, the condition \equ(A6.3.43b) implies that
$$x_{-1} = - \sum_{j'=h+1}^{-1} \left[ \sum_{i=j'}^{-1}
\b_{j',i} x_i + \sum_{i=j'}^{-1} \b'_{j',i} (x_i + \a\l_i) +
\tilde\b^{(T)}_{j'} + \tilde\b^{(\n)}_{j'} - \a \b^{(0)}_{j}
\right]\;,\Eqa(A6.3.40)$$
so that, if $h+1 \le j \le -1$, the $x_j$ satisfy the equation
$$x_j = - \sum_{j'=h+1}^j \left[ \sum_{i=j'}^{-1}
\b_{j',i} x_i + \sum_{i=j'}^{-1} \b'_{j',i} (x_i + \a\l_i) +
\tilde\b^{(T)}_{j'} + \tilde\b^{(\n)}_{j'} - \a \b^{(0)}_{j}
\right]\;.\Eqa(A6.3.41)$$

We want to show that equation \equ(A6.3.41) has a unique solution
satisfying the bound
$$|x_j|\le  c_0 (1+|\a|\bar\l_h) \bar\l_h \g^{\th j}\;,\Eqa(A6.3.41a)$$
for a suitable constant $c_0$, independent of $h$, if $\bar\l_h$
is small enough. Hence we introduce the Banach space $\MMM_\th$ of
sequences $\ux = \{x_j, h+1\le j \le-1\}$ with norm
$||\ux||_\th\defin \sup_j |x_j|\g^{-\th j} \bar\l_h^{-1}$ and look
for a fixed point of the operator $\bT:\MMM_{\th}\to\MMM_{\th}$
defined by the r.h.s. of
\equ(A6.3.41). By using the bounds on the various constants appearing
in the definition of $\bT$, we can easily prove that there are two
constants $c_1$ and $c_2$, such that
$$|(\bT \ux)_j| \le c_1\bar\l_h (1+|\a|\bar\l_h) \g^{\th j} +
c_2\bar\l_h \sum_{j'=h+1}^j  \sum_{i=j'}^{-1} \g^{2\th (j'-i)}
|x_i|\;.\Eqa(A6.3.42)$$
Hence, if we take $c_0=M c_1$, $M\ge 2$, and $\bar\l_h$ small enough,
the ball $\BBB_M$ of
radius $c_0 (1+|\a|\bar\l_h)$ in $\MMM_\th$ is invariant under the
action of $\bT$. On the other hand, under the same
condition, $\bT$ is a contraction in all $\MMM_\th$; in fact, if
$\ux, \ux'\in\MMM_\th$, then, if $\bar\l_h$
is small enough,
$$|(\bT \ux)_j - (\bT \ux')_j| \le c_2\bar\l_h^2 ||\ux - \ux'||
\sum_{j'=h+1}^j  \sum_{i=j'}^{-1} \g^{2\th (j'-i)} \g^{\th i} \le
{1\over 2} ||\ux - \ux'|| \bar\l_h \g^{\th j}\;,\Eqa(A6.3.44)$$
It follows, by the contraction principle, that there is a unique fixed point
in the ball $\BBB_M$, for any $M\ge 2$, hence a unique fixed point
in $\MMM_\th$, satisfying the condition \equ(A6.3.41a) with
$c_0=2c_1$.

To complete the proof, we have to show that $\a$ can be bounded
uniformly in $h$. In order to do that, we insert in the l.h.s. of
\equ(A6.3.40) the definition of $x_{-1}$ and we bound the r.h.s. by
using \equ(A6.3.41a) and \equ(A6.3.42); we get
$$|\tilde\l_{-1}-\a\l_{-1}| \le c_3\bar\l_h + c_4 |\a|
\bar\l_h^2\;,\Eqa(A6.3.40a)$$
for some constants $c_3$ and $c_4$. Since $|\l_{-1}|\ge c_5|\l|$,
$\tilde\l_{-1}\le c_6|\l|$ and $\bar\l_h\le 2|\l|$ by the
inductive hypothesis, we have
$$|\a\l_{-1}| \le |\tilde\l_{-1}| + c_3\bar\l_h + c_4 |\a|
\bar\l_h^2 \Rightarrow |\a| \le (c_6 +2c_3 +2c_4|\a|
\bar\l_h)/c_5\;,\Eqa(A6.3.40b)$$
so that, $|\a| \le 2(c_6 +2c_3)/c_5$, if $4c_4 \bar\l_h \le
c_5$.\Halmos
\\
We want now to discuss the properties of the constants $\tilde
z_j$, $h\le j\le -1$, by comparing them with the constants $z_j$,
which are involved in the renormalization of the free measure, see
\equ(A6.1.32). There is a tree expansion for the $z_j$, which can be
written as
$$z_j = \b_j + \b^{(0)}_j  \;,\Eqa(A6.3.44a)$$
where $\b_j$ is the sum over the trees without endpoints of scale
index $0$, while $\b^{(0)}_j$ is the sum of the others,
satisfying the bound $|\b^{(0)}_j| \le C\bar\l_h^2 \g^{\th j}$.
The tree expansion of the $\tilde z_j$ can be written as
$$\tilde z_j = \tilde\b_j + \tilde\b^{(\n)}_j + \tilde\b^{(0)}_j\;,
\Eqa(A6.3.44b)$$
where $\tilde\b_j$ is the sum over the trees without endpoints of
scale index $+1$, such that the special endpoint is of type
$\tilde\l$, $\tilde\b^{(\n)}_j$ is the sum over the trees whose
special endpoint is of type $T_\pm$, and $\tilde\b^{(0)}_j$ is the
sum over the trees with at least an endpoint of scale index $0$.

Since there is no tree contributing to $\tilde\b^{(0)}_j$ without
at least one $\l$ or $\tilde\l$ endpoint and since all trees
contributing to it satisfy the ``short memory property'', by using
Lemma A6.1 (which implies that $|\tilde\l_j| \le C\bar\l_h$), we
get the bound $|\tilde\b^{(0)}_j| \le C\bar\l_h\g^{\th j}$. In a
similar manner, by using the exponential decay of the 
constants $\n_{j,\o}$, we see that
$|\tilde\b^{(\n)}_j| \le C\bar\l_h^2\g^{\th j}$.

Let us now consider $\b_j$ and $\tilde\b_j$. By an argument
similar to that used in the proof of Lemma A6.1, we can write
$$\tilde\b_j -\a \b_j = \sum_{i=j+1}^{-1} \b_{j,i} (\tilde\l_i -
\a\l_i)\;,\Eqa(A6.3.44c)$$
where $\a$ is defined as in Lemma A6.1 and $|\b_{j,i}|\le
C\bar\l_h\g^{2\th j}$. Hence, Lemma A6.1 implies that
$$|\tilde z_j - \a z_j|\le C\bar\l_h \g^{\th j}\;.\Eqa(A6.3.45)$$
\Halmos

\asub(A6.3.10) 
In this section we conclude the bound for 
$\widetilde G^4_+(\bar\kk_1, \bar\kk_2,
\bar\kk_3, \bar\kk_4)$. If we consider the tree expansion for 
$\widetilde G^4_+$, we realize that 
there are various classes of trees contributing to it,
depending on the type of the special endpoint. Let us
consider first the family $\TT_{\tilde\l}$ of the trees with an
endpoint of type $\tilde\l$. These trees have the same structure
of those appearing in the expansion of $G^4_+(\bar\kk_1,
\bar\kk_2, \bar\kk_3, \bar\kk_4)$, except for the fact that the
external (renormalized) propagator of scale $h$ and momentum
$\bar\kk_4$ is substituted with the free propagator $\hat
g_-(\bar\kk_4)$. It follows, by using the bound $|\tilde\l_j|\le
C\bar\l_h$, that a tree with $n$ endpoint is bounded by
$(C\bar\l_h)^n Z_h^{-1} \g^{-4h}$, larger for a factor $Z_h$ with
respect to what we need.

Let us now consider the family $\TT_{\tilde z}$ of the trees with
a special endpoint of type $\tilde z$. Given a tree $\t\in
\TT_{\tilde\l}$, we can associate with it the class $\TT_{\tilde
z,\t}$ of all $\t'\in \TT_{\tilde\l}$, obtained by $\t$ in the
following way:

\0 1) we substitute the endpoint $v^*$ of type $\tilde\l$ of $\t$
with an endpoint of type $\l$;

\0 2) we link the endpoint $v^*$ to an endpoint of type $\tilde z$
trough a renormalized propagator of scale $h$.

Note that $\TT_{\tilde z} = \cup_{\t\in \TT_{\tilde\l}}
\TT_{\tilde z,\t}$ and that, if $\t$ has $n$ endpoints, any
$\t'\in \TT_{\tilde z,\t}$ has $n+1$ endpoints. Moreover, since
the value of $\bar\kk_4$ has be chosen so that $f_h(\bar\kk_4)=1$,
$\hat g_-^{(h)}(\bar\kk_4)=Z_{h-1}^{-1} \hat g_-(\bar\kk_4)$;
hence it is easy to show that the sum of the values of a tree
$\t\in \TT_{\tilde\l}$, such the special endpoint has scale index
$j^*+1$, and of all $\t'\in \TT_{\tilde z,\t}$ is obtained from
the value of $\t$, by substituting $\tilde\l_{j^*}$ with
$$\L_{j^*} = \tilde\l_{j^*} - \l_{j^*}
{\sum_{j=h}^{-1} \tilde z_j Z_j\over Z_{h-1}}\;,\Eqa(A6.3.45a)$$
see Fig. \graf(5vd).
\midinsert
\*
\insertplotbm{300pt}{100pt}%
{\ins{120pt}{60pt}{$-$}
\ins{72pt}{53pt}{$\tilde\l_{j^*}$}
\ins{222pt}{53pt}{$\l_{j^*}$}
\ins{235pt}{70pt}{$\hat g^{(h)}$}
}%
{lambda}{}
\vskip.5truecm
\line{\vtop{\line{\hskip3.3truecm\vbox{\advance\hsize by -4.0 truecm
\0{\css \eqg(5vd).}
{\ottorm The resummation of \equ(A6.3.45a).}
\hfill} }}}
\*
\endinsert
On the other hand, \equ(A6.3.45) and the bound $Z_j\le
\g^{-C\bar\l_h^2 j}$, see \equ(5.40z), imply that, if $\bar\l_h$ is
small enough
$$\sum_{j=h}^{-1} \left| \tilde z_j Z_j - \a z_j Z_j
\right|\le \sum_{j=h}^{-1} C \bar\l_h \g^{\th j} Z_j \le
C\bar\l_h\;.\Eqa(A6.3.47)$$
It follows, by using also the bound \equ(A6.3.38a), that
$$\L_{j^*} = \a\l_{j^*} \left[1 -
{\sum_{j=h}^{-1} z_j Z_j\over Z_{h-1}} \right] + {O(\bar\l_h)\over
Z_h}\;.\Eqa(A6.3.45b)$$
Moreover, since $Z_{j-1}=Z_j (1+z_j)$, for $j\in[-1,h]$, and
$Z_{-1}=1$, it is easy to check that
$$Z_{h-1} - \sum_{j=h}^{-1} z_j Z_j= 1\;.\Eqa(A6.3.48)$$
This identity, Lemma A6.1 and \equ(A6.3.45b) imply the bound
$$|\L_{j^*}| \le C {\bar\l_h\over Z_h}\;,\Eqa(A6.3.48a)$$
which gives us the ``missing'' $Z_h^{-1}$ factor for the sum over
the trees whose special endpoint is of type $\tilde\l$ or $\tilde
z$.

\* Let us now consider the family $\TT_{\n}$ of the trees with a
special endpoint of type $T_\pm$. It is easy to see, by using
the exponential decay of the $\n_{j,\o}$ and the ``short memory property'', 
that the sum over
the trees of this class with $n\ge 0$ normal endpoints is bounded,
for $\bar\l_h$ small enough, by $(C\bar\l_h)^{n+1} Z_h^{-1}
\g^{-4h} \sum_{j=h}^{-1} Z_j^{-2} \g^{2\th (h-j)}$ $\g^{\th j} \le
(C\bar\l_h)^{n+1} Z_h^{-3} \g^{-(4-\th)h}$, which is even better
of our needs.

We still have to consider the family $\TT_1$ of the trees with a
special endpoint of type $T_1$. There is first of all the trivial
tree, obtained by contracting all the $\psi$ lines of $T_1$ on
scale $h$, but its value is $0$, because of the support properties
of the function $\tilde\c(\pp)$. Let us now consider a tree $\t\in
\TT_1$ with $n\ge 1$ endpoints of type $\l$. 
If we call $h_{v_1}=j_1 + 1$ the scale of the vertex $T_1$, then the
dimensional bound of this tree differs from that of a tree with
$n+1$ normal endpoints contributing to $G^4_+(\bar\kk_1,
\bar\kk_2, \bar\kk_3, \bar\kk_4)$ for the following reasons:

\0 1) there is a factor $Z_h^{-1}$ missing, because the external
(renormalized) propagator of scale $h$ and momentum $\bar\kk_4$ is
substituted with the free propagator $\hat g_-(\bar\kk_4)$;

\0 2) there is a factor $|\l_{j_1}| Z_{j_1}^2$ missing, because
there is no external field renormalization in the
$T_1(\psi^{[h,j]})$ contribution to $\bar\VV_J^{(j)}(\psi^{[h,j]})$,
see \equ(A6.3.31b);

\0 3) there is a factor $Z_h^{-1}$ missing, because the 
factor $\tilde Z_{h-1}(\kk^-)$ in the r.h.s. of \equ(A7.4.20) can only
be bounded by a constant, because $\tilde Z_{h-1}(\kk^-)$ is in 
general different from $Z_{h-1}$ on the support of $f^{(h)}$.

It follows that the sum of the values of all trees $\t\in \TT_1$
with $n\ge 1$ normal endpoints, if $\bar\l_h$ is small enough, is
bounded by $(C\bar\l_h)^n \g^{-4h} \sum_{j_1=h}^0 Z_{j_1}^{-2}
\g^{2\th(h-j_1)}$ $\le (C\bar\l_h)^n \g^{-4h} Z_h^{-2}$.

By collecting all the previous bounds, we prove that the bound
\equ(A6.2.26) is satisfied in the case of
$H_+^{4,1}$.\\
\\
{\it Remark.} 
In $T_1$ and in the Grassmannian monomials
multiplying $\n_{j,+}, \n_{j,-}$, an external line is always
associated to a free propagator $\hat g_-(\bar\kk_4)$; this is due
to the fact that, in deriving the Dyson equation, one
extracts a free propagator. Then in the bounds there is a $Z_h$
missing (such propagator is not ``dressed'' in the multiscale
integration procedure), and at the end the crucial identity
\equ(A6.3.48) has to be used to ``dress'' the extracted propagator
carrying momentum $\bar\kk_4$.\\
\\
\asub(A6.3.11) 
We finally describe the modifications to the discussion 
above needed to bound $H_-^{4,1}$.

If we substitute, in the l.h.s. of \equ(A6.3.1) $H_+^{4,1}$ with
$H_-^{4,1}$, we can proceed in a similar way. By using
\equ(A6.2.17a), we get
$$\hat g_-(\kk_4) {1\over L\b}\sum_{\pp} \tilde\chi_M(\pp)
D_+^{-1}(\pp) \hat H^{4,1}_{-}(\pp;\kk_1,\kk_2,\kk_3,\kk_4-\pp) =
$$
$$=\hat g_-(\kk_4) {1\over L\b}\sum_{\pp} \tilde\chi_M(\pp)
{1\over L\b}\sum_{\kk} {C_-(\kk,\kk-\pp)\over D_+(\pp)}
<\hat\psi^+_{\kk,-}\hat\psi^-_{\kk-\pp,-};
\hat\psi^-_{\kk_1,+};\hat\psi^+_{\kk_2,+}; \hat\psi^-_{\kk_3,-};
\hat\psi^+_{\kk_4-\pp,-}>^T+$$
$$-\nu'_-\hat g_-(\kk_4) {1\over L\b}\sum_{\pp} \tilde\chi_M(\pp)
{1\over L\b}\sum_{\kk} {D_-(\pp)\over D_+(\pp)}
<\hat\psi^+_{\kk,-}\hat\psi^-_{\kk-\pp,-};
\hat\psi^-_{\kk_1,+};\hat\psi^+_{\kk_2,+}; \hat\psi^-_{\kk_3,+};
\hat\psi^+_{\kk_4-\pp,-}>^T-$$
$$-\nu'_+\hat g_-(\kk_4) {1\over L\b} \sum_{\pp} \tilde\chi_M(\pp)
 {1\over L\b}\sum_{\kk} <\hat\psi^+_{\kk,+}\hat\psi^-_{\kk-\pp,+};
\hat\psi^-_{\kk_1,+};\hat\psi^+_{\kk_2,+}; \hat\psi^-_{\kk_3,-};
\hat\psi^+_{\kk_4-\pp,-}>^T\;. \Eqa(A6.3.1a)$$
We define $\widetilde G^4_-(\kk_1, \kk_2,\kk_3, \kk_4)$ as in
\equ(A6.3.2) with $\widetilde\WW$ replaced by $\widetilde\WW_-$ given by
$$\eqalign{&
\widetilde\WW_- = \log \int P(d\hat\psi)e^{-T_2(\psi) + \n'_+T_+(\psi)
+ \n'_-T_-(\psi)} e^{-V(\hat\psi)+ \sum_\o\int d\xx
[\phi^+_{\xx,\o}\hat\psi^{-}_{\xx,\o}+
\hat\psi^{+}_{\xx,\o}\phi^-_{\xx,\o}]} \;,\cr
&T_2(\psi) = {1\over M^2} \sum_{\pp} \tilde\chi_M(\pp)
{1\over M^2} \sum_{\kk} {C_-(\kk, \kk-\pp) \over D_+(\pp)}
(\hat\psi_{\kk,-}^+ \hat\psi_{\kk-\pp,-}^-)
\hat\psi^+_{\kk_4-\pp,-} \hat J_{\kk_4} \hat
g(\kk_4)\;,\cr}\Eqa(A6.3.4aa)$$
$T_+, T_-$ being defined as in \equ(A6.3.6). By the
analogues of \equ(A6.3.7b), \equ(A6.3.7c) we obtain
$$\widetilde G^4_-(\bar\kk_1,\bar\kk_2,\bar\kk_3,\bar\kk_4)= 
\hat g_-(\bar\kk_4){1\over M^2}
\sum_{\pp} \tilde\chi_M(\pp)
{\hat H_-^{4,1}(\pp;\bar\kk_1,\bar\kk_2,\bar\kk_3,\bar\kk_4-\pp)\over
D_+(\pp)}\;.\Eqa(A6.3.7ca)$$
The calculation of $\widetilde G^4_-(\bar\kk_1,\bar\kk_2,\bar\kk_3,
\bar\kk_4)$ is done
via a multiscale expansion essentially identical to the one of
$\widetilde G^4_+(\bar\kk_1,\bar\kk_2,\bar\kk_3,\bar\kk_4)$, 
by taking into account
that $\d\r_{\pp,+}$ has to be substituted with
$$\d\r_{\pp,-}={1\over M^2}\sum_\kk{ C_-(\pp,\kk)\over D_+(\pp)}
(\hat\psi_{\kk,-}^+ \hat\psi_{\kk-\pp,-}^-)\;.\Eqa(A6.3.8a)$$
Let us consider the first step of the iterative integration
procedure and let us call again $\bar\VV_J^{(-1)}(\psi^{[h,-1]})$
the contribution to the effective potential of the terms linear in
$J$. Let us now decompose $\bar\VV_J^{(-1)}(\psi^{[h,-1]})$ as in
\equ(A6.3.19) and let us consider the terms contributing to
$\bar\VV^{(-1)}_{J,a,1}(\psi^{[h,-1]})$. The analysis goes exactly
as before when no one or both the fields $\hat\psi_{\kk,-}^+$ and
$\hat\psi_{\kk-\pp,-}^-$ of $\d\r_{\pp,-}$ are contracted. This is
not true if only one among the fields $\hat\psi_{\kk,-}^+$ and
$\hat\psi_{\kk-\pp,-}^-$ in $T_2(\psi)$ is contracted, since in
this case there are marginal terms with two external lines, which
before were absent. The terms with four external lines can be
treated as before; one has just to substitute $D_+(\kk^-) \hat
g_+^{(0)}(\kk^+)$ with $D_-(\kk^-) g_-^{(0)}(\kk^+)$ in the r.h.s.
of \equ(A6.3.24), but this has no relevant consequence. The terms
with two external lines have the form
$$\eqalign{
& \int d\kk^- \hat\psi^+_{\bar\kk_4,-} \hat g_-(\bar\kk_4) \hat
J_{\bar\kk_4} \tilde\chi_M(\bar\kk^4-\kk^-) G_1^{(0)}(\kk_-)
\left\{ {[C_{h,0}^\e(\bar\kk_4)-1] D_{-}(\bar\kk_4) \hat
g^{(0)}_-(\kk^-)\over D_+(\bar\kk_4-\kk^-)} - {u_0(\kk^-) \over
D_+(\bar\kk^4-\kk^-)} \right\}\;, \cr}\Eqa(A6.3.24b)$$
where $G_1^{(0)}(\kk_-)$ is a smooth function of order $0$ in
$\l$. However, the first term in the braces is equal to $0$, since
$|\bar\kk_4|=\g^h$ implies that $C_{h,0}^\e(\bar\kk_4)-1=0$. Hence
the r.h.s. of \equ(A6.3.24b) is indeed of the form
$$\int d\kk^- \hat\psi^+_{\bar\kk_4,-}
\hat g_-(\bar\kk_4) \hat J_{\bar\kk_4} \tilde\chi_M(\bar\kk_4 -
\kk^-) G_1^{(0)}(\kk_-){u_0(\kk^-) \over D_+(\bar\kk^4-\kk^-)}
\;,\Eqa(A6.3.24c)$$
so that it can be regularized in the usual way.

The analysis of $\bar\VV^{(-1)}_{J,a,2}(\psi^{[h,-1]})$ can be done
exactly as before. Hence, we can define again $\tilde\l_{-1}$ and
$\tilde z_{-1}$ as in \equ(A6.3.25a), with $\tilde\l_{-1}=O(\l)$ and
$\tilde z_{-1}=O(1)$.

Let us consider now the terms contributing to
$\bar\VV^{(-1)}_{J,b,1}$, that is those where
$\hat\psi^+_{\bar\kk_4-\pp}$ is not contracted and there is a
vertex of type $T_2$. Again the only marginal terms have four
external lines and have the form
$$\eqalign{
&\sum_{\tilde \o} \int d\pp \tilde\chi_M(\pp)
\hat\psi^+_{\kk^+,\tilde\o} \int d\kk^+
\hat\psi^+_{\kk^+-\pp,\tilde\o} \hat\psi^+_{\bar\kk_4-\pp,-} \hat
g_-(\bar\kk_4) \hat J_{\bar\kk_4}\;\cdot\cr
&{D_-(\pp) \over D_+(\pp)} \left[ F^{(-1)}_{2,-,\tilde\o}(\kk^+,
\kk^+-\pp) + F^{(-1)}_{1,-}(\kk^+, \kk^+-\pp) \d_{-,\tilde\o}
\right]\;,\cr} \Eqa(A6.3.28aa)$$
where we are using again a definition analogue to \equ(A6.132). 
The analysis of the terms
$F^{(-1)}_{1,-}(\kk^+, \kk^+-\pp)$ is identical to the one in
\sec(A6.3.3), while, the symmetry property of the propagator
under the replacement $\kk\to\kk^*$ implies now that, if we define
$$F^{-1}_{2,-,\tilde\o}(\kk^+, \kk^-) = {1\over D_-(\pp)}
\left[ p_0 A_{0,-,\tilde\o}(\kk^+,\kk^-) + p_1
A_{1,-,\tilde\o}(\kk^+,\kk^-) \right]\;,\Eqa(A6.3.29aa)$$
and
$$\LL F^{-1}_{2,-,\tilde\o}={1\over D_-(\pp)}
\left[ p_0 A_{0,-,\tilde\o}(0,0) +p_1 A_{1,-,\tilde\o}(0,0)
\right]\;,\Eqa(A6.3.29ab)$$
then
$$\LL F^{-1}_{2,-,+}= Z_{-1}^{3,-} {D_-(\pp) \over D_+(\pp)}
\virg \LL F^{-1}_{2,-,-}= Z_{-1}^{3,+}\;,\Eqa(A6.3.30aa)$$
where $Z_{-1}^{3,+}$ and $Z_{-1}^{3,-}$ are the same real
constants appearing in \equ(A6.3.30). Hence, the local part of the
marginal term \equ(A6.3.28aa) is, by definition, equal to
$$Z_{-1}^{3,+} T_+(\psi^{[h,-1]}) + Z_{-1}^{3,-}
T_-(\psi^{[h,-1]})\;.\Eqa(A6.3.30ab)$$
The analysis of $\bar\VV^{(-1)}_{J,b,2}$ can be done exactly as
before, so that we can write for $\bar\VV_J^{(-1)}$ an expression
similar to \equ(A6.3.31a), with $T_2(\psi^{[h,-1]})$ in place of
$T_1(\psi^{[h,-1]})$ and $\n'_{-1,\pm}$ in place of $\n_{-1,\pm}$.

The integration of higher scales proceed as in \sec(A6.3.6). In fact,
the only real difference we found in the integration of the first
scale was in the calculation of the $O(1)$ terms contributing to
$\tilde z_{-1}$, but these terms are absent in the case of $\tilde
z_j$, $j\le -2$, because the second term in the expression
analogous to \equ(A6.3.24b), obtained by contracting on scale $j<0$
only one of the fields of $\d\r_{\pp,-}$, is exactly zero.
Also in this case, the constants $\n'_\o$ can be
chosen again so that the an exponentially decaying 
bound is satisfied even by the constants $\n'_{j,\o}$.

In the analysis of the constants $\tilde\l_j$ and $\tilde z_j$
there is only one difference, concerning the bound \equ(A6.3.45),
which has to be substituted with $\tilde z_{-1}-\a z_{-1} \le C$,
in the case $j=-1$, but it is easy to see that this has no effect
on the bound \equ(A6.3.48a). It follows that the final considerations
of \sec(A6.3.10) stay unchanged and we get for $\widetilde
G^4_-(\kk_1,\kk_2,\kk_3,\kk_4)$ a bound similar to that proved for
$\widetilde G^4_+(\kk_1,\kk_2,\kk_3,\kk_4)$, so ending the proof of
the bound \equ(A6.2.26).

\pagina
\setcap{appendix a7: The properties of {$D_\o(\pp)^{-1}
C_\o(\kk,\kk-\pp) $}.}
\appindex{A7}{The properties of {$D_\o(\pp)^{-1}
C_\o(\kk,\kk-\pp) $}.}
\vskip1.truecm
\appendix(A7, The properties of {$D_\o(\pp)^{-1}
C_\o(\kk,\kk-\pp) $}.)

In this Appendix we describe and collect a number of properties
of the operator $D_\o(\pp)^{-1}
C_\o(\kk,\kk-\pp) $, useful in the analysis of the correction identities.
We follow the analogue discussion in section \S4.2 of [BM3].
 
Let us consider the quantity
$$\eqalign{& \D^{(i,j)}_\o(\kk^+,\kk^-)={C_\o(\kk^+,\kk^-)\over D_\o(\pp)}
\hat g^{(i)}_\o(\kk^+) \hat g^{(j)}_\o(\kk^-)=\cr
&={1\over Z_{i-1}Z_{j-1}} {1\over D_\o(\pp)}
\Bigg\{ {\tilde f_i(\kk^+)\over D_\o(\kk^+)}
\Big[ {\tilde f_j(\kk^-)\over \chi_{h,0}(\kk^-) }
-\tilde f_j(\kk^-)\Big]- {\tilde f_j(\kk^-)\over D_\o(\kk^-)}
\Big[ {\tilde f_i(\kk^+)\over \chi_{h,0}(\kk^+) }- \tilde f_i(\kk^+)
\Big] \Bigg\}\;,\cr}\Eqa(A7.4.5)$$
where $\pp=\kk^+-\kk^-$. The above quantity appears in the expansion
for $\hat H_{2,1}$ when both the fields of $T_{\xx,\o}$
are contracted.
Note first that
$$ \D^{(i,j)}_\o(\kk^+,\kk^-)=0\;,\qquad \hbox{if\ } 0>i,j>h\;,\Eqa(A7.4.6)$$ 
since $\chi_{h,0}(\kk^\pm)=1$, if $h<i,j<0$. We will see that this property
plays a crucial role; it says that, contrary to what happens for $G^{2,1}$, at
least one of the two fermionic lines connected to $J$ must have scale $0$ or
$h$.

In the the cases in which $\D^{(i,j)}_\o(\kk^+,\kk^-)$ is not identically
equal to $0$, since $\D^{(i,j)}_\o(\kk^+,\kk^-)=\D^{(j,i)}_\o(\kk^-,\kk^+)$,
we can restrict the analysis to the case $i\ge j$.
 
\vskip.5cm
\0 1) If $i=j=0$, \equ(A7.4.5) can be rewritten as
$$ \D^{(0,0)}_\o(\kk^+,\kk^-)={1\over D_\o(\pp)} \left[ {f_0(\kk^+)
\over D_\o(\kk^+)}
u_0(\kk^-)- {f_0(\kk^-)\over D_\o(\kk^-)} u_0(\kk^+) \right]\;,\Eqa(A7.4.7)$$ 
where $u_0(\kk)$ is a $C^\io$ function such that
$$ u_0(\kk)=\cases{0 & if $|\kk|\le 1$ \cr 1-f_0(\kk) & if $1\le |\kk|$ \cr}\;.
\Eqa(A7.4.8)$$

We want to show that
$$ \D^{(0,0)}_\o(\kk^+,\kk^-) = {\pp\over D_\o(\pp)} {\bf 
S}^{(0)}_\o(\kk^+,\kk^-
)=
{p_0 S^{(0)}_{\o,0}(\kk^+,\kk^-) + p S^{(0)}_{\o,1}(\kk^+,\kk^-) \over
D_\o(\pp)}\;,
\Eqa(A7.4.9)$$
where $S^{(0)}_{\o,i}(\kk^+,\kk^-)$ are smooth functions such that
$$ |\dpr_{\kk^+}^{m_+} \dpr_{\kk^-}^{m_-} S^{(0)}_{\o,i}(\kk^+,\kk^-)|\le
C_{m_+ +m_-}\;, \Eqa(A7.4.10)$$
if $\dpr_\kk^m$ denotes a generic derivative of order $m$ with respect to
the variables $\kk$ and $C_m$ is a suitable constant, depending on $m$.
 
The proof of \equ(A7.4.9) is trivial if $\pp$ is bounded away from $0$, for
example $|\pp|\ge 1/2$. It is sufficient to remark that
$\D^{(0,0)}_\o(\kk^+,\kk^-)$, by the compact support properties of $f_0(\kk)$,
is a smooth function and put $S^{(0)}_{\o,0}=-i\D^{(0,0)}_\o$,
$S^{(0)}_{\o,1}=\o\D^{(0,0)}_\o$. If $|\pp|\le 1/2$, we can use the identity
$$\eqalign{&
\qquad\qquad \D^{(0,0)}_\o(\kk^+,\kk^-)= -{f_0(\kk^+) u_0(\kk^+)\over
D_\o(\kk^+) D_\o(\kk^-)} +\cr
&+ {\pp\over D_\o(\pp)} \int_0^1 dt {\kk^+-t\pp\over|\kk^+- t\pp|}
\left[ f'_0(\kk^+- t\pp) {u_0(\kk^+) \over D_\o(\kk^-)}-
u'_0(\kk^+ - t\pp) {f_0(\kk^+)\over D_\o(\kk^+)} \right]\;,\cr}\Eqa(A7.4.11)$$
from which \equ(A7.4.10) follows. 
 
\vskip.5cm
\0 2) If $i=0$ and $h\le j<0$, we get
$$ \D^{(0,j)}_\o(\kk^+,\kk^-)=-{1\over Z_{j-1}}{\tilde f_j(\kk^-) 
u_0(\kk^+)\over
D_\o(\pp) D_\o(\kk^-)} + \d_{j,h} {1\over \tilde Z_{h-1}(\kk^-)}
{f_0(\kk^+) u_h(\kk^-)\over D_\o(\pp) D_\o(\kk^+)}\;,\Eqa(A7.4.12)$$ 
where
$$ u_h(\kk)=\cases{0 & if $|\kk|\ge \g^h$ \cr 1-f_h(\kk) & if $|\kk|\le \g^h$
\cr}\;.\Eqa(A7.4.13)$$ 
If $j<-1$, the first term in the r.h.s. of \equ(A7.4.12) vanishes for 
$|\pp|\le 1-\g^{-1}$, since $u_0(\kk^+)\not=0$ 
implies that $|\kk^+|\ge 1$, so that
$|\kk^-|=|\kk^+-\pp|\ge 1-(1-\g^{-1})=\g^{-1}$ and, as a consequence, $\tilde
f_j(\kk^-)=0$. Analogously, the second term in the r.h.s. of \equ(A7.4.12)
vanishes for $|\pp|\le 1-\g^{-1}-\g^h$, since $f_0(\kk^+)\not=0$ implies that
$|\kk^+|\ge 1-\g^{-1}$, so that $|\kk^-|\ge \g^h$ and, as a consequence,
$u_h(\kk^-)=0$. On the other hand, if $j=-1$, because $\tilde f_{-1}(\kk)
u_0(\kk)=0$, we can write
$$ u_0(\kk^+)\tilde f_{-1}(\kk^-)=-u_0(\kk^+)\; \pp \int_0^1 dt {\kk^+-t
\pp\over|\kk^+-t\pp|}\tilde f'_{-1}(\kk^+-t\pp)\;.\Eqa(A7.4.14)$$ 
It follows that
$$ \D^{(0,j)}_\o(\kk^+,\kk^-)= {\pp\over D_\o(\pp)} {\bf 
S}_\o^{(j)}(\kk^+,\kk^-)
\;,\Eqa(A7.4.15)$$ 
where $S^{(j)}_{\o,i}(\kk^+,\kk^-)$ are smooth functions such that
$$ |\dpr_{\kk^+}^{m_0} \dpr_{\kk^-}^{m_j} S^{(j)}_{\o,i}(\kk^+,\kk^-)|\le
C_{m_0 +m_j} {\g^{-j(1+m_j)}\over \tilde Z_{j-1}(\kk^-)}\;,\quad h\le j<0\;.
\Eqa(A7.4.16)$$

\vskip.5cm
\0 3) If $i=j=h$ we get
$$\eqalign{& 
\D^{(h,h)}_\o(\kk^+,\kk^-)={1\over D_\o(\pp)}{1\over \tilde Z_{h-
1}(\kk^+)
\tilde Z_{h-1}(\kk^-)}\cdot\cr
&\cdot \left[{f_h(\kk^+)u_h(\kk^-)\over D_\o(\kk^+)}-
{u_h(\kk^+)f_h(\kk^-)\over D_\o(\kk^-)}\right]\;.\cr}\Eqa(A7.4.17)$$
Since this expression can appear only at the last integration step, it is
not involved in any regularization procedure. Hence we only need its size for
values of $\pp$ of order $\g^h$ or larger. It is easy to see that
$$ |\D^{(h,h)}_\o(\kk^+,\kk^-)|\le {C\over M}{\g^{-2h}\over
\tilde Z_{h-1}(\kk^+) \tilde Z_{h-1}(\kk^-)}\;,\quad \hbox{if\ }
|\pp|\ge M\g^h\;.\Eqa(A7.4.18)$$ 

\vskip.5cm
\0 4) If $j=h<i<-1$, we get
$$ \D^{(i,h)}_\o(\kk^+,\kk^-)= {1\over \tilde Z_{h-1}(\kk^-) Z_{i-1}}
{\tilde f_i(\kk^+) u_h(\kk^-) \over D_\o(\pp) D_\o(\kk^+)}\;,\Eqa(A7.4.19)$$ 
which satisfies the bound
$$ |\D^{(i,h)}_\o(\kk^+,\kk^-)|\le {C\over M}{\g^{-h-i}\over
\tilde Z_{h-1}(\kk^-) Z_{i-1}}\;,\quad 
\hbox{if\ }|\pp|\ge M\g^h\;.\Eqa(A7.4.20)$$

\pagina
\setcap{appendix a8: Proof of Lemma 7.3.}
\appindex{A8}{Proof of Lemma 7.3.}
\vskip1.truecm
\appendix(8, Proof of Lemma 7.3)

Proceeding as in Chapter 6, we first solve the equations for
$Z_h$ and $\widehat m^{(2)}_h$ parametrically in 
$\underline\p=\{\p_h\}_{h\le h^*_1}$.
If $|\p_h|\le c|\l|\g^{(\th/2)
(h-h^*_1)}$, the first two assumptions of \equ(7.lem5.2) easily follow.
Now we will construct a sequence $\underline\p$ such that 
$|\p_h|\le c|\l|\g^{(\th/2)
(h-h^*_1)}$ and satisfying the 
flow equation $\p_{h-1}=\g^h\p_h+\b^h_\p(\p_h,\ldots,
\p_{h^*_1})$.\\

\asub(7.pi1) {\it Tree expansion for {$\b^h_\p$}.}
$\b^h_\p$ can be expressed
as sum over tree diagrams, similar to those used in \sec(5.5). 
The main difference is that they
have vertices on scales $k$ between $h$ and $+2$. The vertices on scales 
$h_v\ge h^*_1+1$ are associated to the truncated expectations \equ(3.28a); 
the vertices on scale $h_v=h^*_1$ 
are associated to truncated expectations w.r.t. the propagators
$g^{(1,h^*_1)}_{\o_1,\o_2}$; the vertices on scale $h_v<h^*_1$ 
are associated to truncated expectations w.r.t. the propagators
$g^{(2,h_v+1)}_{\o_1,\o_2}$. Moreover the end--points
on scale $\ge h^*_1+1$ are associated to the couplings $\l_h$ or $\n_h$,
as in \sec(5.5); 
the end--points on scales $h\le h^*_1$ are necessarily associated to the
couplings $\p_h$.\\

\asub(7.pi2) {\it Bounds on {$\b^h_\p$}.}
The non vanishing trees contributing to $\b^h_\p$ must have at least 
one vertex on scale $\ge h^*_1$: in fact the diagrams depending only on 
the vertices of type $\p$ are vanishing (they are chains, 
so they are vanishing, because of the compact support property
of the propagator). This means that, by the short 
memory property: $|\b^h_\p|\le c|\l|\g^{\th(h-h^*_1)}$.\\

\asub(7.pi3) {\it Fixing the counterterm.}
We now proceed as in Chapter 6 
but the analysis here is easier, because no $\l$
end--points can appear and the bound $|\b^h_\p|\le c|\l|\g^{\th(h-h^*_1)}$
holds.
As in Chapter 6, we can formally consider the flow equation 
up to $h=-\io$, even if $h^*_2$ is a finite integer. This
is because the beta function is independent of 
$\widehat m^{(2)}_k$, $k\le h^*_1$
and admits bounds uniform in $h$. If we want to fix 
the counterterm $\p_{h^*_1}$
in such a way that $\p_{-\io}=0$, we must have, for any $h\le h^*_1$:
$$\p_h=-\sum_{k\le h}\g^{k-h-1}\b^k_\p(\p_k,\ldots,\p_{h^*_1})\;.\Eqa(7.must)$$
Let $\tilde{\MMM}$ be the space of sequences $\underline\p=
\{\p_{-\io},\ldots,\p_{h^*_1}\}$ 
such that $|\p_h|\le c |\l| \g^{-(\th/2) (h-h^*_1)}$.
We look for a fixed point 
of the operator $\tilde{\bT}:\tilde{\MMM}
\to\tilde{\MMM}$ defined as:
$$(\tilde\bT\underline\p)_h=-\sum_{k\le h}\g^{k-h-1}
\b^k_\p(\p_k;\ldots;\p_{h^*_1})\,.\Eqa(7.5.8c0)$$
Using that $\b^k_\p$ is independent from $\hat m_k^{(2)}$
and
the bound on the beta function, choosing $\l$ small enough and proceeding
as in the proof of Theorem 6.1, we find that $\tilde\bT$ is a contraction
on $\tilde{\MMM}$, so that we find a unique fixed point, and the first
of \equ(7.4.58b) follows.\\

\asub(7.pi4) {\it The flows of $Z_h$ and $\widehat m^{(2)}_h$.}
Once that $\p_{h^*_1}$ is fixed via the iterative procedure of \sec(7.pi3),
we can study in more detail the flows of $Z_h$ and $\widehat m^{(2)}_h$ 
given by \equ(7.flow). Note that $z_h$ and $s_h$ can be again expressed as 
a sum over tree diagrams and, as discussed for $\b^h_\p$, see \sec(7.pi2),
any non vanishing diagram must have at least one vertex on scale $\ge h^*_1$. 
Then, by the short memory property, see \sec(short), we have 
$z_h=O(\l^2\g^{\th(h-h^*_1)})$ and $s_h=O(\l\widehat m^{(2)}_h\g^{\th
(h-h^*_1)})$ and, repeating the proof of Lemma 6.1,
we find the second and third of \equ(7.4.58b).\\

\asub(7.pi6){\it The Lipshitz property {\equ(7.ph)}.}
Clearly,  
$\p_{h^*_1}^*(\l,\s_1,\m_1)-\p^*_{h^*_1}(\l,\s_1',\m_1')$ can be expressed
via a tree expansion similar to the one discussed above;
in the trees with non vanishing value, there 
is either a difference of propagators at scale $h\ge h^*_1$
with couplings $\s_h,\m_h$ and  $\s'_h,\m'_h$, giving
in the dimensional bounds
an extra factor
$O(|\s_h-\s_h'| \g^{-h})$ or $O(|\m_h-\m_h'| \g^{-h})$; or a 
difference of propagators at scale $h\le h^*_1$
(computed by definition at $\widehat m^{(2)}_h=0$)
with the ``corrections'' $a^\o_h,c_h$ associated to $\s_1,\m_1$ or 
$\s'_1,\m'_1$, giving
in the dimensional bounds
an extra factor
$O(|\s_1-\s_1'|)$ or $O(|\m_1-\m_1'|)$. Then, 
$$\eqalign{&\Big|\p_{h^*_1}(\l,\s_1,\m_1)-\p_{h^*_1}(\l,\s_1',\m_1')\Big|
\le c|\l|\sum_{k\le h^*_1}\g^{k-h^*_1-1}\cdot\cr
&\qquad\cdot \Big[ \sum_{h\ge h^*_1}
\left({|\s_h-\s_h'|\over \g^h}+{|\m_h-\m_h'|\over \g^h}\right)+
\sum_{k\le h\le h^*_1}\big(|\s_1-\s_1'|+|\m_1-\m_1'|\big)\Big]\;,\cr}
\Eqa(7.last)$$
from which, using \equ(6.mu) and \equ(6.sigma), we easily get \equ(7.ph).

\pagina
\setcap{appendix a9: Independence from boundary conditions.}
\appindex{A9}{Independence from boundary conditions.}
\vskip1.truecm
\appendix(9, Independence from boundary conditions.)

In this Appendix we prove that the limit $\lim_{M\to\io}{1\over M^2}\log\Xi^{
\g_1,\g_2}_{AT}$ considered in \sec(7.an) is independent of the 
boundary conditions $\g_1,\g_2$, in particular we 
prove that there exist a constants $C,c>0$ such that 
$$\Big|\log{\Xi^{
\g_1,\g_2}_{AT}\over \Xi^-_{AT}}\Big|\le C e^{-c M \g^{h^*_2}}
\;,\Eqa(A9.1)$$
where we recall that $\Xi^-_{AT}$ is the partition function with 
antiperiodic boundary conditions in all directions and $h^*_2$ is the scale
introduced in \sec(7.6.101). Note that, if $\g^{h^*_2}>0$, as we are assuming, 
the propagator $g^{(\le 1)}$ of the $\psi$ field has a mass $O(\g^{h^*_2})$. 
The analysis of this Appendix is based on the analogue analysis in Appendix 
G of [M].
 
By using the construction and the definitions 
in \sec(3.2)--\sec(3.3), we can write
$$\log \Xi^{
\g_1,\g_2}_{AT}= \int P_{\g_1}^{(1)}(d\psi^{(1)},d\c^{(1)})
P_{\g_2}^{(2)}(d\psi^{(2)},d\c^{(2)})
e^{\tilde\l V(\psi,\c)}\,,\Eqa(A9.2)$$
where $P_{\g_j}^{(j)}$ are defined as in \equ(4.26) with 
$P_\s(d\psi)$ in the l.h.s. of \equ(4.26) replaced by $P(d\psi)$
and the $\g_j$--boundary conditions replacing the antiperiodic ones.

Proceeding as in Chapter 4 and 5, we see that 
$\log \Xi^{
\g_1,\g_2}_{AT}$ can be written as sum of terms of the form 
$\sum_{\xx_1,\ldots,\xx_n}
W_{\g_1,\g_2}(\xx_1,\ldots,\xx_n)$, with $\xx_i$ varying in
$[-{M\over 2},{M\over 2}]\times[-{M\over 2},{M\over 2}]$ and the
$W$ are truncated expectations for which a Pfaffian expansion like \equ(4.12)
holds. 
Note that $W(\xx_1,\ldots,\xx_n)$ is periodic with period $M$
in any of its coordinates, for any $\g_1,\g_2$; this follows
from the fact that there is an even number
of $\psi,\chi$ fields associated to any $\xx_i$. Moreover 
$W(\xx_1,\ldots,\xx_n)$ is translation invariant, so that we can fix 
one variable to the origin $\V0$, for instance $\xx_1$:
$$\sum_{\xx_1,\ldots,\xx_n}
W_{\g_1,\g_2}(\xx_1,\ldots,\xx_n)=\sum_{\xx_1,\ldots,\xx_n}
W_{\g_1,\g_2}({\bf 0},\xx_2,\ldots,\xx_n)\;.\Eqa(A9.3)$$
We can write $\sum_{\xx_1,\ldots,\xx_n} W$
as $\sum^*_{\xx_1,\ldots,\xx_n}W+ \sum^{**}_{\xx_1,\ldots,\xx_n}W$,
where $\sum^*_{\xx_1,\ldots,\xx_n}$ is over 
the coordinates $\xx_i$ varying in
$[-{M\over 4},{M\over 4}]\times[-{M\over 4},{M\over 4}]$ and 
$\sum^{**}_{\xx_1,\ldots,\xx_n}W$ is the rest.
Then $\sum^{**}_{\xx_1,\ldots,\xx_n} W $ is $O(e^{-c\g^{h^*_2}M})$,
as in $W$ there is surely a chain of propagators
exponentially decaying connecting the point $\V0$
with a point outside
$[-{M\over 4},{M\over 4}]\times[-{M\over 4},{M\over 4}]$.

On the other hand in $\sum^{*}_{\xx_1,\ldots,\xx_n} W $ 
we can use the Poisson summation formula, stating
that
$${1\over M}\sum_{n=0}^{M-1}f({n2\pi\over M}+{\a\pi\over M})
=\sum_{n\in \zzz}\hat f(nM)(-1)^{\a n}\;,\Eqa(A9.4)$$
where $f$ is any smooth $2\pi$-periodic function and $\a=0,1$. From
\equ(A9.4) we find, if $g_{\L_M,\g_j}(\xx)$
is the propagator corresponding to $P_{\g_j}(d\psi^{(j)},d\c^{(j)})$:
$$\eqalign{&g_{\L_M,\g_j}(\xx)(\xx-\yy)=\sum_{\nn\in\zzz^2}
(\e_j')^n(\e_j)^{n_0}g(\xx-\yy+\nn M)\defin\cr
&\defin g(\xx-\yy)+\d g_{\g_j}(\xx-\yy)\;,\cr}\Eqa(A9.5)$$
where $g(\xx)=\lim_{M\to\io}
g_{\L_M,\g_j}(\xx)$, independen of boundary conditions.
Note that the only dependence 
on boundary conditions in the r.h.s. of \equ(A9.5)
is in $\d g_{\g_j}(\xx-\yy)$
and it holds, if
$|x-y|\le{M\over 2}$, $|x_0-y_0|\le{M\over 2}$:
$$|\d g(\xx-\yy)|\le C e^{-c_2 \g^{h^*_2} M}\;,\Eqa(A9.6)$$
with a proper constant $c_2$. Hence all the terms 
in $\sum^{*}_{\xx_1,\ldots,\xx_n} W $ with at least a 
$\d g(\xx-\yy)$ are exponentially bounded,
while the part with only 
$g(\xx-\yy)$ is independent from boundary conditions (and it 
cancels in the expansion for $\log(\Xi^{\g_1,\g_2}_{AT}/\Xi^{-}_{AT}$).
This proves \equ(A9.1).

\pagina
\setcap{Acknowledgments}
\bibindex{Acknowledgments}
\vskip1.truecm
\centerline{\titolo Acknowledgments}
\acapo\acapo
I would like to thank
Prof. J. L. Lebowitz for his nice invitations to visit Rutgers University
in the last two years. Part of the work I presented here
was done in the stimulating atmosphere of Rutgers University.\\
\\ 
Vorrei ringraziare il Prof. G. Gallavotti e il Prof. G. Benfatto
per la fiducia, la disponibilita' e l'interesse 
con cui mi hanno seguito in questi anni.\\
Ringrazio  
il Prof. G. Gentile e il Prof. F. Bonetto, con i quali ho collaborato 
durante tutto il corso del Dottorato, e che hanno contribuito moltissimo
alla mia formazione scientifica.\\
Un particolare ringraziamento va infine al Prof. V. Mastropietro:
gli ultimi tre anni di stretta collaborazione con lui sono stati per me  
importantissimi. Senza le sue idee ed il suo fondamentale contributo 
non sarebbe stato possibile realizzare questo lavoro. 

\pagina  
\setcap{References}
\bibindex{References}
\vskip1.truecm
\centerline{\titolo References}
\vskip.5truecm
\halign{\hbox to 1.5truecm {[#]\hss} &
        \vtop{\advance\hsize by -1.55 truecm \0#}\cr

AT&{J. Ashkin, E. Teller: Statistics of Two-Dimensional
Lattices with Four Components.
{\it Phys. Rev.} 64, 178-184 (1943). }\cr
Ai80&{M. Aizenman: Translation invariance and instability 
of phase coexistence in the two dimensional
Ising system, {\it Comm. Math. Phys.} 73, 83--94 (1980)}\cr
Ai82&{M. Aizenman: Geometric analysis of $\varphi \sp{4}$ 
fields and Ising models. I, II.  {\it Comm. Math. Phys.}  86  1--48 (1982)}\cr
BG&{G. Benfatto, G. Gallavotti:
Renormalization group. Physics notes 1, Princeton University
Press (1995). }\cr
BG1&{G. Benfatto, G. Gallavotti:
Perturbation Theory of the Fermi Surface in Quantum Liquid. A General
Quasiparticle Formalism and One-Dimensional Systems.
{\it J. Stat. Phys.} {\bf 59}, 541--664 (1990). }\cr
BGPS&{G. Benfatto, G. Gallavotti, A. Procacci, B. Scoppola:
Beta function and Schwinger functions for a Many Fermions System 
in One Dimension. {\it Comm. Math. Phys.} {\bf 160}, 93--171 (1994).}\cr
BM&{G. Benfatto, V.Mastropietro:
Renormalization group, hidden symmetries
and approximate Ward identities in the $XYZ$ model.
{\it Rev. Math. Phys.} 13 (2001), no. 11, 1323--143;
and {\it Comm. Math. Phys.} 231, 97-134 (2002)}\cr
BM1&{G. Benfatto, V. Mastropietro: Ward identities and 
Chiral anomaly in the Luttinger liquid, cond-mat/0409049}\cr
BM2&{G. Benfatto, V. Mastropietro: Ward Identities and 
Vanishing of the Beta Function for d = 1 Interacting Fermi Systems,
{\it Jour. Stat. Phys.} 115, 143--184 (2004)}\cr
BM3&{G. Benfatto, V. Mastropietro: On the density--density
critical indices in interacting Fermi systems, {\it Comm. Math. Phys.}
231, 97--134 (2002)}\cr
BMW&{E. Barouch, B. M. McCoy, T. T. Wu:
Zero--Field Susceptibility of the Two-Dimensional Ising Model near Tc,
{\it Phys. Rev. Lett.} 31, 1409--1411 (1973)}\cr
Ba&{R. J. Baxter: {\it Phys.Rev. Lett.} 26 (1971) 832;
{\it Phys. Rev. Lett.} 26 (1971) 834;
{\it Ann. Phys.} 70 (1972) 193, 323; (with M.N. Barber)
{\it J. Phys.} C 6 (1973) 2913;
{\it J. Stat. Phys.} 15 (1976) 485;
{\it J. Stat. Phys.} 17 (1977) 1}\cr
Ba82&{R. Baxter, Exactly solved models in statistical mechanics,
Academic Press (1982)}\cr
Bad&{M. Badehdah, S. Bekhechi, A. Benyoussef, M. Touzani:
Finite-size-scaling study of the anisotropic spin-${1\over 2}$ 
Ashkin--Teller model,
{\it Physica} B 291, 394--399 (2000)}\cr  
Bak&{P. Bak, P. Kleban, W. N. Unertl, J. Ochab, G. Akinci, N. C. Bartelt,
T. L. Einstein: Phase Diagram of Selenium Adsorbed on the Ni(100) Surface: 
A Physical Realization of the Ashkin-Teller Model, {\it Phys. Rev. Lett.} 
54, 1539--1542 (1985)}\cr
Bar&{N. C. Bartelt, T. L. Einstein, L. T. Wille: Phase diagram and 
critical properties of a two-dimensional lattice-gas model of 
oxygen ordering in YBa$_2$Cu$_3$O$_z$
{\it Phys. Rev. B} 40, 10759 (1989)}\cr
Bez&{C. G. Bezerra, A. M. Mariz, J. M. de Araujo, F. A. da Costa:
The anisotropic Ashkin-Teller model: a Renormalization Group study, 
{\it Physica} A 292, 429--436 (2001)}\cr
Be&{S. Bekhechi, A. Benyoussef, A. Elkenz, B. Ettaki, M. Loulidi:
Phase transitions in the anisotropic Ashkin--Teller model,
{\it Physica} A 264, 503--514 (1999)}\cr 
BeM1&{G. Benfatto, V. Mastropietro: Ward identities and Dyson equations
in interacting Fermi systems. To appear on {\it J. Stat. Phys.}}\cr
BoM&{F. Bonetto, V. Mastropietro: Beta function and Anomaly 
of the Fermi Surface for a d=1 System of interacting Fermions in a 
Periodic Potential. {\it Comm. Math. Phys.} {\bf 172}, 57--93 (1995).}\cr
C&{C. G. Callan: Broken Scale Invariance in Scalar Field Theory,
{\it Phys. Rev.} D 2, 1541--1547 (1970)}\cr
D&{R. L. Dobrushin: Gibbsian Random Fields for lattice systems with 
pairwise interactions, {\it Func. Anal. and Appl.} 2, 292--301 (1968)}\cr
DJ&{C. Di Castro, G. Jona--Lasinio: 
On the Microscopic Foundation of Scaling Laws, {\it Phys. Lett.} 29A, 322-323
(1969)}\cr
DL&{IE Dzyaloshinski, A. I. Larkin: Correlation functions 
for a one--dimensional Fermi system with long--range interaction 
(Tomonaga model), {\it Zh. Eksp. Teor. Fiz.} 65, 411 (1973) [{\it Sov.
Phys. JETP} 38, 202--208 (1974)]}\cr
DM&{C. Di Castro, W. Metzner: Ward identities and the beta function in 
the Luttinger liquid, {\it Phys. Rev. Lett.} 67, 3852--3855 (1991);
Conservation laws and correlation functions in the Luttinger liquid, 
{\it Phys. Rev.} B 47, 16107--16123 (1993)}\cr
DR&{E. Domany, E. K. Riedel: Phase Transitions in Two-Dimensional Systems
{\it Phys. Rev. Lett.} 40, 561--564 (1978)}\cr
F&{C. Fan, On critical properties of the Ashkin-Teller model,
{\it Phis. Lett.}, 6, 136-136 (1972)}\cr
FKG&{C. M. Fortuin, P. W. Kasteleyn, J. Ginibre: Correlation 
inequalities on some partially ordered sets, {\it Comm. Math. Phys.} 22,
89--104 (1971)}\cr
Fr&{J. Froelich: On the triviality of $\l\phi^4_d$ theories and the 
approach to the critical point in $(d\ge 4)$ dimensions, {\it Nucl. Phys.}
200B, 281--296 (1981)}\cr
G&{A. Giuliani: Gruppo di rinormalizzazione per un sistema 
di fermioni interagenti 
in due dimensioni, Diploma thesis, Roma 2001.}\cr
G1&{G. Gallavotti: Renormalization theory and Ultraviolet stability 
via Renormalization Group methods, {\it Rev. Mod. Phys.} 57, 471--569 
(1985)}\cr
GM&{G. Gentile, V. Mastropietro:
Renormalization group for one-dimensional fermions.
A review on mathematical results. {\it Phys. Rep.} 352 (2001), no. 4-6,
273--43}\cr
GM1&{A. Giuliani, V. Mastropietro: Anomalous universality in the 
anisotropic Ashkin--Teller model, cond-mat/0404701, 
to appear on {\it Comm. Math. Phys.}}\cr
GM2&{A. Giuliani, V. Mastropietro: Anomalous critical exponents 
in the anisotropic Ashkin--Teller model, cond-mat/0409550,  
to appear on {\it Phys. Rev. Lett.}}\cr
GM68&{G. Gallavotti, S. Miracle--Sole': Correlation functions
of a lattice system, {\it Comm. Math. Phys.} 7, 271--288 (1968)}\cr
GN&{G. Gallavotti, F. Nicolo': Renormalization theory for four 
dimensional scalar fields, I and II. {\it Comm. Math. Phys.} 100, 
545--590 (1985); and {\it Comm. Math. Phys.} 101, 
1--36 (1985)}\cr
GS&{G. Gentile, B. Scoppola: Renormalization Group and 
the ultraviolet problem in the Luttinger model. {\it Comm. Math. Phys.}
{\bf 154}, 153--179 (1993).}\cr
Gr&{R. B. Griffiths: Correlations in Ising ferromagnets, I, {\it Jour.
Math. Phys.} 8, 478--483 (1967)}\cr
H&{C. Hurst, New approach to the Ising problem, 
{\it J.Math. Phys.} 7,2, 305-310 (1966)}\cr
Hi&{Y. Higuchi: On the absence of non traslationally invariant Gibbs states
for the two dimensional Ising system, in {\it Random Fields}, ed. J. Fritz, 
J.L.Lebowitz, D. Szaz, North Holland, Amsterdam (1981)}\cr
ID&{C. Itzykson, J. Drouffe, "Statistical field theory: 1," Cambridge Univ.
Press, 1989.}\cr
KO&{B. Kaufman, L: Onsager: Cristal statistics. III. 
Short range order in a binary Ising lattice, 
{\it Phys. Rev} 76, 1244--1252 (1949)}\cr
KW&{L. P. Kadanoff, F. J. Wegner, {\it Phys. Rev. B} 4, 3989--3993 (1971)}\cr
KWa&{M. Kac, J. C. Ward: 
A Combinatorial Solution of the Two-Dimensional Ising Model, {\it Phys. Rev.}
88, 1332--1337 (1952)}\cr
Ka49&{B. Kaufman: Cristal statistics. II. Partition function evaluation 
by spinor analysis, {\it Phys. Rev} 76, 1232--1243 (1949)}\cr
Ka63&{P. W. Kasteleyn, Dimer Statistics and phase transitions, 
{\it J. Math.Phys.} 4, 287 (1963)}\cr
Ka66&{L. Kadanoff: 
Scaling Laws for Ising Models Near Tc, L.P. Kadanoff, {\it Physics} 2 263 
(1966)}\cr
Ka69&{L. Kadanoff: Correlations along a line in the two--dimensional
Ising model, {\it Phys. Rev.} 188, 859--863 (1969)}\cr
Ka77&{L. P. Kadanoff,
Connections between the Critical Behavior of the Planar Model and That of the
Eight-Vertex Model. {\it Phys. Rev. Lett.} 39, 903-905 (1977)}\cr
L1&{E. H. Lieb: Exact solution of the problem of entropy of two-dimensional
ice,
{\it Phys. Rev. Lett.}, 18, 692--694, (1967); The residual entropy 
of square ice, {\it Phys. Rev.} 162, 162--172 (1967)}\cr
L2&{E. H. Lieb: Exact Solution of the F Model of an Antiferroelectric, 
{\it Phys. Rev. Lett.} 18, 1046--1048 (1967)}\cr
L3&{E. H. Lieb: Exact Solution of the Two-Dimensional Slater KDP Model 
of a Ferroelectric, {\it Phys. Rev. Lett.} 19, 108--110 (1967)}\cr
LP&{A. Luther, I. Peschel. Calculations
of critical exponents in two dimension from quantum
field theory in one dimension.
{\it Phys. Rev. B} 12, 3908-3917 (1975)}\cr
LW&{E. H. Lieb, F. Y. Wu: Two Dimensional Ferroelectric Models, in 
Phase Transitions and Critical Phenomena, edited by C. Domb and M. Green 
(Academic, 1972), Vol. 1, p. 331--490}\cr
Le&{A. Lesniewski:
Effective action for the Yukawa 2 quantum field Theory.
{\it Comm. Math. Phys.} {\bf 108}, 437-467 (1987). }\cr
Le74&{J. L. Lebowitz: GHS and other inequalities, {\it Comm. Math. Phys.}
28, 313--321 (1974)}\cr
M&{V. Mastropietro: Ising
models with four spin interaction at criticality, {\it Comm. Math. Phys}
{\bf 244}, 595--642 (2004)}\cr
ML&{D. Mattis, E. Lieb:
Exact solution of a many fermion system and its associated boson field.
{\it J. Math. Phys.} {\bf 6}, 304--312 (1965). }\cr
MW&{B. McCoy, T. Wu, The two-dimensional Ising model,
Harvard Univ. Press, 1973.}\cr
MPW&{E. Montroll, R. Potts, J.Ward. Correlation
and spontaneous magnetization of the two dimensional
Ising model. {\it J. Math. Phys.} 4,308 (1963)}\cr
N&{M. P. M. den Nijs: Derivation of extended scaling relations between
critical exponents in two dimensional models from the one dimensional
Luttinger model, {\it Phys. Rev. B}, 23, 11 (1981)
6111-6125}\cr
O&{L. Onsager: Critical statistics. A two dimensional
model with an order-disorder transition.
{\it Phys. Rev.}, 56, 117-149 (1944)}\cr
PB&{A. M. M. Pruisken, A. C. Brown.
Universality fot the critical lines of the eight vertex,
Ashkin-Teller and Gaussian models, {\it Phys. Rev.} B, 23, 3 (1981)
1459-1468}\cr
PS&{H. Pinson, T. Spencer: Universality in 2D critical Ising model,
unpublished.}\cr
Pe&{R. Peierls: On Ising's model of ferromagnetism. {\it Proceedings 
of the Cambridge Philosophical Society} 32, 477--481 (1936).}\cr
Po&{J. Polchinski: Renormalization group and effective Lagragians,
{\it Nuc. Phys.} B231, 269--295 (1984)}\cr
Ru63&{D. Ruelle: Correlation functions of classical gases, {\it Annals of 
Physics} 25, 109--120 (1963)}\cr
Ru69&{D. Ruelle: Statistical Mechanics, Benjamin, New York (1969)}\cr
Ru79&{L. Russo: The infinite cluster method in the two dimensional 
Ising model, {\it Comm. Math. Phys.} 67, 251--266 (1979)}\cr
S&{S. Samuel:
The use of anticommuting variable integrals in statistical mechanics'',
{\it J. Math. Phys.} 21 (1980) 2806}\cr
SML&{T. D. Schultz, D. Mattis, E. H. Lieb: 
Two-dimensional Ising model as a soluble problem of many fermions,  
{\it Rev. Mod. Phys.}  36  856--871 (1964)}\cr
So&{J. Solyom: The Fermi gas model of one--dimensional conductors,  
{\it Adv. Phys.} 28, 201--303
(1979)}\cr
Spe&{T. Spencer: A mathematical approach to universality in
two dimensions. {\it Physica A} {\bf 279}, 250-259 (2000). }\cr
Su&{B. Sutherland: Exact Solution of a Two-Dimensional Model for 
Hydrogen-Bonded Crystals,
{\it Phys. Rev. Lett.} 19, 103--104 (1967);
Two-Dimensional Hydrogen Bonded Crystals,
{\it J. Math. Phys.} {\bf 11}, 3183--3186 (1970)}\cr
Sy&{K. Symanzik: Small distance behaviour in field theory and power counting,
{\it Comm. Math. Phys.} 18 227-246 (1970)}\cr
TM&{C. A. Tracy, B. M. McCoy:
Neutron Scattering and the Correlation Functions of the Ising Model near Tc,
{\it Phys. Rev. Lett.} 31, 1500--1504 (1973)}\cr
W&{F. J. Wegner: Duality relation between the Ashkin-Teller
and the eight-vertex model, {J. Phys. C}, 5,L131--L132
(1972)}\cr
W1&{K. G. Wilson: Renormalization Group and Critical Phenomena. 1. 
Renormalization Group and The Kadanoff Scaling Picture, {\it Phys. Rev.} B4, 
3174 (1971)}\cr
W2&{K. G. Wilson: Renormalization Group and Critical Phenomena. 2. 
Phase Space Cell Analysis of Critical Behavior, {\it Phys. Rev.} B4, 3184
(1971) }\cr
WF&{K. G. Wilson, M. E. Fisher: Critical exponents in 3.99 dimensions,
{\it Phys. Rev. Lett.} 28, 240-243 (1972) }\cr
WL&{F. Y. Wu, K. Y. Lin: Two phase transitions in
the Ashkinh-Teller model. {J. Phys. C}, {\bf 5}, L181--L184
(1974).}\cr
WMTB&{T. T. Wu, B. M. McCoy, C. A. Tracy, E. Barouch:
Spin-spin correlation functions for the two-dimensional Ising model: 
Exact theory in the scaling region, {\it 
Phys. Rev. B} 13, 316--374 (1976)}\cr
Wu&{F. W. Wu: The Ising model with four spin interaction.
{\it Phys. Rev.} B 4, 2312-2314 (1971).}\cr
Ya&{C. N. Yang: The spontaneous magnetization of a two--dimensional 
Ising model, {\it Phys. Rev.} 85, 808--816 (1952)}\cr
ZJ&{J. Zinn Justin: Quantum field theory and critical phenomena
(Oxford University Press, 2002)}\cr
}

\fineindicecorrente
\bye